\documentclass[12pt,twoside]{report}

\usepackage{ifpdf}

\usepackage{graphicx}

\usepackage{diss}

\usepackage{journals}

\usepackage{txfonts}

\usepackage{mynatbib}
\bibpunct[, ]{(}{)}{;}{a}{}{,}

\usepackage{stmaryrd}
\usepackage{amssymb}

\usepackage[figuresright]{rotating}

\ifpdf
 \usepackage[pdftex, plainpages=false, colorlinks=true, pdfstartview=FitV,
             linkcolor=red, citecolor=red, urlcolor=blue, pagebackref]{hyperref}
 \def\ads#1{\href{#1}{ADS}}
\else
 \def\href#1#2{{#2}}
 \def\ads#1{}
 \def\phantomsection{}
\fi

 \usepackage{backrefx}
 \renewcommand{\backrefpagesname}{}
 \renewcommand{\backrefpagesnames}{}
 \renewcommand{\backreflist}{,\space}
 \renewcommand{\backrefformat}[1]{[#1]}
 \renewcommand{\backrefnocite}{}

\ifpdf
\else
\fi
\ifpdf
\else
\fi
\newcommand{\powerlaw}{{\fontfamily{ptm}\fontseries{m}\fontshape{sc}\selectfont{powerlaw}}}
\newcommand{\phabs}{{\fontfamily{ptm}\fontseries{m}\fontshape{sc}\selectfont{phabs}}}
\newcommand{\zphabs}{{\fontfamily{ptm}\fontseries{m}\fontshape{sc}\selectfont{zphabs}}}
\newcommand{\tbabs}{{\fontfamily{ptm}\fontseries{m}\fontshape{sc}\selectfont{tbabs}}}
\newcommand{\pexrav}{{\fontfamily{ptm}\fontseries{m}\fontshape{sc}\selectfont{pexrav}}}
\newcommand{\pexriv}{{\fontfamily{ptm}\fontseries{m}\fontshape{sc}\selectfont{pexriv}}}
\newcommand{\diskbb}{{\fontfamily{ptm}\fontseries{m}\fontshape{sc}\selectfont{diskbb}}}
\newcommand{\xstar}{{\fontfamily{ptm}\fontseries{m}\fontshape{sc}\selectfont{xstar}}}

\newcommand{\zgauss}{{\fontfamily{ptm}\fontseries{m}\fontshape{sc}\selectfont{zgauss}}}
\newcommand{\reflion}{{\fontfamily{ptm}\fontseries{m}\fontshape{sc}\selectfont{reflion}}}
\newcommand{\kyrline}{{\fontfamily{ptm}\fontseries{m}\fontshape{sc}\selectfont{kyrline}}}
\newcommand{\laor}{{\fontfamily{ptm}\fontseries{m}\fontshape{sc}\selectfont{laor}}}
\newcommand{\kyconv}{{\fontfamily{ptm}\fontseries{m}\fontshape{sc}\selectfont{kyconv}}}
\newcommand{\ky}{{\fontfamily{ptm}\fontseries{m}\fontshape{sc}\selectfont{ky}}}
\newcommand{\smedge}{{\fontfamily{ptm}\fontseries{m}\fontshape{sc}\selectfont{smedge}}}
\newcommand{\refsch}{{\fontfamily{ptm}\fontseries{m}\fontshape{sc}\selectfont{refsch}}}

\newcommand{\textscown}[1]{{\fontfamily{ptm}\fontseries{m}\fontshape{sc}\selectfont{#1}}}

\newcommand{\ad}{accretion disc}

\newcommand{\textposown}{\textit}

\usepackage[bf]{caption}
\begin{document}

\thispagestyle{empty}
\pagestyle{empty}
\begin{center}
 
\vspace{1cm} 
%{\LARGE 
{\Large

CHARLES UNIVERSITY IN PRAGUE
%Charles University in Prague\vspace{.3cm}

Faculty of Mathematics and Physics \vspace{.3cm}

%{\large Astronomical Institute}
\vspace{.6cm}

ACADEMY OF SCIENCES OF THE CZECH REPUBLIC \vspace{.3cm}
%ACADEMY OF SCIENCES of the Czech Republic \vspace{.3cm}
%Academy of Sciences of the Czech Republic \vspace{.3cm}

Astronomical Institute
}

\vspace{1.5cm}

\begin{figure}[hbtp]
\hspace*{5.1cm}
\includegraphics[width=4.5cm]{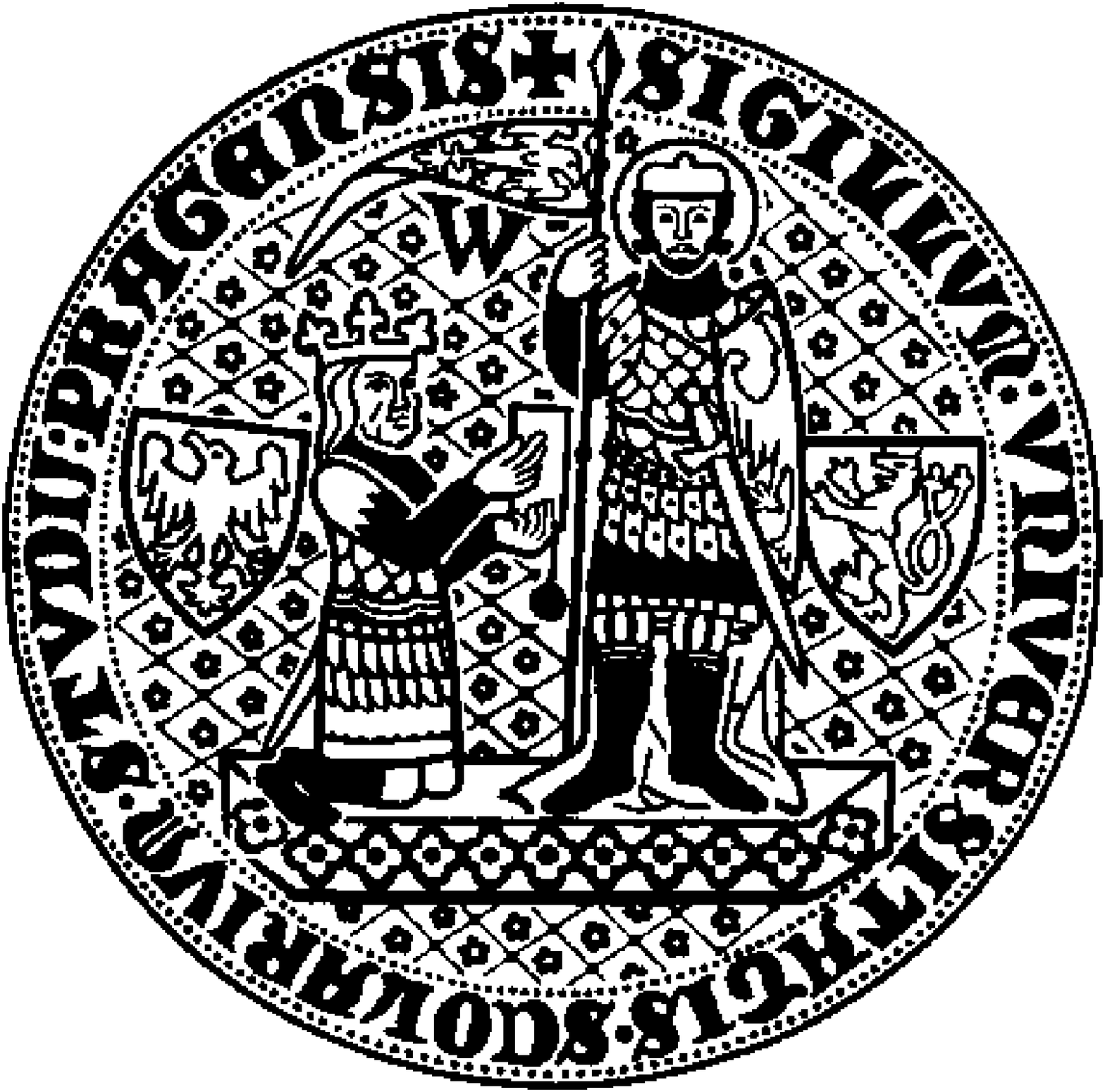}
\end{figure}

\vspace{1.5cm}

{\bf\LARGE Looking into the inner black hole accretion disc\\[5pt] 
with relativistic models of iron line}

\vspace{1cm}
{\large Thesis submitted for the degree of Doctor Philosophiae}

\vspace{1.5cm}
{\Large Mgr. Ji\v{r}\'{i} Svoboda}

\vfill
{\large Supervisor: doc.\ RNDr.\ Vladim\'{\i}r Karas, DrSc.}

\rule{140mm}{0.1mm}

\vspace{1.5cm}
{\large Prague, April 2010}
\end{center}

% ##########################################################################

%\clearpage
%\pagestyle{empty}
%\input{referees}

% ##########################################################################

% I will use roman numbering for the first few pages (contents,...)
\renewcommand{\thepage}{\roman{page}}

\thispagestyle{empty}
\noindent
This doctoral thesis was done at the Astronomical Institute of the Academy
of Sciences of the Czech Republic during doctoral studies at the Faculty
of Mathematics and Physics, Charles University in Prague, in the years 2006-2010.
\vspace{0.8cm}

\noindent
\begin{tabular}{p{4.5cm}p{10.5cm}}
\textsc{Phd student:} & Mgr. Ji\v{r}\'{i} Svoboda\\&\\
\textsc{Study branch:} & 4F1 Theoretical physics, astronomy and astrophysics
\\&\\
\textsc{Supervisor:} & doc.\ RNDr.\ Vladim\'{\i}r Karas, DrSc.\\&\\
\textsc{Address:}
            &Astronomical Institute \\
            & Academy of Sciences of 
             the Czech Republic\\
            %&Institute of Physics\\
            &Bo\v cn{\'\i} II/1401a, 141 31 Praha 4\\
            &Czech Republic\\&\\
\textsc{Referees:}
            &Prof. Giorgio Matt\\
            &Roma Tre University\\
	    &Faculty of Mathematics, Physics and Natural Sciences\\
            &Department of Physics\\
            &Italy\\
            &\\
            &\\
            &Prof. RNDr. Zden\v{e}k Stuchl\'{i}k, CSc.\\
            &Silesian University in Opava\\
            &Faculty of Philosophy and Science\\
            &Department of Physics\\
            &Czech Republic\\

\end{tabular}
\vspace{5.6cm}

%\noindent This summary was distributed on 21st May 2010.

%\vspace{0.4cm}

\noindent The defence was held on 23rd June 2010 in front of the committee of
the study branch 4F1 Theoretical physics, astronomy and astrophysics 
at MFF UK, Ke Karlovu 3, 121 16 Praha 2, Czech Republic. %, room No. 105 (252).
\vspace{0.4cm}

%\noindent The thesis is at disposal in the Department of doctoral studies 
%of MFF UK, Ke Karlovu 3, 121 16 Prague 2, Czech Republic.
%\vspace{1.0cm}

%\newpage

\cleardoublepage
\pagestyle{myheadings}
\chapter*{Abstract}
\thispagestyle{empty}
%\phantomsection\addcontentsline{toc}{chapter}{Abstract}
%\addtocontents{toc}{\vspace*{\baselineskip}}
X-ray observations of active galactic nuclei and black hole binaries
offer a unique laboratory for testing the general relativity
in strong gravity regime, for studying accretion physics around black holes,
and for constraining properties of accreting black holes.
In this Thesis, we discuss black hole spin measurements 
employing the relativistic iron line profiles in the X-ray domain. 

We investigate the iron line band for two representative sources -- 
MCG\,-6-30-15 (active galaxy) and GX 339-4 (X-ray binary).
We compare two relativistic models of the broad 
iron line, {\laor} and {\kyrline}.
%which employ different ways to measure the spin.
%/nebo ???: 
In contrast to {\laor},
the {\kyrline} model has the spin value as a variable parameter.
However, the {\laor} model can still be
used for evaluation of the spin if one identifies the inner 
edge of the disc with the marginally stable orbit.
%ma jednu hodnotu spinu a urcuje to jen s rin
We realise that the discrepancies in the results obtained with the 
two models are within general uncertainties of the spin determination 
using the skewed line profile when applied to the current data.
This implies that the spin is currently
determined entirely from the position of the marginally stable orbit
while the effect of the spin on the overall line shape would be resolvable
with higher resolution X-ray missions like IXO (International
X-ray Observatory).

We show that the precision of the spin measurements
depends on an unknown angular distribution of the disc emission.
Often a unique profile is assumed, 
invariable over the entire range of radii in the disc 
and energy in the spectral band.
However, an improper prescription for the directionality profile
affects the parameters inferred for the relativistic broad line model.
We study how sensitive the spin 
determination is to the assumptions about the intrinsic angular distribution of the emitted photons. 
We find that the uncertainty of the directional emission distribution translates to 20\% 
uncertainty in the determination of the radius of marginally stable orbit.

By assuming a rotating black hole in the centre of an accretion disc,
we perform radiation transfer computations of an X-ray irradiated disc atmosphere 
(NOAR code) to determine the directionality of outgoing X-rays in the 2-10 keV energy band. 
Based on these computations, we find that from the simple formulae
for the directionality, the isotropic case reproduces the simulated 
data with the best accuracy. 
The most frequently used limb darkening law favours higher values of
spin and, in addition, a steeper radial emissivity profile. 
We demonstrate our results on the case of XMM-Newton observation of MCG\,-6-30-15,
for which we construct confidence levels of chi-squared  statistics, 
and on the simulated data for the future X-ray IXO mission. 

Furthermore, we present a spectral analysis 
of an XMM-Newton observation of a Seyfert 1.5 galaxy IRAS~05078+1626 being the 
first X-ray spectroscopic study of this source.
The lack of the significant relativistic blurring of the reflection model component
suggests the accretion disc to be truncated at a farther radius (inner disc radius $R_{\rm in} \geq 60\,R_{g}$).

As a by-product of our reduction of the XMM-Newton data, 
we find that a careful treatment of the raw
instrumental data is necessary to obtain the highest quality data.
As a crucial step, we consider the correct re-binning of the data
reflecting the energy resolution of the used instrument.
Photon pile-up is another problem which may occur in the data
of very bright sources, and it may significantly affect the spectral shape. 

In summary, we found relativistic iron line
models to be a feasible method for measuring the
spin of black holes at all scales -- from solar-mass
microquasars to giant black holes of billions solar masses 
in distant quasars. Some useful constraints on spin are achievable 
already from X-ray spectra of currently operating instruments.
However, our simulations with the tentative 
IXO response show a significant improvement 
in the accuracy of spin measurements in the future.\\[5pt]

\textbf{Keywords}: black holes - accretion disc - active galaxies - Galactic X-ray binaries
%\newpage
%\chapter*{Abstrakt}
%\input{abstrakt.tex}
% ##########################################################################

%\cleardoublepage
%\pagestyle{empty}
%\input{dedication.tex}

% ##########################################################################

\cleardoublepage
\pagestyle{myheadings}
\chapter*{Acknowledgements}
\thispagestyle{empty}
%\phantomsection\addcontentsline{toc}{chapter}{Acknowledgements}
%\addtocontents{toc}{\vspace*{\baselineskip}}
  
%\vspace{-0.15cm}

This Thesis would not be completed without a big help from my colleagues,
friends and family to whom I would like to express my hearted thanks.

%pracovni

First of all, I would like thank to my Supervisor, doc.\,Vladim\'{i}r Karas,
for giving me many ideas for the work presented in this Thesis, 
and for introducing me to an international scientific community. 
%starting with allowing me to be present at General Assembly of the International Astronomical
%Union held in Prague 2006 and continuing with introducing me
%to the international collaboration group working in research
%of relativistic astronomical objects (``FERO'' = Finding Extreme Relativistic Objects). 
I owe very much to my collaborators of our two articles
in the Astronomy and Astrophysics journal. Hence, after Vladim\'{i}r Karas,
I would also like to thank Michal Dov\v{c}iak (from the
Astronomical Institute of the Academy of Sciences),
Matteo Guainazzi (from the European Space Astronomy Centre of ESA in Madrid), 
and Ren\'{e} Walter Goosmann (currently at Strasbourg University but sharing
the office with me in Prague 2006-2008) for their scientific
ideas.

%pratele a rodina

I am indebted to my parents who supported me in the education
from my childhood and helped me to establish my new settlement 
in Prague. I owe very much to my mother Jana who regrettably
deceased two years ago. I dedicate this work to memory of her.
I am grateful to my father Anton\'{i}n to encourage me to have 
astronomy as my hobby and to study it at the University.
I would also like to thank my brother Anton\'{i}n 
and all my friends who have shared my enthusiasm for astronomy with me.
I am grateful for all the discussions which we had about astronomy,
black holes or theory of relativity,
because the genuine interest of my friends and relatives 
encouraged me in this work very much.

My warmest thanks belong to Lucie, my future wife, for her tireless psychological
support and for her patient listening to me when I was bothered about any problem.
She helped me significantly to overcome the most difficult
moments of my life. Without her support, I would perhaps never finish
my Thesis. In a similar way, I would like to thank my recently found new family.

% kolegove v praci

I appreciate discussions with my colleagues at the Prague's department of the
Astronomical Institute of the Academy of Sciences, especially the members of 
the Prague Relativistic Astrophysical Group and ``pidi''-seminar group.
I owe a lot to Michal Dov\v{c}iak and Mirek K\v{r}\'{i}\v{z}ek for helping
me with the technical computational stuff. I am very grateful to Michal Dov\v{c}iak,
Ivana Stoklasov\'a\,--\,Orlitov\'a, Jirka Hor\'{a}k and Ond\v{r}ej Kop\'{a}\v{c}ek 
for careful reading of some parts of the Thesis.

% formalni - ustavy, granty

I acknowledge the Astronomical Institute of the Academy of Sciences of Czech Republic
for providing me a stimulating work environment 
and appropriate computational equipment.
Further, I acknowledge the financial support of the student 
research grant of the Charles University (ref.\ 33308),
the doctoral student program of the Czech Science Foundation (ref.\ 205/09/H033), 
the ESA Plan for European Cooperating States (project No.\ 98040),
the Centre for Theoretical Astrophysics (project No.\ LC06014),
and the grant of the Czech Ministry of Education, Youth and Sports (project No.\ ME09036).

% ##########################################################################

\cleardoublepage
\pagestyle{myheadings}
\markboth{PREFACE}{PREFACE}
\chapter*{Preface}
\thispagestyle{empty}
\phantomsection\addcontentsline{toc}{chapter}{Preface}
\addtocontents{toc}{\vspace*{\baselineskip}}

\begin{quote}
In questions of science the
authority of a thousand is not
worth the humble reasoning
of a single individual.
\begin{flushright}
 Galileo Galilei, 1632 \\ 
\end{flushright}
\end{quote}

Black holes are objects which were first created in minds
of theoretical astrophysicists, and for a long time, they were not 
supposed to exist in the real Universe, %in the endless Universe, 
in the endless world of stars and planets.
Even Albert Einstein who developed the theory, which allowed the existence
of black holes, did not believe that the nature would be
so crazy to give permission to such objects to form.
However, theoretical astrophysicists calculating the details
of the stellar collapse, such as Chandrasekhar, 
Oppenheimer, Snyder etc., predicted that the collapse
of too massive stars could not be stopped by any means
and that it must go on to create space-time singularities,
which were later named by Wheeler as black holes. 
Meanwhile, with no idea of connection,
astronomers classified radio bright galaxies as a peculiar group
distinguished from the standard galaxies by extremely strong radio power.
Presently, we generally believe that accretion on a rapidly rotating 
super-massive black hole is the process behind such an enormous power.

The first suggestions of black holes as real celestial objects
came with the development of X-ray astronomy in 1960s. 
Riccardo Giaconni, one of the pioneers of X-ray astronomy, 
won the Nobel prise in physics in 2002 for opening 
the X-ray window to the Universe. 
The discoveries of pulsars, quasars and X-ray binaries started 
a fruitful life of a new branch of astronomy,
astronomy of ultra-compact objects.
Since that time, black holes have fascinated many people
around the world, not only scientists. 
Studying these objects brings light to the 
physics of stellar collapse, galaxy formation, accretion physics
and behaviour of matter in the strong gravitational field.
%Properties of radiation coming from black hole accretion discs
%represent a unique test of revealing physics of accreting black holes.

Astrophysical black holes are actually very simple objects 
being characterised only by their mass and angular momentum (spin).
In the current knowledge, it seems that there are two populations
of black holes according to their mass -- stellar black holes
of mass of several solar masses and super-massive black holes
of mass of millions to billions solar masses. The distribution
of their spin is still unknown. Measuring the spin is 
difficult because the effect of the spin quickly decreases with
the growing distance from the black hole. Matter feels the black hole spin
only within several gravitational radii.
However, the black hole spin plays
an important role in black hole energetics and evolution. 
The spin is assumed to be responsible for generation and 
up-keeping of the powerful relativistic jets.
The information about the spin value on a statistically significant
sample of black holes is important in the understanding of 
formation and growth of black holes. It can significantly help
to answer the question if the observed spin value is natal 
or if the rotating black holes are spun-up via accretion.

The innermost parts of black hole accretion discs can be uncovered
in high energetic radiation, such as X-rays or $\gamma$-rays.
This is for two reasons.
First, such an energetic radiation can originate only under the extreme
conditions close to a black hole.
Second, any weaker radiation is often efficiently absorbed by surrounding matter
and cannot reach a distant observer. Fast development of X-ray 
detectors in the last two decades has allowed astronomers to provide spectra with
an unprecedented sensitivity, and so constrain accretion flows within a few
gravitational radii and measure the black hole spin. There are
currently several methods of spin measurements. 
Modelling of relativistic iron line represents
one of them, and it is particularly suitable because it is applicable
to black holes at all mass scales.

In this Thesis, %we will try to put a small crock in the 
%big but far not complete mosaic of the knowledge about 
%black holes and radiation from their accretion discs.
%We 
we will look into the inner black hole accretion disc
with relativistic models of iron line.
In Section~\ref{rellinemod}, we summarise the basic concepts
of this method, and compare two models which employ different ways
to determine the spin value.
In Section~\ref{directionality}, we study in detail the role of
the angular emissivity on the spin value measurement.
Both analyses are provided on the current XMM-Newton data
whose reduction is described in Section~\ref{xmm_analysis},
and on the simulated data of a next generation X-ray mission
with a significantly higher energy resolution.
In Section~\ref{xmm_analysis}, we also present the results
of our X-ray spectroscopic study
of the Seyfert~1.5 galaxy IRAS~05078+1626, which represents
the first X-ray spectroscopy measurement of this source.
The achieved results are discussed at the end of each Chapter.
Main conclusions of the Thesis  
are summarised in Section~\ref{concl}.
Future perspectives are pronounced in Section~\ref{futureperspectives}.

Most of the results presented in the Thesis were published
in two papers in Astronomy and Astrophysics journal
\citep{2009A&A...507....1S, 2010A&A...512A..62S},
and several proceedings \citep{SvobodaWDS, 2008mqw..confE..85S, 
2009hrxs.confE..42S}.

\vspace{2cm}
In Prague, April 2010 \\
\begin{flushright}
 Ji\v{r}\'{i} Svoboda
\end{flushright}

% ##########################################################################

\cleardoublepage
\pagestyle{myheadings}
\markright{CONTENTS}
\tableofcontents
%\addcontentsline{tocw}{chapter}{Contents}
%\addtocontents{toc}{\protect\thispagestyle{empty}}
\addtocontents{toc}{\protect\thispagestyle{myheadings}}
\addtocontents{toc}{\vspace*{-\baselineskip}}

\ifpdf
 \clearpage
 \thispagestyle{empty}
\fi

% ##########################################################################

\cleardoublepage
% I will use arabic pagenumbering from now
\renewcommand{\thepage}{\arabic{page}}
\setcounter{page}{1}
%\thispagestyle{empty}
%\phantomsection\addcontentsline{toc}{chapter}{Preface}

% ##########################################################################
\clearpage
%\thispagestyle{empty}
%\phantomsection\addcontentsline{toc}{chapter}{Introduction}
%\markboth{INTRODUCTION}{INTRODUCTION}
\chapter{Introduction}
 \thispagestyle{empty}
 
%\section{Black hole accretion}

%literatura:
% Frank,King,Raine: 2002apa..book.....F

\section{Black holes}
\label{blackholes}

Black holes are regions of space-time in which the gravitational well is so deep
that no particle or even light can escape. First ideas of existence
of such objects from which a distant observer cannot get any signal
due to the escape velocity greater than speed of light
were formulated by \citet{michell1784} and \citet{laplace1799}
(recent English translation can be found in \citealp{2003gr.qc.....4087S}).
However, real foundations for black hole theory were laid by
discovery of general theory of relativity by \citet{1916AnP...354..769E}.
The first exact solution of Einstein's equations 
describing a black hole was found by \citet{1916AbhKP......189S} 
for a point mass assuming spherical symmetry. 
The radius of the event horizon is accordingly called 
Schwarzschild radius:
\begin{equation}
 r_{\rm s} = 2r_{\rm g} = \frac{2GM}{c^2},
\label{rs}
\end{equation}
where $r_{\rm g}$ is the gravitational radius which is equivalent 
to the mass if the geometrised units with $c=G=1$ are used. 
The Schwarzschild metric in Schwarzschild coordinates is:
\begin{equation}
\label{schwarzschild}
ds^{2}=-\left(1-\frac{r_s}{r}\right)dt^{2}+\frac{1}{\left(1-\frac{r_s}{r}\right)}dr^{2}+r^{2}d\Omega^{2},
\end{equation}
where $d\Omega^{2}\equiv d\theta^{2}+\sin^{2}\theta\, d\varphi^{2}$, $t$ is
the time measured by an observer at infinity at rest, 
$r$ is the radial coordinate, $\theta$ is latitude,
and $\varphi$ is the azimuthal angle.

The solution for rotating black holes was found almost half a century
later by \citet{1963PhRvL..11..237K}. The Kerr's metric in
the Boyer-Lindquist spheroidal coordinates ($t,r,\theta,\varphi$) 
and geometrised units is \citep[e.g.][chap.~33]{1973grav.book.....M}:

\begin{equation} \label{metric} 
  {\rm d}s^{2} =
  -\frac{\Delta\Sigma}{A}\,{\rm d}t^{2} 
  +\frac{A\,\sin^{2}\theta}{\Sigma}\; 
   \big({\rm d}\varphi-\omega\,{\rm d}t\big)^2
  +\frac{\Sigma}{\Delta}\,{\rm d}r^{2} 
  +\Sigma\,{\rm d}\theta^{2}, 
\end{equation}
where the metric functions are: 
$\Delta(r)=r^{2}-2r+a^2$,
$\Sigma(r,\theta)=r^{2}+a^{2}\cos^{2}\theta$,
$A(r,\theta)={\left(r^{2}+a^{2}\right)}^{2}-\Delta(r)\,a^{2}\sin^{2}\theta$,
and $\omega(r,\theta)=2ar/A(r,\theta)$;
$a$ denotes the specific rotational angular momentum (spin) of the central
body. 

The conversion factor from the angular momentum $J_{\rm{}phys}$ 
(in physical units) to the angular momentum $J$ (in geometrical units)
reads: $J=(G/c^3)J_{\rm{}phys}$. The geometrised dimension of $J$ is 
the square of the length [cm$^2$].
It is convenient to make all geometrised quantities dimensionless by 
scaling them with the appropriate power of mass $M$.
The dimensionless specific angular momentum, $a\equiv J/M^2$, spans the
range $-1\leq a\leq1$, where the positive/negative value refers 
to the motion co/counter-rotating with respect to the $\varphi$-coordinate.
We will further assume co-rotational motion only ($a \geq 0$).
The magnitude of $a$ is thought to be less than unity
in order to have a regular horizon and avoid the case of naked 
singularity. The value of the outer horizon is:
\begin{equation}
r_+=1+{\left(1-a^2\right)}^{1/2}.
\end{equation}

For the Keplerian angular velocity of the orbital motion, 
we obtain \citep{1972ApJ...178..347B}:
\begin{equation}
\label{kep}
\Omega_{_{\rm{}K}}(r)=\frac{1}{r^{3/2} + a}.
\end{equation}
For the linear velocity with respect to a locally 
non-rotating observer, we have:
\begin{equation}
 {\cal V}(r) = \frac{r^2-2ar^{1/2}+{a}^2}{\Delta^{1/2}\left(r^{3/2}
 +{a}\right)}.
\label{vr}
\end{equation} 

\begin{figure}[tbh]
\begin{center}
\includegraphics[width=0.75\textwidth]{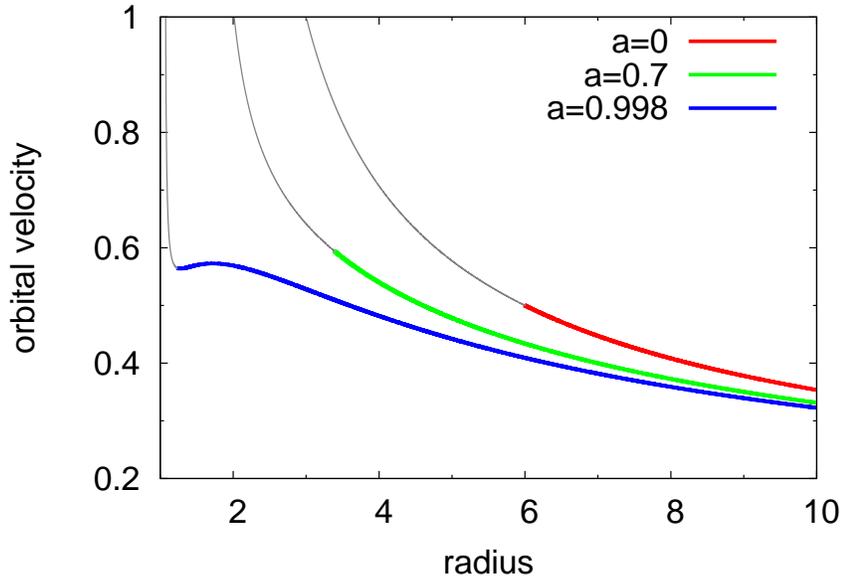}
\caption{Orbital velocity 
${\cal V}(r)$ of co-rotational motion near a rotating black hole, 
as given by formula~(\ref{vr}) for three values of the
black-hole dimensionless angular momentum parameter $a$ (spin). The thick coloured part 
of each curve indicates the range of radii above the marginally stable
orbit, $r\geq r_{\rm ms}(a)$, where the circular motion is stable. The thin curve
indicates an unstable region at small radii.}
\label{v_orbit}
\end{center}
\end{figure}

The velocity at the marginally stable orbit reaches a considerable
fraction of the speed of light $c$ (=1) and has a similar value 
${\cal V}(r) \approx \left(0.5-0.6\right)\,c$ for any value of the angular momentum, see Figure~\ref{v_orbit}.
For large spin, a small dip develops in the velocity profile 
near the horizon. Although it is an interesting feature \citep[see][]{2005PhRvD..71b4037S},
its magnitude is far too small to be recognised with current observational facilities.

The Schwarzschild and Kerr metric represent solutions of Einstein's equations
relevant for astrophysical black holes which are assumed to be electrically 
neutral. Nevertheless, they can be further generalised by taking of 
the electric charge into the consideration. The corresponding solutions
are Reissner-Nordstr{\"o}m metric for a non-rotating charged black hole,
and Kerr-Newman metric for a rotating charged black hole, respectively.
Besides the mass, angular momentum and electric charge, black holes do not
have any other parameters, which is often called ``no-hair theorem''
(this statement gets its name from a comment by the famous astrophysicist 
John~A.~Wheeler 1968, see also \citet{1973grav.book.....M}).

\section{Marginally stable orbit}

The marginally stable orbit ($r_{\rm ms}$), 
sometimes also called innermost stable circular orbit (ISCO), 
is the closest orbit to the centre of a black hole
where the orbit of a test particle is stable.
Below this orbit, only unstable or unbound orbits coming from infinity
may occur.
%around black hole with the velocity equal to the speed of light. 

\begin{figure}[tbh]
\begin{center}
\includegraphics[width=0.7\textwidth]{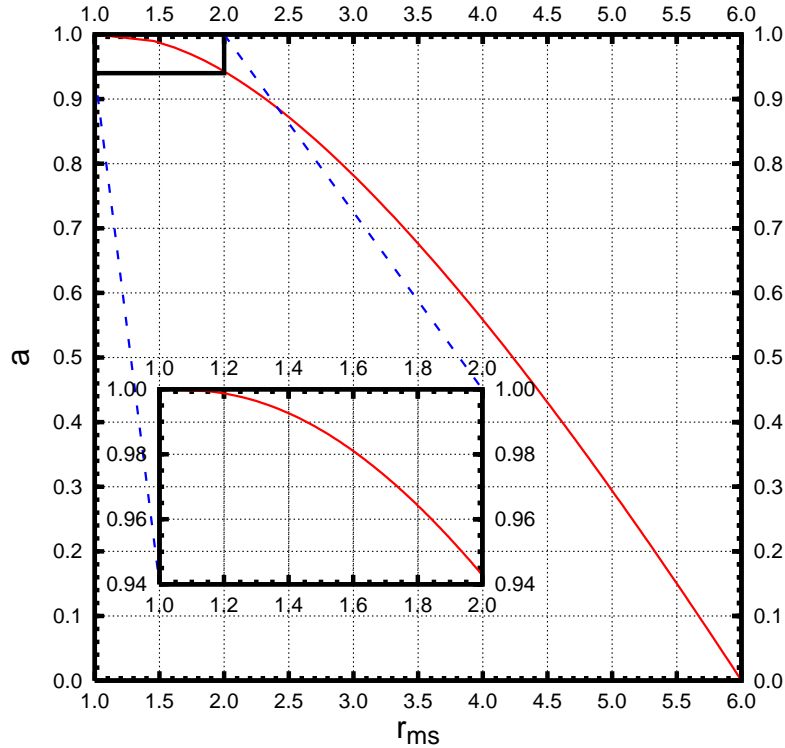}
\caption{Relation between spin $a$ and marginally stable orbit $r_{\rm ms}$.}
\label{a_rms}
\end{center}
\end{figure}

The position of ISCO depends on the
value of the spin \citep{1972ApJ...178..347B}:
\begin{equation}
r_{\rm ms} = 3+Z_2-\big[\left(3-Z_1)(3+Z_1+2Z_2\right)\big]^{1 \over 2},
\label{rms} 
\end{equation}
$Z_1 = 1+(1-a^2)^{1 \over 3}[(1+a)^{1 \over 3}+(1-a)^{1 \over 3}]$ and
$Z_2 = (3a^2+Z_1^2)^{1 \over 2}$. Notice that $r_{\rm ms}(a)$ spans the
range of radii from $r_{\rm ms}=1$ for $a=1$ (the case of a maximally
co-rotating black hole) to $r_{\rm ms}=6$ for $a=0$ (static black hole).
Figure~\ref{a_rms} illustrates the relation~(\ref{rms}) graphically.
It is generally supposed that the rotation of the astrophysical black 
holes is limited by an equilibrium value, $a \doteq 0.9982$, because of
capture of photons from the disc \citep{1974ApJ...191..507T}. 
This translates to $r_{\rm{}ms} \doteq 1.23$.
For a hypothetically higher value of the spin than $a=1$,
the radius of the marginally stable orbit 
decreases to $r_{\rm ms}=2/3$ and then increases again \citep{1980BAICz..31..129S}.

The ISCO is an important quantity in the 
standard accretion disc theory because the inner edge of the 
accretion disc is assumed to coincide with it. However, 
this may not be satisfied precisely under realistic circumstances
\citep{2008MNRAS.390...21B}. The magnitude of the resulting 
error on spin measurements (see Section~\ref{spin_meas})
was constrained recently by \citet{2008ApJ...675.1048R} 
who applied physical arguments about
the emission properties of the inner flow. 
It is very likely that this discussion will have
to continue for some time until the emission properties of
the general relativistic MHD flows are fully understood.

\section{Black hole accretion}

\subsection{Accretion power}

Black holes are interesting objects for the theoreticians
since they represent a natural laboratory for testing
the theory of general relativity, but they are also important
objects for astronomers due to their interaction with
the surrounding matter. Nowadays, black holes cannot be observed
directly as the sensitivity of gravitational wave detectors
is still insufficient. However, there is an increasing amount
of observations of the electromagnetic radiation of the matter
accreting on black holes. These observations can constrain
the black hole parameters as well as the accretion physics.

The accretion power is due to conversion
of the gravitational potential energy into radiation.
For a body of mass \textit{M} and radius \textit{$R$} 
we can estimate the energy released by the accretion
of a particle with mass $m$ as: 
\begin{equation}
\label{accr1}
\Delta E_{\rm{acc}} = \frac{GMm}{R}.
\end{equation}
We can compare the power of the accretion process 
with the energy output by nuclear fusions. 
In the case of hydrogen burning, we obtain $\Delta E \approx 6\times10^{18}\rm{erg\,g}^{-1}$.
For a neutron star with $M\approx M_{\odot}$
and $R\approx 10$\,km, the energy output by accretion on the stellar surface is
$\Delta E_{\rm {acc}} \approx 10^{20}\rm{erg\,g}^{-1}$.
This means that more energy per a mass unit is released 
by the accretion on a compact object than from the nuclear synthesis.
Accretion onto super-massive black holes in quasars powers
the most luminous sources in the Universe.

The maximal luminosity produced by the spherical accretion on a star
is restricted by the Eddington limit:
\begin{equation}
\label{eddl}
L_{\rm{Edd}} = \frac{4\pi GMm_{p}c}{\sigma_{T}},
\end{equation}
where $m_{p}$ is the mass of a proton (i.e. a nucleus of hydrogen) 
and $\sigma_{T}$ is Thomson cross section.
The Eddington luminosity represents the highest possible
luminosity on the conditions of spherically symmetric
accretion $\dot{M}$ and totally ionised accreting material.
In this case, the opacity is dominated by electron scattering
and we suffice only with Thomson cross section.
\footnote{The 
radiation influences mostly electrons which are of the less
mass than protons, vice-versa the gravity more attracts the massive particles. 
Electrons and protons hold together due to electromagnetic interaction.}
For accretion powered objects the Eddington limit
implies a limit on the steady accretion rate.

If all the kinetic energy of infalling matter is
conversed to radiation at the surface of the central body,
then the accretion luminosity $L_{\rm{acc}}$ is given by:
% \ref{lacc}
\begin{equation}
\label{lacc}
L_{\rm{acc}} = \frac{GM\dot{M}}{R}.
\end{equation} 
In the case of black hole accretion, 
the radius $R$ does not refer to the surface
but a natural choice is a Schwarzschild radius. 
The uncertainty of the value of the luminosity can be
parametrised by a dimensionless efficiency $\eta$ 
\citep{1973A&A....24..337S, 1982MNRAS.200..115S},
which measures how efficiently the rest mass energy 
of the accreted material is converted to radiation:
\begin{equation}
\label{lacc-bh}
L_{\rm{acc}} = \eta\frac{GM\dot{M}}{R} \approx \eta\dot{M}c^{2}.
\end{equation}
Comparing eq. (\ref{lacc-bh}) with energy released by the burning of hydrogen we get $\eta=0.007$
for nuclear synthesis. 
\citet{1973A&A....24..337S} mentioned $\eta \approx 0.06$
in the case of Schwarzschild's metric, and in Kerr's metric 
$\eta$ can achieve $40\%$.
The estimation of realistic value for $\eta$
in black hole accretion is an important problem,
a reasonable guess for it would appear to be $\eta \approx 0.1$ 
\citep{2002MNRAS.335..965Y, 2002ApJ...565L..75E, 2004MNRAS.351..169M}.

\subsection{Accretion discs}
\label{ad}

The dominant accretion process for compact objects
involves disc accretion. Most of the accreting matter, 
which is gas supplied by a donor star in binary systems
or the host galaxy in quasars, 
possesses sufficient angular momentum to go into 
the orbit around the black hole, forming an
accretion disc. The previous relations 
calculated for spherically symmetric case will serve
as convenient approximations and estimations.
Accretion disc physics includes many processes, including gravity, 
hydrodynamics, viscosity, radiation and magnetic fields. 
The angular momentum of matter in an accretion disc is gradually 
transported outwards by stresses (turbulent, magnetic, etc.).
The time scale for redistributing angular momentum
is long compared with either the orbital or radiative time scales,
which allows matter to gradually spiral inwards.

\subsubsection{Steady thin accretion discs (Shakura-Sunyaev)}

Steady thin accretion discs represent the standard disc solution
found by \citet{1973A&A....24..337S}. The model is applicable
when the characteristic values of luminosities are sub-Eddington, 
$L/L_{\rm Edd} \ll 1$, i.e. 
when the vertical component of the thermal disc radiation alone 
cannot support the matter against gravity at substantial 
altitudes above the disc plane.
This condition may be violated if the thermal motion is predominant.
The thermal criterion for the Shakura-Sunyaev thin disc is:
\begin{equation}
\label{thincrit_temperature}
\frac{k_{\rm B}T}{\mu c^2}\,\frac{r}{r_g} \ll 1,
\end{equation}
where $k_{\rm B} \doteq 1.38\times 10^{-23}$\,J\,K$^{-1}$
is the Boltzmann constant, $T$ is the gas temperature,
and $\mu$ is the mean mass per particle.
When the condition (\ref{thincrit_temperature}) is not fulfilled, 
luminosities may reach Eddington values, $L/L_{\rm Edd} \approx 1$, 
and the radiation-pressure force becomes
comparable to that of the gravity. The height of the disc
ceases to be small and the thick-disc solution
needs to be considered. 

When the assumption of geometrical thinness is justified,
only radial advection is dominant and the equations 
for the conservation of mass and angular momentum $\Omega$
can be written as:
\begin{equation}
\label{mass_cons}
r\frac{\partial \Sigma}{\partial t}+\frac{\partial}{\partial r}(r\Sigma v_{r})=0,
\end{equation}
and
\begin{equation}
\label{mom_cons}
r\frac{\partial}{\partial t}(\Sigma r^{2}\Omega)+\frac{\partial}{\partial r}(\Sigma v_{r}r^{3}\Omega))=\frac{1}{2\pi}\frac{\partial G}{\partial r},
\end{equation}
where 
\begin{equation}
\label{torque}
G(r,t)=2\pi r^{3}\nu\Sigma\frac{\partial \Omega}{\partial r}
\end{equation}
defines the torque, $\Sigma$ is the surface density, $v_{r}$ is the velocity
in the radial direction, and $\nu$ is the kinematic viscosity.
Combining eqs.~(\ref{mass_cons}) - (\ref{torque}) and using eq.~(\ref{kep})
we obtain the equation governing time evolution of surface density
in the Keplerian disc:
\begin{equation}
\label{evol_surf_dens}
\frac{\partial \Sigma}{\partial t}=\frac{3}{r}\frac{\partial}{\partial r}(\sqrt{r}\frac{\partial}{\partial r}(\nu\Sigma\sqrt{r})).
\end{equation}

The kinematic viscosity $\nu$ may be a function of local variables in the disc.
The eq.~(\ref{evol_surf_dens}) has the form of non-linear diffusion equation 
governing the behaviour of $\Sigma(r,t)$.
Given a solution for $\Sigma(r,t)$ the radial velocity is:
\begin{equation}
\label{rad_vel}
v_{r}=-\frac{3}{\Sigma \sqrt{r}}\frac{\partial}{\partial r}(\nu\Sigma\sqrt{r}).
\end{equation}

Finally, we need some prescription for $\nu$ to close the system of equations 
and to fully determine the radial structure of the accretion disc.
All of the qualities currently ignored, such a detailed
micro-physics, enter into the problem via $\nu$.
After radial integration of the equations~(\ref{mass_cons}) and (\ref{mom_cons})
and assuming time-steady discs ($\frac{\partial}{\partial t}=0$), we get:
\begin{equation}
\label{mass_cons_steady}
r\Sigma v_{r}=\rm{const}. 
\end{equation}
and
\begin{equation}
\label{mom_cons_steady}
\nu\Sigma=\frac{\dot{M}}{2\pi}\left[1-\left(\frac{r_{in}}{r}\right) ^{\frac{1}{2}}\right].
\end{equation}
At $r=r_{in}$, the viscous torque $G(r)$ vanishes. It is generally assumed
that this radius coincides with the marginally stable orbit
because below it the matter losses the centrifugal support.

The essential idea of the accretion process is dissipation of energy.
The dissipation is caused by viscous torques and is given by:
\begin{equation}
\label{diss_1}
D(r)=\frac{G\frac{\partial\Omega}{\partial r}}{4\pi r}. 
\end{equation}
Using eq.~(\ref{torque}) we obtain:
\begin{equation}
\label{diss_2}
D(r)=\frac{1}{2}\nu\Sigma\left(r\frac{\partial\Omega}{\partial r}\right)^{2}. 
\end{equation}
For the Keplerian steady disc, we get the relationship for $D(r)$
independent of $\nu$:
\begin{equation}
\label{diss_3}
D(r)=\frac{3GM\dot{M}}{8\pi r^{3}}\left[1-\left(\frac{r_{in}}{r}\right)^{\frac{1}{2}}\right]. 
\end{equation}
The luminosity produced by the disc between radii $r_{1}$ and $r_{2}$ is given by:
\begin{equation}
\label{ldisc_1}
L\left(r_{1},r_{2}\right)=2\int_{r_{1}}^{r_{2}}D(r)2\pi rdr,
\end{equation}
and using eq.~(\ref{diss_3}):
\begin{equation}
\label{ldisc_2}
L\left(r_{1},r_{2}\right)=\frac{3GM\dot{M}}{2}\int_{r_{1}}^{r_{2}}\left[1-\left(\frac{r_{in}}{r}\right)^{\frac{1}{2}}\right]\frac{dr}{r^{2}}. 
\end{equation}
Setting $r_{1}=r_{in}$ and $r_{2}\rightarrow\infty$, we obtain the luminosity
for the whole disc:
\begin{equation}
\label{ldisc_3}
L_{\rm{disc}}=\frac{GM\dot{M}}{2r_{in}}=\frac{1}{2}L_{\rm{acc}}.
\end{equation}

The thin disc is characterised by no motions or accelerations in the $z$-direction
and the relevant Euler's equation has the form:
\begin{equation}
\label{euler_z}
\frac{1}{\rho}\frac{\partial p}{\partial z}=\frac{\partial}{\partial z}\left[\frac{GM}{\left(r^{2}+z^{2}\right)^{\frac{1}{2}}}\right],
\end{equation}
where $p$ is the pressure.
For a thin disc $\left(z\ll r\right)$ we can write:
\begin{equation}
\label{eulerz2}
\frac{1}{\rho}\frac{p}{h}=-\frac{GMh}{r^{3}},
\end{equation}
where $h$ is the typical vertical scale height of the disc.
From eq.~(\ref{eulerz2}) we can estimate the thickness of the disc as:
\begin{equation}
\label{thickness}
h\approx\frac{c_{s}}{\Omega_{K}}=\frac{c_{s}r}{v_{K}},
\end{equation}
where $v_{K}$ is the local Keplerian velocity and $c_{s}=\sqrt{\frac{p}{\rho}}$ is the sound speed. 
Hence, we can write the criterion for the thin disc as:
\begin{equation}
\label{thickness_velocity}
c_{s}\ll v_{K}.
\end{equation}
It means that the accretion disc is geometrically thin 
when the Keplerian velocity is highly supersonic.

The total pressure of the disc is the sum of gas and radiation pressure.
The equation of state has the form:
\begin{equation}
\label{eos}
P = \frac{\rho k_{\rm B} T}{\mu m_{\rm p}} + \frac{4\sigma}{3c}T^4.
\end{equation}

The relationship for the central temperature can be derived from
evaluating the heat loss per unit area by the radiative transport
with the thermal energy by viscous dissipation given by eq.~(\ref{diss_3}):
\begin{equation}
\label{rad_trans}
\frac{4\sigma}{3\tau}T^4 = D(r)=\frac{3GM\dot{M}}{8\pi r^{3}}\left[1-\left(\frac{r_{in}}{r}\right)^{\frac{1}{2}}\right],
\end{equation}
where $\tau$ is the optical depth of the disc, which can be defined as:
\begin{equation}
\label{tau}
\tau = \kappa_{\rm R} \rho H = \kappa_{\rm R} \Sigma,
\end{equation}
where $\kappa_{\rm R}$ is the total Rosseland mean opacity.
For hot discs around compact objects, free-free transitions
and Thomson scattering contribute mostly to the opacity.
The above estimation of the central temperature value is valid
only for the case of an optically thick disc with $\tau \gg 1$.

Up to now, we have considered equations of the mass conservation, the angular momentum
conservation, energy conservation, hydrostatic equilibrium, equation
of state, and equation of radiative transport for the thin steady disc.
In order to study the detailed physical structure of such a disc,
or any aspect of its time-variability, as well as stability,
the knowledge of the viscosity $\nu$ is required.
A simple and very useful model of the viscosity 
is $\alpha$-model \citep{1973A&A....24..337S}: 
\begin{equation}
\label{alpha}
\nu=\alpha c_{s}h,
\end{equation}
where $\alpha$ is a dimensionless parameter, which can take the value from $0$ to $1$.
In the first approximation, $\alpha$ is a constant for a given {\ad}.

The viscous stress $f_{\varphi}$ exerted in the $\varphi$-direction can be defined for a Keplerian disc as:
\begin{equation}
\label{visc_stress_1}
f_{\varphi}=\frac{3}{2}\eta\,\Omega_{K},
\end{equation} 
where $\eta$ is the coefficient of dynamic viscosity. For turbulent motion,
it can be expressed as:
\begin{equation}
\label{dynvisc}
\eta\approx\rho v_{\rm{turb}}l_{\rm{turb}},
\end{equation} 
where $\rho$ is the matter density, $v_{\rm{turb}}\leq c_{s}$ is the velocity of turbulent cells relative to the
mean gas motion and $l_{\rm{turb}}\leq h$ is the size of the 
largest turbulent cells \citep{1959flme.book.....L, 1983JBAA...93R.276S}.
Using eq.~(\ref{thickness}) we can estimate the value of viscosity stress as:
\begin{equation}
\label{visc_stress_2}
f_{\varphi}\leq\left(\rho c_{s}h\right)\Omega_{K}\approx\rho {c_{s}}^{2}=P.
\end{equation}
In general, we may write:
\begin{equation}
\label{visc_stress_3}
f_{\varphi}=\alpha P,
\end{equation}
where $\alpha$ is the same parameter as in eq.~(\ref{alpha}).
From eq.~(\ref{rad_vel}), the radial velocity can be expressed by this model as:
\begin{equation}
\label{rad_vel_alpha}
v_{r} \approx \frac{\nu}{r} \approx \frac{\alpha c_{s}h}{r} \ll c_{s}.
\end{equation}
Thus, the radial inflow is very subsonic.

The steady thin disc solution (or also called Shakura-Sunyaev's solution)
allows to express the central density $\rho(r)$, the surface density $\Sigma(r)$,
the central pressure $P(r)$, the disc height $h(r)$, the radial drift $v_r(r)$,
the central temperature $T(r)$, and the optical depth $\tau(r)$ as functions
of the parameters $M, \dot{M}$, and $\alpha$.

\subsubsection{Accretion flows in sub- and super-Eddington regime}
\label{accretion_flows}

The condition for the steady thin disc accretion is not always accomplished.
The accretion rate is the main factor which constrains the shape of the accretion 
flow. In the low (sub-Eddington) accretion regime, the cooling mechanism via
radiation ceases to be sufficiently efficient and advection mechanism takes 
place instead \citep[and references therein]{1977ApJ...214..840I, 
1994ApJ...428L..13N, 1995ApJ...438L..37A, 2008NewAR..51..733N}.
In the super-Eddington regime, the gas pressure is strong enough to expand 
significantly the disc vertically and the height of the disc needs
to be taken into account
\citep{1988ApJ...332..646A, 2001MNRAS.324..119Y, 2009ApJS..183..171S}.

Generally, we may write the energy conservation equation per unit volume as:
\begin{equation}
\label{entequ}
\rho T \frac{dS}{dt} = q_+ - q_-,
\end{equation} 
where $\rho$ is the density, $T$ is the temperature, $S$ is the entropy per
unit mass, $t$ is the time, and $q_+$ and $q_-$ are the heating and cooling
rates per unit volume. Since all the entropy stored in the gas is advected 
with the flow, the left-hand of eq.~(\ref{entequ}) may be replaced by a quantity
$q_{\rm adv}$ which represents cooling rate via advection. The heat
energy released by viscous dissipation is partially lost by radiative
cooling $q_-$ and partially by advective cooling $q_{\rm adv}$:
\begin{equation}
\label{entequ2}
q_+ = q_- + q_{\rm adv}.
\end{equation} 
For the Shakura-Sunyaev thin disc is $q_- \gg q_{\rm adv}$.
If, oppositely, $q_- \ll q_{\rm adv}$, the gas
is radiatively inefficient and the accretion flow is under-luminous
($L \ll 0.1 \dot{M}c^2$). Such an accretion flow is known as
ADAF (advection-dominated accretion flow), or RIAF (radiatively
inefficient accretion flow). 

There are two distinct regimes
of advection-dominated accretion flow. The first one occurs
when the cooling time scale is much larger than the accretion 
time scale, $t_{\rm cool} \gg t_{\rm acc}$. It is the standard
case of the ADAF, a self-similar solution of which is described 
by \citet{1994ApJ...428L..13N}.
The second regime corresponds to very high scattering optical depth when
the radiation is unable to diffuse out of the system, i.e.
the photon diffusion time is much larger than the accretion 
time scale, $t_{\rm diff} \gg t_{\rm acc}$. This radiation-trapped
regime was briefly discussed by \citet{1979MNRAS.187..237B}
and then developed in detail by \citet{1988ApJ...332..646A}
as the ``slim disc'' model.

% some relations for ADAFs

\begin{figure}
\begin{center}
\includegraphics[width=0.7\textwidth]{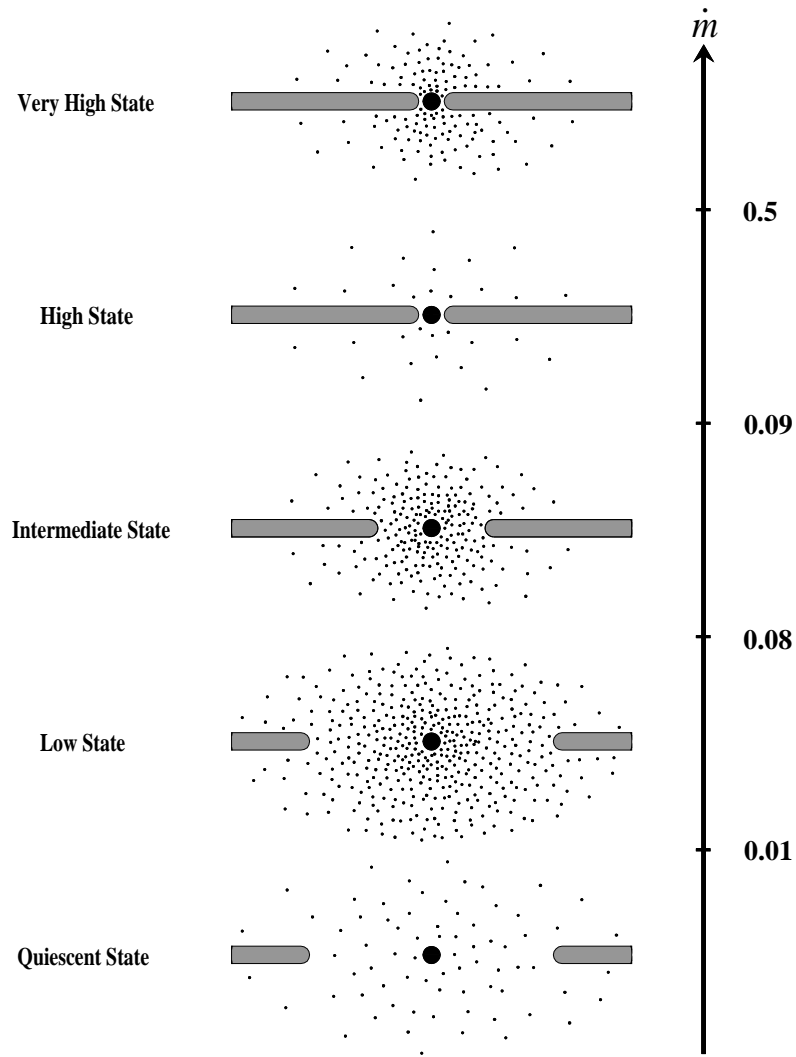}
\caption{Schematic sketch of the configuration of the accretion flow
in different spectral states shown as a function of the mass accretion
rate in Eddington units. The ADAF is indicated by dots. The thin disc
(horizontal bars) extends to the innermost stable orbit only
in the high state. In lower states, the truncation of the disc occurs,
and the transition radius depends on the accretion rate. 
Figure is adopted from \citet{1997ApJ...489..865E}.}
\label{uni_accretion}
\end{center}
\end{figure} 

Accretion state of an accretion disc evolves in time
because the accretion rate varies
and the accretion flow regimes interchange accordingly.
We know from the observations that the black hole binaries
change the spectral properties rapidly and several different
X-ray states were defined 
\citep[for review see][and references therein]{2006ARA&A..44...49R}.
A close connection between the states and accretion flow regimes
was discussed by \citet{1997ApJ...489..865E}. The value of 
the accretion rate is the main criterion for appearance of 
a certain X-ray state. Their schematic sketch is shown in
Figure~\ref{uni_accretion}, which illustrates the link between the accretion states
and the mass accretion rate. However, this simple
picture is rather far from the complete description of the spectral 
states behaviour, not taking into account dynamic properties
of the corona including strong flares, hysteresis of the states
when interchanging etc.

The main conclusion of the unification scheme is that below
some critical value of the accretion rate, such as
$\dot{m}_{\rm crit} \approx 0.08\,\dot{m}_{\rm Edd}$ \citep{1997ApJ...489..865E},
the accretion disc does not extend to the innermost stable orbit,
but, instead, it is truncated at some further radius whose
position is inversely proportional to the accretion rate.
The standard thin (Shakura-Sunyaev) accretion disc occurs
only in the high state. In the very high state, the radiation 
pressure inflates the disc in the close neighbourhood to the black hole, 
and consequently, the advection plays again an important role
in the inner flows (slim disc solution).

%spectral energy distribution for ssd (Camenzind)

\subsubsection{Thermal radiation of accretion discs}

The radiation temperature follows from thermodynamical considerations:
\begin{equation}
\label{kT}
kT_{\rm{rad}} \approx h\nu,
\end{equation}
where % $k \doteq 1.38\times 10^{-23}\rm{J.K}^{-1}$ is Boltzmann constant,
$h\doteq 6.626\times 10^{-34}\rm{J\,s}$ is the Planck constant and $\nu$ is the mean
frequency. The value of this temperature can be estimated by comparing it with 
the black-body temperature $T_{\rm{b}}$ and the thermal temperature $T_{\rm{th}}$:
\begin{equation}
\label{bbT}
T_{\rm{b}}=\left(\frac{L_{\rm{acc}}}{4\pi R^{2}\sigma}\right)^{\frac{1}{4}},
\end{equation}
\begin{equation}
\label{ThermT}
T_{\rm{th}}=\frac{GMm_{p}}{3kR},
\end{equation}
where $\sigma \doteq 5.67\times10^{-8}\,\rm{J\,s}^{-1}\,\rm{m}^{-2}\,\rm{K}^{-4}$ is the Stefan-Boltzmann constant.
These temperatures are two extreme cases for $T_{\rm{rad}}$. If the accretion flow is optically thick,
the radiation reaches thermal equilibrium before escaping and $T_{\rm{rad}} \approx T_{\rm{b}}$.
On the other hand, if the conversion to radiation is direct (from optically thin material),
then $T_{\rm{rad}} \approx T_{\rm{th}}$.
It gives the limits for the radiation temperature:
\begin{equation}
\label{Temperatures}
T_{\rm{b}}\leq T_{\rm{rad}}\leq T_{\rm{th}}.
\end{equation}
For a solar-mass neutron star we will give the range for photon energy:
\begin{equation}
\label{energies}
1\rm{keV}\leq h\nu \leq 50\rm{MeV}.
\end{equation}
Thus, we can expect accreting neutron stars and black holes
to appear as X-ray emitters, or possibly $\gamma$-ray sources. 

If we suppose that the viscously dissipated energy (eq. \ref{diss_3}) is radiated 
as a black-body spectrum (Shakura-Sunyaev disc), using eq.~(\ref{bbT}) we get the relationship 
for the surface temperature of the disc:
\begin{equation}
\label{tempAD}
T(r)=\left(\frac{3GM\dot{M}}{8\pi\sigma r^{3}}\left[1-\left(\frac{r_{in}}{r}\right)^{\frac{1}{2}}\right]\right)^{\frac{1}{4}}. 
\end{equation}
For a fixed ratio between the source luminosity and Eddington limit
and a scaled radius $r/M$ the temperature of the disc depends on the mass as:
\begin{equation}
\label{tempAD_2}
T(r)\propto M^{-\frac{1}{4}}. 
\end{equation}
%??? obrazek profilu termalniho zareni pro ruzny polomery
The temperature of the innermost region of an accretion disc 
surrounding a solar-mass black hole is $T \approx 10^7$\,K.
Using eq.~(\ref{kT}), the corresponding spectral energy is 
$E \approx 1$\,keV. Thus, the thermal component of the 
black hole accretion disc in an X-ray binary occurs in the
soft X-rays. For a super-massive black hole, the accretion
disc temperature is $T \approx 10^5-10^6$\,K for the mass
range $M \approx 10^6-10^9\,M_{\odot}$. The thermal disc
component has its maximum in the ultraviolet energy band and 
the spectral energy is $E \lesssim 0.1$\,keV for $M \approx 10^6\,M_{\odot}$,
and $E \lesssim 0.01$\,keV for $M \approx 10^9\,M_{\odot}$.

\subsubsection{Time dependent discs}

Accretion discs are fueled by a variable amount of the accreting matter,
and thus, the time variability of accretion discs is a natural consequence.
Let us estimate different time scales: dynamical, thermal and viscous. 
The dynamical (or orbital) time scale can be defined as:
\begin{equation}
\label{tdyn}
t_{\rm{dyn}}=\Omega^{-1}.
\end{equation}
The thermal time scale, on which the dissipated energy is radiated from the {\ad}, is given by:
\begin{equation}
\label{ttherm}
t_{\rm{th}} \approx \frac{t_{\rm{dyn}}}{\alpha} = \frac{1}{\alpha \Omega}.
\end{equation}
The viscous time scale which is linked with the redistribution of the angular momentum is defined as:
\begin{equation}
\label{tvisc}
t_{\rm{visc}} \approx \frac{r^{2}}{\nu} ~ \left(\frac{r}{h}\right)^{2}t_{\rm{th}}.
\end{equation}
Comparing the time scales we get for a thin disc:
\begin{equation}
\label{tvisc_2}
t_{\rm{visc}} \gg t_{\rm{th}} > t_{\rm{dyn}}.
\end{equation}

On dynamical time scales we can well consider the temperature 
and $\alpha$ parameter independent of time.
Although the $\alpha$-prescription is very successful in many applications,
there are some limitations of this approach. Especially, the presence
of the magnetic field plays an important role 
in the accretion disc stability.
The present-day theoretical models assume that 
the turbulence, which is responsible for the angular momentum transport,
may be due to a magneto-rotational instability \citep{1991ApJ...376..214B}.

%\subsubsection{Stability of an accretion disc}

%The stability of a differentially rotating disc with angular
%velocity $\Omega(r)$ is given by Rayleigh criterion:
%\begin{equation}
%\label{rayleigh}
%\frac{d}{dr}[r^2\Omega(r)] > 0.
%\end{equation}
%For geometrically thin discs with Keplerian angular velocity
%(eq.~\ref{kep}), this condition is fulfilled and the disc
%is hydrodynamically stable.  However, the presence
%of a weak magnetic field destabilizes the disc \citep{1991ApJ...376..214B}.

%MRI

\subsection{Accretion disc atmosphere and disc reflection}
\label{atmosphere}

X-ray emission of both, black hole binaries and active galaxies,
is characterised by a power-law component with an exponential
cutoff at high energy ($E \approx 300$\,keV). Its origin is suggested
to be due to multiple inverse Compton scatterings of the ``seed photons''
from the accretion disc (UV photons in the case of AGN, soft X-ray photons
in the case of black hole binary) in the optically thin 
accretion disc atmosphere, so called ``corona'' \citep{1975ApJ...195L.101T, 1991ApJ...380L..51H}.
The corona is believed to consist of hot relativistic electrons
which are possibly heated up by the magnetic dissipation processes.
These processes may be caused by amplification of 
the magnetic field due to convective motions 
and differential rotation within a hot inner region of the accretion disc,
resulting into flaring events in the places of magnetic re-connections
\citep{1979ApJ...229..318G,1994ApJ...432L..95H,2004A&A...428..353C}.

Geometrical properties of the corona are still uncertain, 
as well as the distribution of the Comptonising electrons 
(thermal, non-thermal, or mixed). 
The corona might be ``sandwiching'' the accretion disc \citep{1991ApJ...380L..51H},
or rather be locally centralised with a typical size of a few of $r_{g}$. 
In this geometry (``sphere + disc''), the corona is irradiated 
by soft photons from the cooler outer parts of the accretion disc
\citep{1976ApJ...204..187S, 1994ApJ...432L..95H,1995ApJ...449L..13S,1997ApJ...487..759D}. 

%An early model for the origin of a power-law component
%is origin in a hot optical thin inner accretion flow which is 
%formed when the inner edge of the accretion disc does not refer 
%to the last stable orbit around a black hole. The first idea
%of such a two-temperature accretion disc was formulated by
%\citet{1976ApJ...204..187S}, but it was later shown that this
%specific disc model was thermally unstable \citep{1976MNRAS.177...65P}.

Some X-ray photons produced in the corona may escape directly to the observer
and then be detected as the primary power-law radiation, but some of them may illuminate
the disc and be reflected from its surface before reaching the observer
\citep{1974A&A....31..249B}. The illuminating radiation is partly absorbed
in the disc medium and partly re-radiated from the accretion disc.
The reprocessed spectrum is characterised mainly by the Compton hump
at $E \geq 10$\,keV and the fluorescent iron line at $E \geq 6-7$\,keV, 
as seen in Figure~\ref{reflection_cold_ionized} or Figure~\ref{XraySpec}.

The Compton hump is the result of the increased importance of Compton scattering
compared to the bound-free absorption. The cross-section of the bound-free 
absorption $\sigma_{\rm bf}$ decreases namely with increasing energy 
\citep{1983ApJ...270..119M}, except for absorption edges,
while Compton cross-section $\sigma_{\rm C}$ is significant up to
$E \geq 50$\,keV \citep{1988ApJ...335...57L}. 
This effect makes the overall spectral hardening
and forming of the Compton hump \citep{1988ApJ...335...57L, 1988MNRAS.233..475G}.

\begin{figure}
\centering
\includegraphics[width=0.7\textwidth]{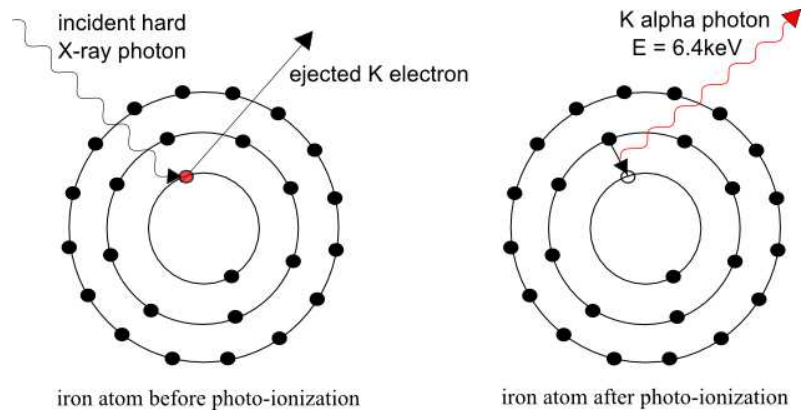}
\caption{A schematic sketch of the photoionisation of a neutral iron atom
followed by production of a K$\alpha$ photon. The three atomic shells
are shown, from inside: K, L, M. Two electrons of the N shell are grouped
with the M shell electrons in the picture for simplicity.
%The K$\alpha$ photon originates when an electron from the L shell
%fills the vacancy produced by X-ray photoionisation. When the
%vacancy filling electron comes from the M shell, K$\beta$ photon
%is released. Similarly, if the vacancy by photoionisation is produced
%in the L shell, L$\alpha$ photon is radiated if M shell electrons
%fills the vacancy.
}
\label{xrf}
\end{figure} 

\begin{figure}
  \includegraphics[width=0.99\textwidth]{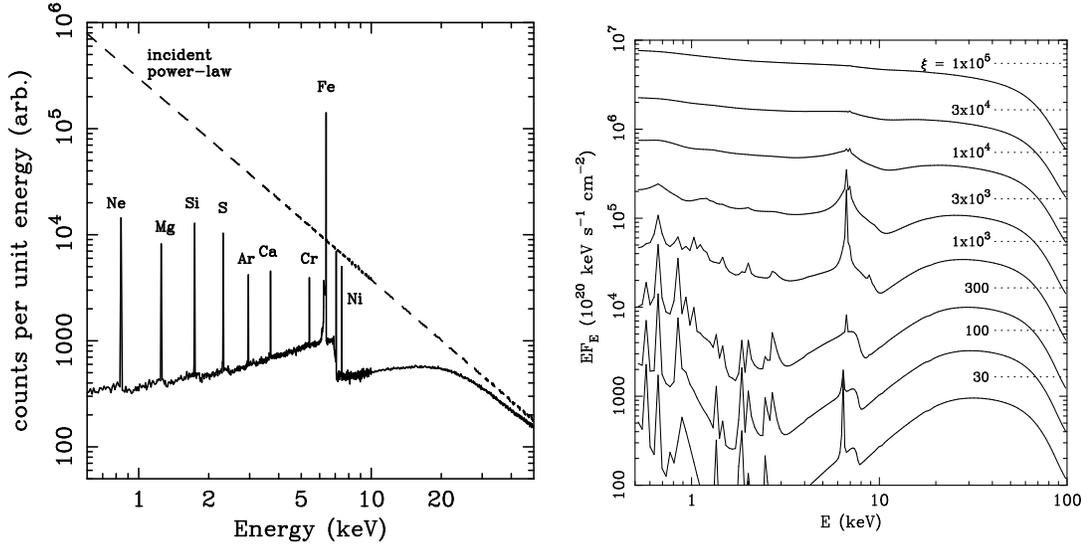}
  \caption{\textposown{Left:} Disc reflection spectrum originated by irradiation
of a neutral slab of gas by an incident power law radiation (dashed line).
Individual fluorescent lines are apparent the iron line being most prominent.
Figure adopted from \citet{1997MNRAS.286..513R}.
\textposown{Right:} Reflection spectra for different ionisation states
(from bottom to top $\xi=30 - 10\,000$).
Figure adopted from \citet{1999MNRAS.306..461R}.}
 \label{reflection_cold_ionized}
\end{figure}

In addition to the reflection continuum, intrinsically narrow features,
absorption edges and fluorescent emission lines, 
are other significant imprints of the reflection from an accretion disc.
These spectral features have a great potential in investigation
of the properties and localisation of the original source of radiation.
They are defined with specific energies and so, detecting of energy shifts
and broadening of the features is an important observational tool
which allows us to investigate the innermost region of a black hole
accretion disc.

A schematic picture of photoionisation followed by production
of an X-ray photon is shown for the case of a neutral
iron atom in Figure~\ref{xrf}. The incident
radiation must have high energy (hard X-ray) to be able to kick
off an inner electron. The minimal energy of the incident
photon is equal to the binding energy of the electron in the atom,
and it corresponds to the absorption edge in the reprocessed spectrum.
The originally neutral atom is photoionised
and the electron vacancy is immediately filled up by an electron dropping down
from a higher atomic level, L shell or M shell.
This is accompanied by a release of the energy equivalent to the
difference between the energy levels either in the form of radiation
(fluorescence), or by ejecting of an outer electron (Auger effect).
\footnote{The Auger effect may be interpreted as a further, inner,
photoionisation. A photon released by dropping down of an
electron from higher to lower level (from L to K shell)
is absorbed by an outer electron. The photon has enough energy
to eject the electron from the atom.}

The fluorescent emission line is named according to
the shell from which the electron was ejected by the photoionisation,
K line if the ejected electron is from the K shell,
L line if it is from the L shell etc. Further division follows
the origin of an electron filling the vacancy
in the inner shell. If it comes from a neighbouring shell
a Greek letter $\alpha$ is added to the name, and gradually
with higher levels ($\beta$, $\gamma$,...).
The K$\alpha$ line, shown in Figure~\ref{xrf}, means that 
a K electron was ejected by the photoionisation, and an L-shell
electron jumped into the vacancy. This is a more probable transition
than filling the vacancy by an M-shell electron. The probability
ratio of K$\alpha$ to K$\beta$ appearance is well defined
quantity and it is about 7.4 : 1. % \citep{1993MNRAS.263..314L}. 
Similarly, if the vacancy by photoionisation is produced
in the L shell, L$\alpha$ photon is radiated if an M shell electron
fills the vacancy.

The probability of the fluorescence occurrence, i.e. 
probability that an X-ray photon is radiated
from the atom after photoionisation, is characterised by
the fluorescence yield $\omega$. This quantity depends strongly on the
atomic number, approximately as $\omega \approx Z^4$
\citep[e.g.][]{1972RvMP...44..716B}. Taking into account cosmic
abundances as well, the iron fluorescent lines are expected
to be particularly strong. Moreover, the iron K$\alpha$ and K$\beta$
lines occur at the energies ($6.4$\,keV and $7.06$\,keV, respectively)
where both, the thermal radiation of an accretion disc 
and the reflection continuum (represented mainly by the Compton hump)
are minimal, and where only a few absorption lines of lighter 
elements occur. This all makes the iron K lines to be relatively 
easily observable spectral features (see left panel of Figure~\ref{reflection_cold_ionized}). 
Properties of the iron K$\alpha$
line in the X-ray illuminated cold accretion disc picture were 
extensively studied by \citet{1991MNRAS.249..352G,1991A&A...247...25M}
and \citet{1992A&A...257...63M}.

The energy and the intensity of the fluorescent line depends significantly
on the ionisation state \citep{1993MNRAS.261...74R,
1993MNRAS.262..179M, 2005MNRAS.358..211R}.
The ionisation parameter may be defined as:
\begin{equation}
\label{ionizpar}
\xi(r)=\frac{4\pi F_{x}(r)}{n(r)},
\end{equation} 
where $F_{x}$ is the flux received per unit area of the disc at a radius $r$,
and $n(r)$ is the co-moving electron number density.

The intrinsic energy of the fluorescent line monotonically increases
with higher ionisation state because the binding
energies of the inner shells increase with ionisation.
In the case of iron, the energy of the K$\alpha$ is very close to $6.4$\,keV
up to Fe\,XVII, then the energy value increases up to $6.7$\,keV for Fe\,XXV
(helium-like iron atom), and finally reaches the value $E=6.97$\,keV for
Fe XXVI (hydrogen-like iron atom) when the last electron is ejected
by photoionisation and another electron is caught up by the 
fully ionised atom and drops down to the innermost K shell.
The energy of the K~photon depends also on the sub-level of the
dropping down electron, which makes the K emission line
to be a doublet with energies $E_{\rm K\alpha1} = 6.404$\,keV
and $E_{\rm K\alpha2} = 6.391$\,keV. This difference is, however,
very small and beyond the resolution abilities of the detectors 
on-board the current X-ray satellites.

The intensity of the fluorescent line is maximal for $\xi \approx 1000 - 3500$.
This is due to two effects. First, the fluorescent yield is higher for 
ionised iron atoms than for neutral ones. Second, the incident
radiation which is strong enough to ionise iron atoms is not so much 
photoabsorbed by dissociated lighter elements.
The fluorescence does not occur if all the matter is too highly ionised,
so that all electrons are unbound. The iron fluorescent line cannot be produced
if $\xi~>~5000\,\rm{erg\,cm\,s^{-1}}$.
Reflection spectra for different ionisation are shown in
the right panel of Figure~\ref{reflection_cold_ionized}.

%importance of fluorescent lines - broad features
%bud tady nejakou zminku nebo az dal...

%tady by se mohl dat appendix pro popis comptonova rozptylu...
%
%\section{X-ray reflection}

%A model of such a X-ray reflected spectrum is presented in Fig.~\ref{XraySpecAGN}, \cite{fa05}.
%There are four prominent features of the spectrum. The underlying power-law component
%was already discussed. The soft excess corresponds with a change of the continuum slope. 
%The nature of this bump is not well understood.
%It can be due to the thermal Comptonization or emission by bremsstrahlung.
%Approximately $50\%$ of all Seyfert galaxies have characteristic absorption lines
%in the soft excess of their spectra. They are created by photo-absorption
%in a highly ionised medium, the so-called warm absorber.
%An emission line around 6.4keV$^3$ and a hardening of the power-law continuum
%above 10keV are the imprints of the reflected spectrum \cite{gefa91,po90,rofa93}. 
%It is produced by Compton scattering, fluorescence of iron atoms and any other emission
%from the disc associated with the hard X-ray illumination. 
%\footnotetext[3]{The conversion between energies and frequencies is given by the
%Planck's equation $E=h\nu$, $\nu\left[\rm{Hz}\right]\doteq 0.24\times 10^{18}E\left[\rm{keV}\right]$.}

\section{Observational evidence of accreting black holes}

\subsection{Stellar-mass black hole binaries}

The first black hole candidates to be identified were celestial
bodies of small size and mass only a few solar masses,
in close orbits to ordinary companion stars emitting
intense and rapidly flickering X-rays. This emission
is attributed to the radiation of inward-spiralling matter
in the form of accretion disc (see Section~\ref{ad}). The first
established black hole binary was Cygnus X-1 \citep{1972Natur.235...37W, 1972Natur.240..124B}.
This object is persistently bright in X-rays because its 
companion is a blue super-giant of spectral type O9.7Iab \citep{1973ApJ...179L.123W}
which fuels the accretion onto black hole by large amount.
These types of objects are classified as High-Mass X-ray Binaries (HMXBs)
and next to the Cygnus X-1, two black hole binaries in the Large Magellanic
Cloud, LMC X-1 and LMC X-3, belong to this category.

More frequently observed black hole binaries are, however,
Low-Mass X-ray Binaries (LMXBs), or also called X-ray novae,
which are transient and change rapidly the spectral state
according to the accretion rate of the infalling matter.
Two of them are especially remarkable, GRS~1915+105 belongs to the
brightest X-ray objects on the sky (excluding the Sun),
and since its eruption in August 1992 it remained very
bright for more than one decade; and GX~339-4, which undergoes
frequent outbursts followed by very faint states.
%, but it has never been observed to reach the quiescent state.
In total, about two dozens of black hole binaries are
confirmed and two other dozens are the candidates. 
For review about X-ray properties of the observed black hole 
binaries see \citet{2006ARA&A..44...49R}.

%spectral states

\subsection{Active galactic nuclei}
\label{section_agn}

Another class of the astrophysical black holes are super-massive
black holes (SMBH) which settle in the dynamical centres of galaxies.
The immense nuclear activity of some galaxies was detected already 
at 1930s by radiotelescopes, but the 
spatial resolution was poor to measure the position of the radiation
source. In the early 1940s, Carl Seyfert discovered intense nuclear
activity together with the presence of highly ionised and 
extremely broad (up to 8500\,km\,s$^{-1}$)
optical emission lines in a sample of galaxies \citep{1943ApJ....97...28S}.
Consequently, a new subclass of galaxies, Seyfert galaxies, was established.

Another windows to the Universe, especially X-ray and infrared, 
revealed that some galactic nuclei are enormously bright compared 
to the rest of galaxies in the whole spectral energy range. 
These objects are commonly denoted as Active Galactic Nuclei (AGN). 
Some AGNs are faint radio sources like M31, but other nuclei, 
like the one in a quasar 3C~273, belong to radio-loud AGNs, 
which are characterised by collimated jets 
of energetic particles spanning millions of light years into the space.

Such an activity of galactic nuclei is attributed to the 
accretion on the super-massive black hole with the mass 
of several millions to billions of solar masses,
$M_{\rm BH} \approx 10^6-10^9 M_{\odot}$, \citep{1984ARA&A..22..471R}.
The precise value of the central black hole mass is measured 
from the velocity dispersion of the stars orbiting near 
to the centre (see Section~\ref{mass_meas} for more details).

\begin{figure}[tbh]
\begin{center}
\includegraphics[width=0.45\textwidth]{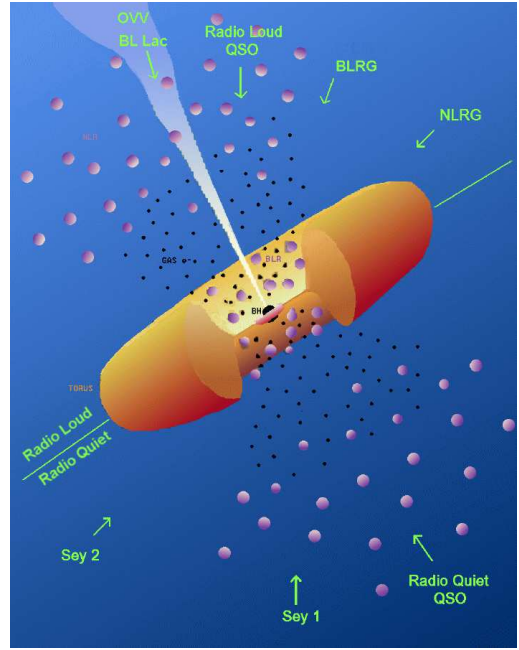}
%\includegraphics[width=0.49\textwidth]{elvis}
%\caption{\textposown{Left:} Standard picture of the AGN unified scenario
%\citep{1993ARA&A..31..473A, 1995PASP..107..803U}.
%\textposown{Right:} Proposed structure of quasar by \citet{2000ApJ...545...63E}.}
\caption{Standard picture of the AGN unified scenario
\citep{1993ARA&A..31..473A, 1995PASP..107..803U}.
All the observed active galaxies have the same
intrinsic structure but are viewed under a different inclination.
(QSO = quasar, Sey = Seyfert galaxy, BLRG = Broad Line Region
Galaxy, NLRG = Narrow Line Region Galaxy, OVV = Optically
Violent Variables, BL Lac = BL Lac Objects)}
\label{unification}
\end{center}
\end{figure} 

\begin{figure}[tbh]
\begin{center}
\includegraphics[angle=-90,width=0.7\textwidth]{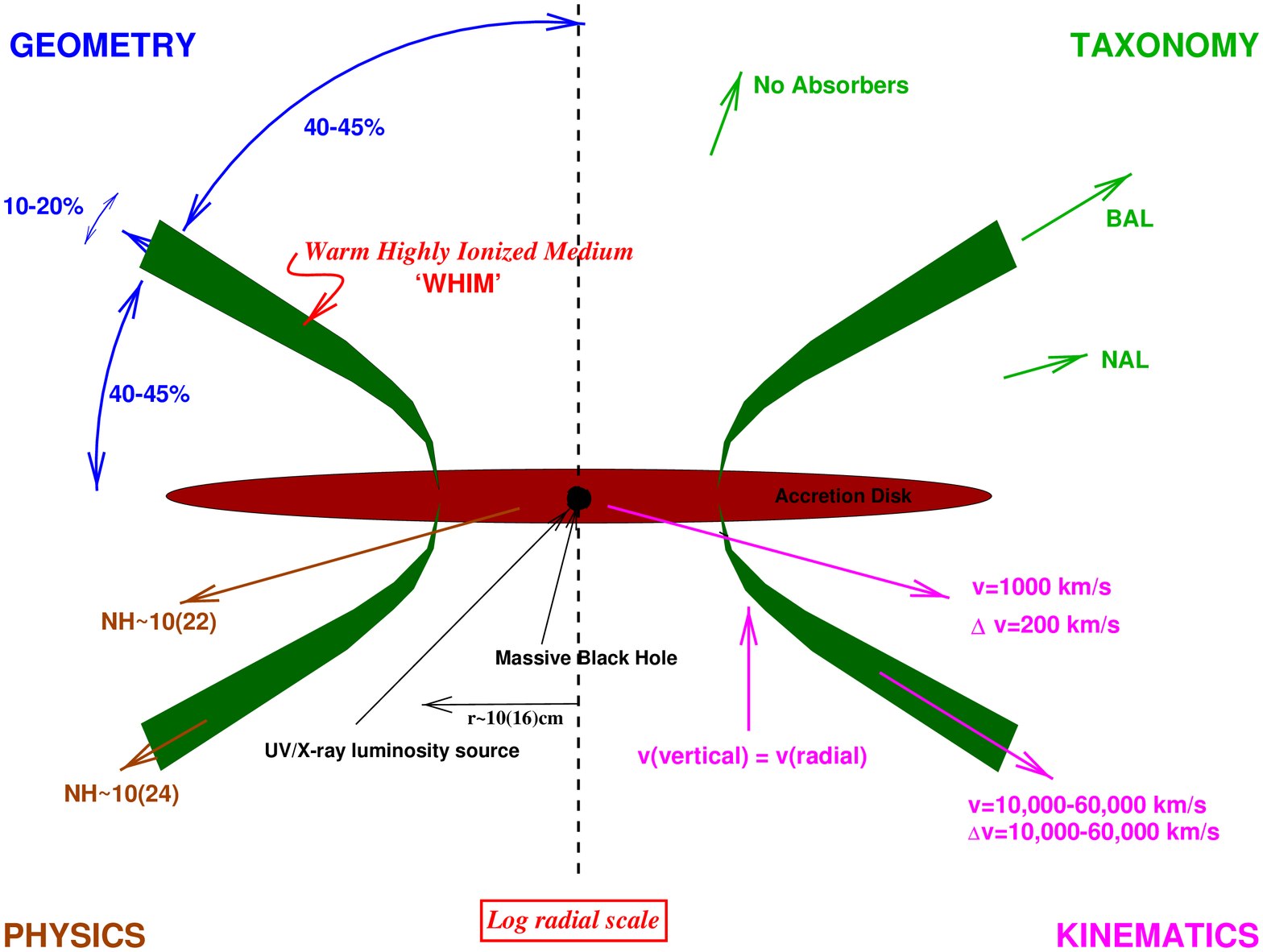}
\caption{Proposed structure of quasar by \citet{2000ApJ...545...63E}.
The figure is divided into four quadrants which illustrate the following
(clockwise from top left): the opening angles of the structure,
the spectroscopic appearance to a distant observer at various angles
(BAL = broad absorption lines, NAL = narrow absorption lines),
the outflow velocities along different lines-of-sight, and the typical
column densities of the absorber.}
\label{elvis}
\end{center}
\end{figure} 

Extensive mapping of properties of the individual AGNs across
the  whole electromagnetic spectrum lead to the origin 
of several different empirical AGN subclasses, which gradually
evolved into a realisation that a unification into a single
family of intrinsically similar active galaxies may be possible
\citep{1985ApJ...297..621A,1993ARA&A..31..473A,1995PASP..107..803U}.
The different appearance is mainly due to the orientation effect,
see Figure~\ref{unification}. The activity in the radio spectral energy
range, so called ``radio loudness'', is the only large distinction,
probably connected with the presence or absence of the relativistic
jet. 

Particularly interesting subclass of the AGNs are Seyfert galaxies
which belong to the low-luminosity radio-quiet AGNs. 
Its significant role in the unification scheme
is mainly due to the relatively frequent occurrence at low redshift
(close to our Galaxy), which enables a spatial resolution
unachievable for the distant quasars. The Seyfert galaxies are
divided into two main groups, Seyfert 1 and Seyfert 2, according
to the presence or absence of broad lines, especially Balmer lines
of hydrogen. According to the unification scenario, as seen in the
Figure~\ref{unification}, the broad line region of Seyfert 2s
is obscured by a torus surrounding the central region 
as the nucleus is seen under a high value of the 
inclination.\footnote{Standard convention is that the inclination
angle is zero when we see the disc/torus along the symmetric axis.
We say that we see the disc ``face-on''. Oppositely, if the observer
is in the disc plane, the inclination is 90 degrees and the disc is ``edge-on''.}
The torus does not enter the line-of-sight between us and the 
central broad line region when we see the nucleus under a low value 
of the inclination. The typical value of the inclination angle of Seyfert~1s
is around 30 degrees. More sensitive detectors constructed
in 1980s enabled fainter division of Seyferts into spectral 
subclasses \citep{1989agna.book.....O}. 
The spectral type of a Seyfert galaxy may be expressed as
\citep{1990agn..conf.....B}:
\begin{equation}
{\textrm{Spectral type}} = 1 + \left[{\frac{\textrm{narrow-line flux}}{\textrm{total flux in lines}}}\right]^{0.4}.
\label{seyfert_type}
\end{equation}

In spite of the unquestionable success of the standard unification scheme
by \citet{1993ARA&A..31..473A},
some observational facts are not explained within this standard AGN picture. 
Especially, broad absorption lines
($\approx$\,0.1$c$) in about 10$\%$ of quasars and highly ionised outflows
($v \approx 1000$\,km\,s$^{-1}$) in narrow absorption lines in a half
of Seyfert galaxies belong to the issues which are beyond the
standard picture. Hence, \citet{2000ApJ...545...63E} proposed
a different structure for quasars which can be also applied to other AGNs,
see Figure~\ref{elvis}. The warm highly ionised medium, WHIM, is 
in the form of an outflow from the disc and has a conical configuration.
The opening angle of the cone is about 70-90~degrees and its width
is at maximum about 10-20~degrees. If we translate it to the probability
of observation of a quasar through the conical sheath then broad absorption
lines of the WHIM should be detected in about 10-20\,\% quasars.
This corresponds to the proportional representation of the so 
called BAL quasars in the observations. The spectra are characterised
with high values of column density of the ionised absorber
$N_{\rm H} \approx 10^{24}$\,cm$^{-2}$.

\subsection{Black hole binaries versus active galaxies: similarities and differences}
\label{similarity}

\begin{figure}
\centering
	\includegraphics*[width=0.7\textwidth]{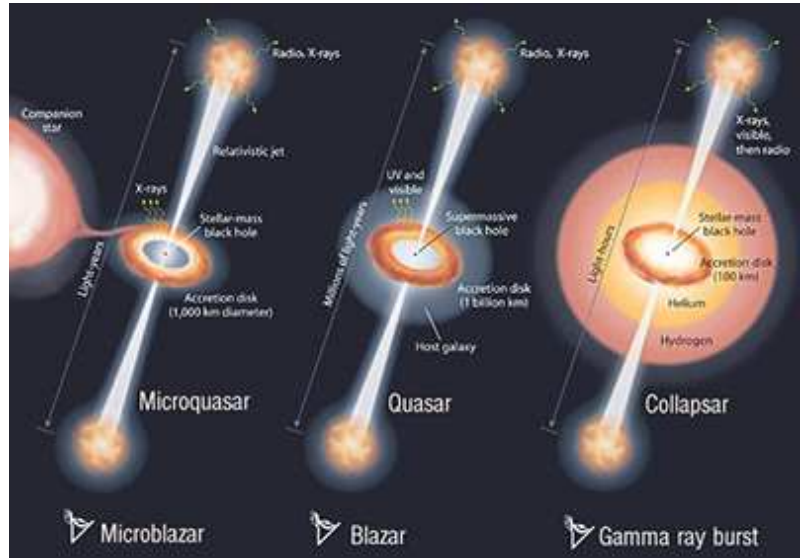}
\caption{The same physical mechanism can be responsible for three 
different types of objects: microquasar (\textposown{left}), quasar (\textposown{middle}), 
and collapsar (\textposown{right}). 
Figure is adopted from \citet{2006IAUS..230...85M}.}
\label{mirabel}
\end{figure}

\begin{figure}
	\centering
		\includegraphics*[width=0.95\textwidth]{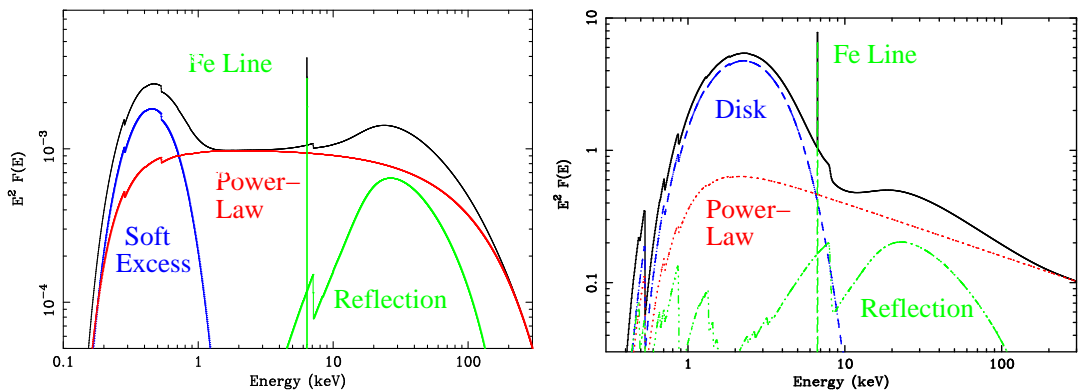}
	\caption{Model of intrinsic X-ray spectrum of a Seyfert 1 AGN (\textposown{left}) 
and a stellar-mass black hole binary (\textposown{right}). 
Relativistic smearing is not included
in the figure. The figure illustrates similarity of the AGN and BHB 
X-ray spectrum. Hard X-rays are dominated by the power-law component 
(red curve) which partially reflects on the disc (green curve). 
The reflection component is dominated by the Compton hump at 
$E \approx 20-40$\,keV and the iron fluorescent line. In the figure, 
the accretion disc of the AGN is assumed to be cold while the disc 
of BHB is ionised, and so, the shape of the reflection component 
differs accordingly. The soft excess (blue curve) 
observed in several Seyfert galaxies is shown here as 
a black-body component but it also may arise through disc 
reflection (for more details, see Section~\ref{similarity}).
Figure is adopted from \citet{2007ARA&A..45..441M}.
}
	\label{XraySpec}
\end{figure}

Both, black hole binaries and active galaxies, are powered
by black hole accretion. The masses of black holes in binaries
and galactic nuclei are very different (see previous Sections).
The typical length and time scale differ accordingly (eq.~\ref{rs}).
However, the physical mechanism of the energy balance seems to be the same,
and so, the AGNs can be interpreted as scaled-up Galactic black holes,
or vice-versa, Galactic black holes are often
called as ``microquasars'' \citep{1998Natur.392..673M}. 
The unified picture of both types of objects
is illustrated in Figure~\ref{mirabel} where another object,
collapsar, is added to the same family of accretion powered objects, as well.
The collapsar occurs when a massive star 
undergoes the final gravitational collapse, and 
it appears only in a short flash. If the jet is aligned
with the line-of-sight the objects appear as blazars, microblazars,
and gamma ray bursts, respectively. 

Besides the similar appearance of AGNs and BHBs the temporal
behaviour seems to be the same, only proportionally scaled 
\citep{2006Natur.444..730M,2008Natur.455..294U}.
The minimal time scale for the spectral variability is limited by the
black hole horizon (eq.~\ref{rs}):
\begin{equation}
 T_{\rm variability} \gtrsim \frac{r_g}{c} = \frac{GM}{c^3}.
\end{equation} 
The typical time scale for stellar-mass black hole binaries
is of order of milliseconds, 
for super-massive black holes in AGNs
it is of order of minutes to hours.
%Thus, while the black hole binaries evolve very rapidly and may change 
%their spectral states during a single observation, AGNs can 
%be assumed to occur always in the same accretion regime.

The very rapid flux variability is a characteristic feature 
of both kinds of objects and was constrained in several observations.
The RXTE (Rossi X-ray Timing Explorer) satellite is suitable
for studying millisecond variability of BHBs 
\citep[see e.g. the review by][and references therein]{2006csxs.book...39V}.
The variability of AGNs was also reported in several works
\citep[see e.g.][]{1999MNRAS.306L..19I,1999ApJ...524..667T,
2004MNRAS.349.1435M,2004MNRAS.348.1415V,
2005MNRAS.363..586U,2006A&A...445...59T,
2006MNRAS.368..903P,2008Natur.455..369G,2009A&A...507..159D}
although more sensitive instruments or longer exposure times
would be eligible to increase the accuracy.

The time-averaged spectrum of a black hole binary evolves
as the spectral states change, but in general, it may be interpreted
as a composition of power-law radiation, thermal multi-colour black-body 
radiation of the accretion disc \citep{1984PASJ...36..741M}, 
and reflection radiation from the disc,
all absorbed by a local (possibly ionised) matter and the interstellar
Galactic gas. The model spectrum of BHB is shown
in the right panel of Figure~\ref{XraySpec}. The model spectrum
of an AGN is very similar (left panel of Figure~\ref{XraySpec}),
but the thermal component is dominated in the UV energy range,
and, thus, it is obscured by the interstellar gas, which is opaque
to UV photons. Nevertheless, the high energy tail of a black-body
spectrum can reach soft X-rays if the temperature of the disc 
is high enough. Indeed, a ``soft excess'' is observed in several
Seyfert galaxies which may be interpreted by the thermal radiation
of the disc (as shown in the left panel of Figure~\ref{XraySpec}).

However, \citet{2006MNRAS.365.1067C} and \citet{2006MNRAS.371L..16G}
argued that the thermal radiation interpretation is doubtful
since the temperature would be the same for a big sample of AGN
with distinct masses, and therefore, they suggested alternative explanations.
The ``soft excess'' may be due to reflection by the ionised surface
of the accretion disc \citep{2006MNRAS.365.1067C}. 
For a cold disc, the reflection is dominated
by a plenty of emission lines in soft X-rays.
In this case, the ``soft excess'' may also occur when
the reflection radiation comes from the innermost regions
of an accretion disc and the individual lines are blended
due to relativistic smearing.
Another interpretation of the ``soft excess'' is that it appears
due to the partially ionised and Doppler smeared absorption \citep{2006MNRAS.371L..16G}.

%The reflection component of AGN is shown as neutral while ionised is supposed to occur on a cold disc 
%although some works suggest a significant role of the ionisation 
%\citep{}

\section{Measuring black hole parameters}

Astrophysical black holes are characterised only by their mass and angular momentum.
Although electrically charged black holes also represent the solution of Einstein's equations,
their existence in the Universe is very unlikely since the electromagnetic repulsion
is about 42 orders of magnitude stronger than the gravitational attraction. Thus,   
any charged black hole will be immediately neutralised by an accreting matter.

\subsection{Mass}
\label{mass_meas}

Masses of astrophysical black holes are known with relatively
good precision and customarily categorised in three groups 
of stellar-mass black holes ($\lesssim30M_{\odot}$), intermediate mass
black holes ($10^2$--$10^5\,M_{\odot}$), and super-massive black
holes ($\gtrsim10^6\,M_{\odot}$) in galactic nuclei
\citep[][]{2007IAUS..238....3C,2008ChJAS...8..273Z,
2009ApJ...699..800V,2009arXiv0910.0313C}. 
The intermediate mass black holes 
may occur in centres of globular star clusters
and represent a plausible explanation of so called ultra-luminous
X-ray sources (ULXs) \citep{2001MNRAS.321L..29K, 2002MNRAS.330..232C, 
2004IJMPD..13....1M, 2009ApJ...703.1386S}, 
but their existence has not yet been unambiguously accepted
\citep[e.g.][]{2004MNRAS.347L..18K, 2009ApJ...696..287S}.

The mass of the black hole $M_{\rm BH}$ in a binary system
can be determined from the knowledge of the inclination $i$ 
of the system and the mass ratio between two components
of the binary (the mass of the companion star can be estimated
from the spectral type of the star). The mass function is given by:
\begin{equation}
f(M) = \frac{PK_{\rm C}^{3}}{2\pi G} = 
\frac{M_{\rm BH}\sin^{3}i}{ \left( 1+ \frac{M_{\rm C}}{M_{\rm BH}}\right)^{2} },
\label{mf_binary}
\end{equation}
where $P$ is the orbital period, $K_{\rm C}=v \sin i$ is
the maximal line-of-sight Doppler velocity of the companion
star, and $M_{\rm C}$ is the mass of the companion star, 
which is usually lower than the black hole mass (if it is not a super-giant as
in the case of the Cyg X-1 system). 
 
%Jeansova rovnice, magorian relation
%korelace 

The mass of black holes in galactic nuclei can be derived
from the dynamical properties of the stellar neighbourhood.
The first velocity moment of the collisionless Boltzmann equation,
the Jeans equation, gives the mass as the function of radius
\citep[see e.g.][]{2008gady.book.....B, 2007coaw.book.....C}:
\begin{equation}
 M(r) = \frac{rv^2}{G} + \frac{r\sigma_{r}^2}{G} \left[-\frac{d \ln \rho_*}
{d \ln r} - \frac{d \ln \sigma_{r}^2}{d \ln r} - \left( 1 - 
\frac{\sigma_{\theta}^2}{\sigma_{r}^2} \right) - \left( 1 - 
\frac{\sigma_{\varphi}^2}{\sigma_{r}^2} \right) \right],
\label{jeans}
\end{equation} 
where $v$ is the rotational velocity, $\sigma_i^2$ are velocity
dispersions and $\rho_*$ is the density of stars. All these quantities
are measurable, but several problems exist. First, we observe
these quantities projected on the sky, and de-projection is rather
complicated process. Second, the most galaxies are not spherically 
symmetric. The high quality of the data are required. Observations performed 
by the Hubble Space Telescope can serve to extract the line-of-sight
velocity dispersions (LOSVD). Then, an orbit based approach, known
as Schwarzschild's method is applied to de-project the quantities.
Finally, maximum entropy models are used to account for axisymmetry 
instead of the spherical symmetry, and the potential is derived
from the profile of the surface brightness. 
The velocity maps can be achieved with higher precision 
from microwave maser emission of water molecules
\citep{1995Natur.373..127M,2002ApJ...565..836G}.

In order to estimate the black hole mass one can use the fact
that the stars around more massive nucleus orbit with higher 
velocities. The empirical relation between the black hole mass
and velocity dispersion of the nearby stars, ``$M-\sigma$'' relation,
was found on the sample
of galaxies with relatively known values of the central mass
\citep{2000ApJ...539L...9F,2000ApJ...539L..13G}:
\begin{equation}
% \log \frac{M_{\rm BH}}{M_{\odot}} = \alpha + \beta \log \frac{\sigma_*}{200 {\rm km\,s}^{-1}}
\log\left(M_{\rm BH}/M_{\odot}\right) = \alpha + \beta \log \left(\sigma_*/ 200\,{\rm km\,s}^{-1}\right).
\label{msigma}
\end{equation} 

Earlier, the correlation between the central mass and the bulge luminosity
was discovered \citep{1995ARA&A..33..581K, 1998AJ....115.2285M}.
This relation is often called ``Magorrian'' relation.
It has larger scatter than the ``$M-\sigma$'' relation, 
and hence it is used less often at present.
However, both correlations have important implications for theories
of the galaxy and bulge formation and the interactions of the stars
with the central super-massive black hole.

In fact, the value of the mass determined using such relations is
rather an order-of-magnitude estimation than the precise measurement
due to the relatively large scatter of the correlations.
Our own Galaxy is an exception in the mass determination
of black holes in galactic nuclei because it is so close ($\approx 25000$\,ly) 
that the individual stars orbiting around the central 
black hole are observed \citep{1996Natur.383..415E, 
1998ApJ...509..678G, 2005ApJ...620..744G, 2009ApJ...692.1075G}.
The latter one presents the black hole mass to be
$M_{\rm BH} = (4.3 \pm 0.2_{\rm stat} \pm 0.3_{\rm sys}) \times 10^6 M_{\odot}$,
where the uncertainty is given by the statistical
and systematical error, in which the uncertainty of the distance
is included. The suitable view to measure dynamical properties
of the closest stars to the Galactic centre is due to the
most sensitive infrared detectors at the largest telescopes
(VLT, Keck's telescopes). 

Though the infrared observations reveal some accretion
activity in the form of the flares \citep{2003Natur.425..934G},
there is no observational evidence for a standard 
Shakura-Sunyaev accretion disc in the Galactic centre.
The advection dominated accretion flow is likely to
explain the X-ray spectrum and also the luminosity 
of the Galactic centre \citep{1995Natur.374..623N}.

\subsection{Angular momentum (spin)}
\label{spin_meas}

The mass of a black hole is relatively easy to measure
because the attractive gravitational force reaches to the
large distance and affects therefore orbital movement of the
companion star in a binary system or the surrounding stars and gas
in the central region of a galaxy, respectively.
More challenging is to measure the value of the spin of a black hole
because the spin causes curvature
of the space-time in a detectable level only within a 
few gravitational radii around the black hole.
Despite this small outreach, the black hole spin plays
an important role in the black hole energetics, especially,
it is assumed to be responsible in
generating and up-keeping of the powerful relativistic jets
\citep{1969NCimR...1..252P,1977MNRAS.179..433B}.
The information about the spin value on a statistically significant
sample of black holes is important in the understanding of 
the formation and the growth of black holes. It can significantly help
to answer the question if the observed spin value is natal or if the 
black hole rotation is accelerated via the accretion 
\citep[see e.g.][]{1999MNRAS.305..654K, 2005ApJ...620...69V}.

There are several observational methods which provide 
a good opportunity to explore the innermost region of an accretion
disc, and thus, to constrain the spin value. The summary of them
is listed and briefly discussed:

\subsubsection{Continuum fitting}

In the thin Shakura-Sunyaev disc, the temperature decreases 
with the distance as given by eq.~(\ref{tempAD}).
When the innermost edge of the disc corresponds to the
last stable circular orbit, which should be the case
in at least one accretion state (high/soft), its position can be
determined from the spectral fitting of the thermal component.
This is the basic concept of the method which is called
X-ray continuum fitting method and which was first carried out 
by \citet{1997ApJ...482L.155Z} to measure black hole spin.

The temperature does not depend only on the inner edge
of the disc, but also on the accretion rate. The thermal
spectrum is further distorted by the rotational and gravitational
frequency shift, and therefore, the fully relativistic model is
acquired for this purpose \citep[\textsc{kerrbb} model]{2005ApJS..157..335L}.
The spectral hardening factor $f_{\rm col} = T_{\rm col}/T_{\rm eff}$
as a function of the Eddington 
scaled disc luminosity \citep{2005ApJ...621..372D}
and a switch parameter for zero/nonzero torque condition at the 
inner edge are included in the \textsc{kerrbb} model.
The spectral hardening factor 
(or also called colour correction) $f_{\rm col}$ plays an important 
role especially in higher accretion rates \citep{1995ApJ...445..780S}.

\subsubsection{Iron K$\alpha$ line profile}
\label{intro_ironline}

The iron K$\alpha$ line profile originates by the reflection of hard X-rays
on the accretion disc surface. % (see Section~\ref{atmosphere}). 
The combination of three effects makes the iron K$\alpha$ line 
easily detectable in X-ray spectra of black hole binaries 
as well as active galaxies. First, the fluorescent yield is 
higher for heavier elements ($\approx Z^4$, see Section~\ref{atmosphere}).
Second, the relative cosmic abundances of iron are high compared to other 
heavy elements. And last, the energy of iron K$\alpha$ line occurs 
in the spectral range of a simple continuum (see Figure~\ref{XraySpec}).

The profile of the intrinsically narrow emission line is 
distorted by thermal motion, Compton scattering,
Doppler broadening, and relativistic effects on the radiation 
including relativistic Doppler shift, lensing, gravitational redshift, 
and time delay (if the radiation source is not steady in time). 
The Doppler broadening due to the rapid orbital motion and the relativistic 
effects are much more significant than the thermal or Compton broadening
if the emitted radiation comes from the inner region
of an accretion disc (within a few hundreds of $r_g$),
making the line profile extremely smeared
even for a non-rotating Schwarzschild black hole \citep{1989MNRAS.238..729F}.
The effects are amplified in the case of a rotating black hole
since the marginally stable orbit shifts closer to the black hole
(see Sec.~\ref{blackholes} and Fig.~\ref{a_rms}), and are maximal
for a maximally rotating black hole \citep{1991ApJ...376...90L}.

The advantage of this method is that the line 
profile is completely independent of black hole mass, and is widely
applicable to BHBs as well as to  AGNs \citep[for reviews see][]{2003PhR...377..389R,
2007ARA&A..45..441M, 2009Ap&SS.320..129G}. 
The line profile depends on 
the source properties -- its geometrical position, which is influenced
by the black hole spin (the innermost edge of a disc), and orientation. 
This makes this technique
a suitable tool for investigation of the nature of the innermost region
of the accretion disc, and also for measurement of the inclination angle of the disc.
This method is more widely investigated and discussed in the following
Sections of the Thesis.

\subsubsection{Quasi-periodic oscillations}

X-ray quasi-periodic oscillations (QPOs) are transient phenomena
associated with the non-thermal states and state transitions.
There are two kinds of QPOs - low frequency quasi-periodic
oscillations (LFQPOs) at roughly 0.1-30\,Hz in the power
density spectra (PDS), and high frequency quasi-periodic
oscillations (HFQPOs) at roughly 40-450\,Hz in PDS.
The typical frequency of HFQPO corresponds approximately
to the orbital frequency at the ISCO, and are thus relevant
for the spin measurement. Although several models were suggested
\citep[e.g.][]{1998ApJ...492L..59S,1998ApJ...499..315T,2001A&A...374L..19A,
2003MNRAS.344L..37R,2005A&A...436....1T,2006A&A...451..377H}, 
the satisfactory description of QPOs is not well established.

\subsubsection{Variability and reverberation}

Small size of a ``hole'' in the inner accretion disc implies the
variability on the relatively short time-scales. 
Temporal changes in the primary radiation are translated
to the reflection radiation with a certain time lag $\tau$.
The characteristic time lag is approximately $\tau \approx r_{\rm ms}$/$c$.
A clear evidence for such a lag, a 30\,s reverberation lag between
direct X-ray continuum and Fe\,L emission accompanying the relativistic
reflection, was reported by \citet{2009Natur.459..540F}. This measurement
was possible thanks to a high overabundance of iron in this particular galaxy
and a very long exposure time by XMM-Newton satellite.

\subsubsection{Polarimetry}

The last method of the spin measurement is via X-ray polarimetry.
The polarisation of X-rays from accretion discs around black
holes was studied by \citet{1975ApJ...198L..73L}. The thermal
emission is polarised due to Thomson scattering in a disc atmosphere.
\citet{1980ApJ...235..224C} showed that the polarisation
features are strongly affected by general relativistic effects.
Other authors have considered effects of magnetic fields on the resulting
polarisation \citep[e.g.][]{1998MNRAS.293....1A, 2008A&A...481..217S}.
Especially, rotation of the polarisation angle is a sensitive
quantity. Recent studies \citep{2008MNRAS.391...32D, 2009ApJ...701.1175S}
illustrate how the polarisation features depend on the spin value
in the thermal state of an accreting black hole. Although
the models capable to compute Stokes parameters 
of a polarised accretion disc spectrum are ready for use
\citep{2004ApJS..153..205D}, any X-ray polarimeter useful
for this type of measurement has not yet been launched.
This method is hence promising in the future.

% ##########################################################################

\chapter{Relativistic lines}
 \chaptermark{Relativistic lines}
 \thispagestyle{empty}
 \label{rellinemod}

\section{Line profiles in strong gravity regime}

In classical astronomy, broad emission lines are useful diagnostic tool 
for measuring of the temperature
(thermal broadening) or velocity dispersion (Doppler broadening) of an observed system. 
However, X-ray spectroscopy of some X-ray binaries and active galaxies has revealed iron 
fluorescent lines so broad and asymmetric that their profiles cannot be explained
in terms of classical physics and instead, a complex fully general-relativistic
approach needs to be taken into account \citep{1989MNRAS.238..729F,1995Natur.375..659T,
2003PhR...377..389R,2007MNRAS.382..194N,2007ARA&A..45..441M}.
The broad iron lines are supposed to originate in close neighbourhood 
of a black hole where
the strong gravitational redshift occurs and the orbital velocities
reach a considerable fraction of the speed of light (see Fig.~\ref{v_orbit}).
%They are part of reflection of the primary radiation 
%(coming from a corona) on the accretion disc

\subsection{Frequency shift}
\label{frequency_shift}

The frequency shift, $g$, is defined as the ratio of the observed frequency
$\nu_{\rm{obs}}$ to the intrinsic emitted frequency $\nu_{\rm{em}}$:
\begin{equation}
\label{g-shift}
\nu_{\rm{obs}}=g\nu_{\rm{em}}.
\end{equation}
In high-energy astronomy, energies are usually used instead of frequencies:
\begin{equation}
\label{g-shift_energy}
E=h\nu=\frac{h\,c}{\lambda}, 
\end{equation}
where $h \doteq 6.626\times10^{-34}$\,J\,s is the Planck constant,
and $\lambda$ is the wavelength. Typically used
units are kiloelectronvolts ($1$\,keV$\doteq 1.602 \times 10^{-16}$\,J) for energies,
or Angstr\"{o}ms ($1 \AA = 10^{-10}$\,m) for wavelengths. The conversion relation is:
\begin{equation}
\label{conversion_keV_A}
E\left[{\rm keV}\right] \doteq \frac{12.4}{\lambda\left[\AA\right]}. 
\end{equation}

In the case of a rotating accretion disc, the g-factor is a function of the position 
on the disc and the emission angle $g=g\left(R,\varphi,\theta_e \right)$. 
In Newtonian approach:
\begin{equation}
\label{g-shift_class}
g_{\rm{class}}=1-\frac{V_{\rm los}}{c},  
\end{equation}
where $V_{\rm los}$ is the line-of-sight velocity. For Keplerian orbital velocity,
$V_{\rm los}=\sqrt{\frac{GM}{R}}\cos{\varphi}\sin{\theta}$.

In special relativistic approach:
\begin{equation}
\label{g-shift_str}
g_{\rm{STR}}=\frac{1}{\gamma\left(1-\frac{V_{\rm los}}{c}\right)},
\end{equation}
where $\gamma=\frac{1}{\sqrt{1-\frac{v^{2}}{c^{2}}}}$ is the Lorentz factor.
The observed frequency is shifted even when the line-of-sight velocity is zero
($V_{\rm los}=0$, ``face-on'' disc). Since $g_{\rm{STR}} = \gamma^{-1} < 1$
in this case, the frequency is shifted to lower energy.
This effect is called as the transverse Doppler shift.

In general relativity, g--factor may be expressed within the approximation 
of geometrical optics, in terms of four-momentum of photons $p_{\mu}$ 
and four-velocities $u^{\mu}$ \citep{1975ApJ...202..788C}:
\begin{equation}
\label{g-factor}
g=\frac{{\left(p_{\mu}u^{\mu}\right)}_{\rm{obs}}}{{\left(p_{\mu}u^{\mu}\right)}_{\rm{em}}}.
\end{equation}
Using ${u^{\mu}}_{\rm{obs}}=[-1,0,0,0]$ (four-velocity of the observer): 
\begin{equation}
\label{g-factor2}
g=\frac{-{\left(p_{t}\right)}_{\rm{obs}}}{{\left(p_{\mu}u^{\mu}\right)}_{\rm{em}}}.
\end{equation}

The frequency shift $g$ and the emission angle $\theta_{\rm e}$ 
may be expressed using constants of motion as:
\begin{equation} 
 g  =  \frac{{\cal{C}}}{{\cal{B}}-r^{-3/2}\xi}, 
 \quad 
 \theta_{\rm e}  =  \arccos\frac{g\sqrt{\eta}}{r}, 
\label{gtheta}
\end{equation}
where ${\cal{B}}=1+{a}r^{-3/2}$, ${\cal{C}}=1-3r^{-1}+2{a}r^{-3/2}$;
$\xi$ and $\eta$ are constants of motion connected with symmetries of
the Kerr space-time \citep{1973blho.conf..343N, 2006AN....327..961K}. 

The computation of g--factor can be provided numerically \citep[e.g.][]{2004ApJS..153..205D, 
2004MNRAS.352..353B, 2006ApJ...652.1028B}, or some feasible approximations
were developed for Schwarzschild metric to perform fast calculations 
\citep[e.g.][]{2002ApJ...566L..85B,2005A&A...441..855P}.

\begin{figure}[tbh!]
\begin{center}
\includegraphics[width=0.49\textwidth]{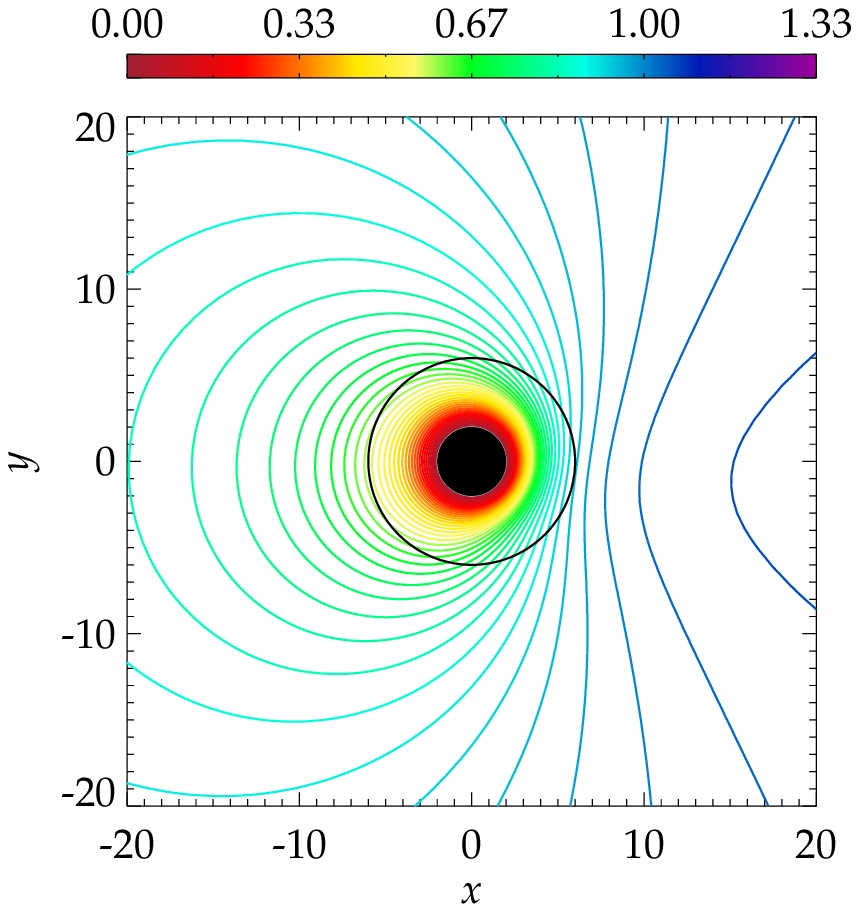}
\includegraphics[width=0.49\textwidth]{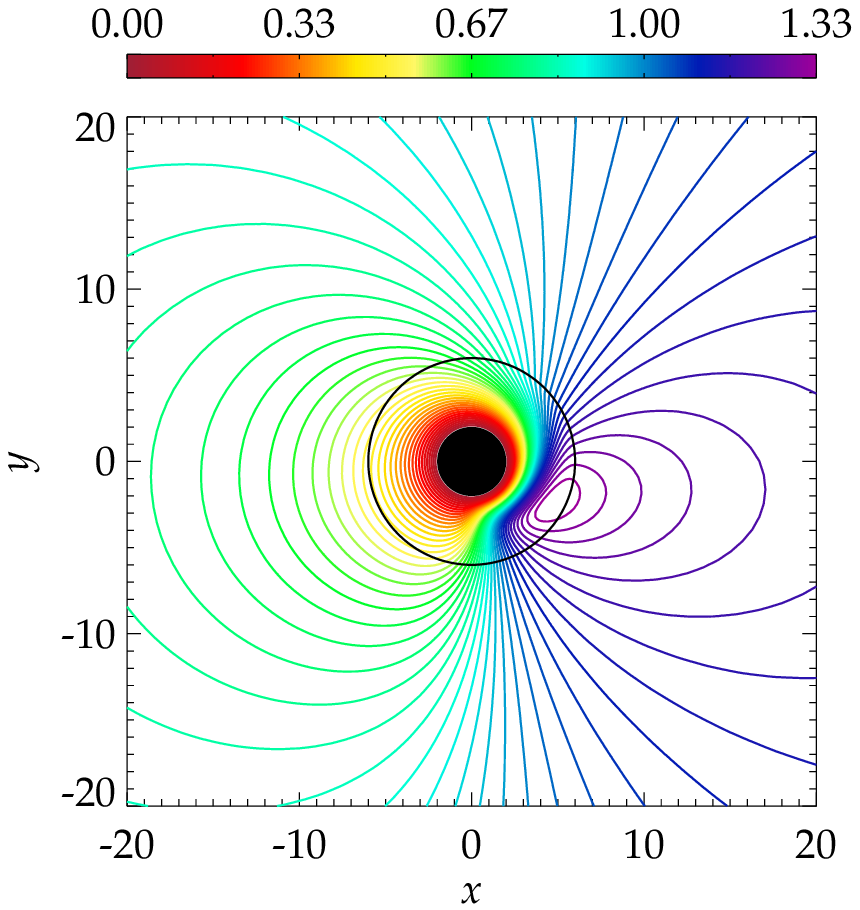}
\includegraphics[width=0.49\textwidth]{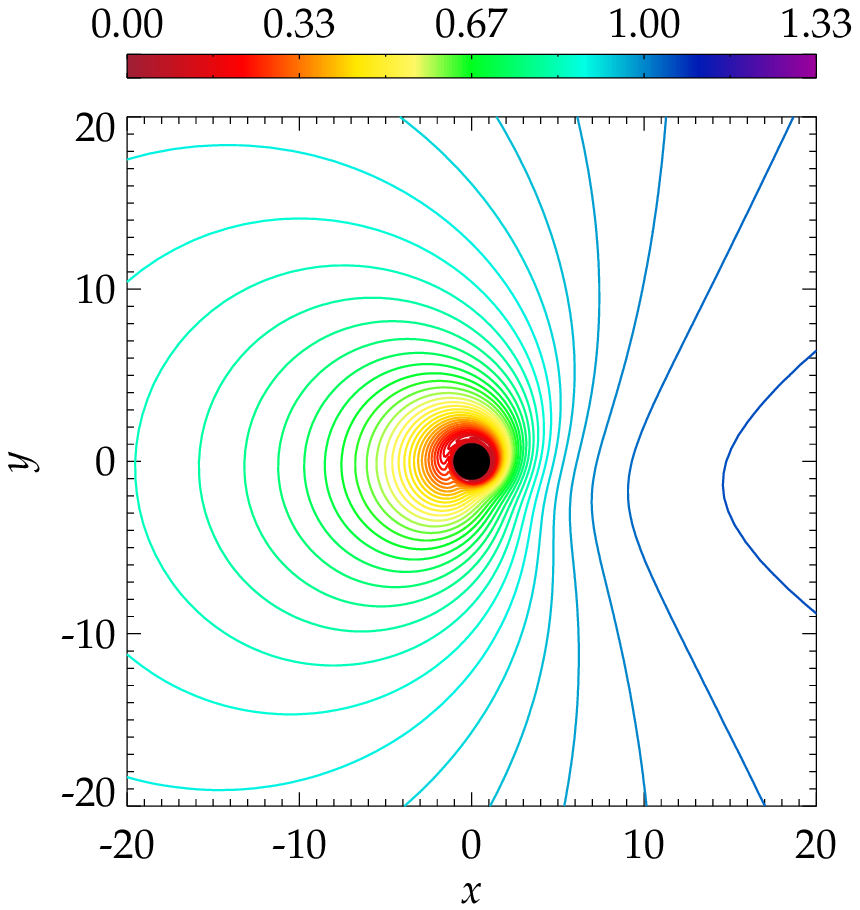}
\includegraphics[width=0.49\textwidth]{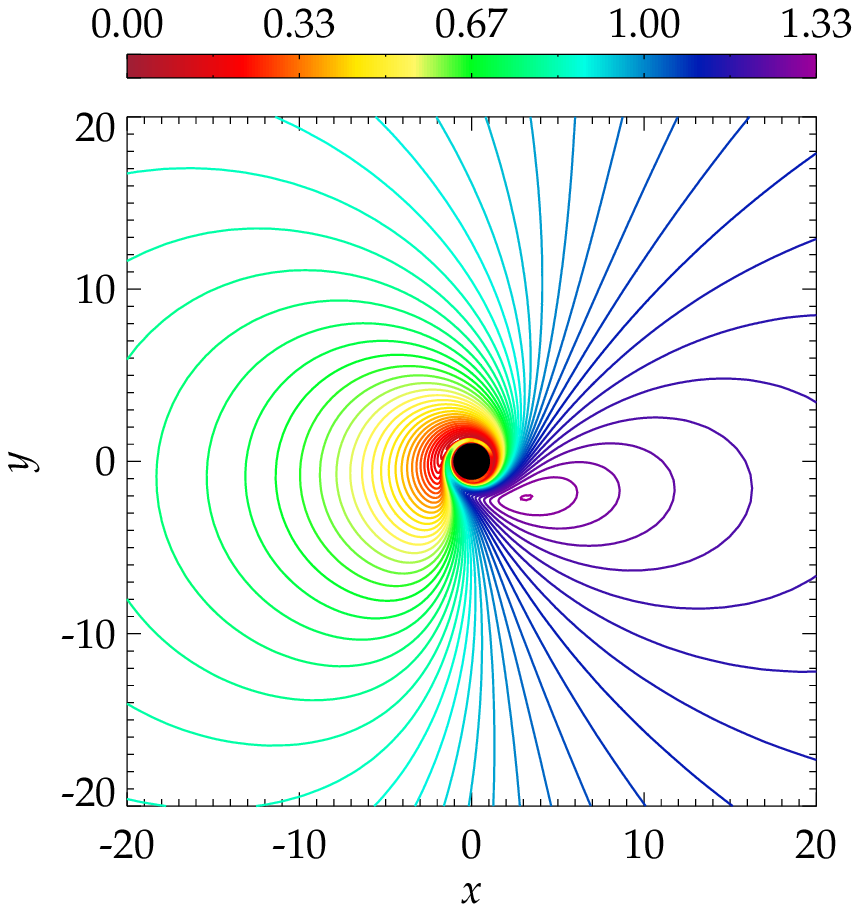}
\caption{Contours of redshift factor $g(r,\varphi)$ near a 
non-rotating Schwarzschild black hole, $a=0$, (\textposown{top})
and a maximally rotating black hole, $a=1$, (\textposown{bottom}), 
depicted in the equatorial plane $(x,y)$. 
The black hole and the accretion disc rotate counter clock-wise.
A distant observer is located towards the top of the figure.
The inner region is shown up to $r=20$ gravitational radii
from the black hole. The black hole is denoted by a dark filled 
circle around the centre, for Schwarzschild black hole
the circle around represents the marginally stable orbit.
Two cases of different observer inclinations
are shown.
\textposown{Left:} $\theta_{\rm o}=30$~deg. \textposown{Right:} $\theta_{\rm o}=70$~deg.
The colour bar encodes the range acquired by $g(r,\varphi)$, 
where $g>1$ corresponds to blueshift (approaching side of the disc),
while $g<1$ is for redshift.}
\label{fig2}
\end{center}
\end{figure}

The contours of constant frequency shift $g$ around a
non-rotating, $a=0$, and a maximally
rotating, $a=1$, black hole are shown for two different values
of the inclination angle in Figure~\ref{fig2}
\citep[an atlas of $g$--factor for different values of spin
and inclination may be found in][]{2004astro.ph.11605D}.
The value of the frequency shift $g$ results from two effects,
gravitational redshift and Doppler shift.
The effect of gravitational redshift quickly increases when
approaching the black hole, and it is infinite at the black hole horizon ($g=0$).
Near the black hole,
the gravitational redshift clearly dominates over the Doppler shift.

The magnitude of the Doppler shift depends on the line-of-sight velocities.
Its effect is also maximal at the closest orbit to the black hole where the orbital speed
is maximal (see Figure~\ref{a_rms}). 
Therefore, the maximal redshift always occurs at the innermost orbit.
However, the position of the maximal blueshift is farther from
the black hole due to the gravitational redshift, 
and its exact position depends on the inclination angle.
For lower inclinations, the Doppler effect is weaker
and hence, it overcomes the gravitational redshift at a farther radius.
Sufficiently far from the black hole, two distinct regions of blueshift and redshift 
are separated according to the rotation and the direction of the 
line-of-sight velocity to the observer. The level of the frequency shift 
decreases with the radius as the orbital velocity decreases as well.

\subsection{Intensity}

The specific intensity $I$ is defined as energy emitted at 
some given frequency $\nu$ into the element of solid angle $\Omega$
in unit of time $t$:
\begin{equation}
\label{intensity_def}
I=\frac{Nh\nu}{{\rm{d}}t{\rm{d}}\nu{\rm{d}}\Omega},
\end{equation}
where $N$ is the total number of photons.

The value of the intensity detected by an observer depends on the frequency shift
of the emitted radiation. This relation may be derived from the Liouville's theorem
which states that the phase-space volume $\Gamma$ is invariant to the
canonical transformations representing the time evolution of the system:
\begin{equation}
\label{liou}
\int_{\Gamma} {\rm{d}}^{n}p\,{\rm{d}}^{n}q=C,
\end{equation}
where $n$ is the dimension, $q$ expresses coordinates and $p$ the conjugated momenta.
In our application, the Liouville's theorem states that the phase-space
density:
\begin{equation}
\label{phase-space_density}
 n=\frac{N}{\Gamma},
\end{equation}
is a constant of the Lorentz transformation. 
The element of the phase-space volume is given by:
\begin{equation}
\label{phase-space_volume}
\Gamma={\rm{d}}^{3}p{\rm{d}}^{3}x=4\pi p^{2}{\rm{d}}p{\rm{d}}\Omega cS{\rm{d}}t.
\end{equation}
Thus,
\begin{equation}
\label{phase-space_density_2}
n=\frac{N}{4\pi p^{2}\rm{d}p\rm{d}\Omega cS\rm{d}t}.
\end{equation}

Substituting from (eq.~\ref{phase-space_density_2}) into (eq.~\ref{intensity_def})
and using $p=\frac{h\nu}{c}$:
\begin{equation}
\label{intensity_const}
I=\frac{4\pi h^{4}cSn\nu^{3}}{c^{3}},
\end{equation}
and
\begin{equation}
\label{Inu3}
\frac{I}{\nu^{3}}=const.
\end{equation}
Thus, using eq.~(\ref{g-shift})
\begin{equation}
\label{Ig3}
\frac{I_{\rm{obs}}}{I_{\rm{em}}}=\frac{{\nu_{\rm{obs}}}^{3}}{{\nu_{\rm{em}}}^{3}}=g^{3}.
\end{equation}

This fact implies that the radiation from a matter approaching to the observer
(blueshift) is boosted while the radiation from a matter receding from the observer
(redshift) is diminished. The level of the intensity amplification (resp. diminution)
depends on the values of the line-of-sight velocity.

\subsection{Line profiles from accretion discs}

For line emission from an accretion disc, 
we can assume an axial symmetry and separability of the radial and 
angular emissivities in some cases. 
Then, the line emission from the disc can be written in the form of product:
\begin{equation}
I_{\rm em}(r_{\rm
e},\mu_{\rm e},E_{\rm e})\equiv {\cal R}(r_{\rm e})\,\mu_{\rm e}\,{\cal M}(\mu_{\rm
e}, E_{\rm e})\,\delta(E_{\rm e}-E_{0}), 
\label{iloc}
\end{equation}
where  $r_{\rm e}$ is the disc radius (distance from the centre), $\mu_{\rm e}=\cos\theta_{\rm e}$
is the cosine of the emission angle measured from the disc normal direction to the equatorial plane,
in the disc co-moving frame, i.e.\ in the local Keplerian frame orbiting
with the angular velocity $\Omega_{_{\rm{}K}}(r)$. Likewise, the
intrinsic energy $E_{\rm e}$ is measured with respect to the local frame.
$E_{0}=E_{0}(\xi)$ is the intrinsic energy of the fluorescent line depending
on the ionisation state.

The radial part is being approximated by a power law, 
\begin{equation}
{\cal R}(r_{\rm e})=r_{\rm e}^{-q}\qquad (q=\mbox{const}),
\label{rloc}
\end{equation}
or by a broken power law:
\begin{equation}
{\cal R}(r_{\rm e})= \left\{
\begin{array}{l}
{\cal R}(r_{\rm e})=r_{\rm e}^{-q_1}\qquad \left(r_{\rm e} \leq r_{\rm b}\right) \\ 
\rule{0pt}{1.4em}
{\cal R}(r_{\rm e})=r_{\rm e}^{-q_2}\qquad  \left(r_{\rm e} > r_{\rm b}\right)
\end{array}
\right.
\label{rloc2}
\end{equation}
The $q$ parameter is typically $q \geq 2$ which means that the intrinsic intensity
of the radiation decreases with the distance.
The standard value is $q=3$, larger values may occur only under certain conditions
in the innermost parts of the disc \citep[see e.g.][]{2008MNRAS.386..759N}.

The angular emissivity law, ${\cal M}(\mu_{\rm e},E_{\rm e})$,
defines the distribution of the intrinsic intensity outgoing from each
radius $r_{\rm e}$ of the disc surface with respect to the perpendicular
direction. The limb darkening law in the form 
${\cal M}(\mu_{\rm e},E_{\rm e})= {\cal M}(\mu_{\rm e}) = 1+2.06\,\mu_{\rm e}$
\citep{1960ratr.book.....C,1991ApJ...376...90L} is most frequently
used. However, the choice
is somewhat arbitrary in the sense that the physical assumptions behind
this law are not satisfied at every radius over the entire surface of 
the accretion disc. 
This aspect is studied in detail in Section~\ref{directionality} of the Thesis.

%$\equiv r_{\rm e}^{-\alpha}\,{\cal M}(\mu_{\rm e},r_{\rm e},E_{\rm e})$,

The observed radiation flux from an accretion disc is obtained by integrating the
intrinsic emission over the entire disc surface, from the inner edge ($r=r_{\rm in}$) to the outer
edge ($r=r_{\rm out}$), weighted by the transfer function
$T(r_{\rm e},\varphi_{\rm e},\theta_{\rm o},a)$ determining the impact of relativistic
energy change (Doppler and gravitational) as well as the lensing effect
for a distant observer directed along the inclination angle $\theta_{\rm o}$
\citep[see][]{1975ApJ...202..788C,1989PASJ...41..763A,1992MNRAS.259..569K,2006AN....327..961K}:
\begin{equation}
F_{\rm obs}\left(\theta_{\rm o},a,E_{\rm o}\right)=\int \! T(r_{\rm e},\varphi_{\rm e},\theta_{\rm o},a)\, I(r_{\rm e},\mu_{\rm e},E_{\rm e})\, {\rm d}g\, r_{\rm e}\, {\rm d}r_{\rm e},
\label{flux_cunningham}
\end{equation}
where the index `e' denotes quantities related to the disc and `o' observed quantities.
%specifically from def $E_{\rm o} = g\,E_{\rm e}$.
The integration is carried out over all possible values of the frequency shift $g$
and the whole surface of an accretion disc.

\begin{figure}
\begin{center}
  \includegraphics[width=0.6\textwidth]{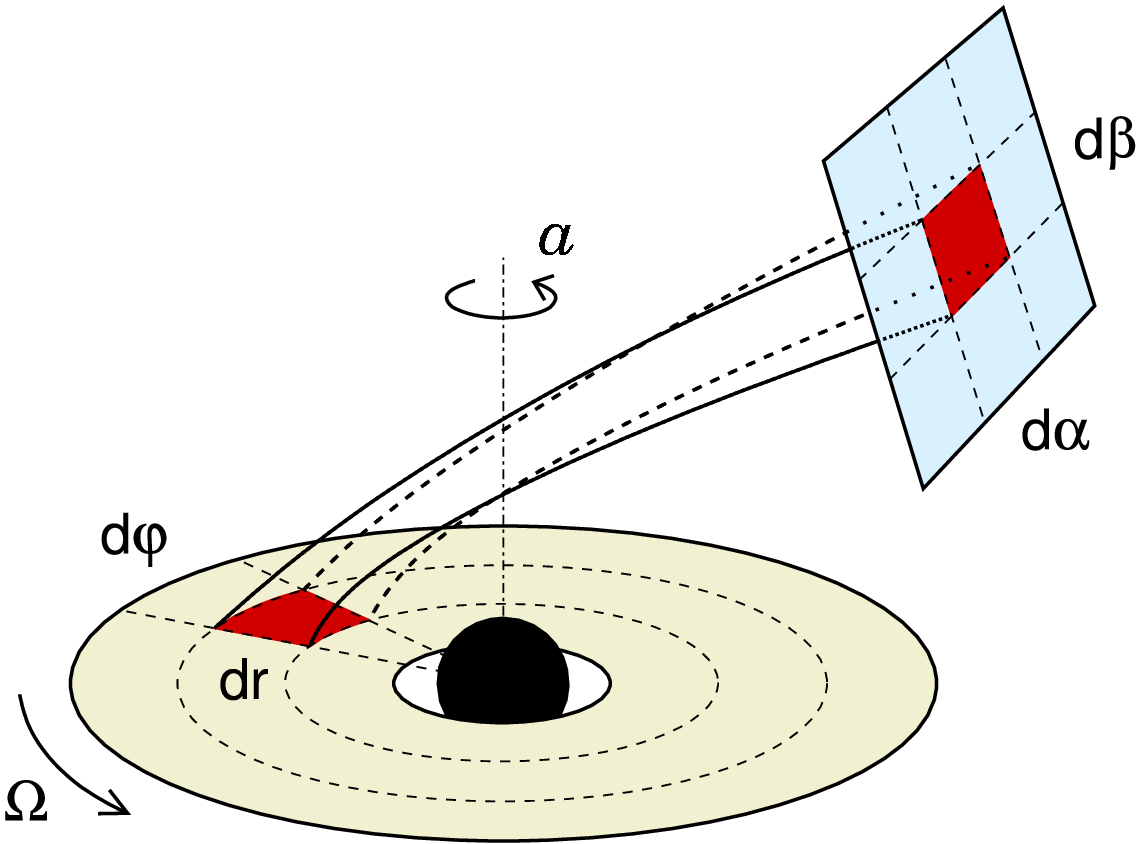}
  \caption{Schematic sketch of the transformation from the Boyer-Lindquist
coordinates of the disc $r$,$\varphi$ to the detector coordinates $\alpha$, $\beta$.
Credit by M.\,Dov\v{c}iak.}
\label{fig_rphi_ab}
\end{center}
\end{figure}

Another approach is employed in \citet{2004ApJS..153..205D}
where the integration is carried out over the coordinates
and $g$-factor is computed at each place.
There are two possibilities how the observed
flux is expressed: 
\begin{enumerate}
 \item in detector coordinates $\alpha$, $\beta$: \\
\begin{equation}
F_{\rm obs}\left(\theta_{\rm o},a,E_{\rm o}\right)=\int \! G_1(r_{\rm e},\varphi_{\rm e},\theta_{\rm o},a)\, I(r_{\rm e},\mu_{\rm e},E_{\rm e})\, {\rm d}\alpha\, {\rm d}\beta,
\label{flux_dovciak_ab}
\end{equation}
 \item in Boyer-Lindquist coordinates of the disc $r$, $\varphi$: \\
\begin{equation}
F_{\rm obs}\left(\theta_{\rm o},a,E_{\rm o}\right)=\int_{0}^{2\pi} \! \int_{r_{\rm in}}^{r_{\rm out}} G_2(r_{\rm e},\varphi_{\rm e},\theta_{\rm o},a)\, I(r_{\rm e},\mu_{\rm e},E_{\rm e})\, r_{\rm e}\,{\rm d}r_{\rm e}\, {\rm d}\varphi_{\rm e},
\label{flux_dovciak_rphi}
\end{equation}
\end{enumerate}
where $G_1(r_{\rm e},\varphi_{\rm e},\theta_{\rm o},a)$ and
$G_2(r_{\rm e},\varphi_{\rm e},\theta_{\rm o},a)$ 
are another transfer functions.
If $I$ is the specific intensity given by eq.~(\ref{intensity_def})
then from eq.~(\ref{Ig3}): $G_1 \equiv g^3$. The relation for $G_2$
may be found from the transformation relation between the $\alpha$, $\beta$ coordinates
and the Boyer-Lindquist coordinates of the disc $r$,$\varphi$ (see Figure~\ref{fig_rphi_ab}):

\begin{equation}
 \frac{{\rm d}\alpha\,{\rm d}\beta}{r\,{\rm d}r\,{\rm d}\varphi} 
= \frac{{\rm d}S_{\rm o}}{{\rm d}S^\perp_{\rm loc}} \times 
  \frac{{\rm d}S^\perp_{\rm loc}}{{\rm d}S_{\rm loc}} \times 
  \frac{{\rm d}S_{\rm loc}}{{\rm d}S} = \frac{\ell\,\mu_{\rm e}}{g},
\label{rel_coord_transform}
\end{equation}
where $S_{\rm o}$ is the observed element surface (i.\,e. in the detector frame),
$S_{\rm loc}$  is the local element surface of the disc, and
$\ell \equiv \frac{{\rm d}S_{\rm o}}{{\rm d}S^\perp_{\rm loc}}$
is the lensing factor defined as the ratio of the cross-section of the flux tube
at the detector to the cross-section of the same flux tube at the disc. 
Using eqs.~(\ref{flux_dovciak_ab})--(\ref{rel_coord_transform}): $G_2 = g^2 \mu_{\rm e} \ell$.

\begin{figure}
  \includegraphics[width=0.49\textwidth]{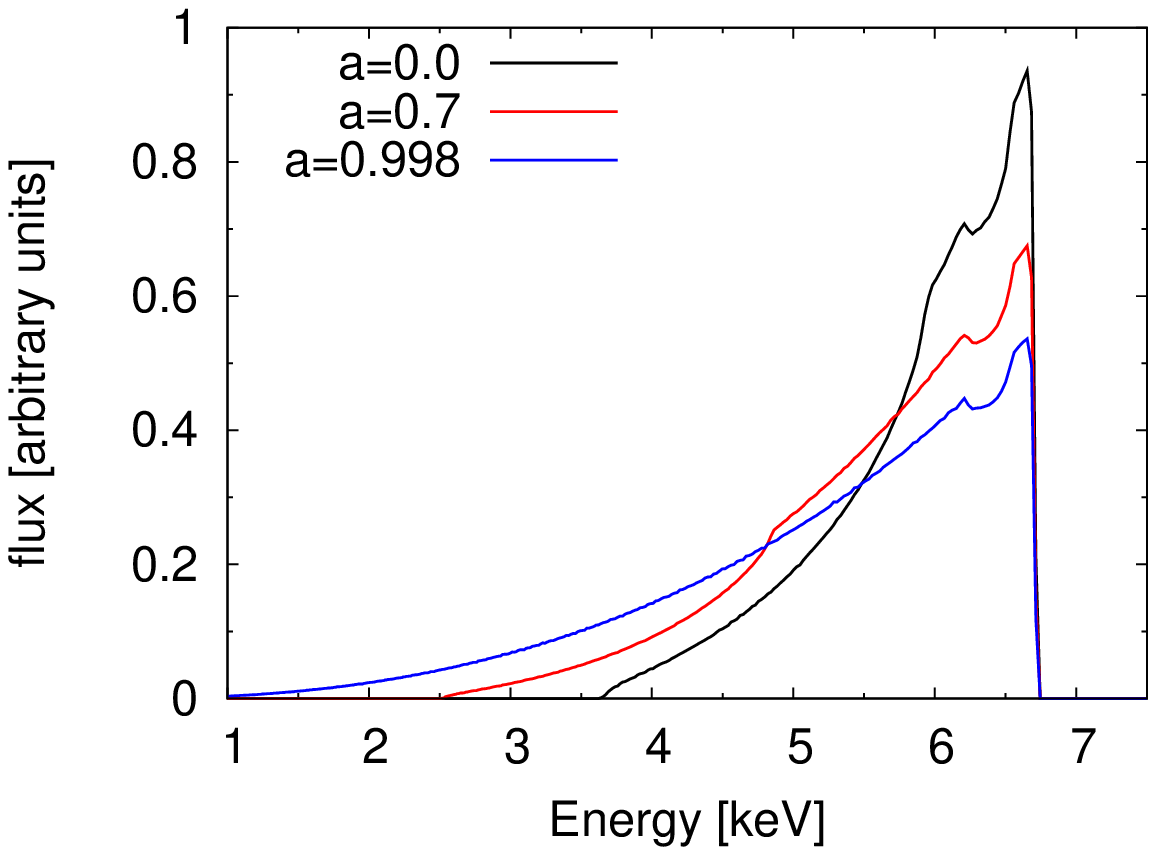}
  \includegraphics[width=0.49\textwidth]{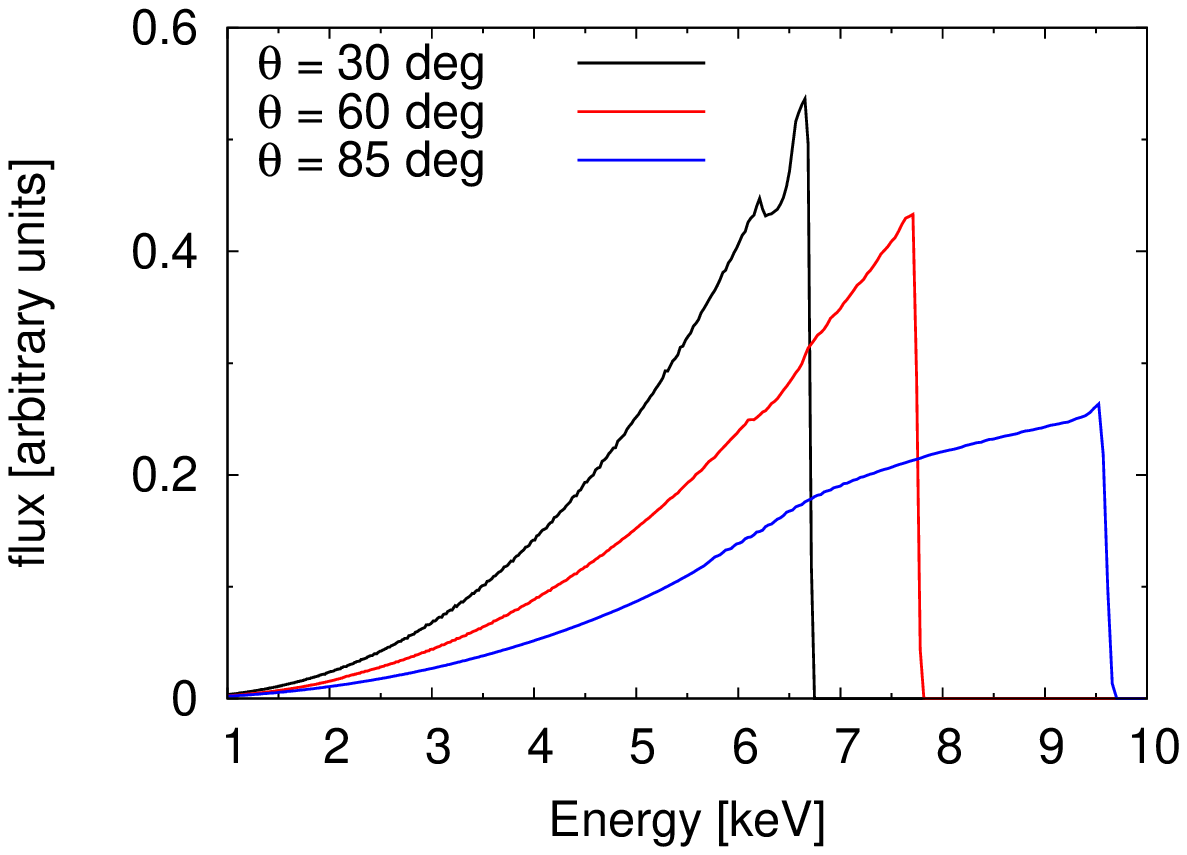}
\caption{Comparison of iron line models produced with the {\kyrline} model
for different set of parameters.
If not further specified the model parameters are $a=0.9982$, $q=3$, $r_{\rm in}=r_{\rm ms}(a)$, 
$r_{\rm out}=400$, $\theta_{\rm o}=30\deg$, $E_0=6.4$~keV
\textposown{Left}: Three curves correspond to the spin $a=0$ (black), $a=0.7$ (red),
and $a=0.998$ (blue), respectively.
\textposown{Right}: Three curves correspond to the inclination $\theta=30$\,deg
(black), $\theta=60$\,deg (red), and $\theta=85$\,deg (blue), respectively.}
\label{compare_ky}
\end{figure}

As a summary, the resulting line profile from an axisymmetric accretion disc is shaped by
these parameters (see eqs.~\ref{iloc} and \ref{flux_dovciak_rphi}):
\begin{enumerate}
  \item inclination angle $\theta_{\rm o}$ 
  \item inner disc radius $r_{\rm in}$
  \item outer disc radius $r_{\rm out}$
  \item radial dependence parameter $q$
  \item angular emissivity law ${\cal M}$
  \item spin $a$
\end{enumerate}

Figure~\ref{compare_ky} shows the theoretical line profiles for different
values of the parameters. The {\kyrline} model \citep{2004ApJS..153..205D}
in the XSPEC fitting package \citep{1996ASPC..101...17A} was used to produce the lines.
The inner disc was fixed to the radius of the marginally stable orbit
$r_{\rm in} = r_{\rm ms}(a)$ which is linked with the spin value by eq.~(\ref{a_rms}).
The line profile has a more extended red wing for higher values of the spin
because the inner edge of the disc shifts to lower radii where more
extreme gravitational redshift occurs.
The line profile is broader for higher values of the inclination angle
because the line-of-sight velocities are higher accordingly
implying a stronger Doppler effect. The stronger Doppler effect
overreaches the gravitational redshift in the region where
the matter moves towards the observer. The position of the maximal
blueshift is closer to the black hole proportionally to the strength
of the Doppler shift and thus, to the inclination angle (see Figure~\ref{fig2}).
The peak of the line is shifted to
higher energies for higher inclinations.
All the lines shown in the figure are normalised to the same flux.

\section{Observational evidence of relativistic iron lines}
\label{rel_lines_observed}

Broad iron lines have been detected in several X-ray spectra of black hole binaries
as well as active galaxies. The first detection of a moderately broad iron line
is attributed to the EXOSAT observations of Cygnus~X-1 \citep{1985MNRAS.216P..65B}.
The redshift and broadening of the line was explained by \citet{1989MNRAS.238..729F}
as a result of relativistic 
smearing. They constructed {\textsc{diskline}} %model
model\footnote{This model is embedded in the 
XSPEC fitting package and may still be used for spectral analysis
of non-rotating or slowly rotating black holes or neutron stars. However,
be aware that the effect of lensing is not included in this model.}
for the relativistic line around a non-rotating black hole. 
A model of the relativistic
line for a maximally rotating Kerr black hole 
was developed independently by \citet{1991MNRAS.250..629K} 
and \citet{1991ApJ...376...90L}.
A strongly asymmetric and redshifted line profile was predicted
for radiation coming from an inner accretion disc around a black hole.
This indicated that the interesting insights on the geometrical properties
of accretion discs in the closest vicinity of a black hole
may be derived from the shape of the line.

The first X-ray spectrometer capable for resolving the shape of the line was
SIS detector (Solid-state Imaging Spectrometer) on-board the ASCA satellite
(Advanced Satellite for Cosmology and Astrophysics).
The first resolved broad iron line was detected 
in the spectrum of a Seyfert galaxy MCG\,-6-30-15
by \citet{1995Natur.375..659T}. This source exhibited an extremely broad iron
line in X-ray spectra of all succeeding satellites (see Section~\ref{mcg}).
The one of the most suitable and still operating missions 
for investigation of broad iron lines is the XMM-Newton satellite 
thanks to its large effective collecting area in the iron line band 
and also a few keV above (see Section~\ref{xmm_newton}). 
Recently, the Suzaku satellite (launched on 10 July 2005)
exceeds XMM-Newton with its 
broad-band coverage (0.4-600\,keV) which allows better constraining of
the continuum (both, primary and reflection).

\begin{figure}[tbh!]
\begin{center}
\begin{minipage}{0.5\textwidth}
 \includegraphics[width=1.\textwidth]{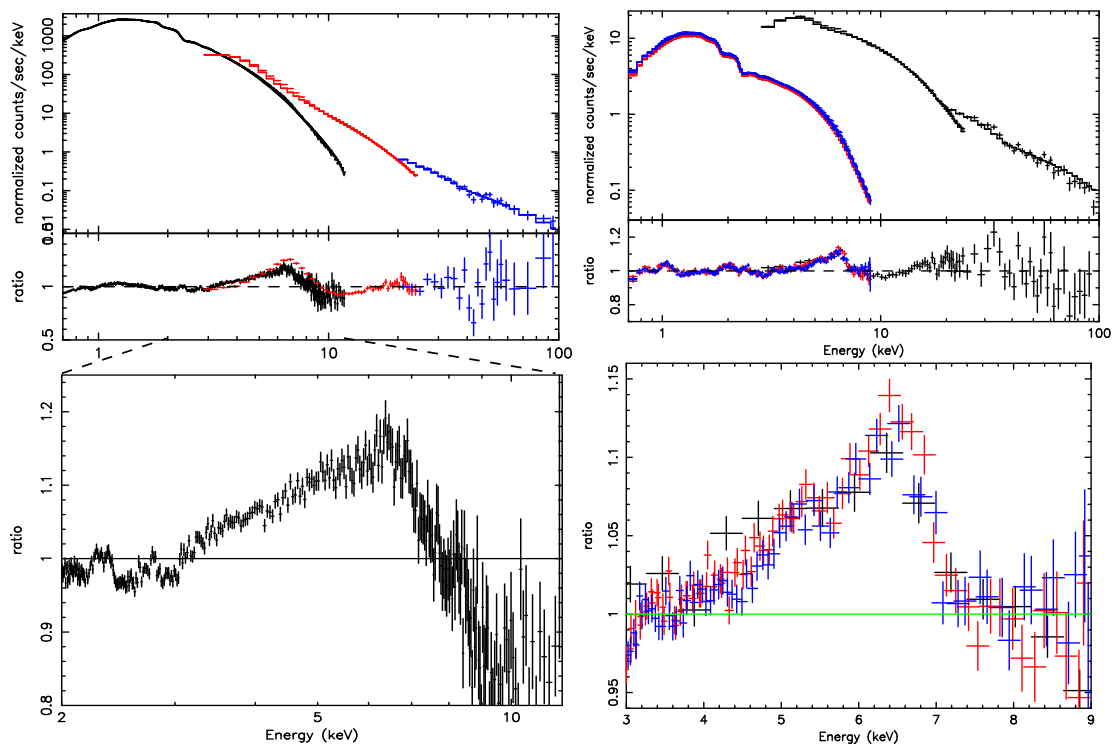}
\end{minipage}
\begin{minipage}{0.49\textwidth}
  \includegraphics[angle=270,width=0.96\textwidth]{mil06_f4.eps} \\
  \includegraphics[angle=270,width=0.93\textwidth]{f5.eps}
\end{minipage}
\caption{\textposown{Left:} X-ray spectrum of GX\,339-4 observed in the high/soft state
by XMM-Newton/EPN (black), RXTE/PCA (red) and HEXTE (blue) \citep{2004ApJ...606L.131M}.
The spectrum is modelled by a multi-colour temperature disk black-body 
and a power law revealing a broad asymmetric iron line. 
Fe K$\alpha$ line is shown in detail in the bottom panel.
\textposown{Right:} X-ray spectrum of GX\,339-4 in the low/hard state  
\citep{2006ApJ...653..525M}. The XMM-Newton data observed during 
the revolution No.\,782 are shown in blue, No.\,783 in red 
and the data obtained by RXTE are black.
Fe K$\alpha$ line is shown in detail in the bottom panel.}
\label{fig_miller}
\end{center}
\end{figure}

\begin{figure}[tbh!]
\begin{center}
  \includegraphics[width=0.5\textwidth,angle=-90]{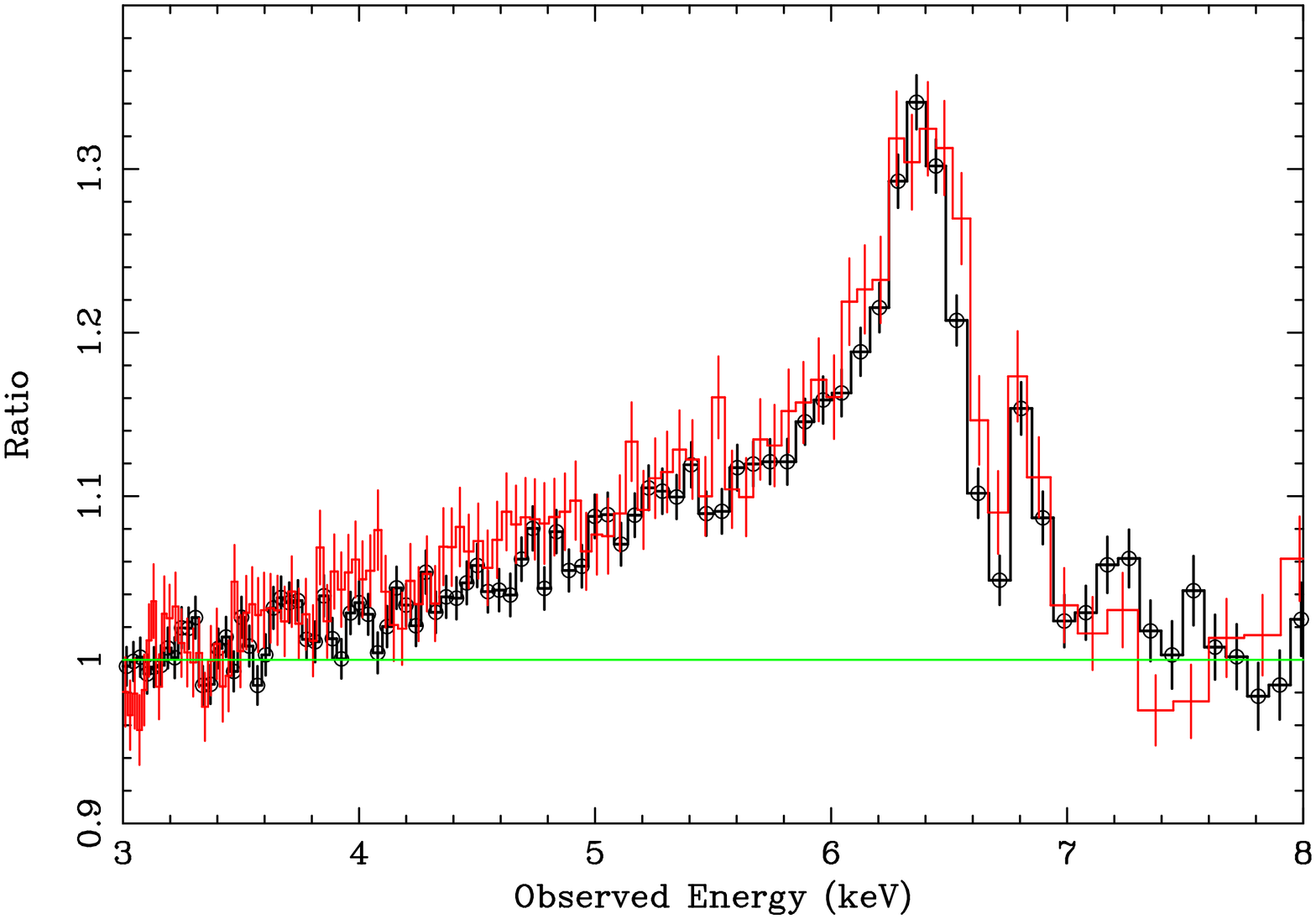}
  \caption{Relativistic iron line in MCG\,-6-30-15 observed by
XMM-Newton (red) and Suzaku (black).
Figure adopted from \citet{2007PASJ...59S.315M}.}
  \label{miniutti_mcg}
\end{center}
\end{figure}

Broad iron lines shown in Figure~\ref{fig_miller} for the case of 
a black hole binary GX\,339-4,
and in Figure~\ref{miniutti_mcg} for the case of a Seyfert~1 galaxy MCG\,-6-30-15
belong to the best examples of the extremely relativistically broadened lines.
We re-analysed the XMM-Newton observations of these sources.
For a brief description of both sources and details of our re-analysis,
see Sections~\ref{gx339} and \ref{mcg}.
Other examples of active galaxies with broad iron lines may be found in 
e.g. in \citet{1997ApJ...477..602N, 2003A&A...401..903G, 2005A&A...432..395S,
2006AN....327.1079R, 2006A&A...445...59T, 2006A&A...453..839P, 
2007MNRAS.382..194N, 2008MmSAI..79..259L, 2008PASJ...60S.277M, 
2009ApJ...691..922M, 2009ApJ...702.1367B, 2009Natur.459..540F, 2010ApJ...713.1256S}.
A statistical distribution of AGN with broad iron lines was studied
by \citet{2006AN....327.1032G} who estimated a fraction of $42 \pm 12\,\%$
of well exposed AGNs that exhibit a relativistically broadened iron line.
Works by \citet{2008MmSAI..79..259L} and de la Calle {\textit{et al.}}
(2009, submitted to {\textit{A}\&\textit{A}}) represent the continuation of this effort. 
Different samples of Seyfert galaxies observed by the XMM-Newton satellite
were studied by \citet{2007MNRAS.382..194N} and \citet{2009ApJ...702.1367B}.
Both groups concluded that the most X-ray spectra of their samples possess a relativistically 
broadened iron line.
Other examples of black hole binaries may be found in \citet{2007A&A...462..657D,
2009MNRAS.394.2080H, 2009ApJ...697..900M}.
Relativistically broadened iron lines were also detected in several X-ray spectra
of neutron stars \citep[see][and references therein]{2010AAS...21536002C}.

%\section{Relativistic line models}

\section{Comparison between the {\laor} and {\ky} models}
\label{laky}

There are several numerical codes for the relativistic disc spectral 
line, some of them were already mentioned.
The most widely used model, over almost two decades, 
has been the one by \citet{1991ApJ...376...90L},
which includes the effects of a maximally rotating Kerr black hole.
In other words, the {\laor} model sets the dimensionless angular momentum $a$ 
to the canonical value of $a=0.9982$ -- so that it cannot be subject of the data fitting procedure. 
\citet{2004ApJS..153..205D} have relaxed this limitation and allowed $a$ 
to be fitted in the suite of {\textscown{ky}} models. 

Other numerical codes have been 
developed independently by several groups \citep{2000MNRAS.312..817M,
1993A&A...272..355V,2004AdSpR..34.2544Z,2004A&A...424..733F,
2004MNRAS.352..353B, 2005MNRAS.363..177C, 2006ApJ...652.1028B}
using different techniques.
The last three equipped their codes with similar functionality
as KY to be used in X-ray spectra modelling.
\citet{2006ApJ...652.1028B} performed useful tests
demonstrating that {\textscown{ky}} and their {\textscown{kerrdisk}} model 
give compatible results
when they are set to equivalent parameter values. 

Although the {\laor} model does not have the spin value as a variable parameter,
it can still be used for evaluation of the spin if one identifies the inner 
edge of the disc with the marginally stable orbit (eq.~\ref{rms}, see also Figure~\ref{a_rms}).
The evaluation of the spin from extension of the red wing only is, however, 
not precise because the spin also modulates the value of the $g$ factor and thus, it affects 
the overall shape of the line by itself. 
This systematic error of spin measurements obtained with the {\laor} model
has not yet been constrained. Further, we compare 
the spin estimation by the {\laor} and {\kyrline} model
to constrain this error.

\begin{figure}[tbh!]
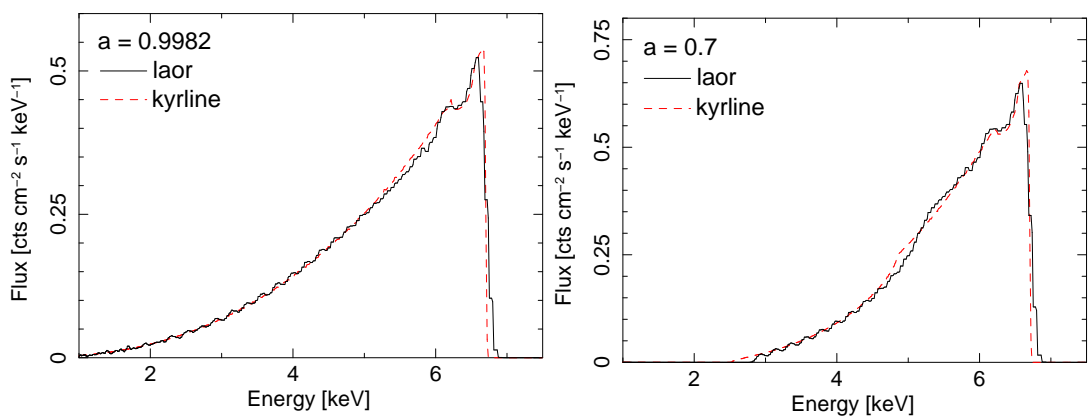

\begin{center}
%\begin{tabular}{ccc}
%\epsfbox{fm_line_ratio.eps}
\includegraphics[angle=270,width=0.475\textwidth]{compare_laorky.eps}
\includegraphics[angle=270,width=0.475\textwidth]{compare_laorky_a07.eps} 
%\end{tabular}
\end{center}
\caption{Comparison of the {\laor} (black, solid) and {\kyrline} (red, dashed)
model for two values of the spin $a = 0.9982$ (left) and $a = 0.7$ (right). 
The other parameters of the line are $E=6.4$\,keV, $q=3$, $i=30^{\circ}$.}
\label{laky_compare}
\end{figure}

The theoretical line profiles of the 
{\laor} and {\kyrline} models are compared in Figure~\ref{laky_compare}
for two values of the spin, $a=0.9982$ (left panel) and $a=0.7$ (right panel).
The inner radius of the disc is set to coincide with the marginally stable orbit
for a given value of the spin, i.e. $r_{\rm in} \doteq 1.235$
and $r_{\rm in} \doteq 3.39$, respectively.
The angular dependence of the emissivity in this example is
given by the limb darkening law ($I\left(\mu_{\rm e}\right)\propto 1 + 2.06\,\mu_{\rm e}$)
which is the only prescription for the directionality in the {\laor} model.
The {\kyrline} model enables to switch between different emission laws. 
Further in this Section, we used two distinct cases and this name convention:
{\kyrline} is equipped with the same limb-darkening law as in the {\laor} model,
and {\kyrline}* uses the limb-brightening law in the form
$I(\mu_{\rm e}) \propto \ln (1+\frac{1}{\cos\theta_{\rm e}})$ \citep{1993ApJ...413..680H}.

In the first comparison (left panel of Fig.~\ref{laky_compare}), 
the difference between the two models 
is noticeable only in the high energy edge of the line profile.
This is caused by poorer energy resolution of the {\laor} model.
In the second comparison (right panel of Fig.~\ref{laky_compare}), an additional discrepancy appears
in the slope of the line profile (look at energies 4.5--6\,keV).
This is the effect of the different frame-dragging taken into account
and it is caused by the fact
that the {\laor} model is calculated only for $a=0.9982$.

Relatively to the total line profile, the difference between the {\laor} 
and {\kyrline} model seems to be rather small. Nevertheless,
we further compare both models applied to the real data with these goals:
\begin{enumerate}
 \item  to investigate the importance of the precise value of spin in the line profile 
formation besides the relation between spin and the marginally stable orbit,
 \item to constrain the uncertainty in the spin determination by the less 
precise {\laor} model.
\end{enumerate}

%(\markcite{{\it Fabian et al.} [2002], {\it Vaughan and Fabian} [2004]} for MCG-6-30-15, and .\markcite{{\it Miller et al.} [2004], {\it Miller et al.} [2006], {\it Reis et al.} [2008]} for GX 339-4). 
%We choose the observations with the highest exposure times for the sources publicly available in the archive.% of the XMM-Newton.
%We reduced the EPIC pn data using the SAS software version 7.1.2 and the most recent calibration files.
%The details of the analysed observations are described in the next section. In the following sections, we present the results
%of our spectral re-analyses and we compare the two relativistic line models on the current and the future simulated data. 
%The analysis on the simulated data for  mission is presented in Section~\ref{future}.
%We did a detailed spectral re-analysis. 

\subsection{Analysis with current data provided by XMM-Newton satellite}

%The XMM-Newton satellite belongs to the instruments with the most
%sensitive current X-ray detectors onboard. ??

For the comparison of the {\laor} and {\kyrline} model
with current data, 
we chose two sources, MCG\,-6-30-15 and GX\,339-4, 
which exhibited extremely broad iron lines 
(see Figures~\ref{fig_miller} and \ref{miniutti_mcg}).
We re-analysed the observations performed by the XMM-Newton satellite
(see Section~\ref{xmm_newton}).

%\subsection{MCG-6-30-15}

%\subsection{GX 339-4}

% and it was not used in the previous analyses. 
%This different approach leads to a significant decrease of the total number of bins and to better statistics -- more independent on the instrument properties. Consequently, the previous fits of GX 339-4 spectrum were not satisfying any more. We found a different fit,
%in which the line strength becomes much weaker. However, the spin value $a \approx 0.7$ enabled us to compare the {\it laor} model with the {\it kyrline} model for an intermediate value of the spin.

%\begin{figure}[tbh!]
%\begin{center}
%\begin{tabular}{c}
%  \epsfxsize=102mm
%  \epsfysize=100mm
%\epsfbox{a_rms2.eps}
%\end{tabular}
%\end{center}
%\caption{Relation between the spin value $a$ and the radius of the marginal stable orbit $r_{\rm ms}$}
%\label{rms}
%\end{figure}

\subsubsection{MCG-6-30-15}

The galaxy MCG\,-6-30-15 is a nearby Seyfert 1 galaxy ($z = 0.008$). 
The skewed iron line has been revealed in the X-ray spectra 
by all recent satellites \citep{1999A&A...341L..27G,2001MNRAS.328L..27W,
2002MNRAS.335L...1F,2002ApJ...570L..47L,2007PASJ...59S.315M}. 
The XMM-Newton observed 
MCG\,-6-30-15 for a long 350\,ks exposure time during summer 2001 
(revolutions 301, 302, 303). We joined
the three spectra into one for further analysis. Details
of the data reduction and spectral re-analysis are described in Section~\ref{mcg}.

We used the same continuum model for the MCG\,-6-30-15 spectrum as presented 
in \citet{2002MNRAS.335L...1F}: the simple power law component absorbed by 
Galactic gas
matter along the line of sight with column density
$n_{\rm H} = 0.41\times 10^{21}$\,cm$^{-2}$.
The employed model is sufficient to fit the 
data above $\approx 2.5$\,keV, which is satisfactory for our goal of the 
comparison of the relativistic line models.
The value of the photon index is $\Gamma =1.90(1)$. The residuals are 
formed by a complex of a broad iron line and two narrow iron lines --
one emission line at $E=6.4$\,keV likely originating in a distant matter 
(torus) and one absorption line at $E = 6.77$\,keV which
can be explained by a blueshifted absorption originating in an outflow.
The rest energy of the broad line is $E=6.7$\,keV, which corresponds to 
the helium-like ionised iron atoms. The spectral complexity in 
the line band allows an alternative explanation -- the model with two 
narrow emission lines at energies $E=6.4$\,keV and $E=6.97$\,keV. 
This alternative model leads to the presence of the broad line component at $E=6.4$\,keV.
See Section~\ref{mcg} for details.
%Both alternatives lead to the presence of the broad line component.
%, but subsisting at different rest energies. The rest energy is $E=6.7$keV for the model with narrow absorption line and $E=6.4$keV for the model with two emission lines. We choose the first model with the absorption feature to the next analysis. The studied broad iron line is shown in the left in Figure~\ref{ratios} and the results of the fits in Table~1.

\begin{table}[tbh!]
\begin{center}
\caption{Results of iron line models for MCG\,-6-30-15 in 2.5--9.5\,keV.}
	\vspace{0.1cm}
\begin{tabular}{c|c|c|c|c}
% 	\multicolumn{5}{c}{\bf Table 1. Results for MCG-6-30-15 in 2.5--9.5\,keV}\\
	\hline \hline
	parameter   &	{\kyrline}&	{\kyrline}*	& {\laor}\,$_{\rm best}$	 & {\laor}\,$_{\rm loc.min.}$\\
	\hline
%	\rule{0cm}{0.35cm}
\rule{0pt}{1.5em}	$a/M$		&	$\textbf{0.94}^{+0.02}_{-0.03}$&	$\textbf{0.95}^{+0.02}_{-0.01}$	&	$\textbf{0.98}^{+0.02}_{-0.01}$	&	$\textbf{0.96}^{+0.02}_{-0.01}$	\\
\rule{0pt}{1.5em}	$i$\,[deg]	&	$26.7(7)$		&	$31.5(7)$		&	$35.7(5)$		&	$26.8(5)$	\\
\rule{0pt}{1.5em}	$E$\,[keV]	&	$6.67(1)$		&	$6.60(1)$		&	$6.48(1)$		&	$6.66(1)$		\\
\rule{0pt}{1.5em}	$q_{1}$		&	$4.9(1)$		&	$3.7(1)$		&	$4.8(1)$		&	$4.7(1)$		\\
\rule{0pt}{1.5em}	$q_{2}$		&	$2.84(4)$		&	$2.11(4)$		&	$2.50(3)$		&	$2.87(3)$		\\
\rule{0pt}{1.5em}	$r_{b}$ 	&	$5.5(2)$		& 	$18.3(5)$		& 	$6.6(2)$		& 	$5.1(2)$		\\
	\hline
\rule{0pt}{1.5em}	$\chi^{2}/v$	&	$175/148$	&	$174/148$ 	&	$170/148$ 	&	$174/148$ 	\\[2pt]
\hline
%	$EW$\,[eV]	&	$761$		&	$757$	  	&	$764$	  	&	$754$	  	\\
%	\multicolumn{4}{c}{Notes: The {\laor} value for the spin was calculated from the value of the inner disc radius assuming the equality with the marginally stable orbit $r_{\rm ms}$. The single-digit errors indicate the error in the last significant digit for the given parameter value. The spin values are constrained with the upper and lower limits. The errors were calculated when other parameters were frozen.}
%\label{mcg_fit}
\end{tabular}
\label{laky_mcg_table}
\end{center}

{\small Notes: The errors in brackets are related to the last significant digit of the number,
and they correspond to 90$\%$ confidence level calculated 
while the other model parameters are fixed.}
\end{table}

%\caption{asfdf}
%\label{gx_fit}
%\end{center}
%\end{table}
%\end{figure}

\begin{figure}[tbh!]
\begin{center}
\begin{tabular}{ccc}
  \includegraphics[angle=270,width=0.31\textwidth]{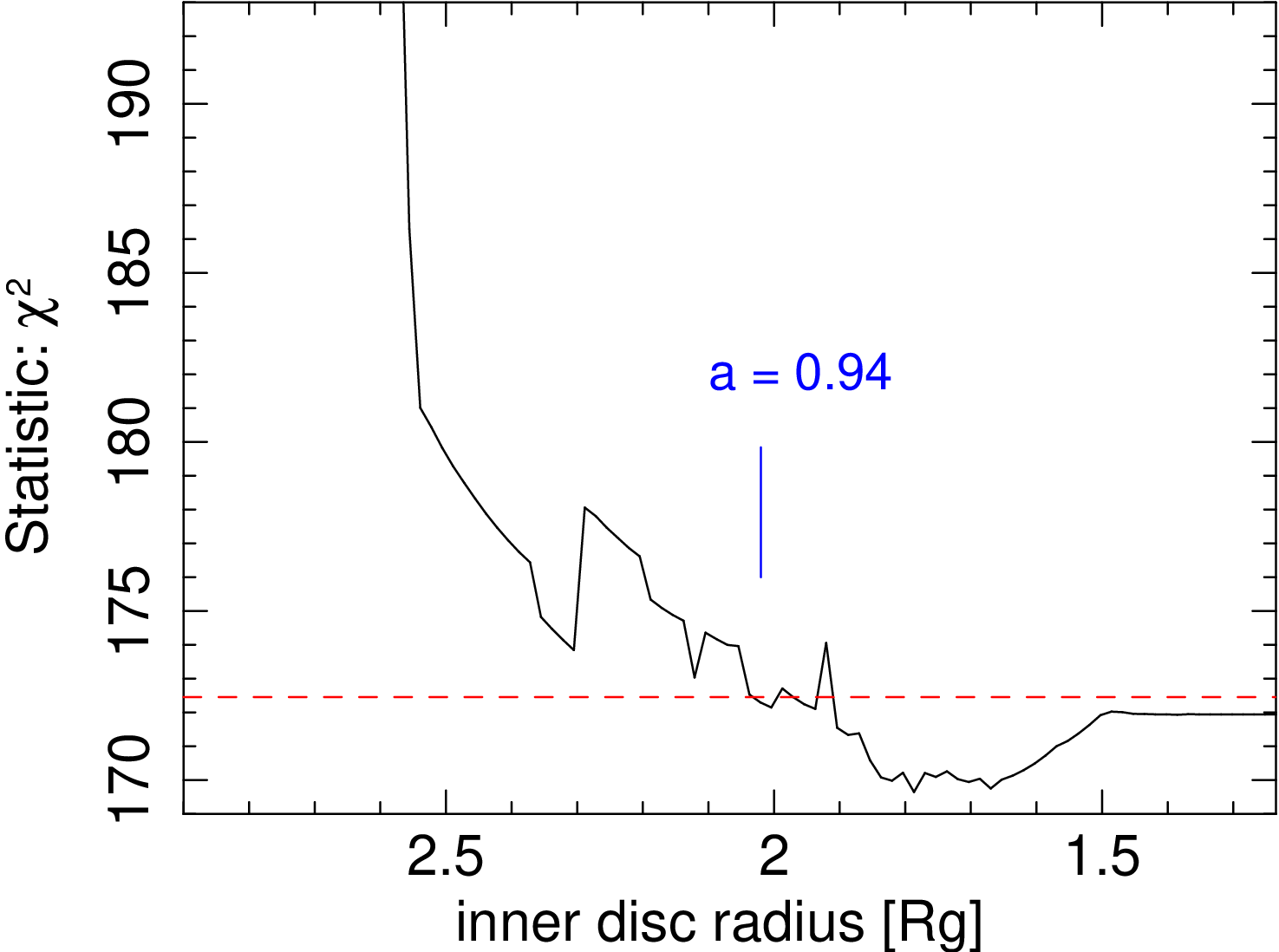} &
  \includegraphics[angle=270,width=0.31\textwidth]{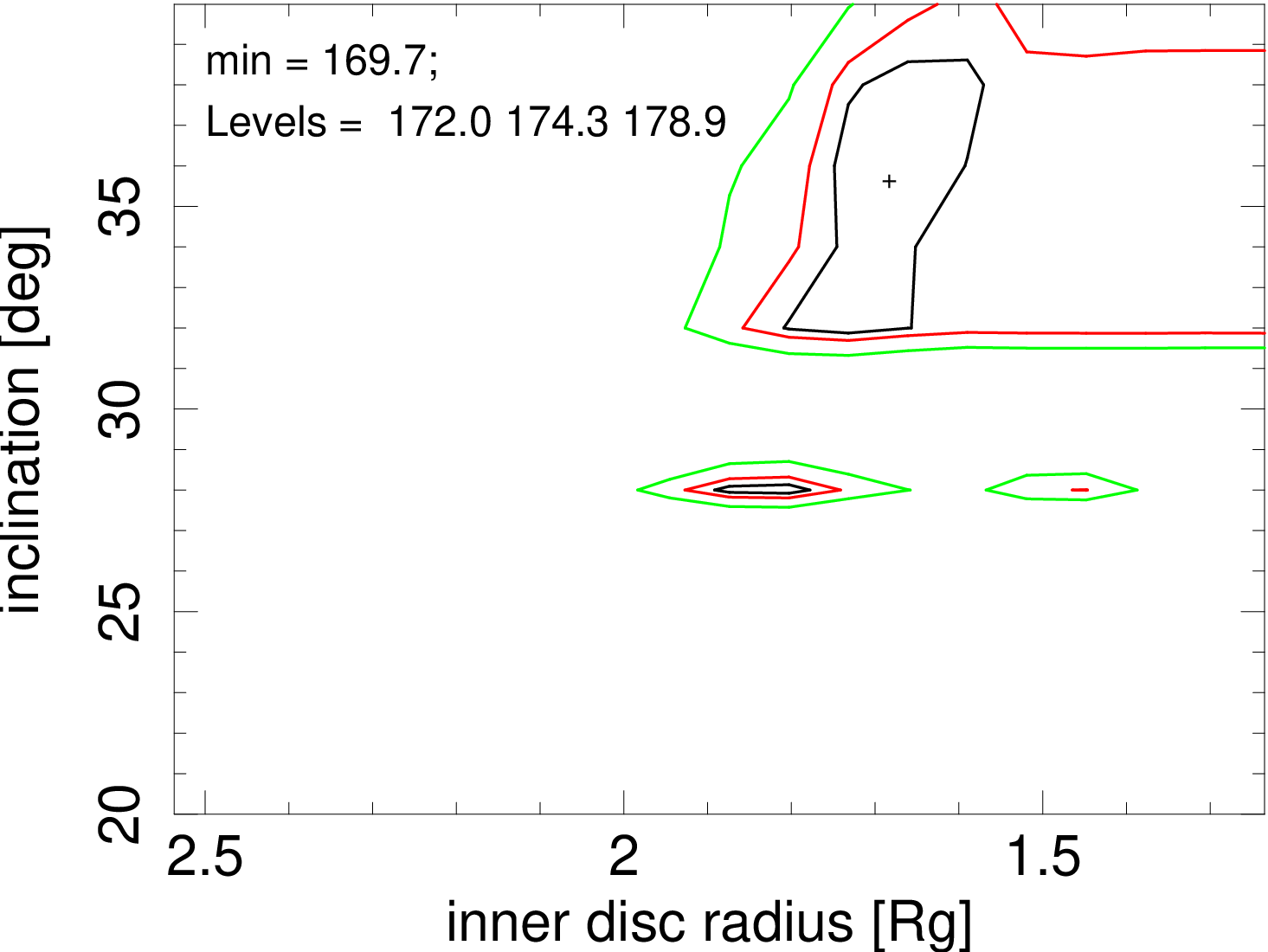} &
  \includegraphics[angle=270,width=0.31\textwidth]{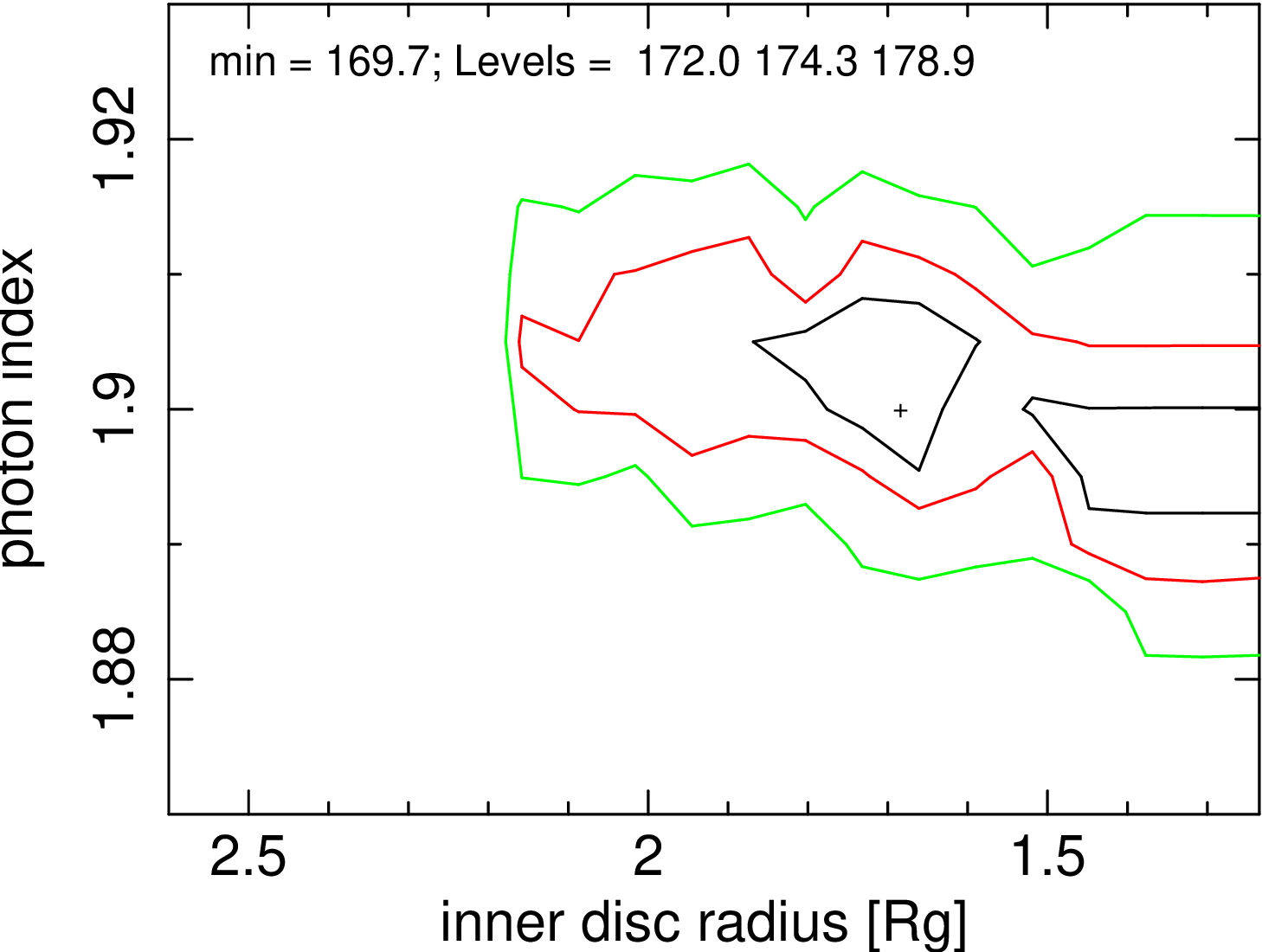}\\
  \includegraphics[angle=270,width=0.31\textwidth]{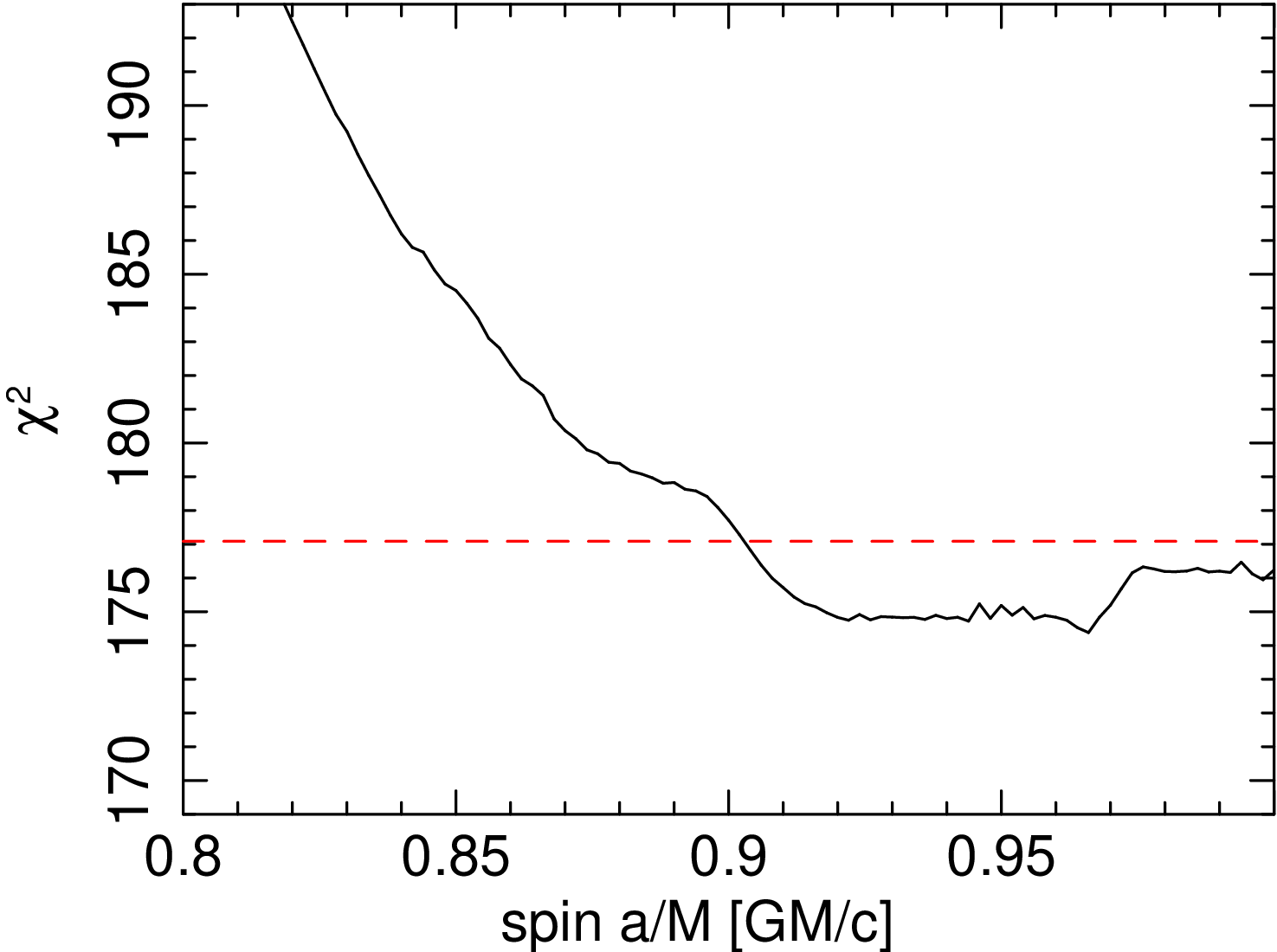} &
  \includegraphics[angle=270,width=0.31\textwidth]{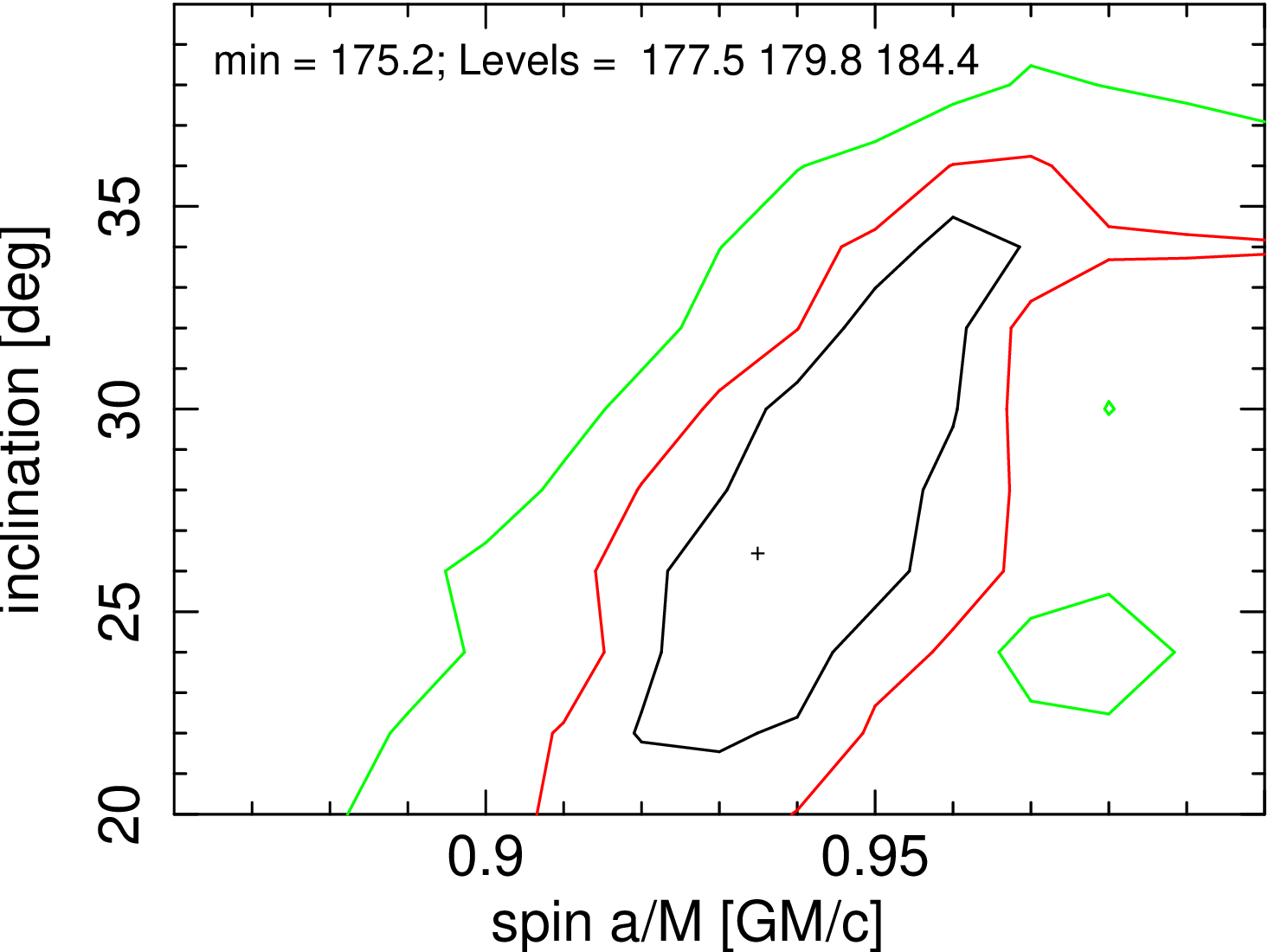} &
  \includegraphics[angle=270,width=0.31\textwidth]{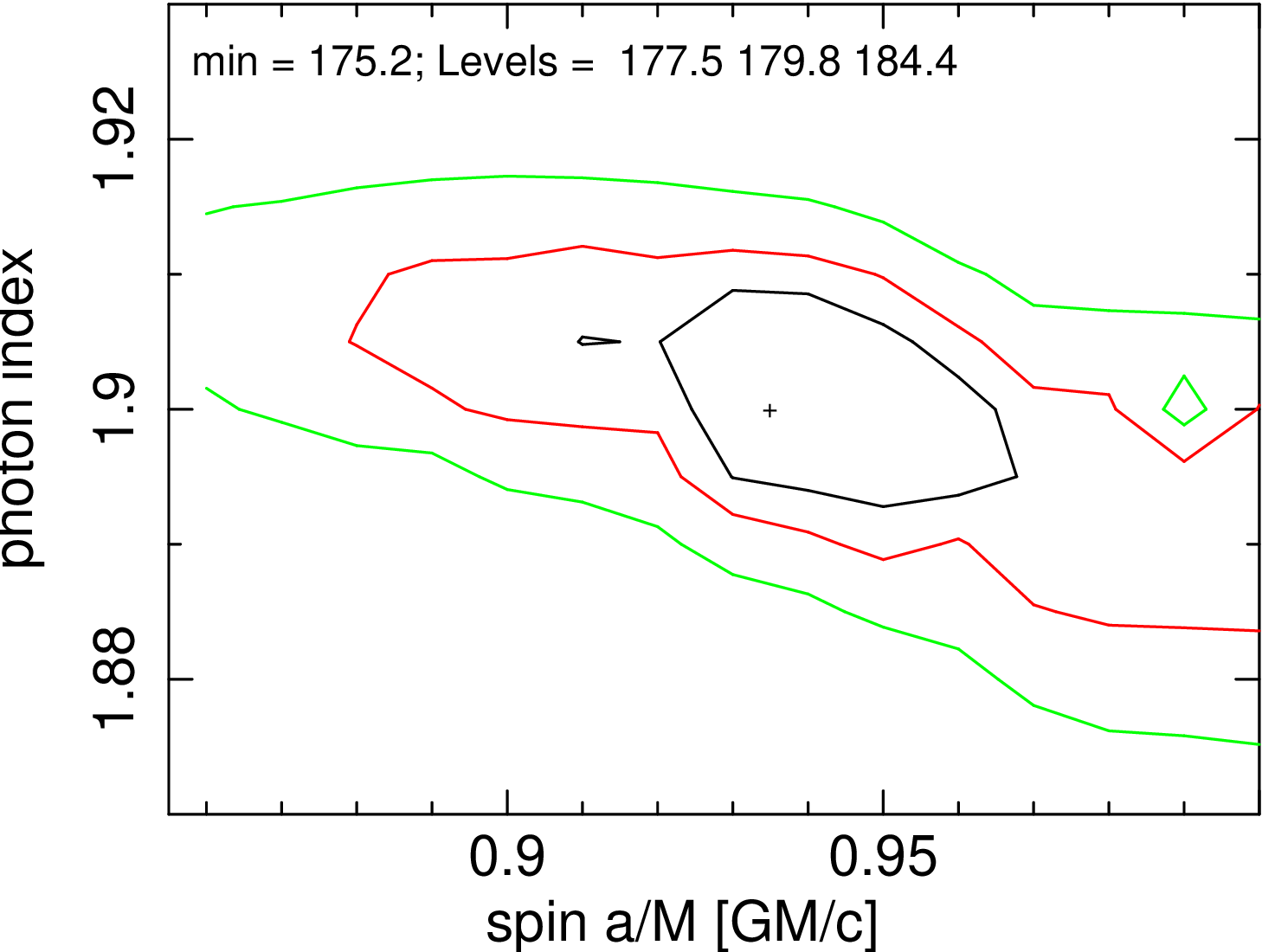}\\
  \includegraphics[angle=270,width=0.31\textwidth]{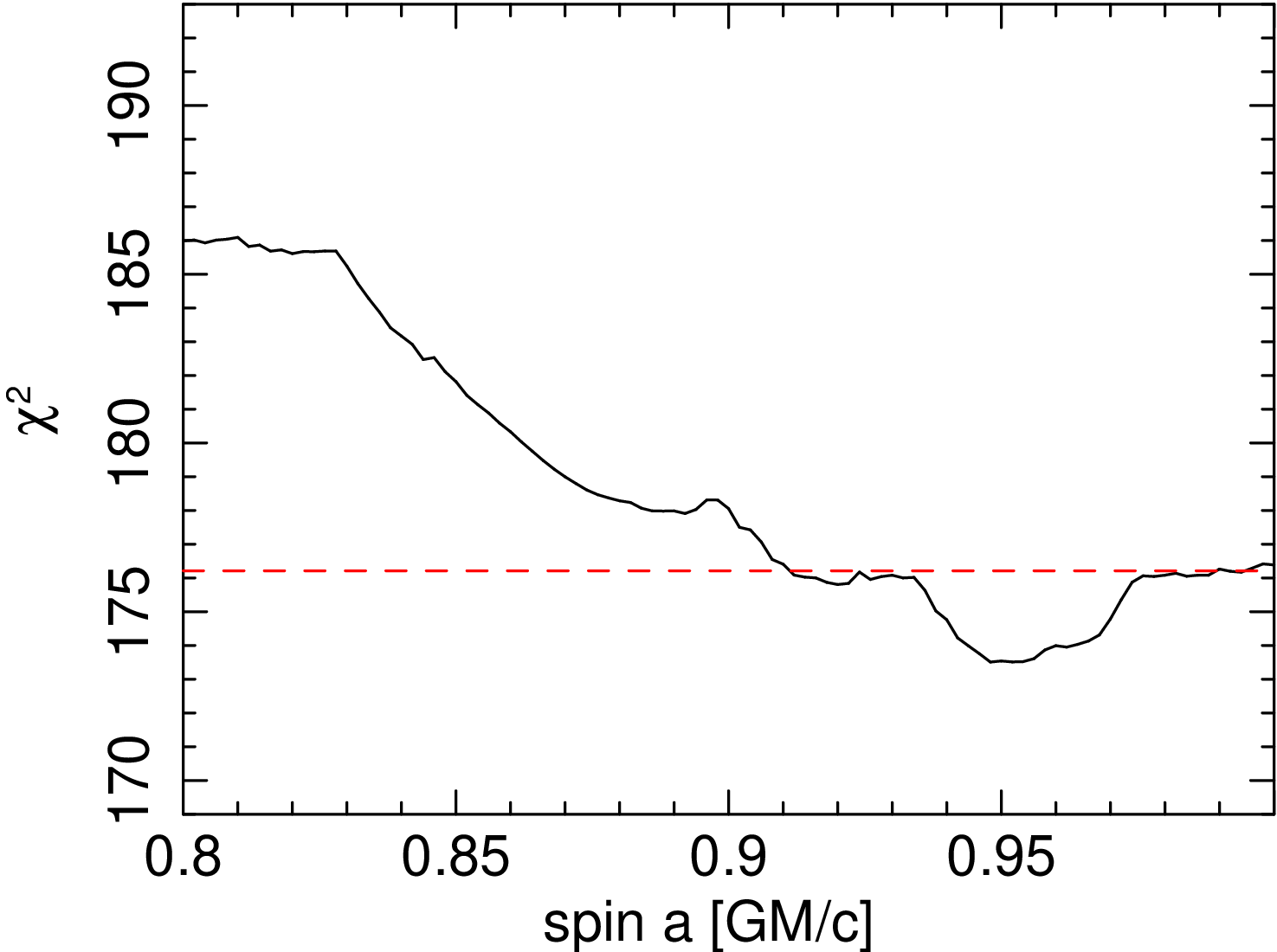} &
  \includegraphics[angle=270,width=0.31\textwidth]{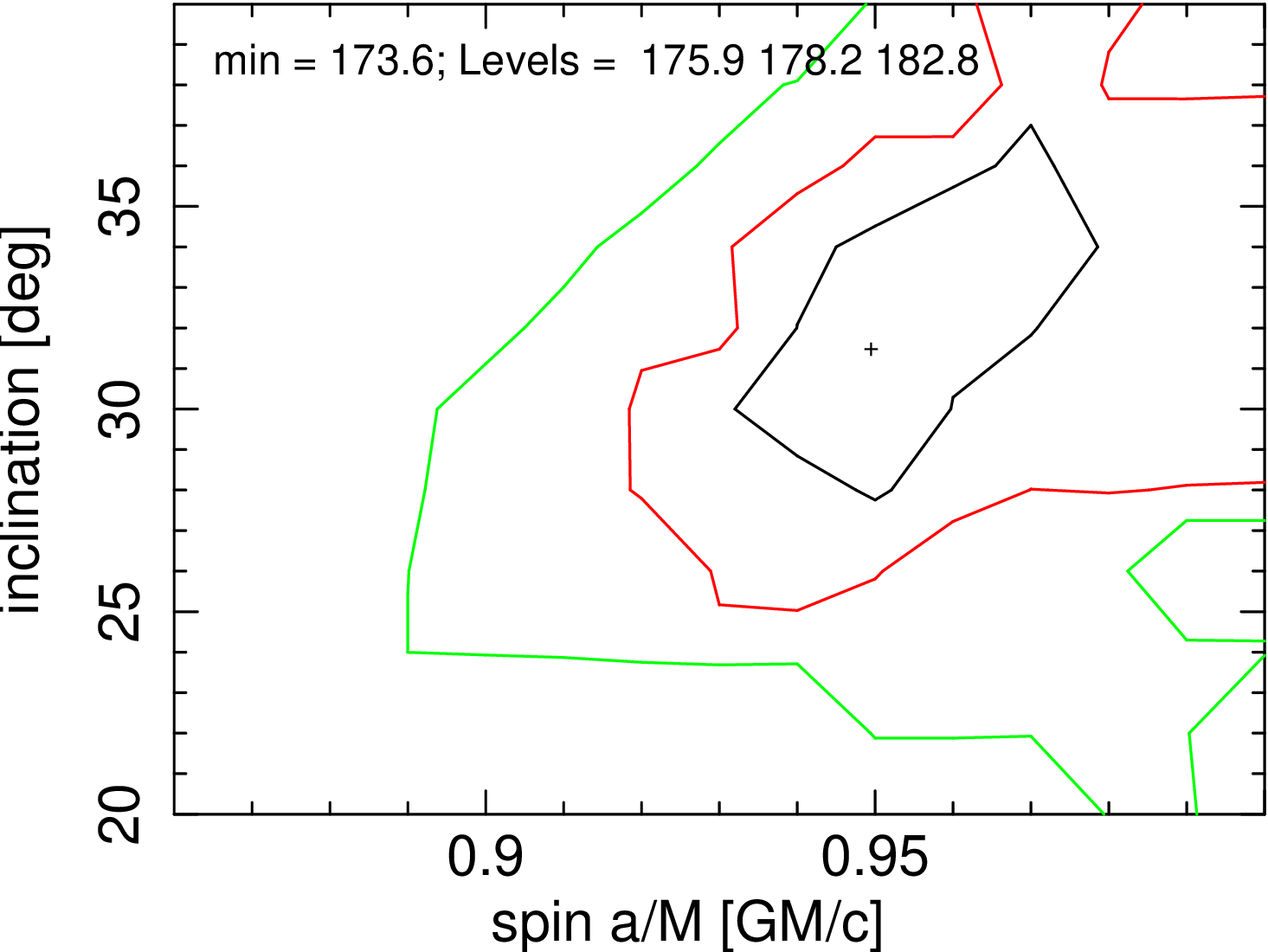} &
  \includegraphics[angle=270,width=0.31\textwidth]{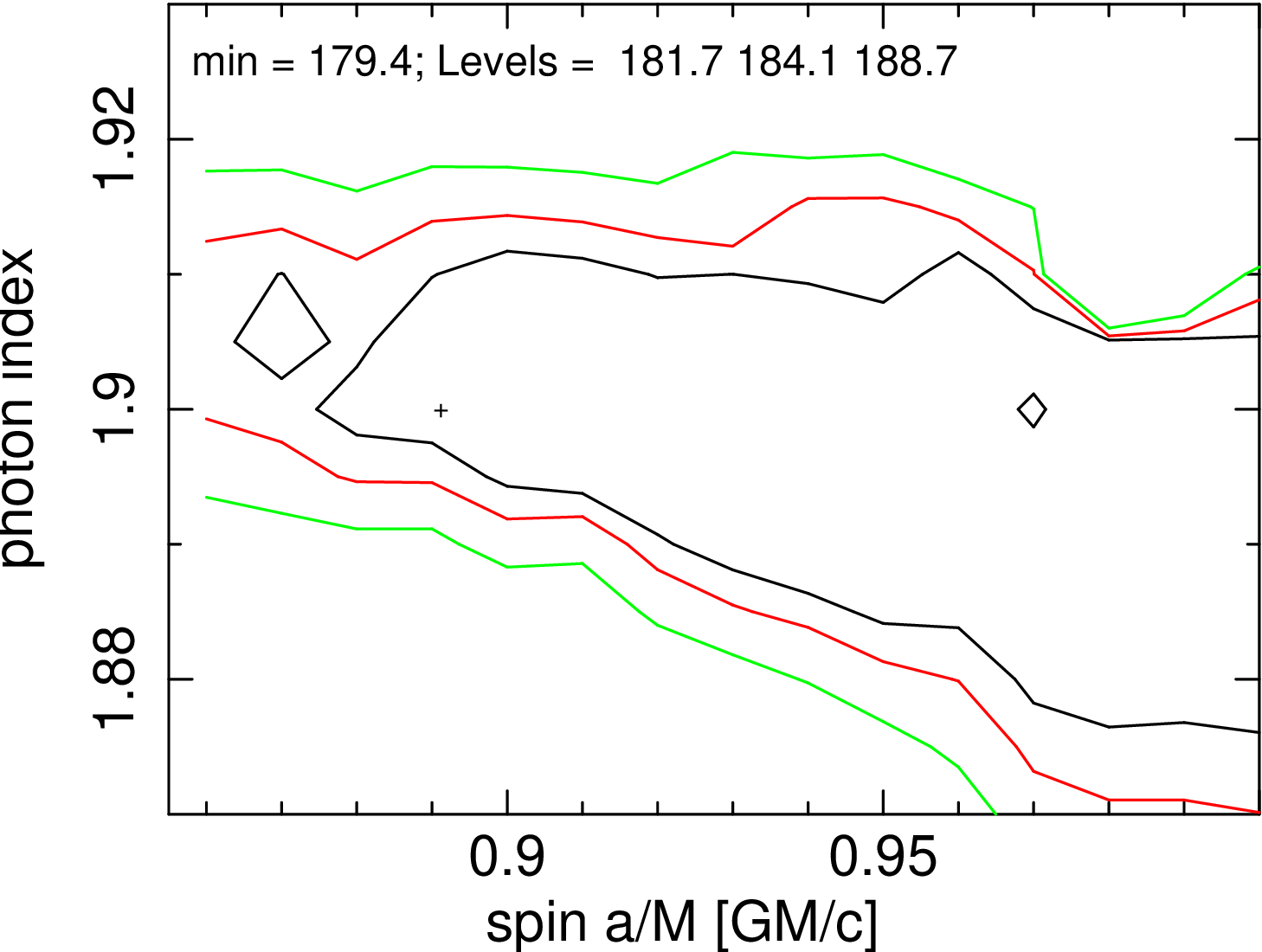}\\
\end{tabular}
\caption{The contour graphs show (from \textposown{left} to \textposown{right}) 
the dependence of the $\chi ^{2}$ value, 
the inclination angle and the power law index on the value of the spin ({\kyrline}) 
or the inner disc radius ({\laor}) for the MCG\,-6-30-15 spectrum in 2.5--9.5\,keV.
The x-axis is oppositely directed in the case of the inner disc radius 
of the {\laor} model as x-variable for an easier comparison with the {\kyrline} results. 
The black, red and green contours correspond to $1\sigma$, $2\sigma$ and $3\sigma$, 
respectively. \textposown{Top}: The results of the {\laor} model.
\textposown{Middle}: The results of the {\kyrline} model with limb darkening. 
\textposown{Bottom}: The results of the {\kyrline}* model with limb brightening.}
\label{mcg_contours}
\end{center}
\end{figure}

A good fit of the broad line was found with a 
broken power law line emissivity with a steeper
dependence on the radius in the innermost region.
%, which suggests a centrally localised corona. 
The goodness of the fit is constrained by the least squared method.
The fit results in 2.5--9.5\,keV  are presented in Table~\ref{laky_mcg_table}. 
The $\chi^{2}$ values give comparable results for all employed models. 
The $\chi^{2}_{\rm red}=\chi^{2}/\nu \approx 1.2$, where $\nu$ 
is the number of degrees of freedom which is related to the total
number of energy bins and model parameters.
The six independent parameters of the {\laor} and {\kyrline} models 
make the global minimum of $\chi ^{2}$ rather wide with several local minima. 
Each model has a different tendency to converge to a different minimum.
Hence, we did not compare only best fits of both models, but also the 
evaluated spin values by the {\kyrline} and {\laor} models
when the other model parameters correspond to each other.
%Firstly, we found the same minimum for both models and then compared their fitting capabilities. 
The equivalent width of the line is $EW \approx 700 \pm 50$\,eV
which is a rather high value but consistent with the result of \citet{2006ApJ...652.1028B}.

The errors presented in the table are evaluated while the 
other parameters of the model are fixed. However,
the realistic errors are higher because the model parameters 
further depend on other parameters of the line and continuum models.
To catch up these relations we produce various contour graphs focusing
on the determination of the spin value, taking into account the other
parameters of the used model, see Figure~\ref{mcg_contours}. 
The dependence of the $\chi ^{2}$ value
on the value of the spin ({\kyrline}) or the inner disc radius ({\laor}) 
is shown in the left column of the figure. 
%We can see from the graph that the {\laor} model enables the extreme 
%spin value while {\kyrline} model gives an upper limit for the spin
%value ($a_{max}=0.965$) for the same statistical relevance. 
The contour graphs for the spin and the inclination angle are shown 
in the middle column. The underlying 
model was fixed in both cases, and it was relaxed
for the third evaluation in the right column 
where the contours for the spin and the power-law index are shown. 
The contour plots reveal quite complicated structure
of the parameter space with several minima, making
the spin estimation rather smeared.
Taking all of these into account, we obtain for the spin value:\\
\[
% a_{KY} = 0.84 - 1.0, 
a_{KY} = 0.94^{+0.06}_{-0.10}
\hspace{0.3cm} \rm{and} \hspace{0.3cm}
% a_{laor} = 0.88 - 0.998. 
a_{laor} = 0.96^{+0.04}_{-0.08}.
\]

Another issue is the smoothness of the contour plots.
Some of the sharp or local features in the presented contour
plots are due to computational problems of the fitting
method but some of them can be smoothed by an improved
choice of the internal computational parameters.
We tried to change some of these parameters.
The smoothest results were generally obtained when
we increased number of fitting iterations together
with wider steps of the value of the examined parameters.
However, one may easily miss a true minimum when stepping 
a parameter value roughly which leads to bad results.

To avoid this problem we always tried first to find a true
minimum and then stepping the parameters in such a way that
the values corresponding to the true minimum are not skipped.
Finding the true minimum is rather difficult problem in such a rich
space of free parameters. Therefore, we usually repeated the fitting
procedure several times with different initial values of the model
parameters and then we compared $\chi^2$ values of the local minima.

\subsubsection{GX\,339-4}
\label{gx339_vhs_spectrum}

The black hole binary GX\,339-4 exhibited a strong broadened iron line 
in the 76\,ks observation in 2002 
when the source was in the very high state \citep{2004ApJ...606L.131M}, 
and also in two 138\,ks observations in spring 2004 
when the source was in the low-hard state \citep{2006ApJ...653..525M}.
See also Figure~\ref{fig_miller} adopted from the two papers.
However, by our re-analysis of the data we found that 
the longer low/hard state observation is 
affected by a pile-up -- see the details in Section~\ref{pileup}.
Hence, we use further only the very high state observation from 2002. 
Details of the data reduction and spectral re-analysis are described in Section~\ref{gx339}.

\begin{table}[tbh!]
\begin{center}
\caption{Results of iron line models for GX\,339-4 in 3--9\,keV.}
 	\vspace{0.1cm}
\begin{tabular}{c|c|c|c}
	\hline \hline
	parameter   &	{\kyrline}&	{\kyrline}*	& {\laor}	 \\
	\hline
\rule{0pt}{1.5em}	$a/M$		&	$\textbf{0.69}^{+0.13}_{-0.12}$&	$\textbf{0.62}^{+0.14}_{-0.14}$	&	$\textbf{0.77}^{+0.08}_{-0.12}$\\
\rule{0pt}{1.5em}	$i$\,[deg]	&	$19 \pm 3$		&	$19 \pm 4$		&	$17 \pm 4$		\\
\rule{0pt}{1.5em}	$E$\,[keV]	&	$6.97_{-0.2}$		&	$6.97_{-0.2}$		&	$6.97_{-0.2}$		\\
\rule{0pt}{1.5em}	$q$		&	$3.45 \pm 0.08$		&	$3.35\pm 0.08$		&	$3.3\pm 0.1$		\\
	\hline
\rule{0pt}{1.5em}	$\chi^{2}/v$	&	$147/125$	&	$148/125$ 	&	$148/125$	\\[2pt]
	\hline
%	$EW$\,[eV]	&	$175$		&	$164$	  	&	$164$	  	\\
%\multicolumn{4}{c}{}\\
%\multicolumn{4}{c}{}\\
\end{tabular}
\label{laky_gx_table}
\end{center}
\end{table}

\begin{figure}[tbh!]
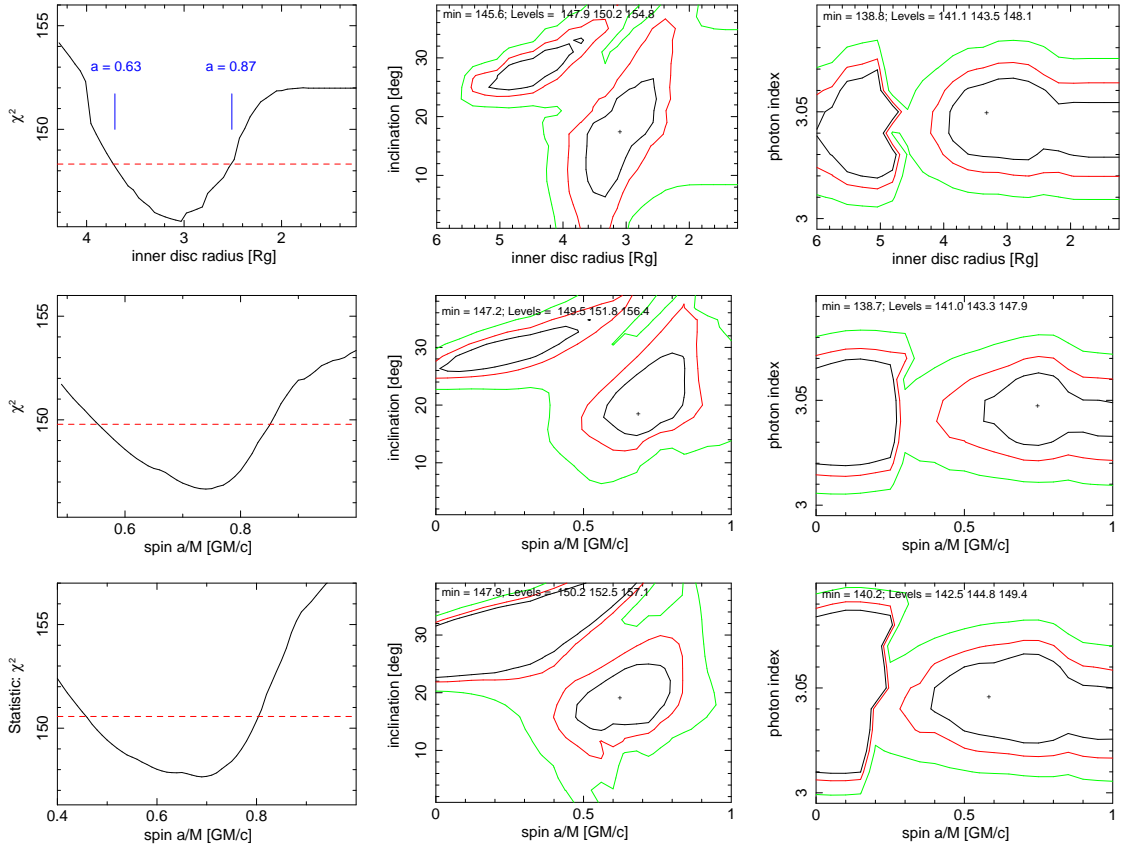

\begin{center}
\begin{tabular}{ccc}
 \includegraphics[angle=270,width=0.31\textwidth]{gxlaor_stepa2.eps} &
 \includegraphics[angle=270,width=0.31\textwidth]{gxlaor_contirin.eps} &
 \includegraphics[angle=270,width=0.31\textwidth]{gxlaor_contpa.eps}\\
 \includegraphics[angle=270,width=0.31\textwidth]{gxkyr_stepa.eps} &
 \includegraphics[angle=270,width=0.31\textwidth]{gxkyr_contia.eps} &
 \includegraphics[angle=270,width=0.31\textwidth]{gxkyr_contpa.eps}\\
 \includegraphics[angle=270,width=0.31\textwidth]{gxkylb_stepa.eps} &
 \includegraphics[angle=270,width=0.31\textwidth]{gxkylb_contia.eps} &
 \includegraphics[angle=270,width=0.31\textwidth]{gxkylb_contpa.eps}\\
\end{tabular}
\caption{The contour graphs show (from \textposown{left} to \textposown{right}) 
the dependence of the $\chi ^{2}$ value, 
the inclination angle and the power law index on the value of the spin ({\kyrline}) 
or the inner disc radius ({\laor}) for the GX\,339-4 spectrum in 3--9\,keV.
The x-axis is oppositely directed in the case of the inner disc radius 
of the {\laor} model as x-variable for an easier comparison with the {\kyrline} results. 
The black, red and green contours correspond to $1\sigma$, $2\sigma$ and $3\sigma$, respectively. 
\textposown{Up}: The results of the {\laor} model.
\textposown{Middle}: The results of the {\kyrline} model with limb darkening. 
\textposown{Down}: The results of the {\kyrline}* model with limb brightening.}
\label{gx_contours}
\end{center}
\end{figure}

The fitting results of the relativistic line models in 3--9\,keV are summarised in 
Table~\ref{laky_gx_table} and Figure~\ref{gx_contours}.
There are two minima found during the fitting procedure. 
We preferred the one which better corresponds to the results
obtained by the independent radio and infrared measurements 
which constrained the inclination angle to be $i < 26^{\circ}$
\citep{2004MNRAS.347L..52G}.
The dependence of the goodness of the fit on the spin value
is shown in the left column of the figure. 
The contour graphs for the spin and the inclination angle 
are depicted in the middle column, and for the spin 
and the power law photon index in the right column of the figure. 
The derived spin value is then:

\[
% a_{KY} = 0.56 - 0.85, 
a_{KY} = 0.69^{+0.16}_{-0.13}
%\]
\hspace{0.3cm} \rm{and} \hspace{0.3cm}
%\[
% a_{laor} = 0.65 - 0.86. 
%a_{laor} = 0.77^{+0.09}_{-0.12}.
a_{laor} = 0.77^{+0.10}_{-0.14}.
\]
%\vspace{0.3cm}

In both cases, MCG\,-6-30-15 and GX\,339-4, the fitting with the {\laor} model
resulted in slightly higher values for the spin, but consistent with the values 
of the {\kyrline} model within the general uncertainties of the spin estimation.
The spin value of MCG\,-6-30-15 is better constrained thanks to its high value
and more counts in the line than in the case of GX\,339-4 (see Table~\ref{table_mcg_gx_cts}).

\begin{table}[tbh!]
\begin{center}
\caption{Count rates of the XMM-Newton observations in $2-10$\,keV.}
%	\vspace{0.35cm}
\begin{tabular}{c|c|c}
	\hline \hline
\rule{0pt}{1.5em}			&	{MCG\,-6-30-15}		&	{\hspace{0.2cm} GX\,339-4 \hspace{0.2cm}} 			\\
	\hline
\rule{0pt}{1.5em}	number of counts $[\times 10^6]$&	$0.97$			&	$3.56$	\\
\rule{0pt}{1.5em}	net counts/s	&	$4.98(1)$			&	$1547(1)$	\\
\rule{0pt}{1.5em}	model counts/s	&	$5.02$				&	$1546$	\\
\rule{0pt}{1.5em}	line counts/s	&	$0.20(1)$			&	$5.1(1)$	\\
\rule{0pt}{1.5em}        line counts	&	$4.37\times10^{4}$		&	$1.15\times10^{4}$	\\[2pt]
\hline

\end{tabular}
\label{table_mcg_gx_cts}
\end{center}
\end{table}

%%%%%%%%%%%%%%%%%%%%%%%%%%%%%%%%%%%%%%%%%%%%%%%%%%%%%%%%%%%%%%%%%%%%%%%%%%%%%%%%%%%%%%%%%%%%%%5555

\subsection{Simulated data of next generation X-ray satellites}
\label{laky_future}

\begin{figure}[tbh!]
\begin{center}
\includegraphics[width=0.45\textwidth]{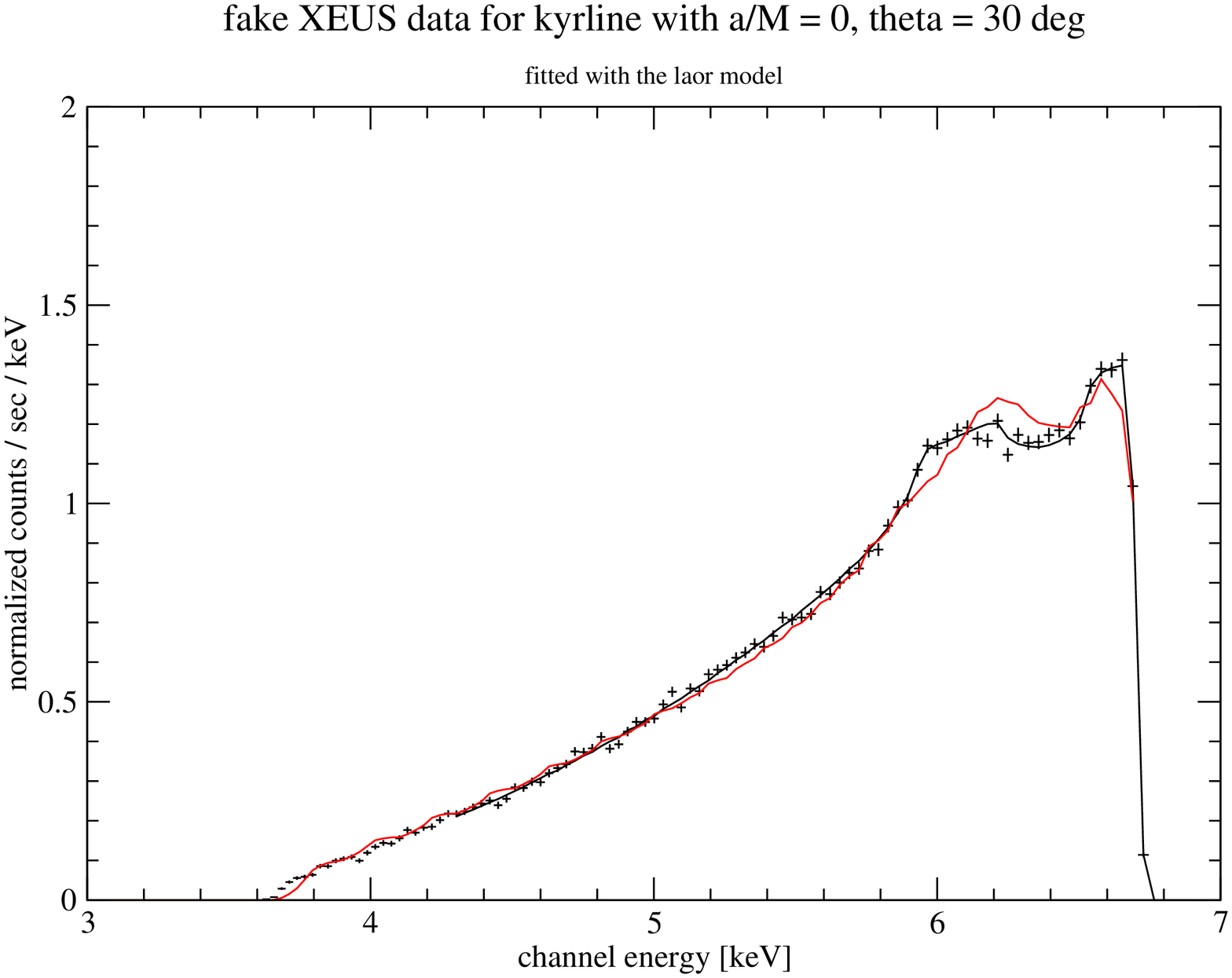} 
\hfill \includegraphics[width=0.45\textwidth]{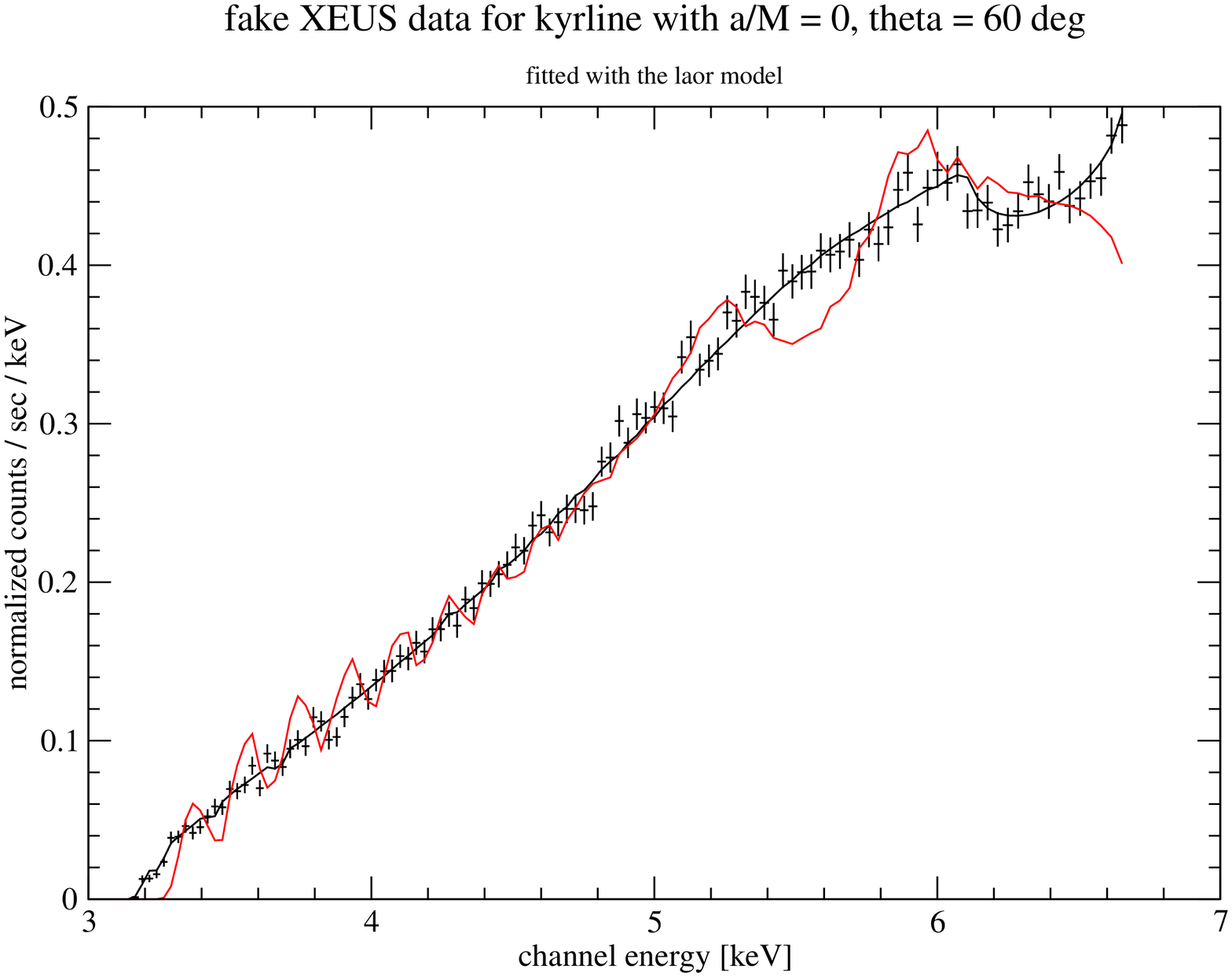} \\[-0.73cm]
\includegraphics[width=0.45\textwidth]{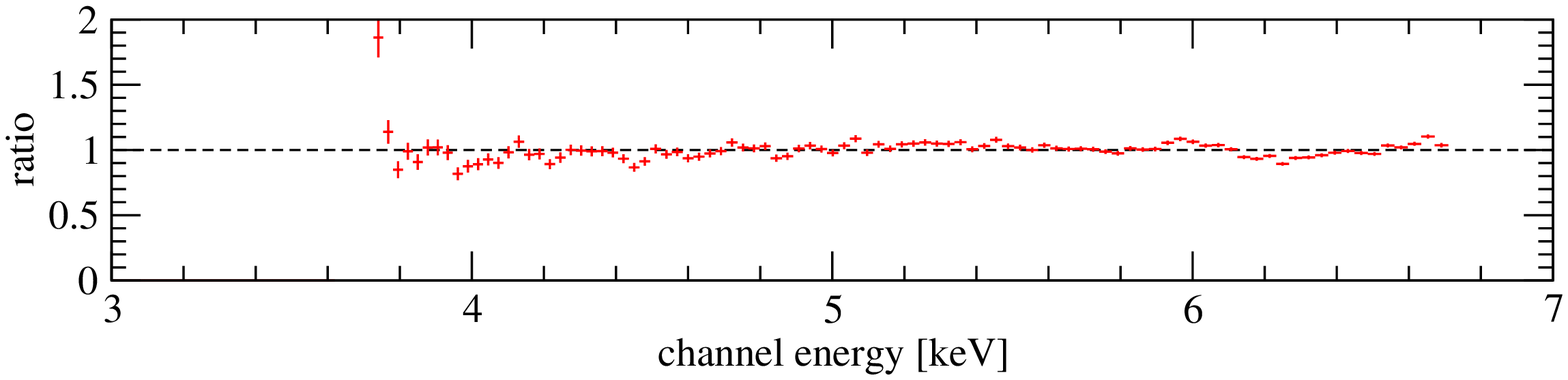} 
\hfill \includegraphics[width=0.45\textwidth]{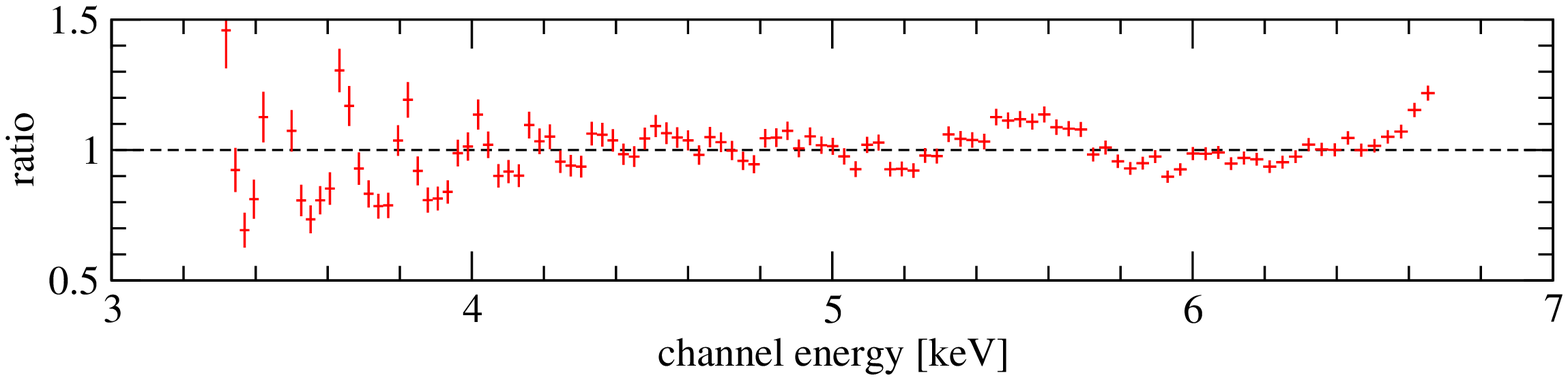} \\[0.1cm]
\includegraphics[width=0.45\textwidth]{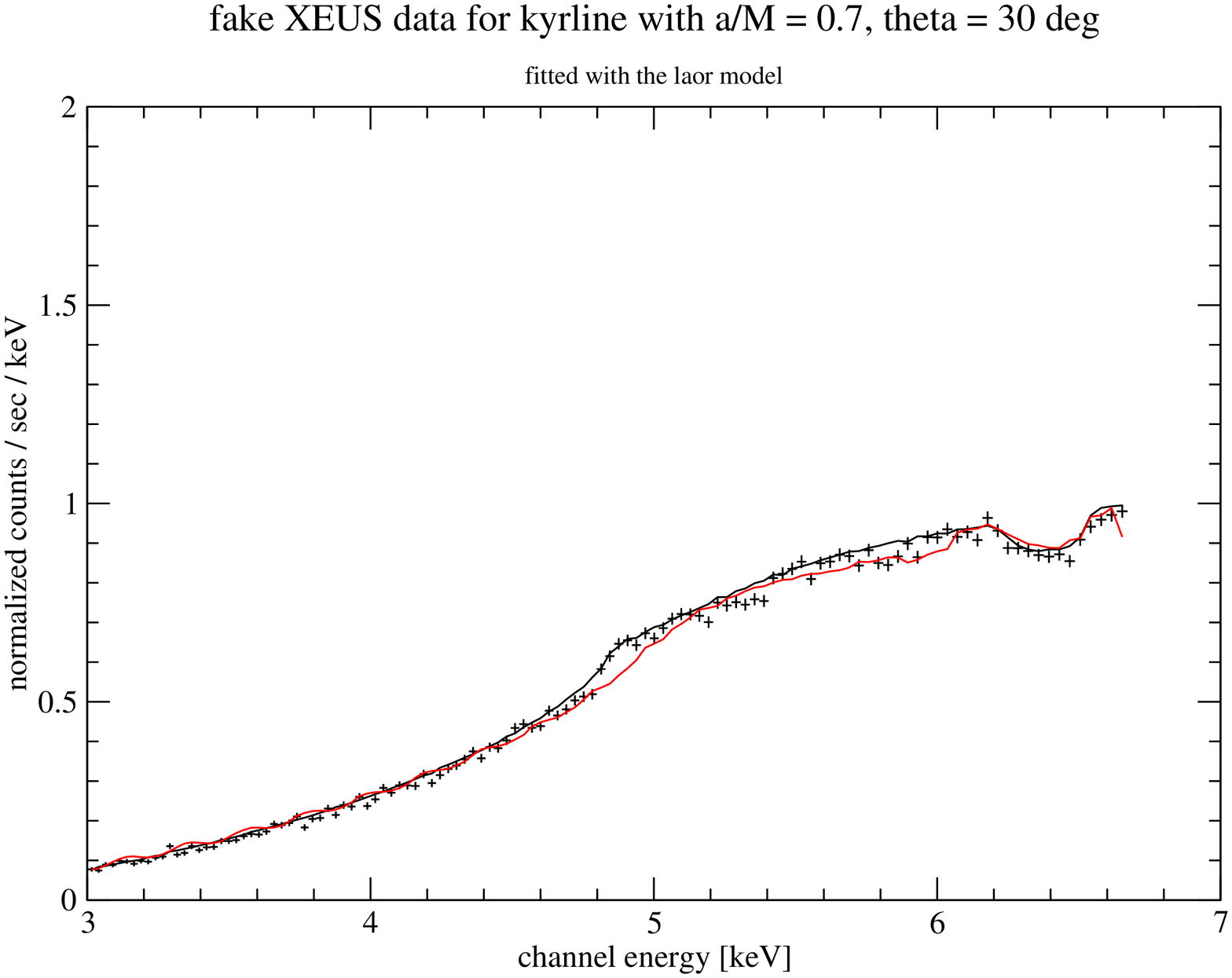} 
\hfill \includegraphics[width=0.45\textwidth]{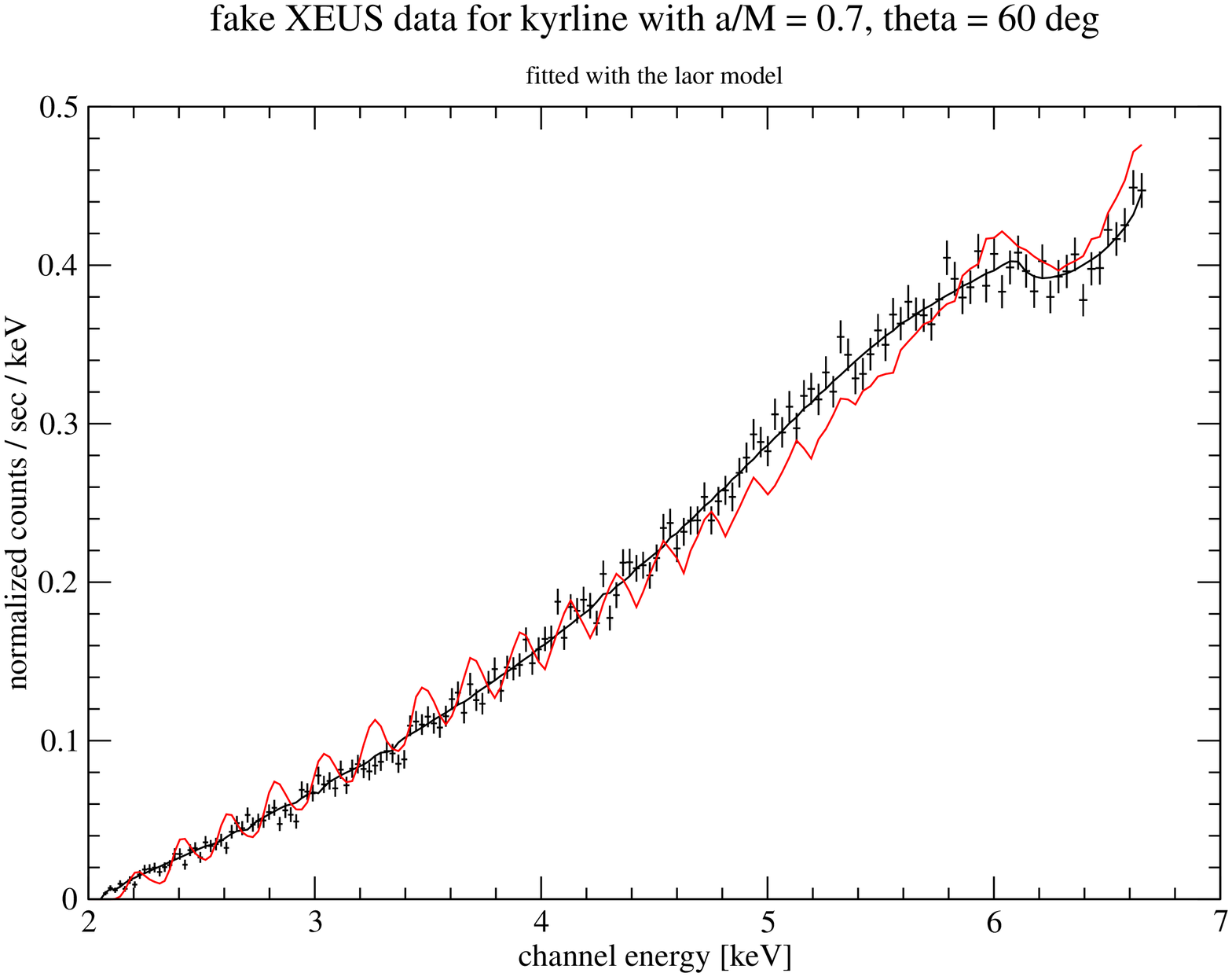} \\[-0.73cm]
\includegraphics[width=0.45\textwidth]{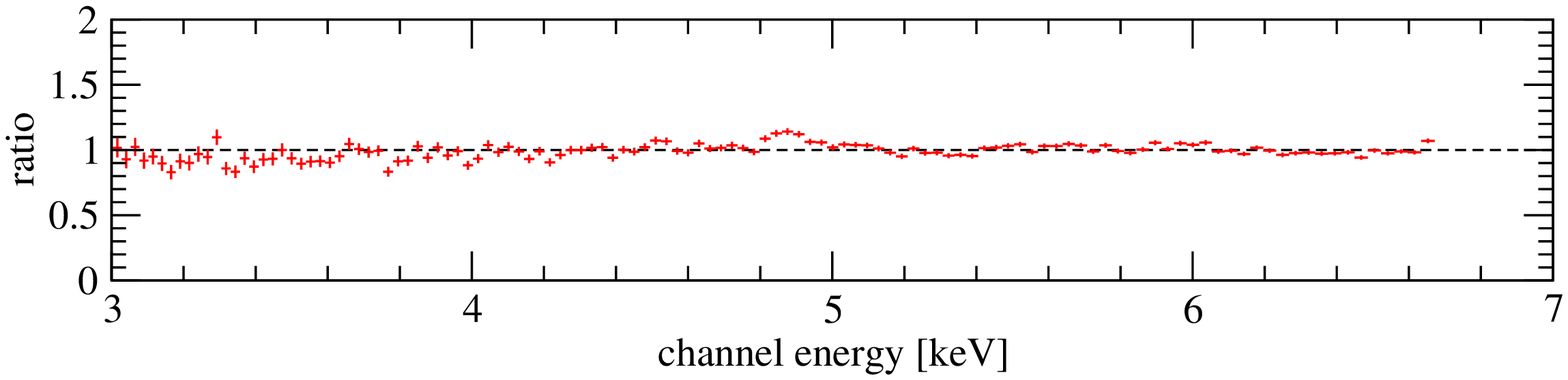} 
\hfill \includegraphics[width=0.45\textwidth]{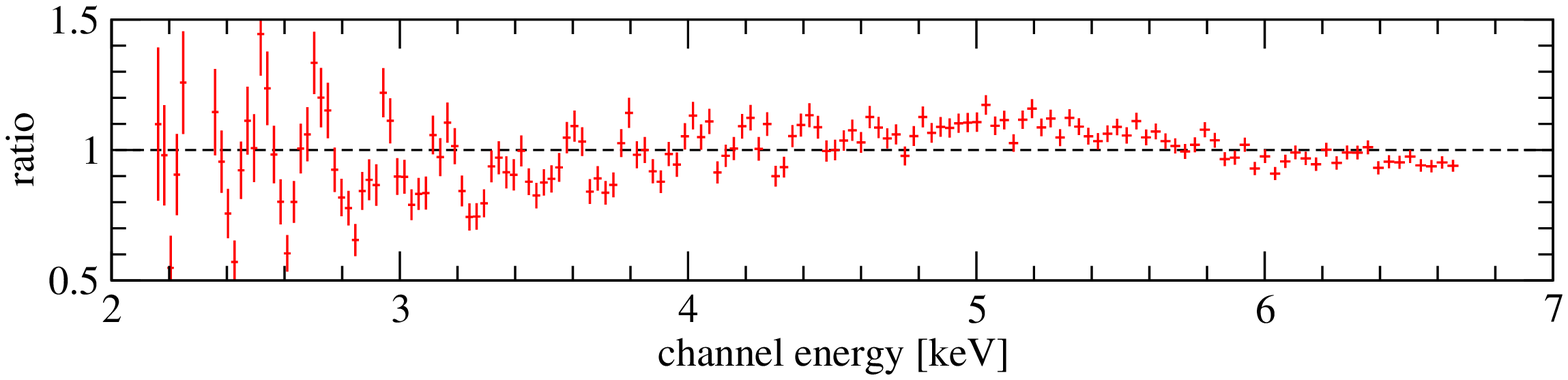} \\[0.1cm]
\includegraphics[width=0.45\textwidth]{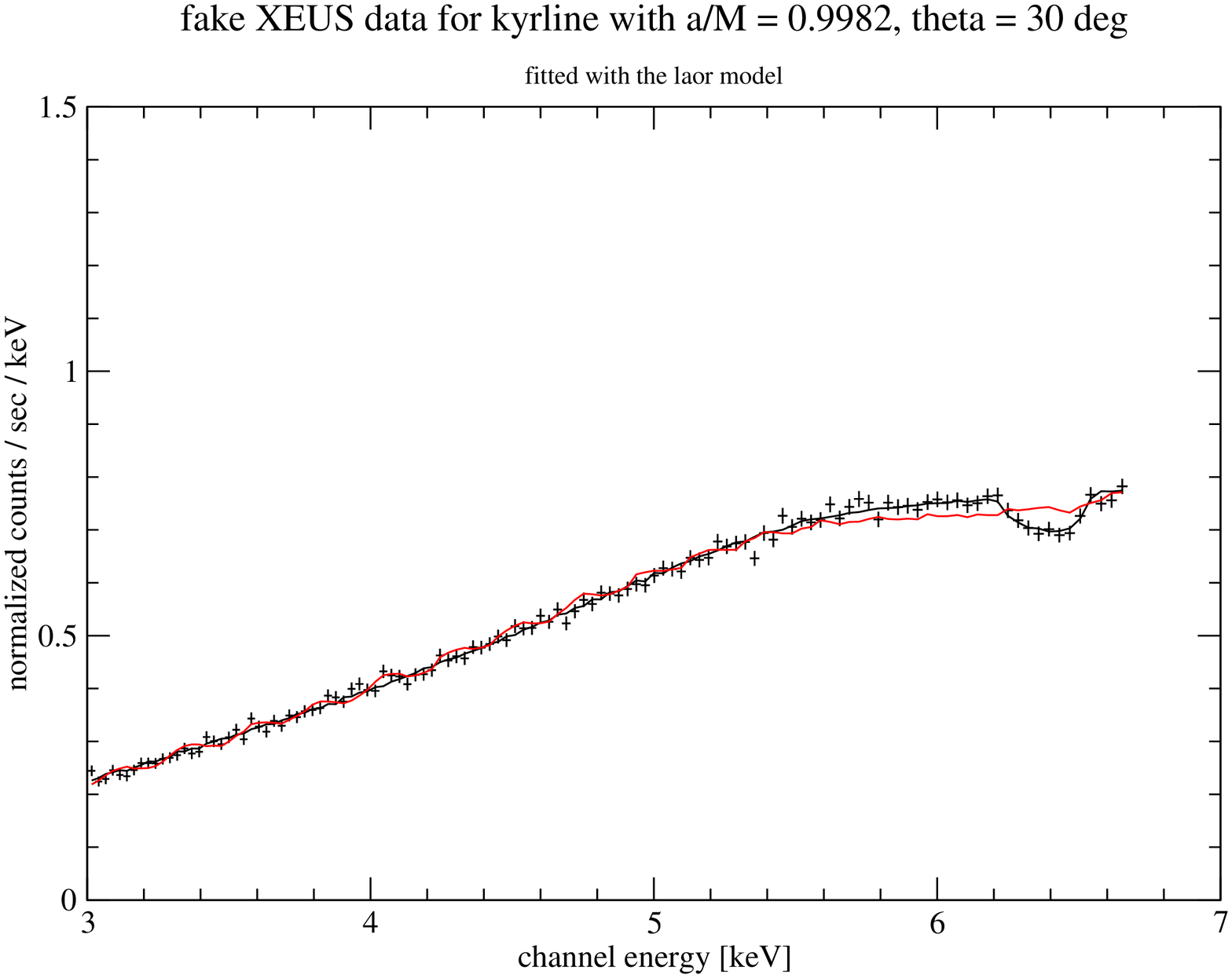} 
\hfill \includegraphics[width=0.45\textwidth]{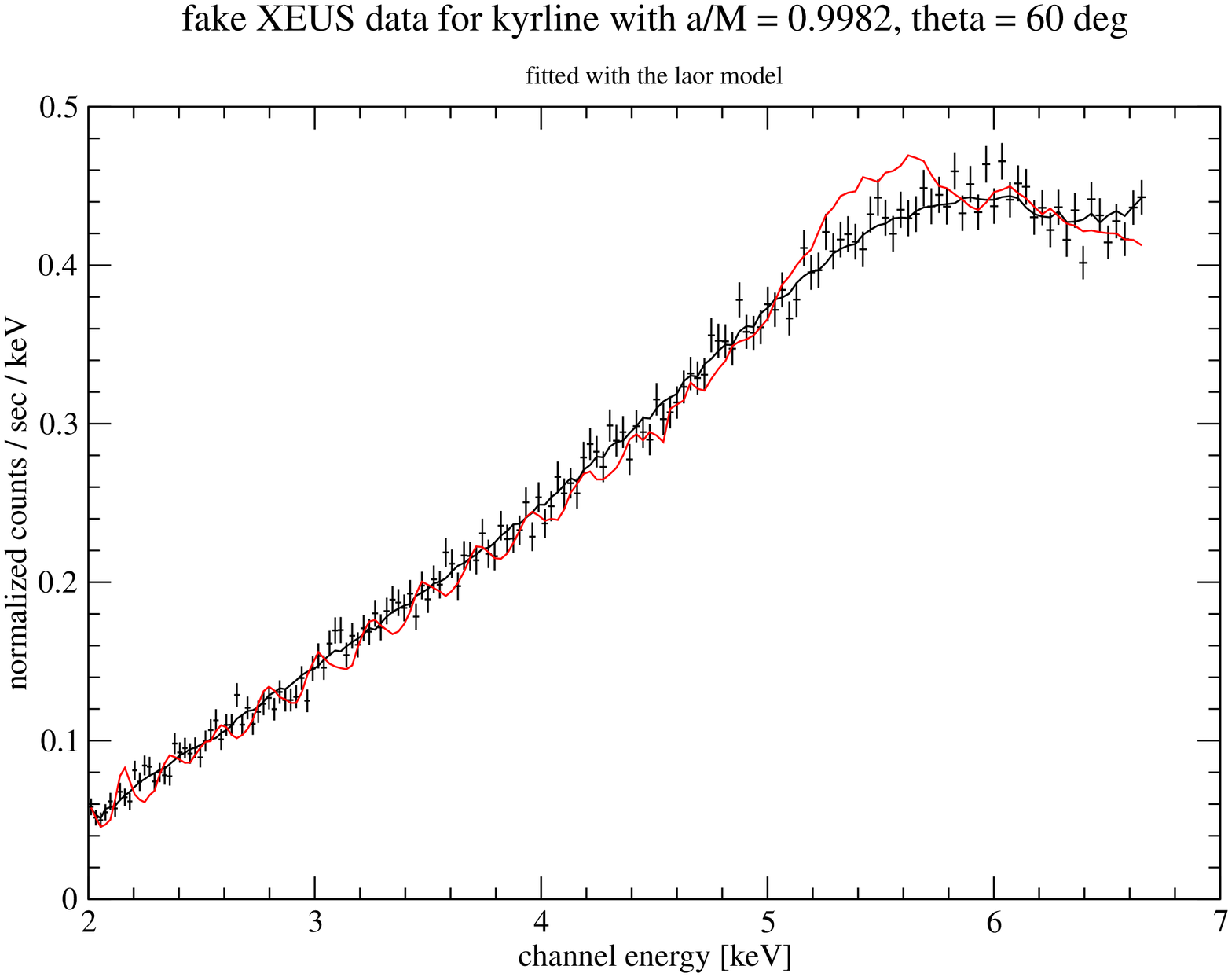}\\[-0.73cm]
\includegraphics[width=0.45\textwidth]{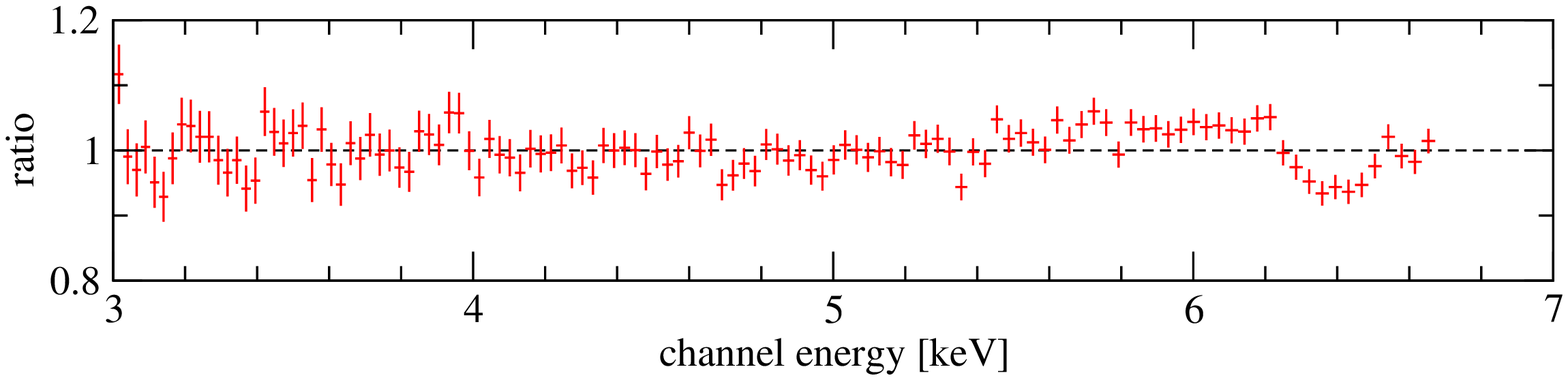} 
\hfill \includegraphics[width=0.45\textwidth]{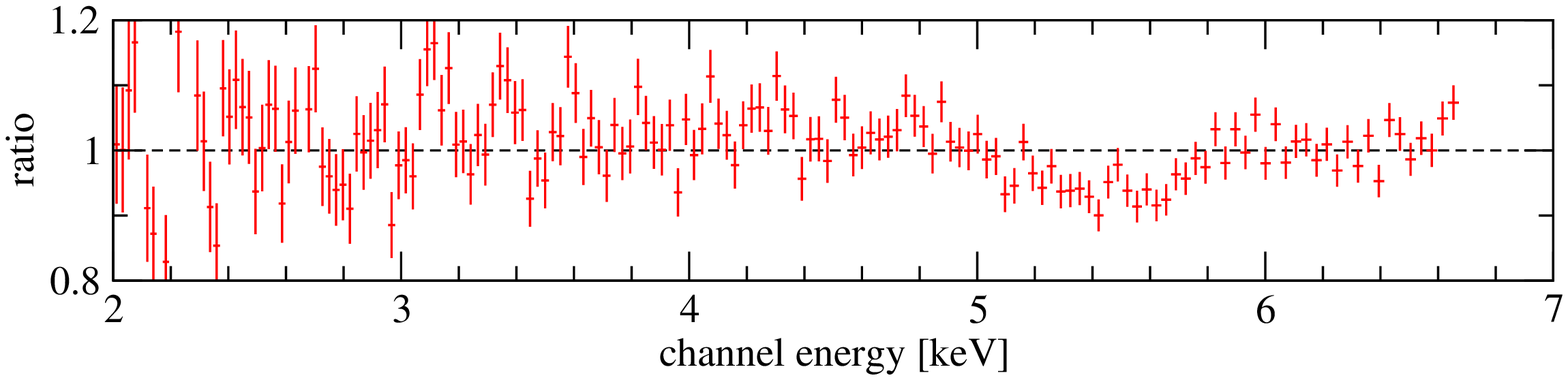} \\[0.1cm]

\caption{Simulated data for three different values of spin 
(from \textposown{top} to \textposown{bottom}: $a=0$, $0.7$, $0.9982$) 
and two different inclination angles $\theta=30$\,deg (\textposown{left}), 
and $\theta=60$\,deg (\textposown{right}). 
The artificial data are shown with the expected errors (black crosses).
Black curve is the seed {\kyrline} model from which the data were generated.
The red curve is the {\laor} model which resulted from the fitting of the
simulated data. Each panel also contains a plot with ratios of the data 
to the {\laor} model.}
\label{xeus_systematic}
\end{center}
\end{figure}

%\begin{figure}[tbh!]
%\begin{center}
%\includegraphics[width=0.47\textwidth]{a00-mcg-6-kyrline-alpha3-theta80-1-9keV-res30eV-laor-fit.eps} \\[-0.73cm]
%\includegraphics[width=0.47\textwidth]{a00-mcg-6-kyrline-alpha3-theta80-1-9keV-res30eV-laor-rat.eps} \\[0.27cm]
%\includegraphics[width=0.47\textwidth]{a07-mcg-6-kyrline-alpha3-theta80-1-9keV-res30eV-laor-fit.eps} \\[-0.73cm]
%\includegraphics[width=0.47\textwidth]{a07-mcg-6-kyrline-alpha3-theta80-1-9keV-res30eV-laor-rat.eps} \\[0.27cm]
%\includegraphics[width=0.47\textwidth]{a09982-mcg-6-kyrline-alpha3-theta80-1-9keV-res30eV-laor-fit.eps} \\[-0.73cm]
%\includegraphics[width=0.47\textwidth]{a09982-mcg-6-kyrline-alpha3-theta80-1-9keV-res30eV-laor-rat.eps}\\ [0.27cm]

%\caption{Continuation of the previous figure
%for the inclination angle $\theta=80$\,deg.}
%\label{xeus_systematic_continue}
%\end{center}
%\end{figure}

\begin{figure}[tbh!]
\begin{center}
\includegraphics[width=0.4\textwidth, angle=270]{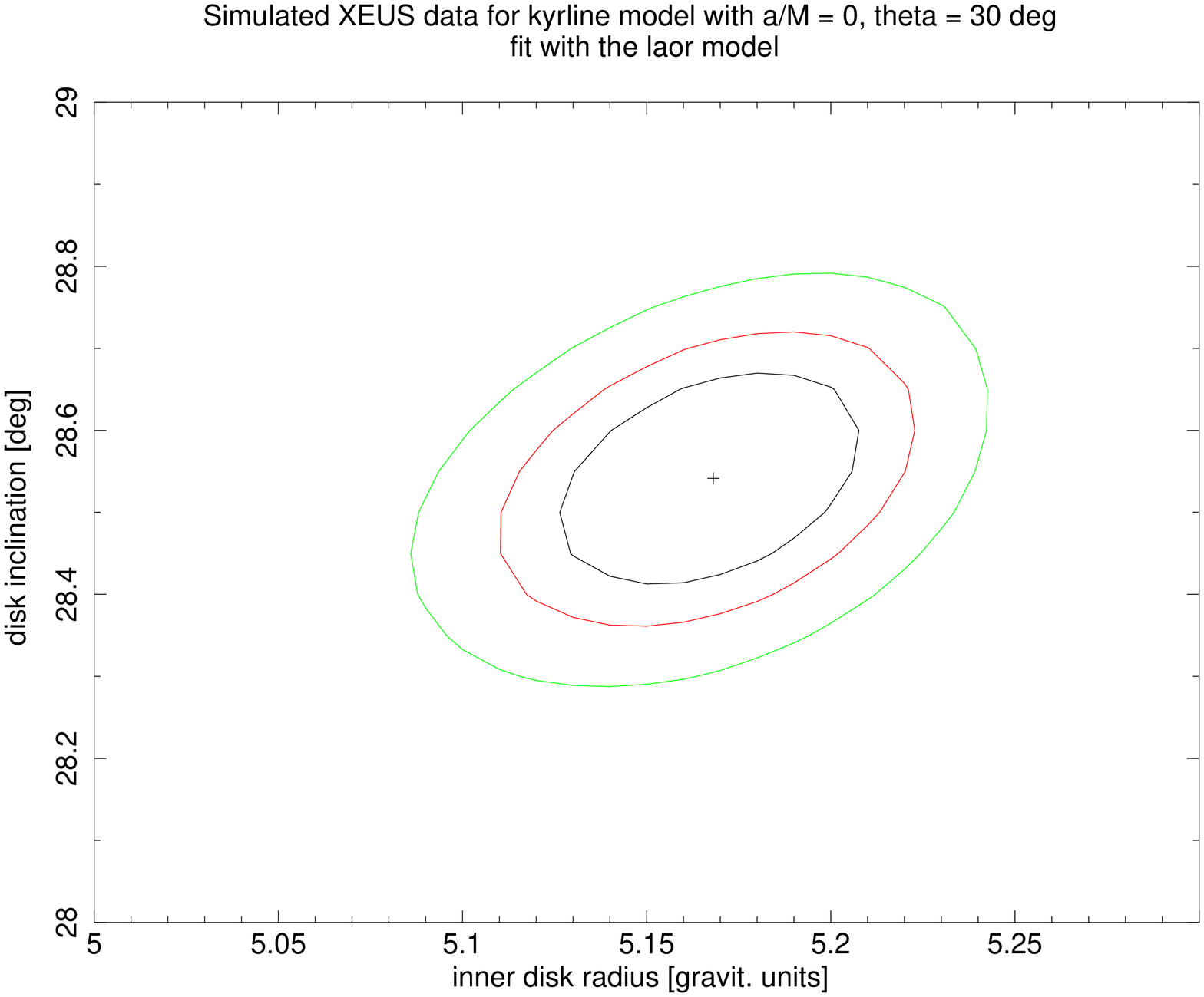} 
\hfill \includegraphics[width=0.4\textwidth, angle=270]{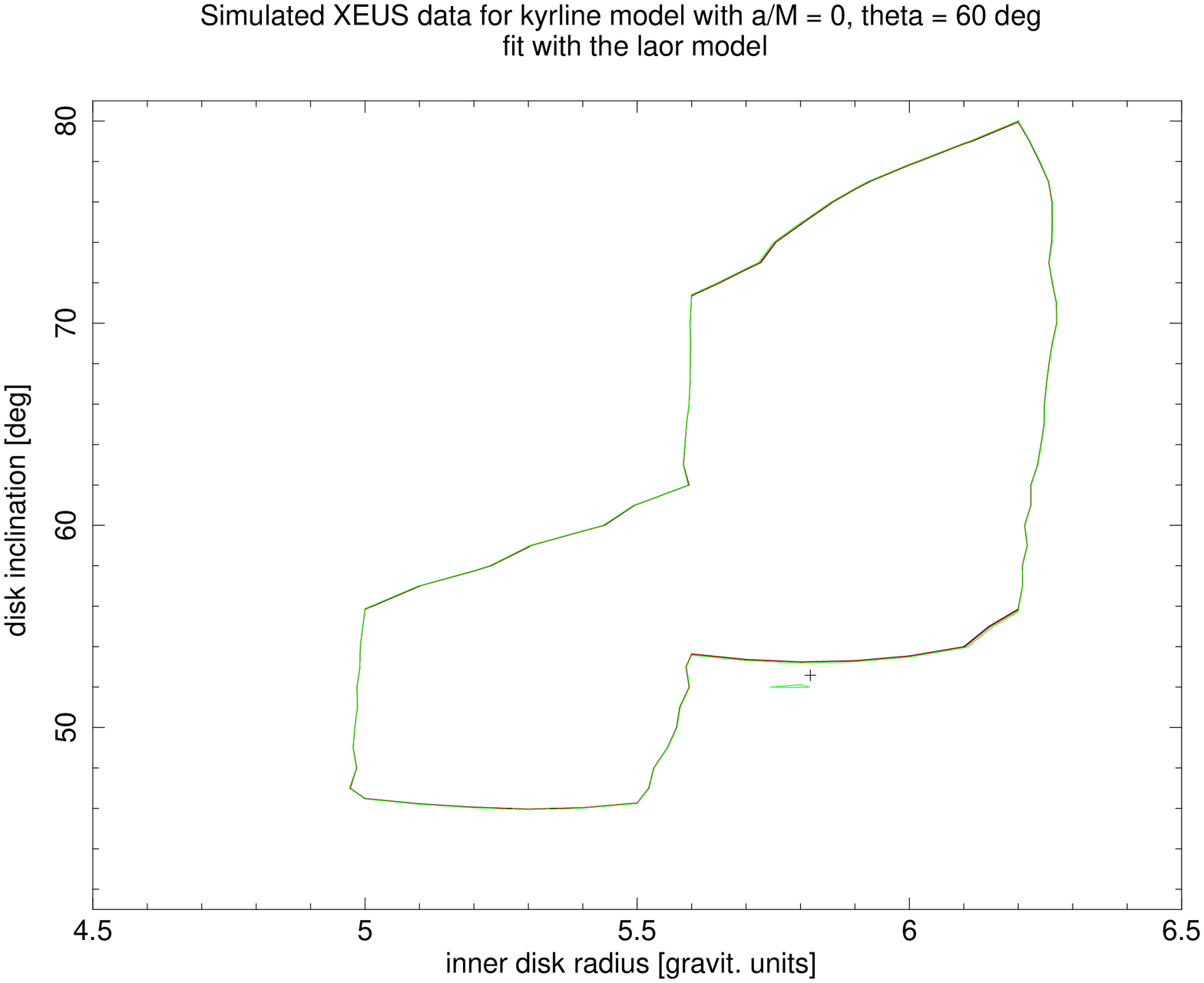} \\[0.2cm]
\includegraphics[width=0.4\textwidth, angle=270]{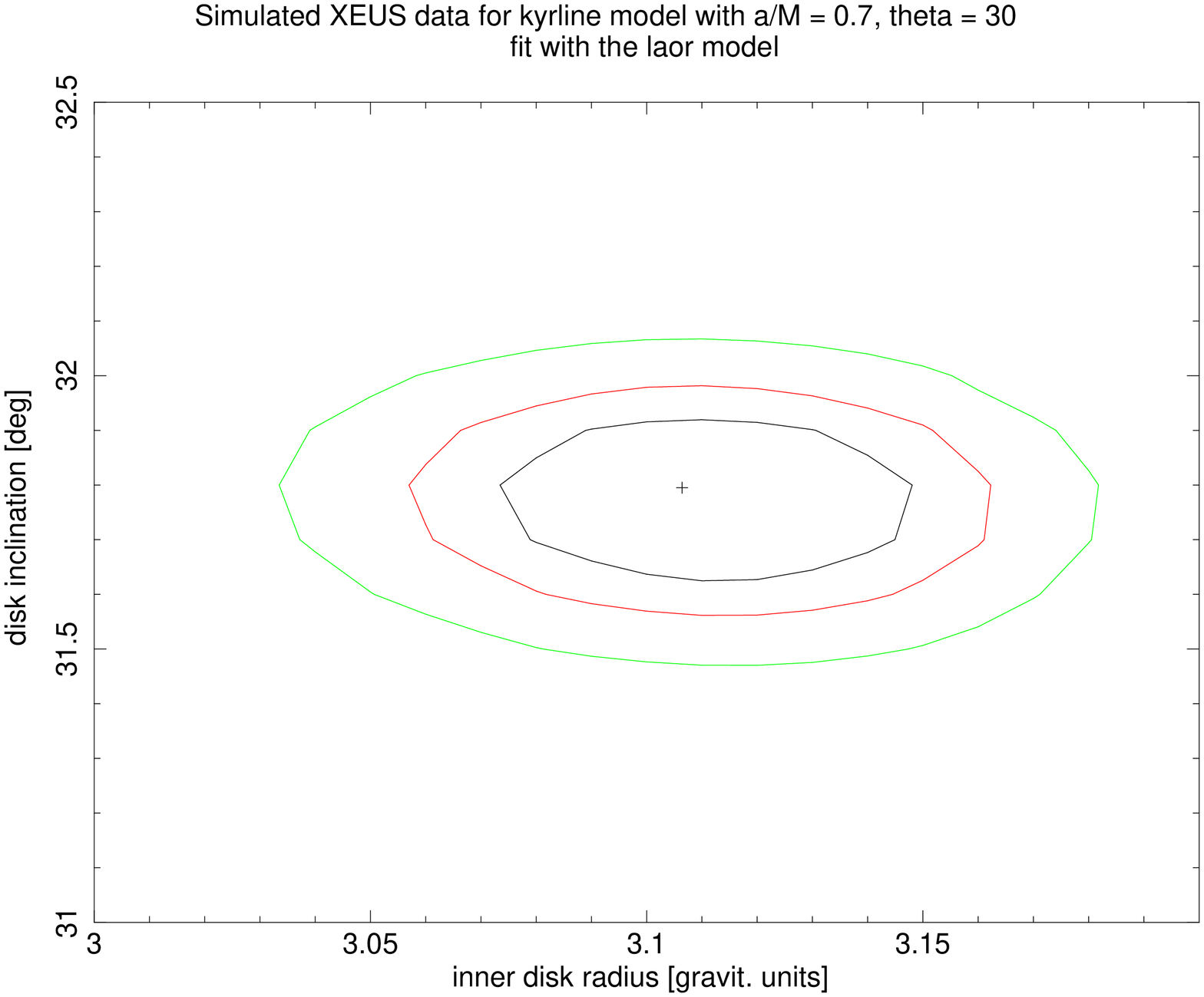} 
\hfill \includegraphics[width=0.4\textwidth, angle=270]{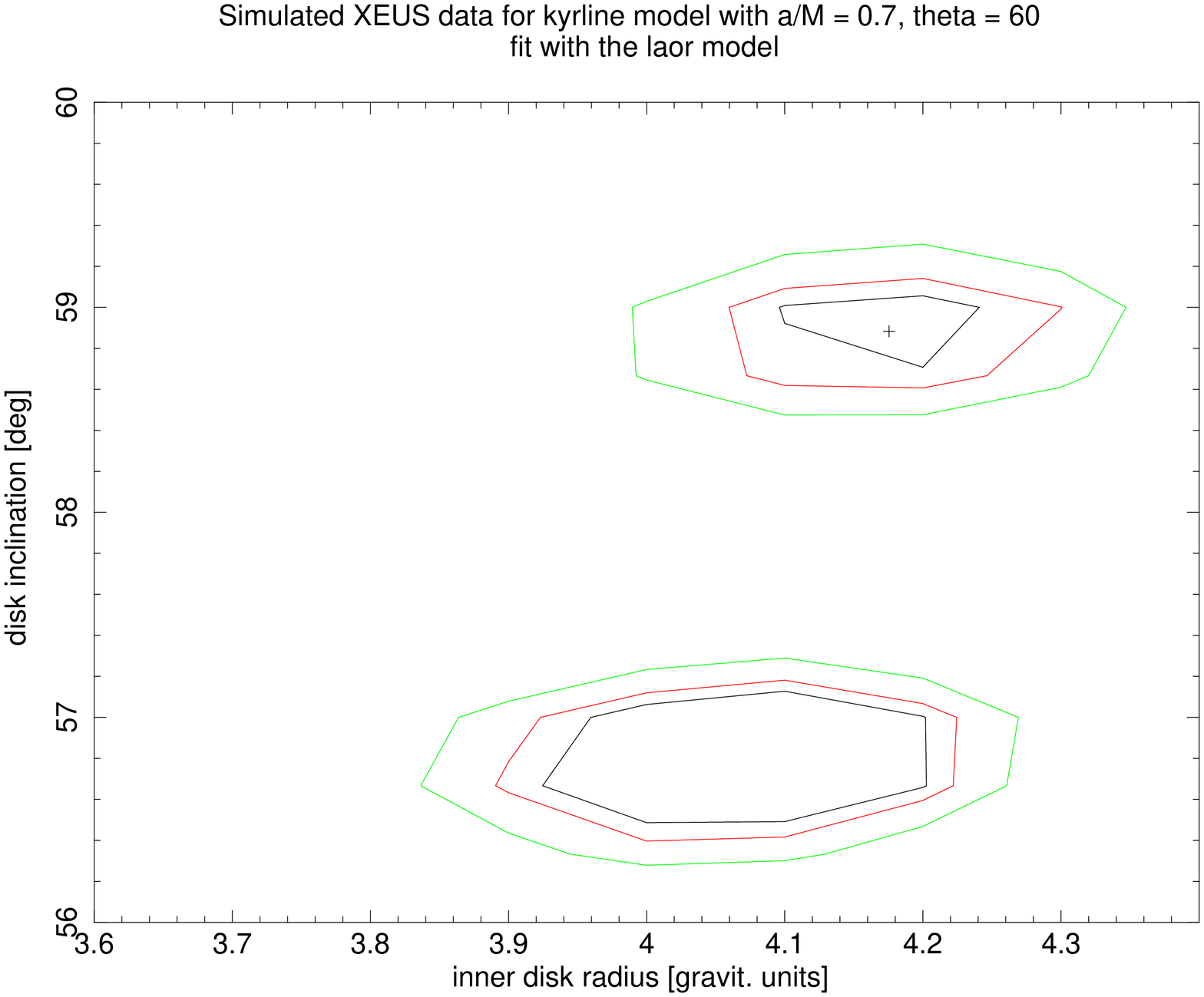} \\[0.2cm]
\includegraphics[width=0.4\textwidth, angle=270]{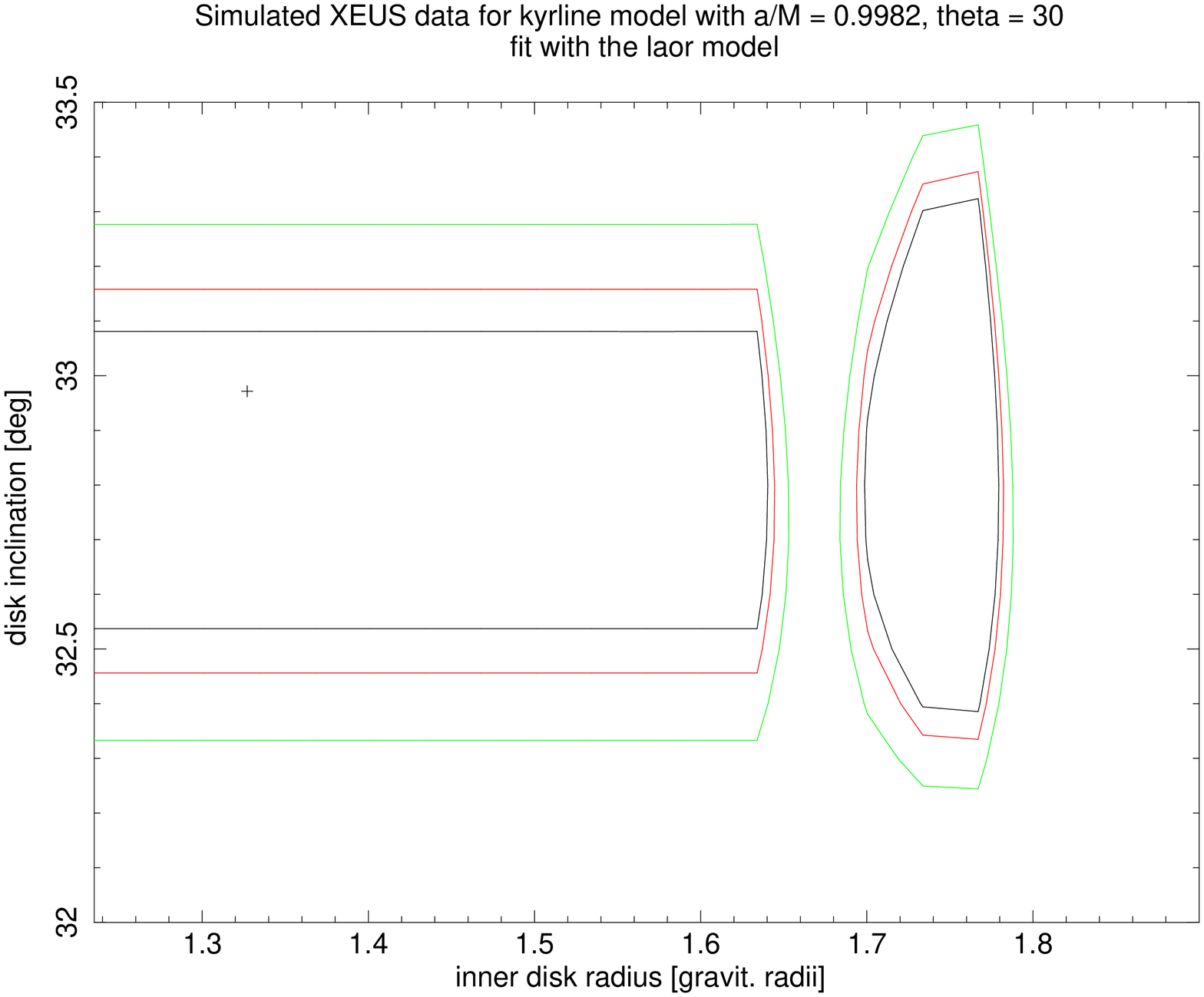} 
\hfill \includegraphics[width=0.4\textwidth, angle=270]{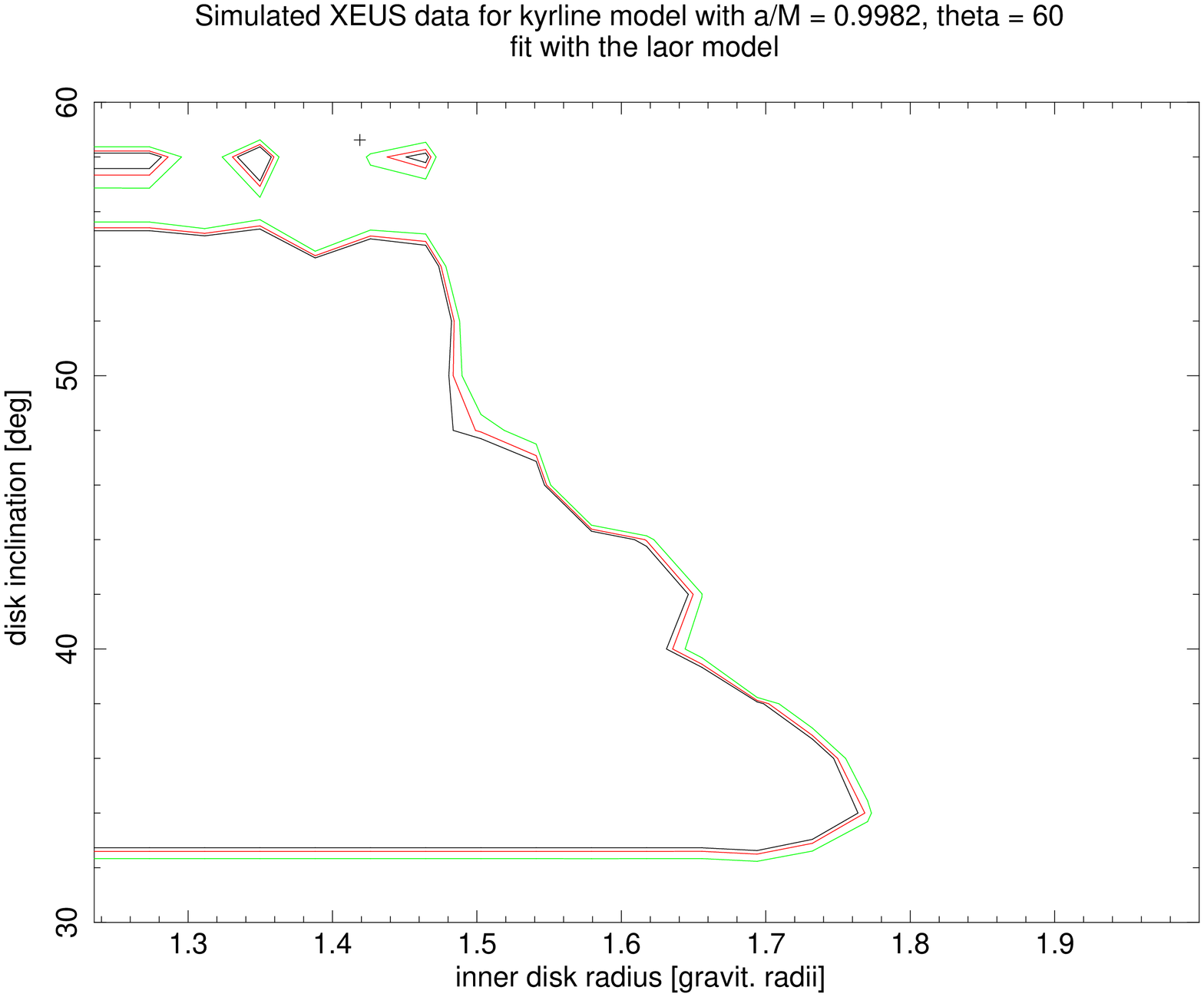} \\[0.2cm]

\caption{Contour graphs of the inclination angle and the inner disc radius
of the {\laor} model applied to the same artificial data as in Figure~\ref{xeus_systematic}.
The fiducial values of the studied parameters are $i=30$\,deg (\textposown{left}), 
$i=60$\,deg (\textposown{right}), and $r_{\rm in}=r_{\rm ms}(a=0)=6\,r_g$ (\textposown{top}),
$3.39\,r_g$ (\textposown{middle}), and $1.235\,r_g$ (\textposown{bottom}).
The individual curves correspond to 1\,$\sigma$, 2\,$\sigma$,
and 3\,$\sigma$, respectively.}
\label{xeus_systematic_cons}
\end{center}
\end{figure}

Further, we applied the {\laor} and {\kyrline} models on the data 
with significantly higher quality supposed to be achieved by on-coming X-ray missions. 
The presently planned International X-ray Observatory 
(IXO, see e.g. \citealp{2010arXiv1001.2329B}) arose from the merging
of the former XEUS and Constellation-X missions. 
Because the details of the IXO mission have not been
fixed at the moment when this work originated, 
we used a preliminary response matrix of the former XEUS mission \citep{2009ExA....23..139A}.
%To this purpose, we simulated data using the presently available response matrix of the XEUS mission.
We generated the data for a {\kyrline} model with a rest energy of the line $E=6.4\,$keV
and radial disc emissivity that follows a power law with the index $q=3$. 
We re-binned the data in order to have a resolution of $30\,$eV per
bin. We then fit the data in the 1--9\,keV energy range with the {\laor} model using the same initial
values of the fitting parameters as for the data simulation. 
Due to insufficient resolution of the {\laor} model, a significant problem appears 
at the high-energy edge of the broad line. This occurs because the next generation instruments
achieve much higher sensitivity in the corresponding energy range.
%Therefore, we excluded the higher-energy drop of the lines from the
%analysis in order to reveal the differences in the overall shape of the line.
 
We examined the artificial data for a grid of values of the angular momentum and the inclination angle.
The results for different values are shown in Figure~\ref{xeus_systematic}.
%are shown in Figures~\ref{xeus_systematic}--\ref{xeus_systematic_continue}.
We excluded the higher-energy drop of the lines 
in order to reveal the differences in the overall shape of the line.
It is clearly seen from the figures that the effect of the spin on the 
shape of the line is sufficiently resolved with the higher quality data. 
The contour graphs of the inner disc radius and the inclination
angle are shown in Figure~\ref{xeus_systematic_cons} for three different
values of the spin, $a=0$, $a=0.7$, $a=0.9982$. 
For lower values of the spin, the results obtained with the {\laor}
model differ more from the fiducial values. The {\laor} model
tends to underestimate the inner disc radius, or, by other words,
to overestimate spin value related to the inner disc radius by 
the relation $a=a\,\left(r_{\rm ms}=r_{\rm in}\right)$.
%The {\laor} model fails for the case of any lower value of the spin than the extremal one.
%The inner disk radius is set to the marginally stable orbit for a given spin ($R_{\rm in}=R_{\rm ms}(a/M)$). 

\begin{table}
\begin{center}
\caption{Results of the {\laor} fit in 3--9\,keV in the simulated spectra.}
\vspace{0.1cm}
\begin{tabular}{c|c|c||c|c}
% \multicolumn{5}{c}{\bf Table 4. Results of {\laor} fit in 3--9\,keV in the simulated spectra}\\
        \hline \hline
\rule{0pt}{1.5em}				& \multicolumn{2}{c||}{\textbf{MCG\,-6-30-15} (simul.)} & \multicolumn{2}{c}{\textbf{GX\,339-4} (simul.)} \\[1.5pt]
\rule{0pt}{1.em}       parameter               &       KY value        &       fitted {\laor} value    &       KY value        &       fitted {\laor} value \\[1.5pt]
\hline
\rule{0pt}{1.5em}        $       R_{\rm in}     $&$     2.00     $&$     1.86^{+0.04}_{-0.02}$   &       $3.39$ 	 	&       $3.20^{+0.05}_{-0.05}$   \\ 
%       &$      a/M             $&$     0.94    $       \\
\rule{0pt}{1.5em}	$       i\,[{\rm deg}]         $&$     26.7    $&$     25.4(3) $ 		&       $19$    	&       $18.6(2)$      \\
\rule{0pt}{1.5em}        $       E_{\rm line}\,[{\rm keV}]        $&$     6.70     $&$     6.71(1) $      	 	&       $6.97$  	&       $6.94(5)$	\\
\rule{0pt}{1.5em}        $       q_{1}           $&$     4.90     $&$     4.51(3) $       	&       $3.45$  	&       $3.2(1)$        \\
\rule{0pt}{1.5em}        $       q_{2}           $&$     2.80     $&$     2.76(2) $       	&	-		&	-		\\
\rule{0pt}{1.5em}        $       r_{\rm b}           $&$     5.5     $&$     6.2(1)  $        	&	-		&	-		\\
\rule{0pt}{1.5em}        $       K_{\rm line}    $&$     8.7\times 10^{-5}$&$ 9.0\times 10^{-5}$ &       $6.5\times 10^{-3}$&$ 6.6\times 10^{-3}   $\\[1.5pt]
\hline
%       $\chi^{2}/v$            &$      1560/873        $&$     1364/873        $               \\
\rule{0pt}{1.5em}        $\chi^{2}/\nu$            & & $1364/873$    & & $2298/873$             \\[2pt]
\hline
\end{tabular}

\label{table_simdata_mcg_gx_laorfit}
\end{center}
{\small \textposown{Note}: Fiducial values of the {\ky} model parameters which
were used to the data simulation are shown in columns ``{\ky} value''.
The inner disc radius is determined
from the relation $r_{\rm in}=r_{\rm ms}(a)$ (eq.~\ref{a_rms}).}
\end{table}

Further, we produced simulated data for the Seyfert galaxy MCG-6-30-15 and the
black hole binary GX\,339-4 using rather simplified models
 which were suitable to fit the current XMM-Newton data. 
For MCG-6-30-15 we used a power law model plus a {\kyrline} model
for the broad iron line, absorbed by Galactic gas
matter along the line of sight: {\phabs} * ({\powerlaw} + {\kyrline}). The
parameters of the continuum are the column density $n_{\rm H} = 0.4\times
10^{21}$\,cm$^{-2}$, the photon index $\Gamma =1.9$ of the power law, and its
normalisation $K_{\Gamma} = 5\times 10^{-3}$. 
The values of the line parameters are summarised in the \textit{KY value} column of 
Table~\ref{table_simdata_mcg_gx_laorfit}.
The exposure time was chosen as $220$\,ks and the flux 
of the source as $1.5\times10^{-11}$erg\,cm$^{-2}$\,s$^{-1}$ in the 2--10\,keV 
energy range (i.e. $1.4\times10^{7}$\,cts).

For GX\,339-4 we used {\phabs} * ({\diskbb} + {\powerlaw} + {\kyrline}) with $n_{\rm H} =
0.6\times 10^{22}$\,cm$^{-2}$, $kT_{\rm in} = 0.87$\,keV ($K_{\rm kT} = 1.4\times 10^{3}$), 
and $\Gamma = 3$ ($K_{\Gamma} = 5.6$). 
The exposure time was chosen as $75$\,ks, the flux of the source as 
$9.3\times10^{-9}$erg\,cm$^{-2}$\,s$^{-1}$ in the 2--10\,keV 
energy range, i.e. $1.5\times10^{9}$\,cts. The number of counts is 
more than two orders of magnitude higher than for the observation of 
the XMM-Newton satellite (see Table~\ref{table_mcg_gx_cts}). 
The reason is due to the loss of 97\% of the photons during the \textit{burst mode}
of XMM-Newton observation which eliminates the pile-up problem (see Section~\ref{xmm_modes}). 
The next generation X-ray missions are supposed to have a calorimeter 
instead of the CCD camera on-board which will get rid of such problems.

\begin{figure}[tbh!]
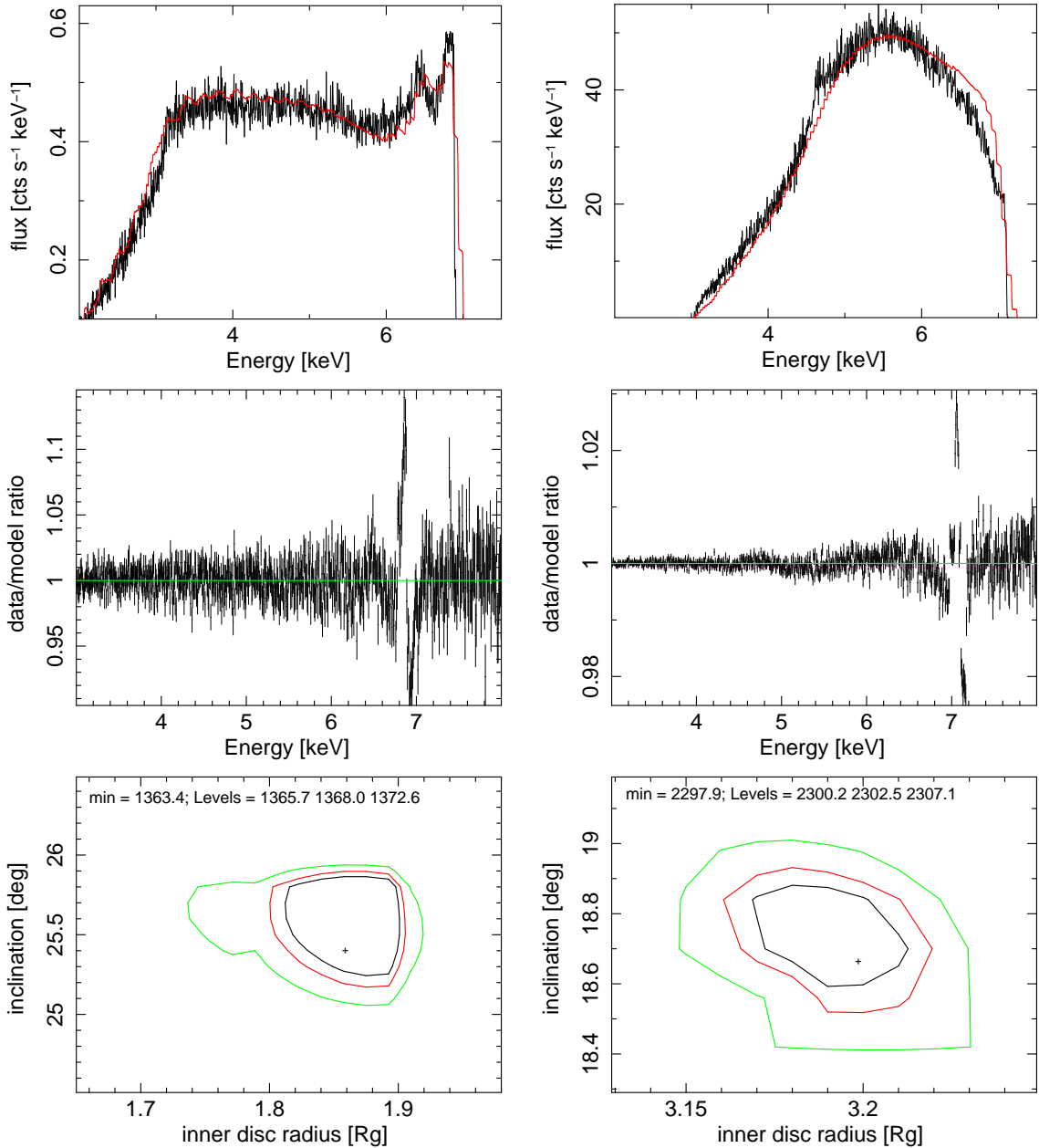

\begin{center}
%\begin{tabular}{ccc}
\includegraphics[angle=270,width=0.48\textwidth]{mcg_sim_line.eps}  
\hfill \includegraphics[angle=270,width=0.48\textwidth]{gx_sim_line.eps} \\[0.2cm]
\includegraphics[angle=270,width=0.48\textwidth]{mcg_sim_ratio.eps} 
\hfill \includegraphics[angle=270,width=0.48\textwidth]{gxf_sim_ratio.eps} \\[0.2cm]
\includegraphics[angle=270,width=0.48\textwidth]{mcg_sim_contirin_best} 
\hfill \includegraphics[angle=270,width=0.48\textwidth]{gx_sim_contirin.eps}\\[0.2cm]
%\end{tabular}
\caption{The simulated spectra for MCG\,-6-30-15 (\textposown{left}) 
and GX\,339-4 (\textposown{right}) 
using the preliminary response matrix of the XEUS mission. 
\textposown{Top}: Broad iron line generated
by the {\kyrline} model (black data) and fitted by the {\laor} model (red curve). 
\textposown{Middle}: Data/model ratio when fitting by the {\laor} model. 
\textposown{Bottom}: Contours for the inclination angle and the inner disc radius
of the {\laor} model. The fiducial values are $r_{\rm in}=2$ and $\theta=26.7$
for the case of MCG\,-6-30-15 (\textposown{left}), and $r_{\rm in}=3.39$ and $\theta=19$
for the case of GX\,339-4 (\textposown{right}).}
\label{xeus_fake}
\end{center}
\end{figure}

%Prior to the spectral analysis we rebinned the data to have approximately a 5eV resolution 
%(as it was planned for the XEUS instrument). 
%We tested different grouping realizing that the discrepancies between the two models increase 
%with larger grouping, but we can see apparent differences already for the most moderate rebinning.
The results of the {\laor} fit are shown in
Table~\ref{table_simdata_mcg_gx_laorfit} and in Figure~\ref{xeus_fake}.
The broad iron line component of the model is plotted in the top panel
of the figure. 
The model continuum components are not displayed there 
to clearly see the deflections of the {\laor} model.
The most prominent discrepancy appears at the higher-energy drop, 
which is clearly seen in the data/model ratio plot
in the middle panel of Figure~\ref{xeus_fake}. 
The model parameters are constrained with small error bars 
(see contours $a$ vs. $i$ in the bottom panel of Figure~\ref{xeus_fake}), 
which reveals a difference between the {\kyrline} and the {\laor} models.

The spin value derived from the analysis using the {\laor} model is:
\[
a_{\rm laor,MCG} = 0.958^{+0.003}_{-0.004} 
\]
for MCG\,-6-30-15, while the fiducial value of the spin was $a_{\rm fid,MCG} = 0.940$, and
\[
 a_{\rm laor,GX} = 0.74^{+0.02}_{-0.02} 
\]
for GX\,339-4, while the fiducial value of the spin was $a_{\rm fid,GX} = 0.70$.\\

\section{Discussion of the results}

We investigated the iron line band for two representative 
sources -- MCG\,-6-30-15 (active galaxy) and GX\,339-4 (X-ray binary).
The iron line is statistically better constrained for 
the active galaxy MCG\,-6-30-15 due to a significantly longer exposure
time of the available observations -- see Table~\ref{table_mcg_gx_cts}
for comparison of count rates of the observations. 
The spectra of both sources are well described by a continuum model 
plus a broad iron line model.

We compared modelling of the
broad iron line by the two relativistic models, {\laor} and {\kyrline}. 
The {\laor} model always assumes an extreme Kerr metric ($a/M=0.9982$), 
and therefore, it is used for the fitting of different spin values only by 
identifying the inner edge of the disc with the marginally stable orbit. 
%The spin is then estimated from the lower boundary of the broad line. 
In the {\kyrline} model, on the other hand, the spin itself is a fitting 
parameter and the metric is adjusted to the actual value of $a/M$.

%and is strictly limited to the extreme Kerr metric. 
%It leads to the predictions of slightly higher values for the spin.
%For all studied cases, the {\laor} model overestimates the spin value.
The discrepancies between the {{\kyrline}} and {{\laor}} results are within 
the general uncertainties of the spin determination using the skewed line profile
when applied to the current data. This means that the spin is currently
determined from the position of the marginally stable orbit rather than
from the overall shape of the line.

Results with both models are apparently distinguishable for higher 
quality data, as those simulated for the next generation X-ray missions. 
We find that the {\laor} model tends to overestimate the spin value 
%(the tendency was already apparent by the analysis of the current XMM-Newton data)
%jako side-product zjistujeme, ze laor ma spatnou rozlisovaci schopnost v energiich, nepresne urcuje horni mez
%As a side-product 
and moreover, it has insufficient energy resolution 
which affects the correct determination of
the high-energy edge of the broad line.
This leads to large $\chi^2$ values by fitting
the high resolution data (see Table~\ref{table_simdata_mcg_gx_laorfit}).

Technically, the {{\kyrline}} model leads to a better defined 
minimum of $\chi^{2}$ for the best fit value.
The confidence contour plots for $a/M$ versus other
model parameters are more regularly shaped.
This indicates that the {{\kyrline}} model 
has a smoother adjustment between the different
points in the parameter space allowing for more reliable constraints on $a/M$. 
The {{\laor}} model has a less accurate grid.
Smoother and more precise results by the {\kyrline} model
are at the expense of the computational speed. Empirically,
we found that the {\kyrline} model is about $10 \times$ slower
than the {\laor} model.

% ##########################################################################

\chapter{Role of the emission directionality in the spin determination}
\chaptermark{Role of the emission directionality \ldots}
 \thispagestyle{empty}
 \label{directionality}

\section{Emission directionality}

\begin{figure}[tbh!]
\begin{center}
\includegraphics[width=0.49\textwidth]{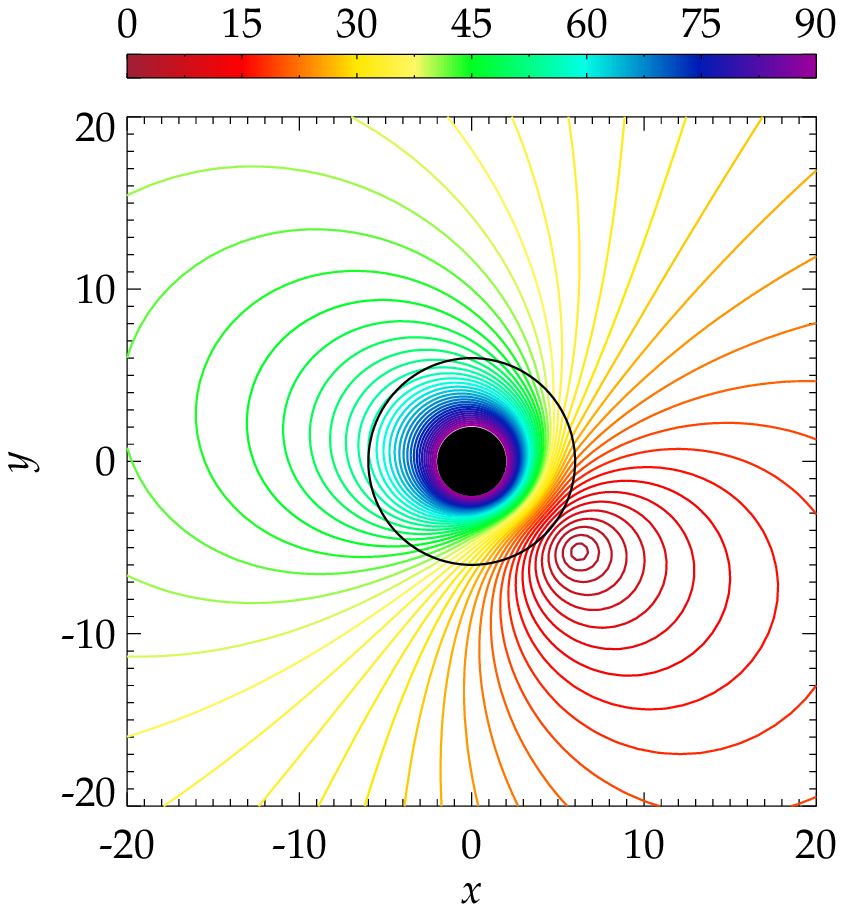}
\includegraphics[width=0.49\textwidth]{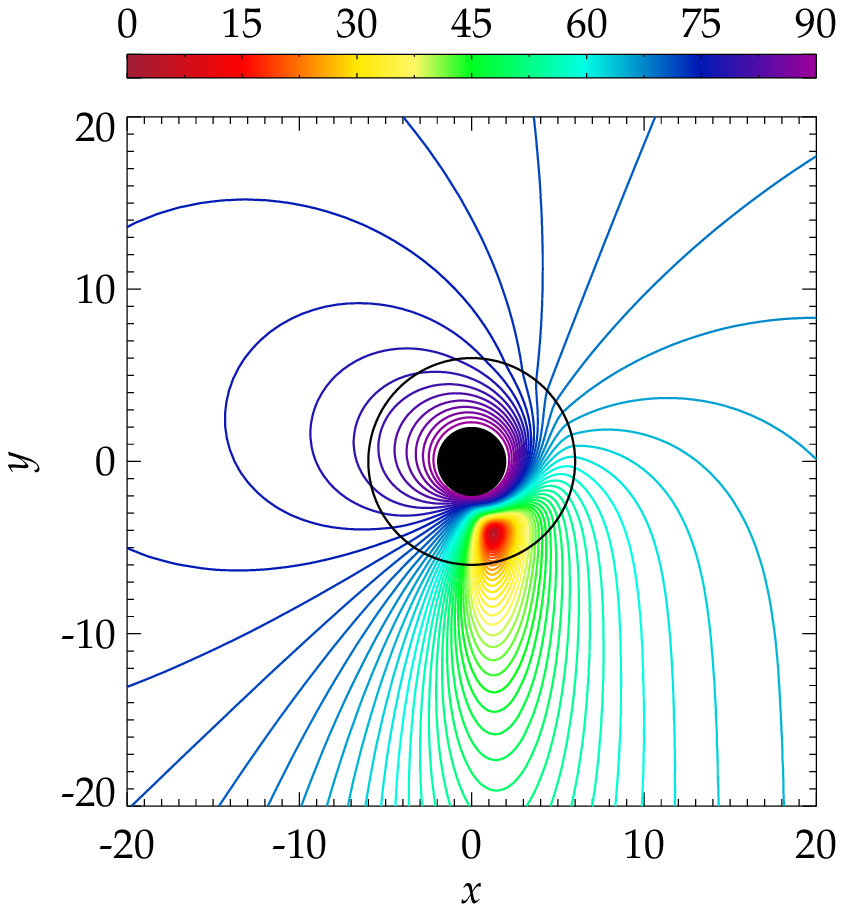}
\includegraphics[width=0.49\textwidth]{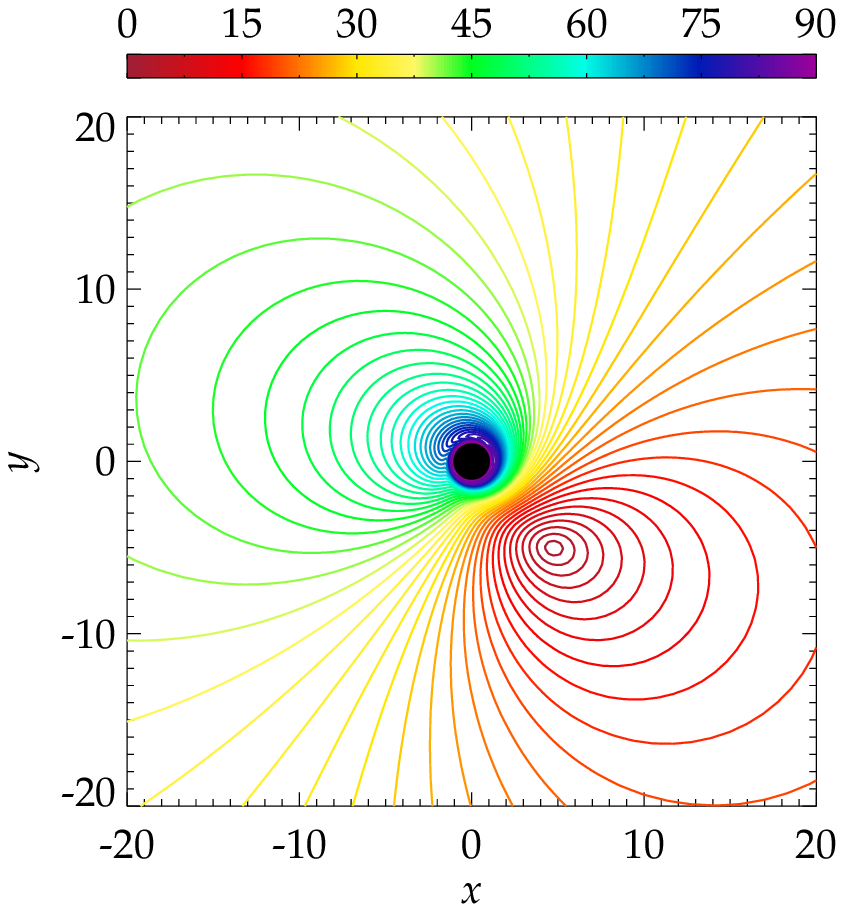}
\includegraphics[width=0.49\textwidth]{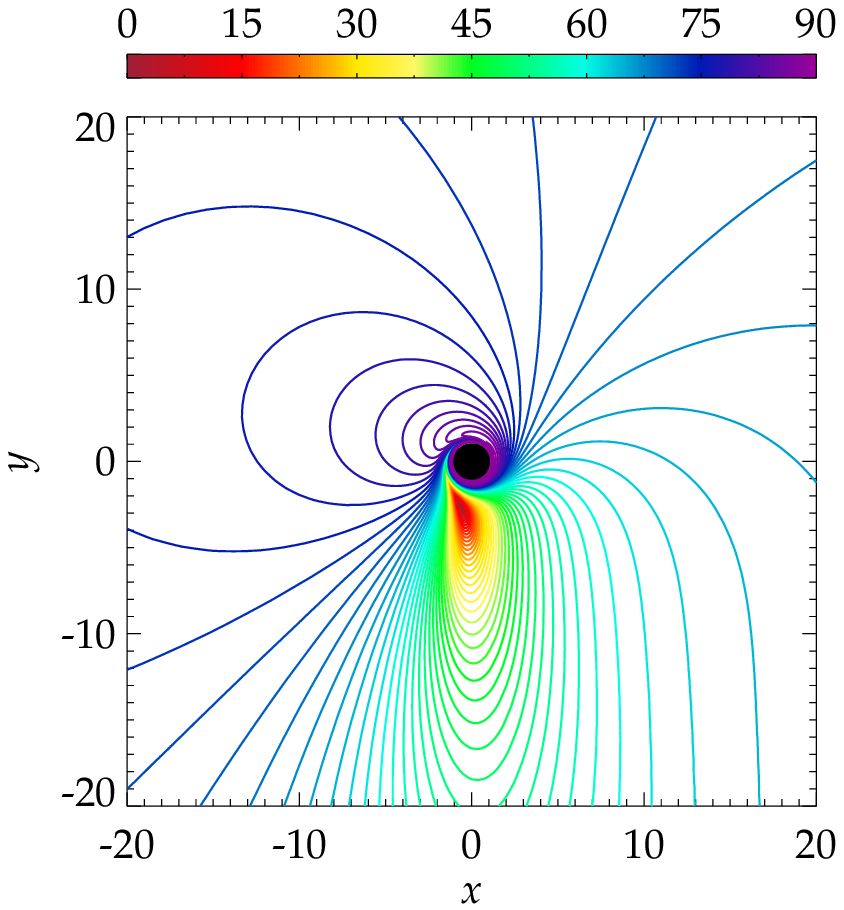}
\caption{Contours of the local emission angle, $\theta_{\rm e}(r,\varphi)$
near a non-rotating Schwarzschild black hole, $a=0$, (\textposown{top}),
and a maximally rotating black hole, $a=1$, (\textposown{bottom}), depicted in the equatorial 
plane $(x,y)$. The black hole and the accretion disc rotate counter clock-wise.
A distant observer is located towards the top
of the figure.
The inner region is shown up to $r=20$ gravitational radii
from the black hole. 
The black hole is denoted by a dark filled 
circle around the centre, for Schwarzschild black hole
the circle around represents the marginally stable orbit.
Two cases of different observer inclinations
are shown, $\theta_{\rm o}=30$~deg (\textposown{left}) and $\theta_{\rm o}=70$~deg
(\textposown{right}).
The colour bar encodes the range of $\theta_{\rm e}(r,\varphi)$ from $0$ to $90$
degrees. The emission angle is
measured from the disc normal direction to the equatorial plane, in
the disc co-moving frame.}%, i.e.\ in the local Keplerian frame orbiting
%with the angular velocity $\Omega_{_{\rm{}K}}(r)$.}
\label{fig3}
\end{center}
\end{figure}

Light rays coming from a black hole accretion disc
are highly curved in strong gravitational field
and the emission angle given by eq.~(\ref{gtheta})
may possess any value, depending
on the position on the disc from which the radiation is reflected.
Figure~\ref{fig3} shows the contours of the constant local
emission angle $\theta_{\rm e}$ from the accretion disc
around a non-rotating, $a=0$,
and a maximally rotating, $a=1$, black hole taking into account their
distortion by the central black hole and assuming that the emitted photons
reach a distant observer at a given view angle $\theta_{\rm o}$.
It clearly illustrates that the local emission angle spans 
the entire range, from $0$ to $90$ degrees for any value of $\theta_{\rm o}$. 
This is due to the combined effects 
of aberration and light bending which grow greatly near the inner rim 
of the disc. In the inner region, the photons are boosted in the direction of 
rotation and they emerge along grazing angles.
Notice for the case of a higher disc inclination, $\theta_{\rm o} = 70$\,deg,
that there is a small region behind the black hole (below, in the picture)
where the local emission angle is very low, i.e. the radiation is emitted
in the direction almost perpendicular to the disc. 
This is the clear effect of the light bending.
Although the contours are most dramatically distorted near the horizon
the emission angle is visibly different from the observer inclination even quite far
from the horizon, at a distance of several tens $r_{\rm{}g}$. This is
mainly due to special-relativistic aberration which decays slowly with
the distance as the disc obeys Keplerian rotation at all radii.
Asymptotically, $\theta_{\rm e}(r,\varphi)\rightarrow \theta_{\rm o}$
for $r\rightarrow\infty$.

As emission directionality, we call the dependence
of the intensity on the emission angle.
Because of the variety of the emission angle values,
the directional distribution of the outgoing radiation is among the important
aspects that must be addressed. 
This functionality depends on how the radiation originates
and also on the surrounding conditions.
In the case
of a black hole accretion disc, the directionality of thermal
and reflection radiation may be significantly different.
In general, the directionality is a function of radius
from the black hole and energy of the emitted radiation.
In practical application, however, a unique profile 
is standardly assumed, invariably over the entire range
of radii in the disc and the energy in the spectral band.
Particularly, the limb darkening law is frequently used
since it is embedded in the currently most widely used {\laor} model (Sect~\ref{laky}).

Limb darkening traditionally refers 
to the gradual diminution of intensity in the image of the surface of 
a star as one moves from the centre of the image to the edge. It is a 
consequence of uneven angular distribution of the radiation flux emerging 
from the stellar surface \citep{1960ratr.book.....C,1978stat.book.....M}. 
Limb darkening results as a combination of two effects: 
(i)~the density of the surface layers decreases in the outward
direction, and (ii)~temperature also drops as the distance from the
centre of the star increases. The outgoing radiance is therefore
distributed in a non-uniform manner. 

The limb darkening law is widely applied also to describe radiation coming 
from an accretion disc around a black hole.
Nevertheless, the actual form of the angular
distribution depends on the physical mechanism responsible for the
emission and the geometrical proportions of the source. 
The limb darkening law is relevant for thermal radiation
of the disc with infinite optical depth. However, already
when the optical depth is assumed to be finite and also internal
heating is taken into account the emission directionality 
significantly changes \citep[see e.g.][]{2006PASJ...58.1073F}. 
The case of reflection is substantially different from the 
thermal radiation because the initial radiation
comes from above the disc and passes through the disc atmosphere.
The higher layers are characterised by higher ionisation and temperature
with the electron scattering as the dominant process there.
The observer looks deeper to the colder 
layers of the atmosphere when the emission angle is low. 
However, the radiation is there more likely to be absorbed  
resulting in the detection
of weaker reflection radiation than when the emission angle is higher
%and the observer looks at higher slabs in the same optical depth
\citep[see discussion in][Sect.~3]{2007A&A...475..155G}.
This effect is just opposite to the limb darkening, and therefore,
it is called as limb brightening.

Given the high velocity of the orbital motion (see Fig.~\ref{v_orbit}) 
and the strong-gravity light bending near the 
black hole, the effect of directional anisotropy of the local emission is enhanced.
As a result of this interplay
between the local physics of light emission and the global effects of the
gravitational field, different processes contribute to the final
(observed) directional anisotropy of the emission.
For this reason it is important to describe the angular
distribution in a correct manner; ad hoc choices of the limb-darkening
law may lead to errors in the determination of the model best fit
parameters, including the inaccuracy in the spin parameter which are
difficult to control, or they may prevent us from estimating the statistical
confidence of the model. %In the case of black hole accretion discs, this
%complication becomes important because the aberration, beaming and
%light-bending effects grow rapidly towards the inner edge of the disc.
%Therefore, even a small discrepancy between the assumed and the correct
%angular emissivity profiles becomes greatly enhanced in the observer
%frame. 

\subsection{Effects of the emission directionality on the iron line profiles}

In optics, Lambert's cosine law describes an emitter producing 
a radiation intensity that is directly proportional to $\mu_{\rm e}\equiv\cos\theta_{\rm e}$.
Lambertian surfaces exhibit the same apparent radiance when viewed from
any angle $\theta_{\rm o}$. Likewise, Lambertian scattering refers to
the situation when the surface radiates as a result of external
irradiation by a primary source and the scattered light is distributed
according to the same cosine law. This is, however, a very special
circumstance; directionality of the emergent light is sensitive to
the details of the radiation mechanism. For example, the classical
result of the Eddington approximation for stellar atmospheres states that
the effective optical depth of the continuum is $\tau=\frac{2}{3}$, and
so the emergent intensity is described by the limb-darkening law,
$I(\mu_{\rm e})\propto\mu_{\rm e}+\frac{2}{3}$. 

In the case of a fluorescence iron line produced by an illuminated plane-parallel 
slab, the angular distribution was investigated by various authors
\citep{1978ApJ...223..268B,1991MNRAS.249..352G,1993ApJ...413..680H,1994MNRAS.267..743G,2000ApJ...540..143R}.
In that case a complicated interplay arises among the angular
distribution of the primary irradiation, reflection and scattering in
the disc atmosphere. Several authors pointed out that it is essential for the reliable
determination of the model parameters to determine the angular directionality of
the broad line emission correctly. \citet{2000MNRAS.312..817M} noted, by
employing the lamp-post model, that ``{\em ...the broadening of the observed
spectral features is particularly evident when strongly anisotropic
emissivity laws, resulting from small $h$ [i.e., the lamp-post elevation
above the equatorial plane], are considered.}`` 

The important role of
the emission angular directionality was clearly spelled out by
\citet{2004MNRAS.352..353B}: ``{\em ...the angular emissivity law (limb
darkening or brightening) can make significant changes to the derived line
profiles where light bending is important``} (see their Fig.~9--13).
Similarly, \citet{2004ragt.meet...33D,2004ApJS..153..205D} and 
\citet{2005ragt.meet...29B} compared the relativistic broad lines produced
under different assumptions about the emission angular directionality. 
However, to verify the real sensitivity of the models to the
mentioned effect of directionality, it is necessary to connect the
radiative transfer computations with the spectral fitting procedure, and to
carry out a systematic analysis of the resulting spectra, taking into
account both the line and the continuum in the full relativistic regime.
We report on our results from such computations in Section~\ref{sec-titan-noar-modelling}.

\citet[][sec. 4.3]{2004MNRAS.352..205R} argue that
the combined effect of photoelectric absorption in the disc and Compton 
scattering in the corona more affect the iron line photons emerging
along grazing light rays than continuum photons. They conclude that
the line equivalent width should be diminished for observers viewing the
accretion disc at high inclination angles. Such a trend can be seen also
in the lamp-post model of \citet{1992A&A...257...63M} and \citet[][see their
Figure~11]{2000MNRAS.312..817M}. However, in the latter work this diminution
is less pronounced when we compare it with the case of intrinsically isotropic
emissivity.

More recently,
\citet{2008MNRAS.386..759N} studied the effect of different
limb darkening laws on the iron line profiles. They pointed out that the
role of emission directionality can be quite significant once the radial
emissivity of the line is fixed with sufficient confidence. However,
this is a serious assumption. In reality, the radial emissivity is not
well constrained by current models.

The angular dependence of the outgoing radiation is determined by the
whole interconnection of various effects. We describe them in more
detail below (Sec.\ \ref{sec-titan-noar-modelling}).
Briefly, the conclusion is such that realistic models require numerical
computations of the full radiative transfer.
We have developed a complete and consistent approach to such radiative
transfer computations in the context of the broad iron-line modelling {\em
together} with the underlying continuum computations. As described 
in considerable detail below, we performed the extensive
computations which are necessary in order to reliably determine the impact
of the emission angular anisotropy on spectral fitting results (namely, on
the determination of the black hole angular momentum). 
In particular, we
describe detailed results from the investigation of the model goodness
(by employing an adequate statistical analysis of the complicated $\chi^2$
parameter space). Such the analysis has not been performed so far in previous
papers because detailed stepping through the parameter space and proper
re-fitting of the model parameters was not possible due to the enormous
complexity of the models and extensive computational costs.

In regular stars and their accretion discs, the relativistic effects 
hardly affect the emerging radiation. The situation is very
different in the inner regions of a black hole accretion disc,
where the energy shift and gravitational lensing are significant.
The observed  signal can be boosted or
diminished by the Doppler effect combined with gravitational redshift:
$I(\mu_{\rm o})/I(\mu_{\rm e})\equiv g^3$ (eq.~\ref{Ig3}), where the $g$ values span 
more than a decade (see Fig.~\ref{fig2}).

\begin{figure}[tb!]
\begin{center}
\includegraphics[width=0.49\textwidth]{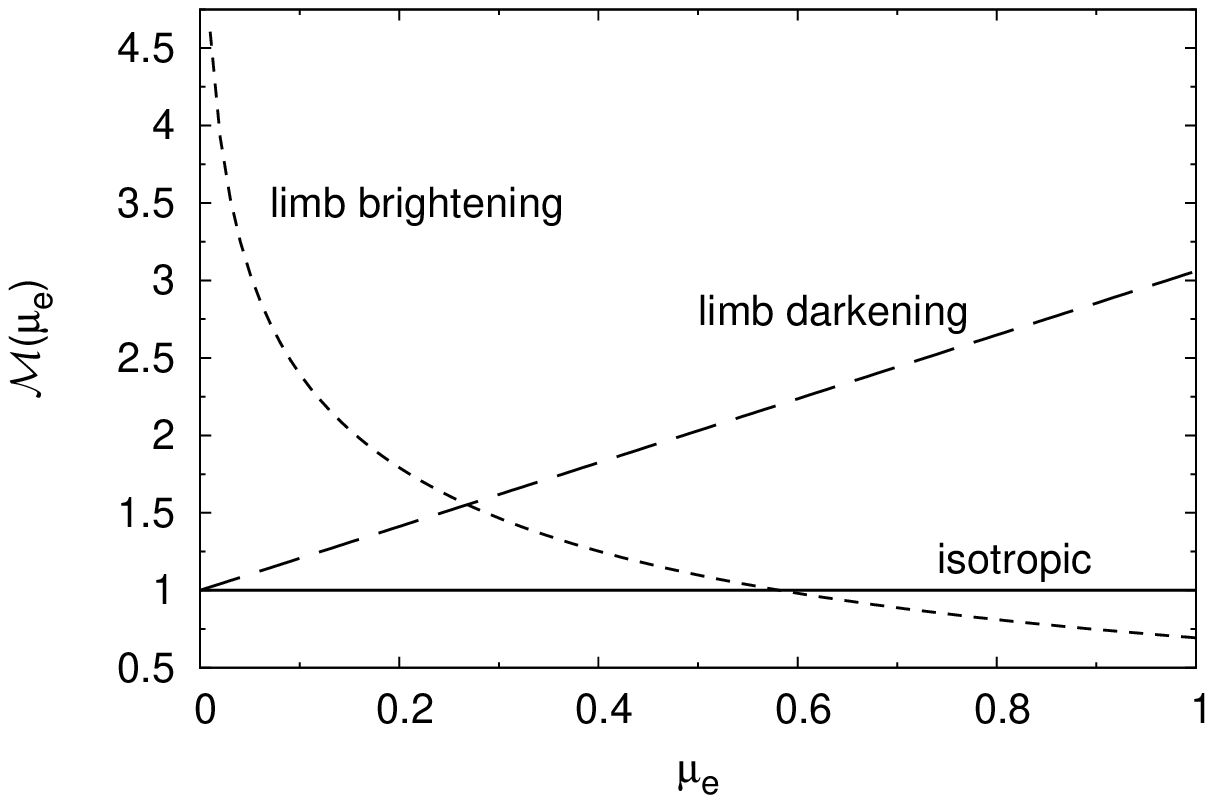}
\hfill~\includegraphics[width=0.49\textwidth]{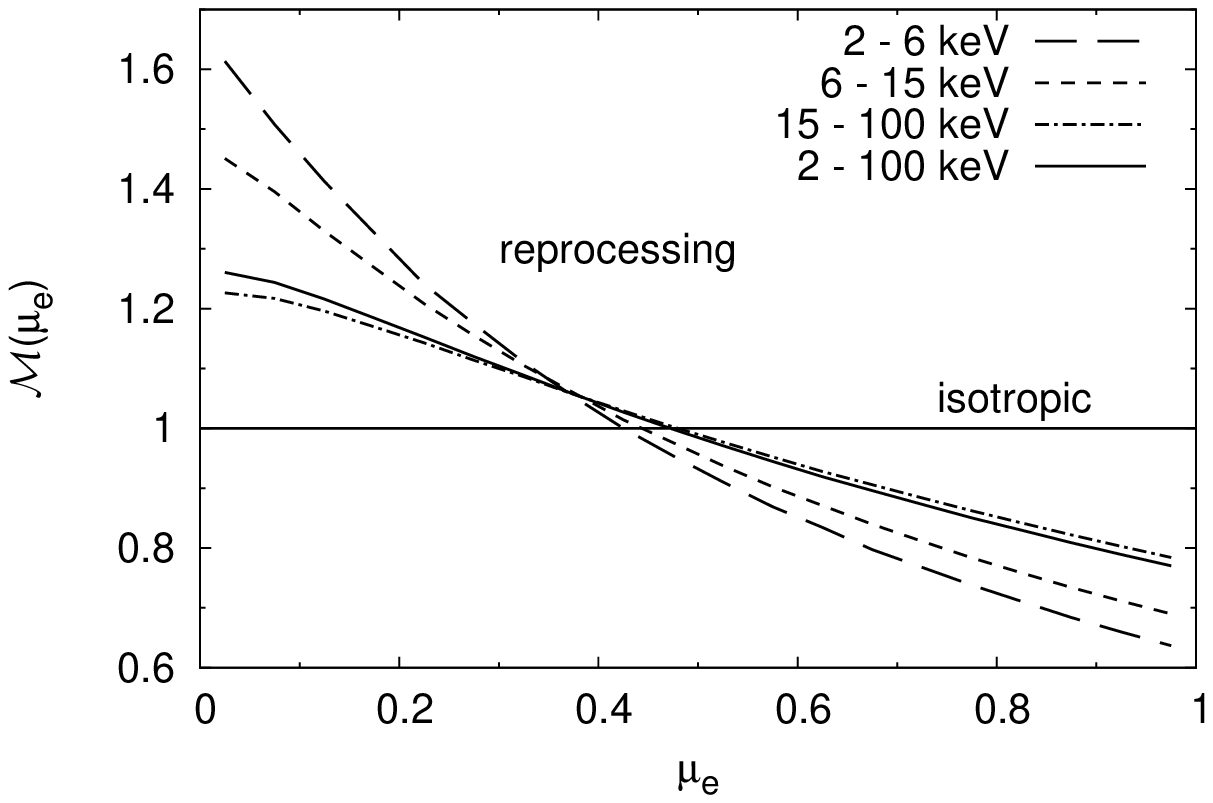}
\caption{\textposown{Left}: directional distribution of the intrinsic emissivity
(see Sec.~\ref{sec:anisotropy31} for details). \textposown{Right}: an example of the 
directional distribution from the numerical computations, showing
results from the reprocessing model with the continuum photon index $\Gamma=1.9$,
integrated over $2$--$100$ keV energy range (solid line), and in several
different energy sub-ranges (line styles are indicated in the inset). 
The latter graph demonstrates the presence of 
limb-brightening effect in comparison with respect to the isotropic emission. 
This effect is clearly visible as a function of the 
emission angle $\mu_{\rm e}$, although its magnitude is smaller than the 
limb-brightening approximation
used in the left panel. See sec.~\ref{sec-titan-noar-modelling}
for details.}
\label{fig4}
\end{center}
\end{figure}

As mentioned above, many authors have adopted the defining choice
\citep{1991ApJ...376...90L} of the cosine profile
for the line angular emissivity: ${\cal M}(\mu_{\rm e},E_{\rm
e})=  1+2.06\,\mu_{\rm e}$. This relation describes the
energy-independent limb-darkening type of profile. However, the choice
is somewhat arbitrary in the sense that the physical assumptions behind
this law are not satisfied at every radius over the entire surface of 
the accretion disc. It has been argued that the limb-darkening
characteristics need to be modified, or even replaced
by some kind of limb brightening in the case of X-ray irradiated disc
atmospheres with Compton reflection
\citep{1993ApJ...413..680H,1994MNRAS.267..743G,1994MNRAS.266..653Z}. The latter should
include the energy dependence, as the Compton reprocessing of the
reflected component plays a significant role. The angle-dependent 
computations of the Compton reflection demonstrate these effects
convincingly \citep{2004A&A...420....1C}.  Indeed, the same effect is
seen also in our computations, as shown in the right panel of Figure~\ref{fig4}.
The increase of emissivity
with the emission angle strongly depends on the ionisation state of the
reflecting material, so the actual situation can be quite complex 
\citep{2007A&A...475..155G}. 

A question arises of whether the current determinations of the black hole
angular momentum might be affected by the uncertainty in the actual
emissivity angular distribution, and to what degree. In fact, this may
be of critical concern when future high-resolution data become available
from the new generation of detectors. We have therefore carried out a
systematic investigation using the \textscown{ky} code to reveal how sensitive
the constraints on the dimensionless $a$ parameter are with respect to
the possible variations in the angular part ${\cal M}(\mu_{\rm e})$ of
the emissivity.

Figures \ref{fig2} and \ref{fig3} demonstrate the main attributes of the
photon propagation in black hole space-time relevant to our problem: the
energy shift and the direction of emission depend on the view angle of
the observer as well as on the angular momentum of the black hole.
Notice that, near the inner rim, the local emission angle is indeed
highly inclined towards the equatorial plane where at the same time the
outgoing radiation is boosted. These effects are further enhanced by
gravitational focusing, which we also take into account in our
calculations.

It should be noted that the energy shifts and emission angles
near a black hole have been
studied by a number of authors in mutually complementary ways. 
The figures shown here have been produced by plotting directly the
content of FITS format files that are encoded in the corresponding
\textscown{ky} routines (see \citet{2004astro.ph.11605D}, where an
atlas of contour plots is presented for different inclinations and spins).
Analogous figures were also shown in \cite{2004MNRAS.352..353B}, who
depicted the dependence of the cosine of the emission angle on the
energy shift of the received photon and the emission radius.

The above quoted papers concentrated mainly on the discussion of the
energy shifts and the emission directions of the individual photons,
or they isolated the role of relativistic effects on the predicted
shape of the spectral line profile. They clearly demonstrated that
the impact of relativistic effects can be very significant. However,
what is still lacking is a more systematic analysis which would
reveal how these effects, when integrated over the entire source,
influence the results of spectra fitting. To this end one needs to
perform an extensive analysis of the model spectra including the
continuum component, match the predicted spectra to the data by
appropriate spectra fitting procedures, and to investigate the
robustness of the fit by varying the model parameters and exploring
the confidence contours.

One might anticipate the directional effects of the local emission
to be quite unimportant. The argument for such an expectation
suggests that the role of directionality should grow with the source
inclination, whereas the unobscured Seyfert 1 type AGNs (where the
relativistically broadened and skewed iron line is usually expected)
are thought to have only small or moderate inclinations. However,
this qualitative trend cannot be used to quantitatively constrain
the model parameters and perform any kind of precise analysis,
needed to determine the black hole spins from current and future
high-quality data. Such an analysis has not been performed so far,
and we embark on it here for the first time.

\section{Iron K$\alpha$ line band examined with different directionalities}
%of the intrinsic emissivity}
\label{sec:anisotropy1}

\subsection{Approximations to the angular emission profile}
\label{sec:anisotropy31}
We describe the methodology which we adopted in order to explore the
effects of the spectral line emission directionality. To this end we
first employ simple approximations, neglecting any dependence on the
photon energy and the emission radius. We set the line intrinsic
emissivity from the planar disc to be described by one of the following
angular profiles,
\begin{equation}
\begin{array}{l}
\mbox{Case 1:}\quad \\ \mbox{Case 2: \rule[-1.3em]{0pt}{3em}}\quad \\ \mbox{Case 3:}\quad
\end{array}
{\cal M}(\mu_{\rm e})= \left\{
\begin{array}{l}
\ln\,(1+\mu_{\rm e}^{-1}) \quad\mbox{\citep{1993ApJ...413..680H}}\\ 1 \rule[-1.3em]{0pt}{3em} \quad \mbox{(locally isotropic emission)}\\ 1+2.06\,\mu_{\rm e}\quad\mbox{\citep{1991ApJ...376...90L}} 
\end{array}
\right.
%\label{cases}
\label{case123}
\end{equation}
%with the radial profiles being a power law ${\cal R}(r_{\rm e})=r_{\rm e}^{-q}$, 
%and ${\cal E}(E_{\rm e})=\delta(E_{\rm e}-E_0)$. 
The three cases correspond, respectively, to
the limb-brightened, isotropic, and limb-darkened angular profiles of the 
line emission. 

The limb-brightening law by \citet{1993ApJ...413..680H} describes the
angular distribution of a fluorescent iron line emerging from an
accretion disc that is irradiated by an extended X-ray source. The
relation was obtained from geometrical considerations and agrees well
with more detailed Monte-Carlo computations \citep{1991MNRAS.249..352G, 1992A&A...257...63M}. 
The physical circumstances relevant for the limb-darkening law are
different, and we include this case mainly because it is implemented
in the {\sc laor} model and frequently used in the data analysis. 
The isotropic case, dividing all limb-brightening and limb-darkening 
emissivity laws, is included in our analysis for comparison.

The radial profile of the emission is set to a unique 
power law, eq.~(\ref{rloc}), over the entire range of radii across the disc.
The directionality formula (\ref{case123}) of the intrinsic
emissivity and the resulting spectral profiles are illustrated in
Fig.~\ref{fig4} (left panel). Naturally, more elaborate and accurate approximations
have been discussed in the literature for some time. For example, \citet{1994MNRAS.267..743G}
in their eq.~(2) include higher-order terms in $\mu_{\rm e}$ to describe
the X-ray reprocessing in the single-scattering Rayleigh approximation. However,
at this stage the first-order terms are sufficient for us to demonstrate 
the differences between the three cases. Later on we will proceed towards
numerical radiation transfer computations that are necessary to derive
realistic profiles of the emission angular distribution and to keep their 
energy dependence.

\begin{figure}[tbh!]
\begin{center}
\includegraphics[width=0.45\textwidth]{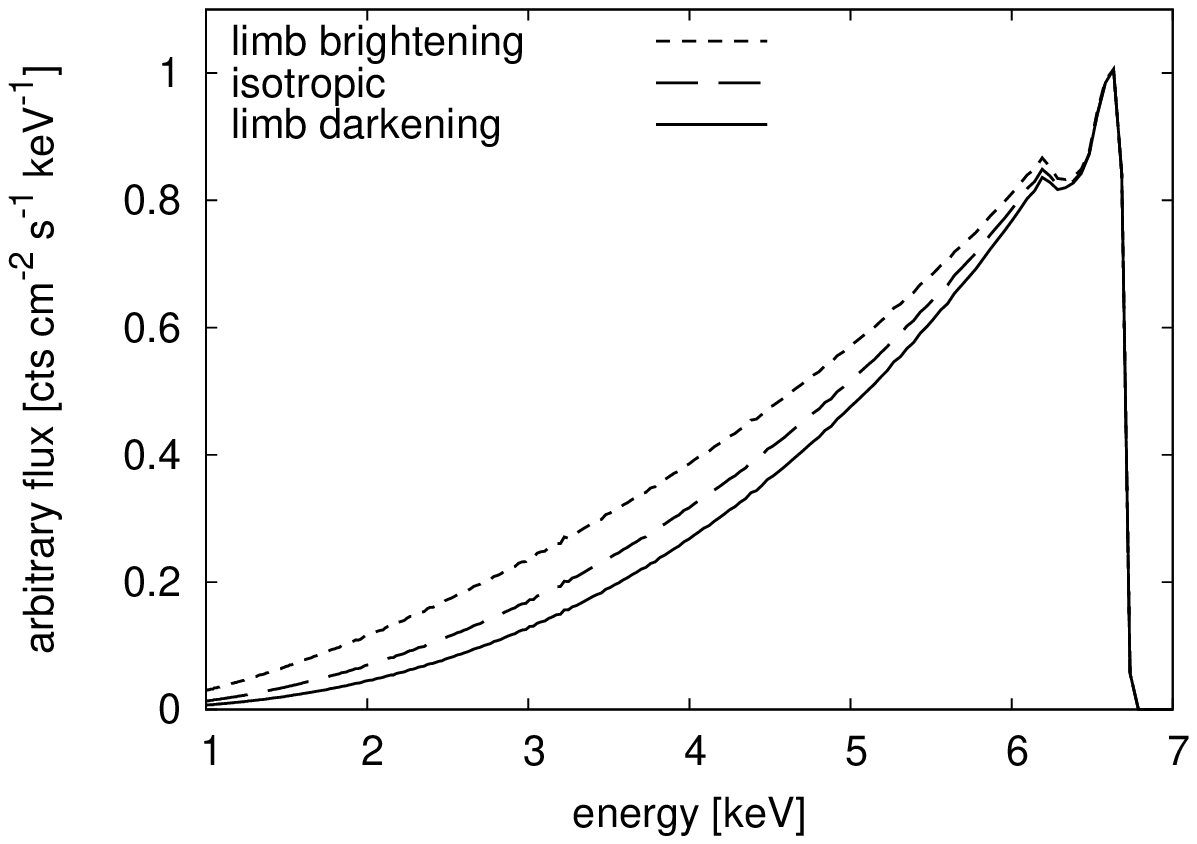}
\hfill
\includegraphics[width=0.45\textwidth]{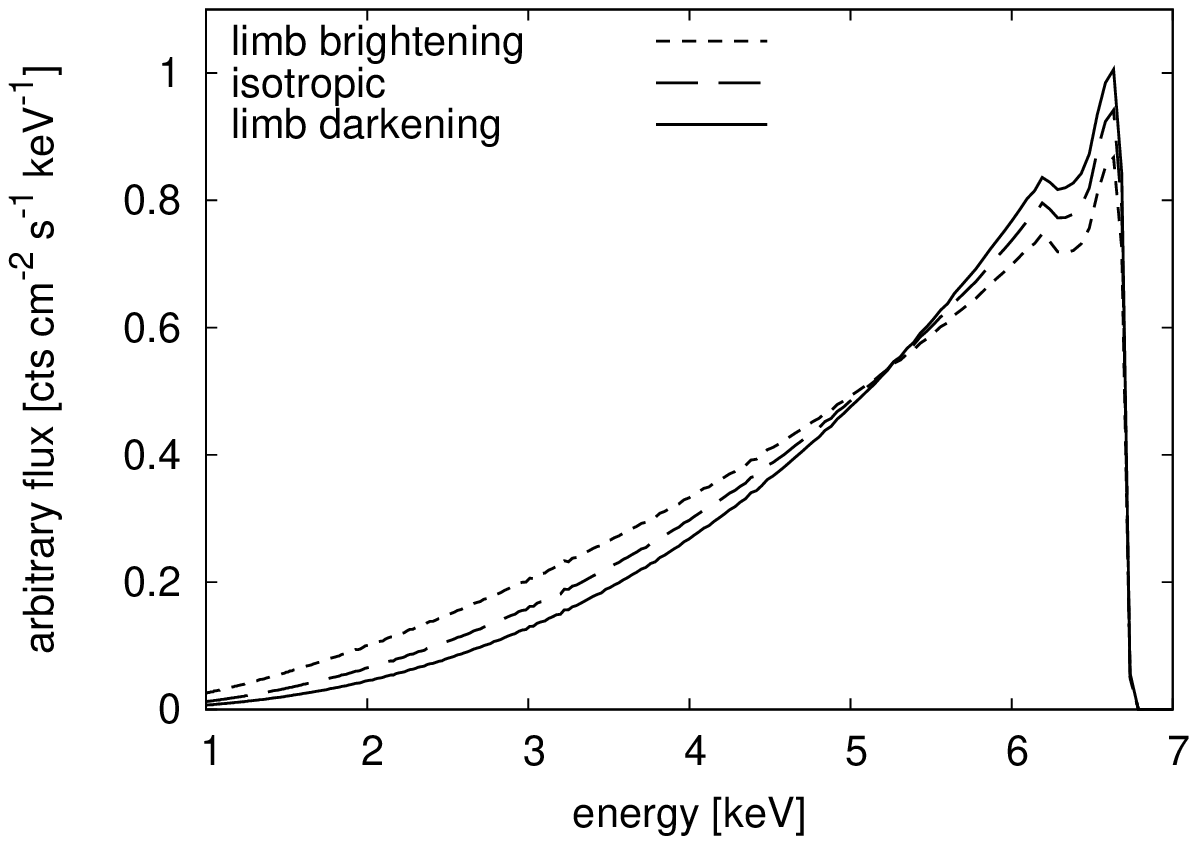}
\caption{\textposown{Left}: theoretical
profiles of the relativistic line (the \textscown{kyrline} model, without continuum),
corresponding to the three cases in the left panel of Fig.~\ref{fig4}.
The lines are normalised with respect to the height of the blue peak.
Model parameters are $a=0.9982$, $q=3$, $r_{\rm in}=r_{\rm ms}(a)=1.23$, $r_{\rm out}=400$,
$\theta_{\rm o}=30\deg$, $E_0=6.4$~keV. 
\textposown{Right}: the same as in the left panel, but with the normalisation set in such a way that
the radiation flux is identical in all three profiles.}
%In this case the most apparent
%difference is the change of the height of the blue peak.}
\label{fig4a}
\end{center}
\end{figure}

\begin{figure}[tbh!]
\begin{center}
\includegraphics[width=0.49\textwidth]{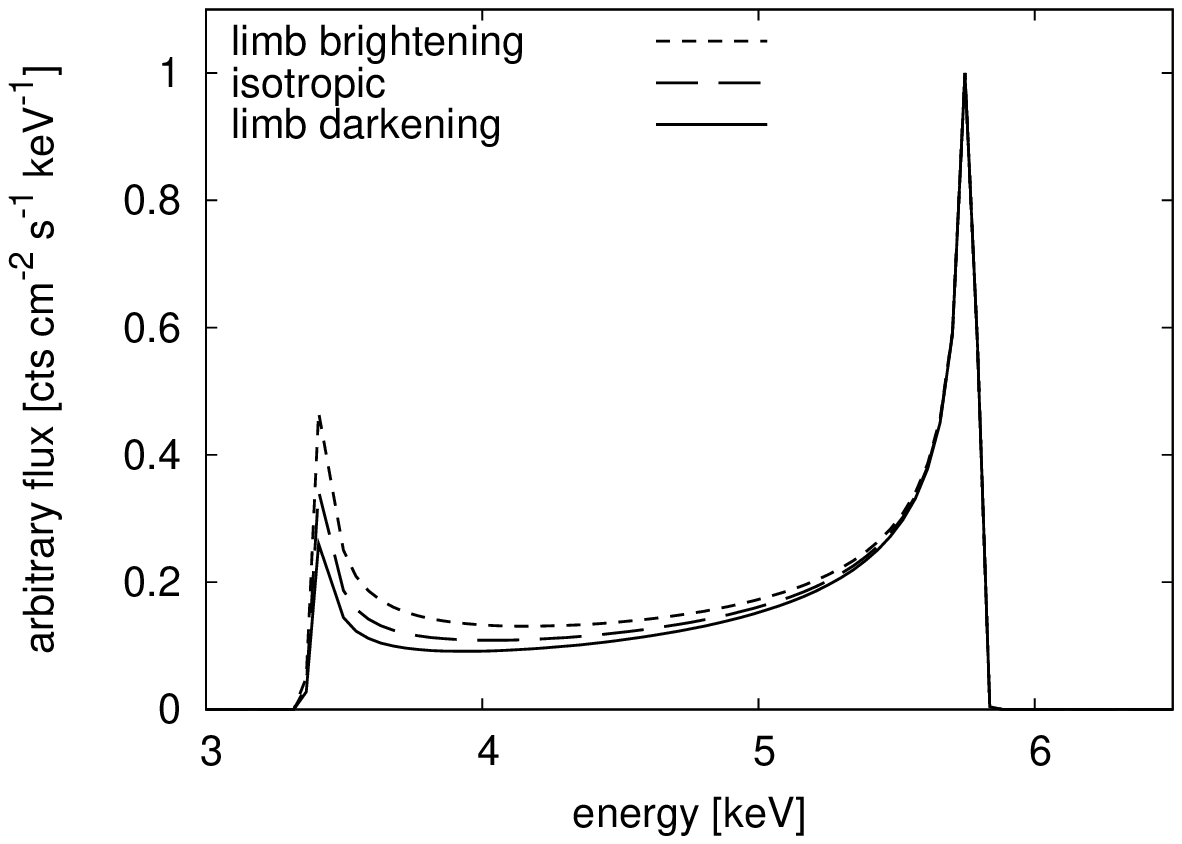}
\hfill
\includegraphics[width=0.47\textwidth]{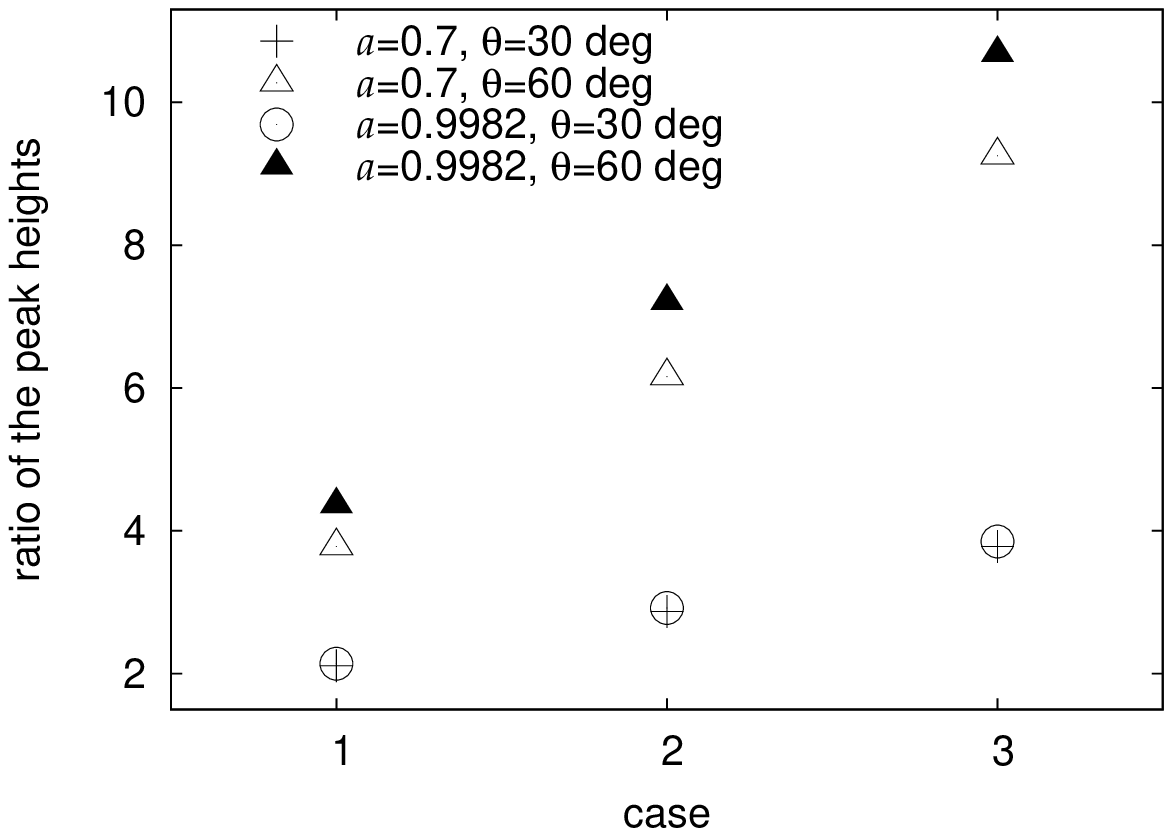}
\caption{\textposown{Left}: The same as in Fig.~\ref{fig4a}, but for a narrow ring 
with $r_{\rm in}=4.7$, $r_{\rm out}=4.8$ (other parameters are
$a=0.9982$,  $q=3$, $\theta_{\rm o}=30\deg$, $E_0=6.4$~keV). 
The lines are normalised with respect to the height of
the blue peak. \textposown{Right}: The ratio of the two peak's heights for the three
cases of the emission angular directionality according to
eq.~(\ref{case123}) and for the two values of the spin and the inclination angle
(crosses and circles are for $\theta_{\rm o}=30\deg$; empty and filled
triangles are for $\theta_{\rm o}=60\deg$).} 
\label{fig4b}
\end{center}
\end{figure}

\begin{figure}[tbh!]
\begin{center}
\includegraphics[width=0.49\textwidth]{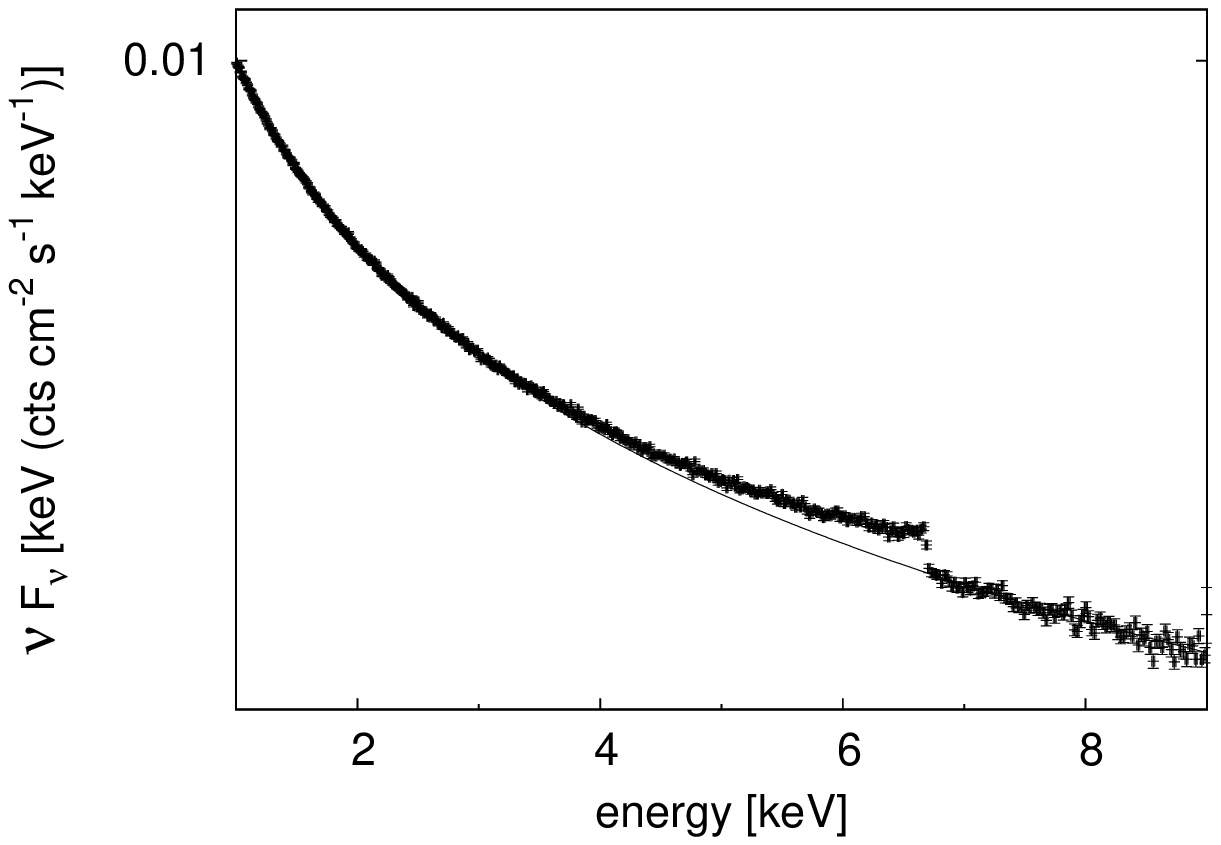}
\hfill
\includegraphics[width=0.49\textwidth]{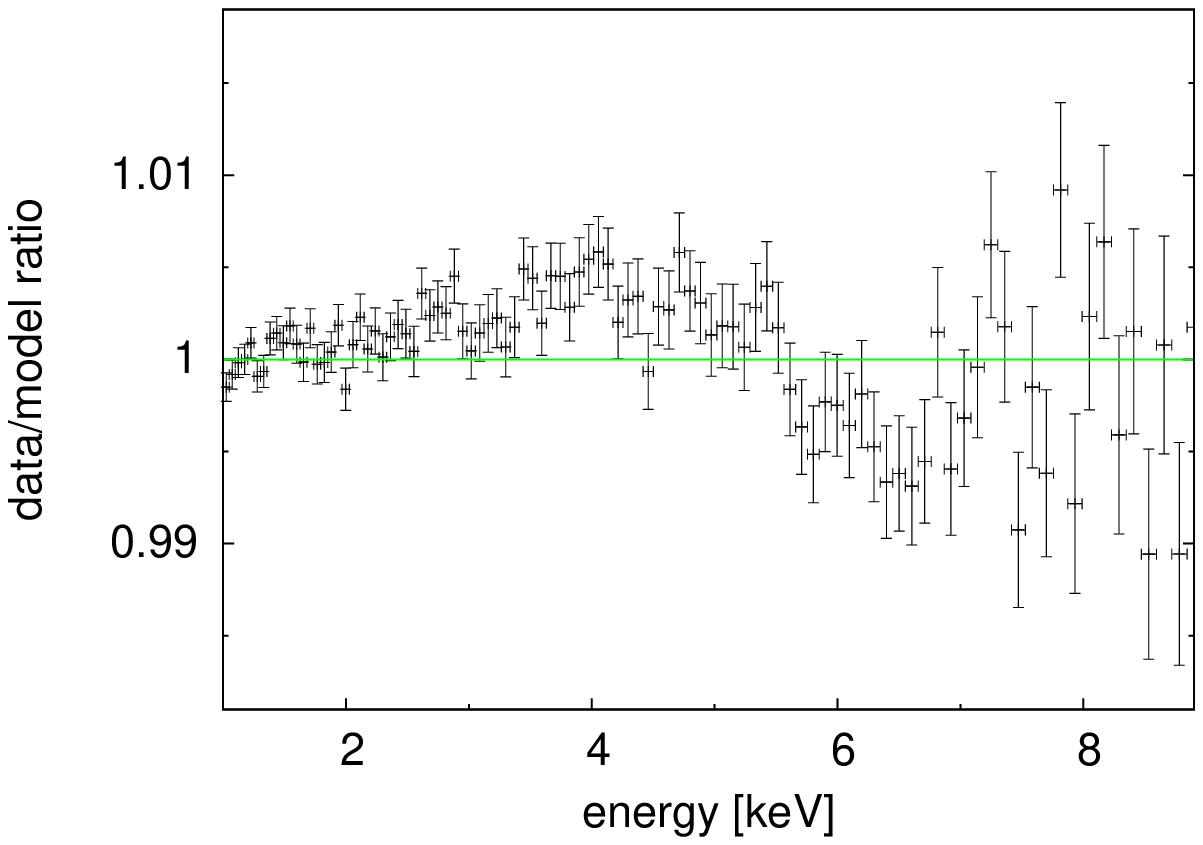}
\caption{\textposown{Left}: Unfolded spectrum generated by the 
\textscown{powerlaw + kyrline} model with parameters of the line identical as
in the Figure~\ref{fig4a}. We assumed the limb-brightening profile, i.e.\ Case~1
of eq.~(\ref{case123}). \textposown{Right}: Ratio of the 
simulated data on the left to
the same model but with the limb-darkening profile (Case~3).
The normalisation of the line was allowed to vary during the fit. The
$\chi^{2}_{\rm red}$ changed from $1.04$ to $1.30$ between the two cases.
The details about the simulation of the data points are the same as described
in the section \ref{simulated}.}
\label{fig5}
\end{center}
\end{figure}

The dependence of the spectral profiles on the angular
distribution ${\cal M}(\mu_{\rm e})$ of the 
intrinsic emission is shown in figures \ref{fig4a}--{\ref{fig4b}.
It is apparent from Figure~\ref{fig4a} that the limb brightening case 
makes the line profile broader (in the left panel)
and the height of the blue peak lower (in the right panel) 
than the limb darkening case for the same set of parameters, 
which can consequently lead to discrepant evaluation of the spin.
While Figure~\ref{fig4a} shows the line profile for an extended
disc, Figure~\ref{fig4b} deals with a narrow ring. In this
case, a typical double-horn profile develops. Although the energy of the
peaks is almost entirely insensitive to the emission angular directionality,
the peak heights are influenced by the adopted limb darkening law. 
In the right panel of Figure~\ref{fig4b}, the ratio of the two peaks
is shown for the three cases of the emission angular directionality according to
eq.~(\ref{case123}). The influence of the directionality is apparent 
and it is comparable to the effect of the inclination angle. 
The spin value has only little impact for large inclinations.

In general, we notice that the extension and prominence of the
red wing of the line are indeed related to the intrinsic emission
directionality. This was examined in detail by \citet{2004MNRAS.352..353B},
who plotted the expected broad-line profiles for different angular emissivity
and explored how the red wing of the line changes as a result
of this undetermined angular distribution.
Similarly, the inferred spin of the black hole
must depend on the assumed profile to a certain degree.
This is especially so because the spin is determined by the extremal
redshift of the red wing of the line (see Section~\ref{rellinemod}). 

However, simple arguments are
insufficient to assess how inaccurate the spin determination might be
in realistic situations, as
the spectral fitting procedure involves several components extending
over a range of energy above and below the iron-line band.
We illustrate this in Figure~\ref{fig5}, noticing that
the different prescriptions produce very similar results outside the line
energy, but they do differ at the broad line energy range. 
The theoretical (background-subtracted) profiles of the relativistic line
cannot alone be used to make any firm conclusion about the error of the best
fit parameters that could result from the poorly known angular emissivity.
To this end one has to study a consistent model of the full spectrum.
With a real observation, the sensitivity to the
problem of directionality (as well as any other uncertainty inherent in
the theoretical model) will depend also on the achieved resolution, 
energy binning and the error bars of the data.

\subsection{Example: directionality effects in MCG\,-6-30-15}
\label{direxammcg}

\begin{figure}[tbh!]
\begin{center}
\includegraphics[width=0.49\textwidth]{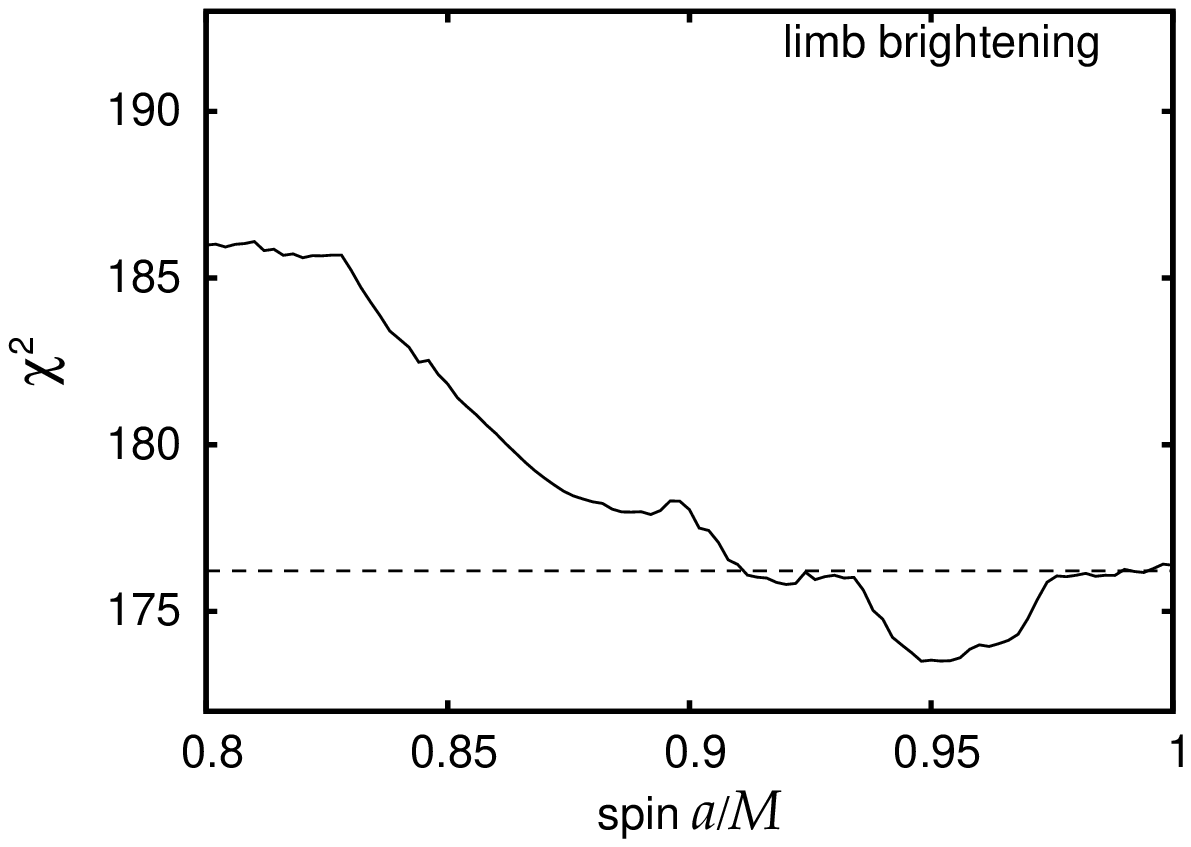}
\hfill
\includegraphics[width=0.49\textwidth]{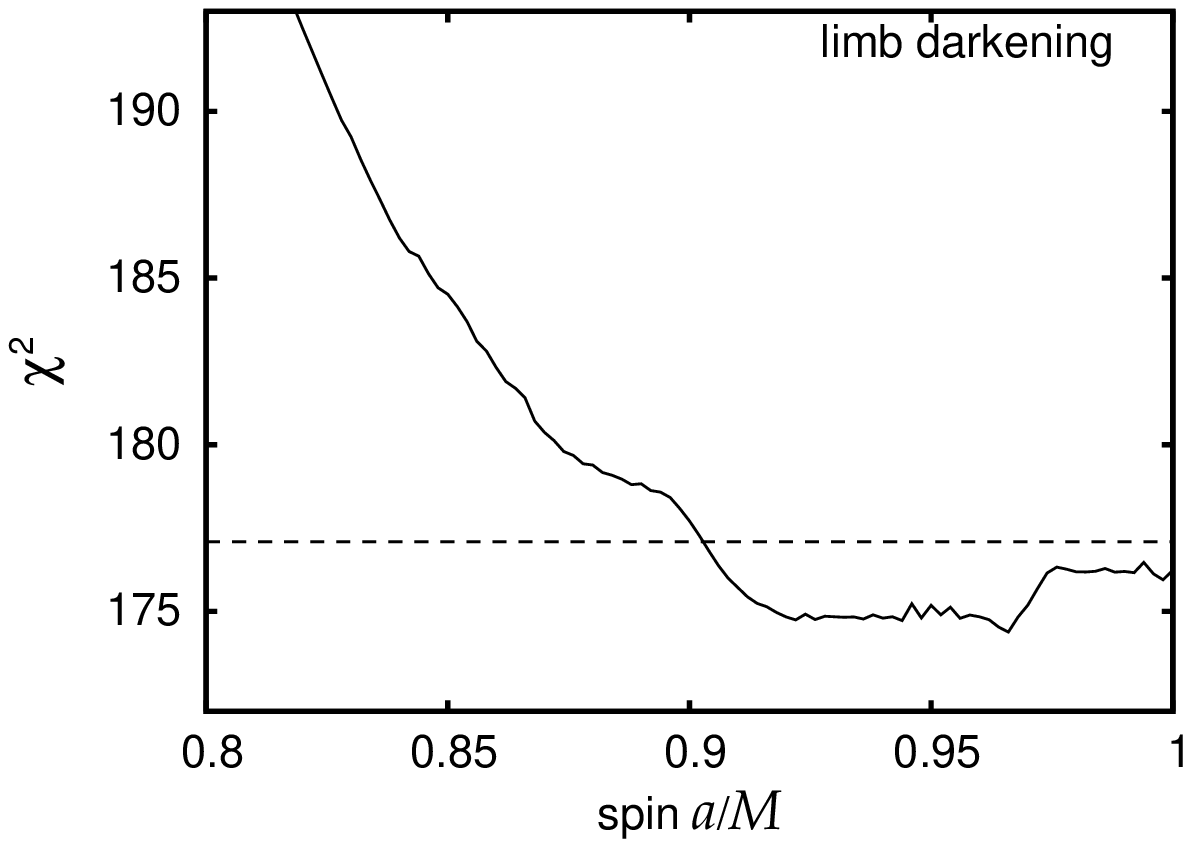}
\caption{The best-fit values of $\chi^2$ statistics for the spin parameter,
which we obtained by gradually stepping $a$ from $0.8$ to $1$.
XMM-Newton data for MCG\,-6-30-15 were employed 
\citep{2002MNRAS.335L...1F}. \textposown{Left}: the limb-brightening
profile (Case~1). \textposown{Right}: the limb darkening profile (Case~3).
The dashed line is the 90\% confidence level
(2 sigma). See the text for a detailed description of the model.}
\label{fig6}
\end{center}
\end{figure}

We re-analysed a long XMM-Newton observation of a nearby Seyfert 1 galaxy MCG\,-6-30-15 
%\citep{2002MNRAS.335L...1F}
to test whether the different directionality approximations can be distinguished 
in the current data. The observation took place in summer 2001 and the acquired exposure time 
was about 350\,ks. 
We reduced the EPN data from three sequential revolutions (301, 302, 303). 
The details of the reduction of the data are described in Section~\ref{mcg}.

We applied the same continuum model
as presented in \cite{2002MNRAS.335L...1F}: the power law component
(photon index $\Gamma =1.9$) absorbed by Galactic gas 
(column density $n_{\rm H} = 0.41\times 10^{21}$\,cm$^{-2}$).
This simple model is sufficient to fit the data above $\approx 2.5$\,keV, 
which is also satisfactory for our goal of reproducing the overall shape of
the broad iron line.
Other components need to be added to the model in order to fully understand 
the spectrum formation and to decide between viable alternatives, including
the presence of outflows and a combination of absorption and reflection effects
\citep{2008MNRAS.483..437,2009A&ARv..17...47T}.
However, our goal here is not to give precedence to any of the particular schemes.
Instead, we rely on the model of a broad line and we test
and compare different angular laws for the emission. 

The residuals from the simple power law model can be explained 
by a complex of a broad line and two narrow lines.
Due to the complexity of the model, it is hard to
distinguish between the narrow absorption line 
at $E\approx6.75$\,keV and the emission line at 
$E\approx6.97$\,keV \citep{2002MNRAS.335L...1F}.
Although the level of $\chi^{2}$ values stays almost the same,
the parameters of the broad iron line do depend on the
continuum model and the presence of narrow lines
(see Section~\ref{mcg}).
By adding an absorption line at 
$E\approx6.75$\,keV, the best fit rest energy of the broad 
component comes out to be $E=6.7$\,keV, and so the broad
line originates in a moderately ionised disc.
This result is consistent with \citet{2003MNRAS.342..239B}.
%but it is rather in the contrast with the analysis 
%of \citet{2002MNRAS.335L...1F} where the rest energy 
%of the broad line was kept fixed to $E=6.4$\,keV.

Figure~\ref{fig6} shows the results of the one-dimensional {\sl steppar} command
performed in XSPEC, 
which demonstrates the expected confidence of $a$-parameter best-fit values
for the two extreme cases of directionality, Case~1 and Case~3.
The model used in the XSPEC syntax is: 
\textscown{phabs}*(\textscown{powerlaw} + \textscown{zgauss} + \textscown{zgauss} + \textscown{kyrline}).
The fixed parameters of the model are the column density 
$n_{\rm H} = 0.41\times 10^{21}$\,cm$^{-2}$, %(from catalogue, frozen during fitting procedure)
the photon index of power law $\Gamma =1.9$, the redshift factor $z = 0.008$, 
the energy of the narrow emission line $E_{\rm em} = 6.4$\,keV,
and the energy of the narrow absorption line $E_{\rm abs} = 6.77$\,keV.
The parameters of the broad iron line were allowed to vary during the fitting procedure. 
Their default values were $E_{\rm broad} = 6.7$\,keV for the energy of the broad iron line,
$\theta_{\rm o} = 30$\,deg for the emission angle, $q_{1} = 4.5$, $q_{2} = 2$ and 
$r_{\rm b} = 10$\,$r_{\rm g}$ for the radial dependence of the emissivity
(the radial part of the intensity needs to be rather complicated
to fit the data and can be expressed as a broken power law:
${\cal R}(r_{\rm e})=r_{\rm e}^{-q_1}$ for $r_{\rm e} < r_{\rm b}$, 
and ${\cal R}(r_{\rm e})=r_{\rm e}^{-q_2}$ for $r_{\rm e} > r_{\rm b}$).

The determined best-fit values for the spin are virtually the same for both cases, 
independent of the details of the limb-brightening/darkening profile. 
However, this result arises on account of the growing complexity of the model.
The differences between the two cases become hidden in different 
values of the other parameters -- especially in $q_1$, $q_2$ and $r_{\rm b}$, 
i.e.\ the parameters characterising the radial dependence of the line
emissivity in \textscown{kyrline} as a broken power law with a break radius $r_{\rm b}$.
We find: (i) $E_{\rm broad} = 6.60(1)$, $\theta_{\rm o} = 31.5(7)$\,deg,
$q_1 = 3.7(1)$, $q_2 = 2.1(1)$, $r_{\rm b} = 18(1)$\,$r_{\rm g}$ for Case~1;
and (ii) $E_{\rm broad} = 6.67(1)$\,keV, $\theta_{\rm o} = 26.7(7)$\,deg,
$q_1 = 5.3(1)$, $q_2 = 2.8(1)$, $r_{\rm b} = 4.9(2)$\,$r_{\rm g}$ for Case~3.
The errors in brackets are evaluated as the 90\% confidence region 
for a single interesting parameter when the values of the other parameters
are fixed.
The combination of three parameters $q_1$, $q_2$, $r_{\rm b}$ thus adjusts 
the best fit in XSPEC. Nonetheless, the clear differences between 
the models occur consistently with theoretical expectations: 
for Case~3 the lower values of the spin,
$a<0.87$, produce larger $\chi^2$  and the best-fit spin can
reach the extreme value within the 90\% confidence threshold.

\subsection{Analysis of simulated data for next generation X-ray missions}
\label{simulated}

\begin{table}
\begin{center}
\caption{The best-fit spin values inferred for the three cases of the
limb darkening/brightening law, eq.~(\ref{case123}), for the \textscown{kyrline} model.}
\vspace{0.2cm}
\begin{tabular}{c|cc|cc}
\hline \hline
\rule[-0.6em]{0pt}{2em}Case & \multicolumn{2}{c|}{$a_{\rm f}=0.7$} & \multicolumn{2}{c}{$a_{\rm f}=0.9982$} \\ 
\rule[-0.6em]{0pt}{2em}no. & $\theta_{\rm f}=30^{\circ}$ & $\theta_{\rm f}=60^{\circ}$ & $\theta_{\rm f}=30^{\circ}$ & $\theta_{\rm f}=60^{\circ}$\\ 
\hline
\rule{0pt}{2em} 1 & $0.56^{+0.04}_{-0.03}$ & $0.69^{+0.03}_{-0.04}$ & $0.92^{+0.03}_{-0.03}$ & $0.981^{+0.013}_{-0.031}$ \\
\rule{0pt}{2em} 2 & $0.66^{+0.05}_{-0.05}$ & $0.70^{+0.02}_{-0.04}$ & $\geq0.966$ & $\geq0.986$\\
\rule[-1em]{0pt}{3em} 3 & $0.74^{+0.05}_{-0.03}$ & $0.70^{+0.03}_{-0.03}$ & $\geq0.991$ & $\geq0.993$\\ 
\end{tabular}
\label{tab1}
\end{center}

The artificial data were generated using the \textscown{kyrline} model with isotropic 
directionality and the fiducial values of parameters (denoted by the subscript ``f'').
See the main text for details.
\end{table}

In order to evaluate the feasibility of determining the spin of a rotating black hole
and to assess the expected constraints from future X-ray data, we produced a set of
artificial spectra. We used a simple model prescription and preliminary
response matrices for the International X-ray Observatory (IXO) mission.\footnote{We used 
the current version of provisional 
response matrices available at {\sf http://ixo.gsfc.nasa.gov/science/responseMatrices.html}
for the glass core calorimeter, dated October 30, 2008.
We used the energy resolution of 5\,eV per bin.} 
Here we limit the energy band in the range $2.5$--$10$~keV. The adopted model
consists of \textscown{powerlaw} for continuum (photon index $\Gamma$ and the
corresponding normalisation $K_\Gamma$), plus \textscown{kyrline} model
\citep{2004ApJS..153..205D} for the broad line. 
The normalisation factors of the model were chosen in such a way that
the model flux matches the flux of MCG\,-6-30-15.
In this section, the simulated flux is around $3.1\times10^{-11}$\,erg\,cm$^{-2}$\,s$^{-1}$
with about 3\% of the flux linked to the broad iron line component.
The simulated exposure time was 100\,ks.

%\footnote{We used XSPEC, ver. 12.2. See {\sf http://heasarc.gsfc.nasa.gov}.}

\begin{figure}[tb!]
\begin{center}
\begin{tabular}{ccc}
%  \multicolumn{3}{c}{\textbf{a=0.6}}\\
  \includegraphics[width=0.315\textwidth]{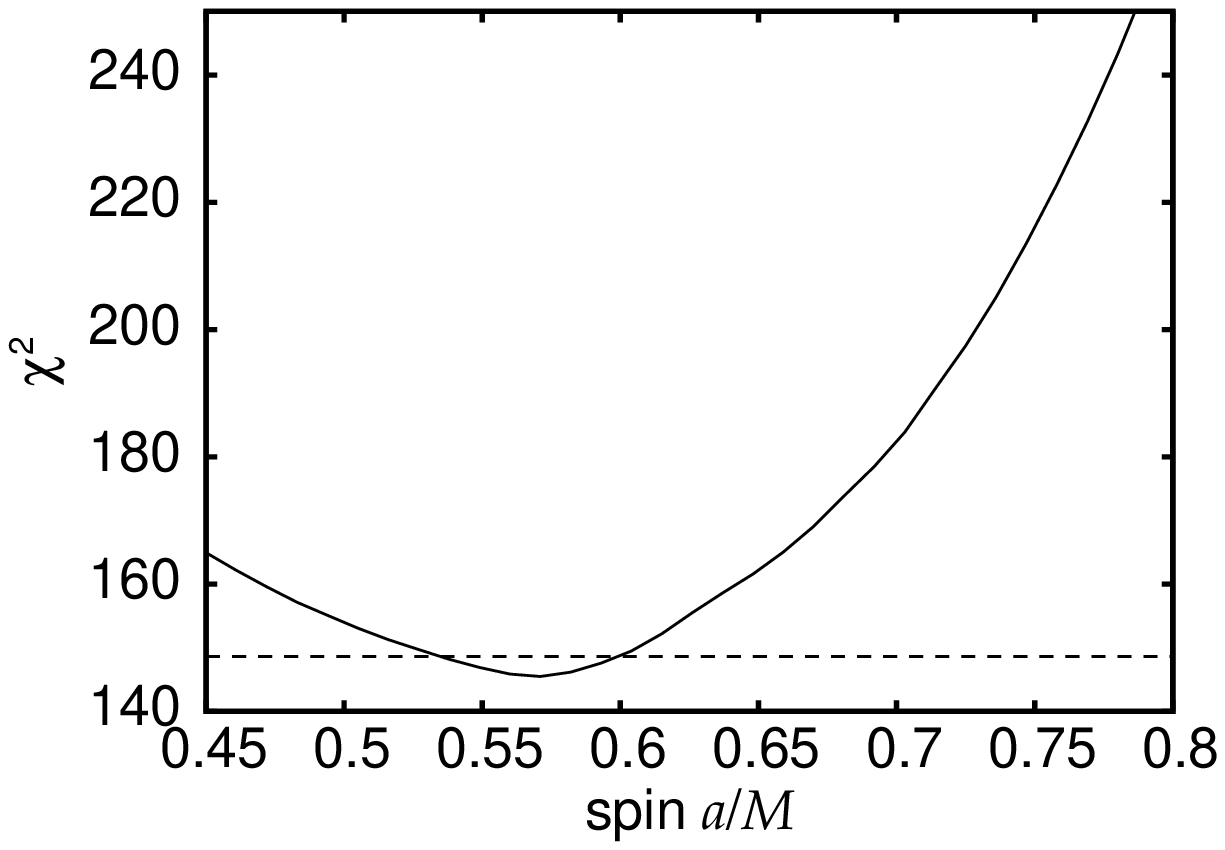} &
  \includegraphics[width=0.315\textwidth]{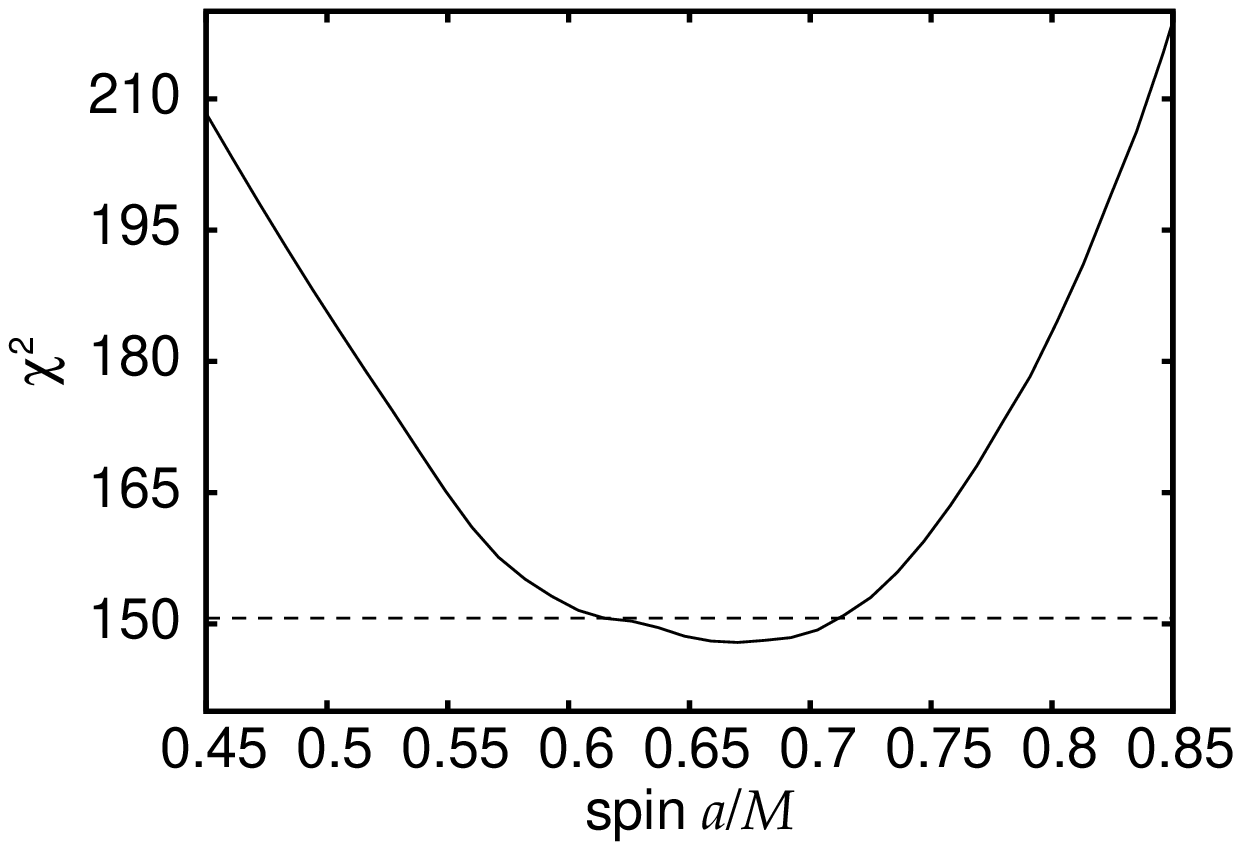} &
  \includegraphics[width=0.315\textwidth]{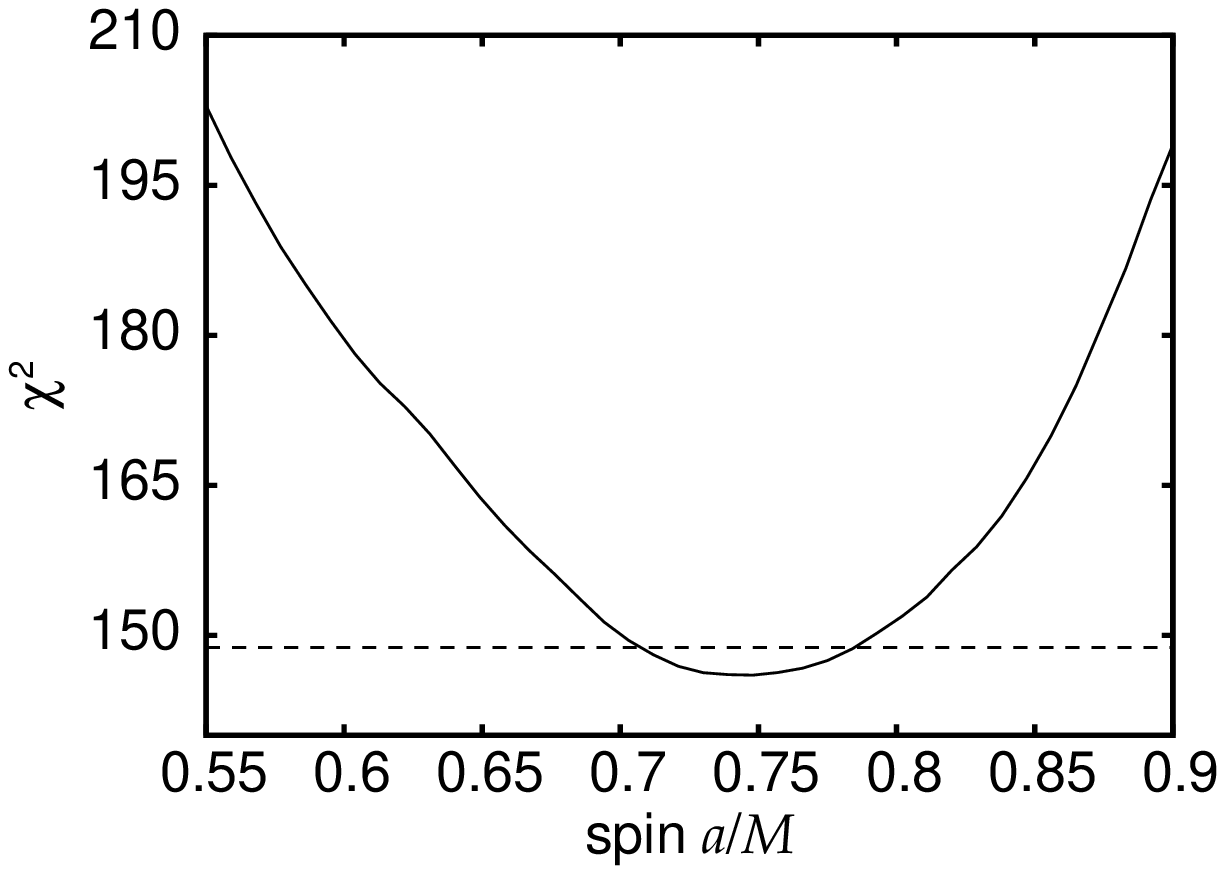}\\
\end{tabular}
\begin{tabular}{ccc}
  \includegraphics[width=0.315\textwidth]{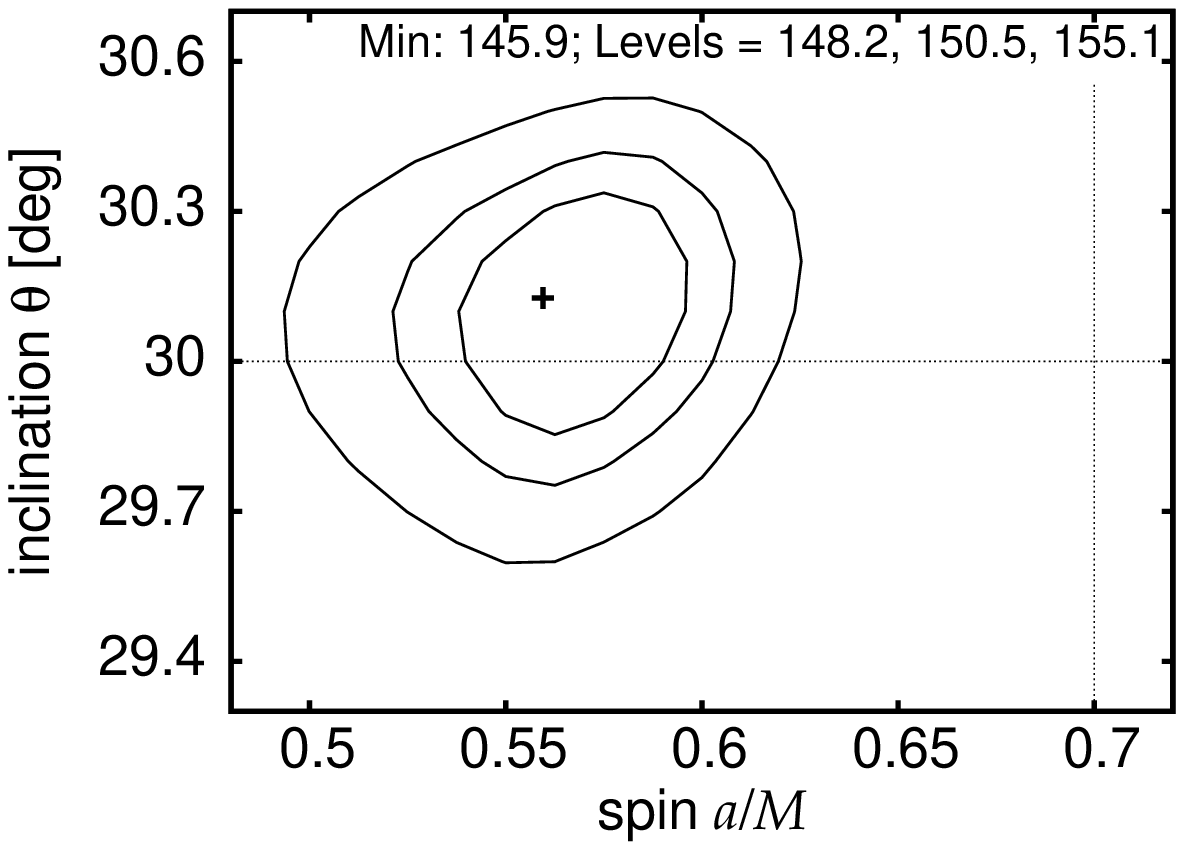} &
  \includegraphics[width=0.315\textwidth]{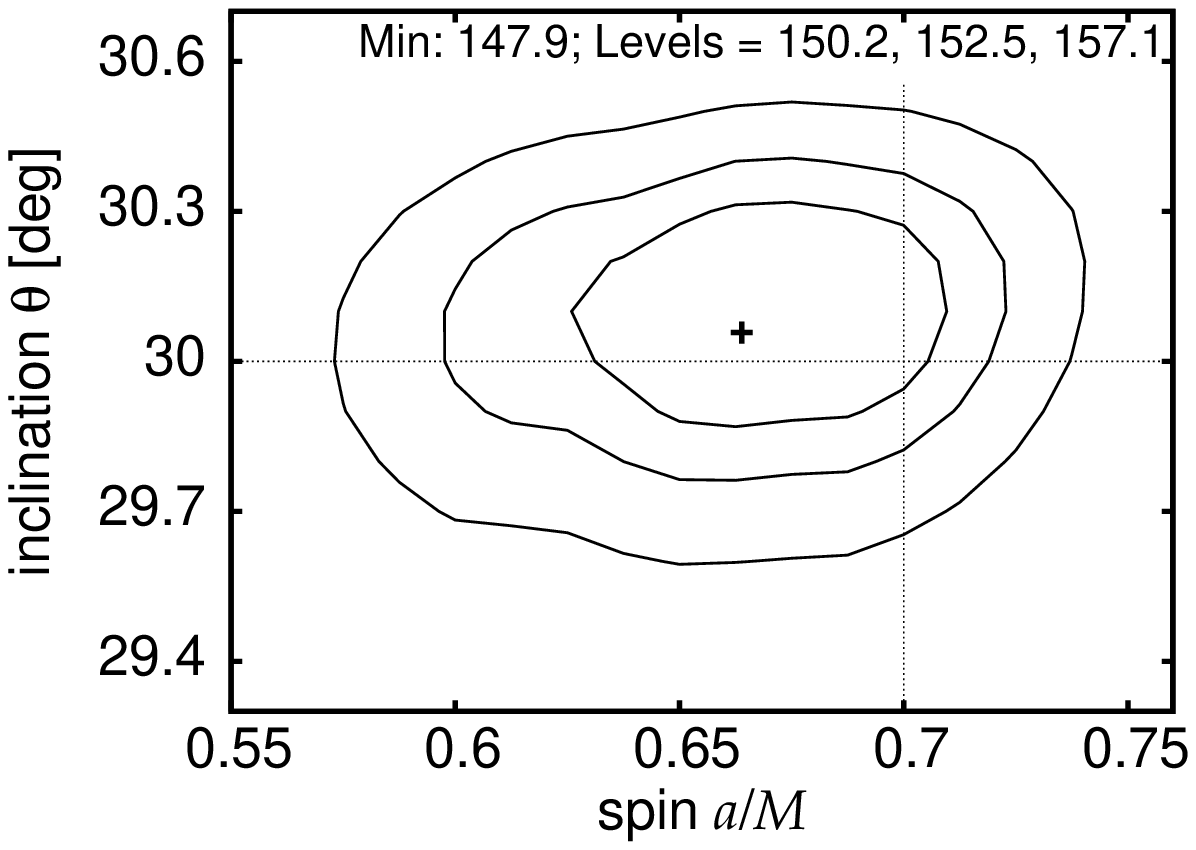} &
  \includegraphics[width=0.315\textwidth]{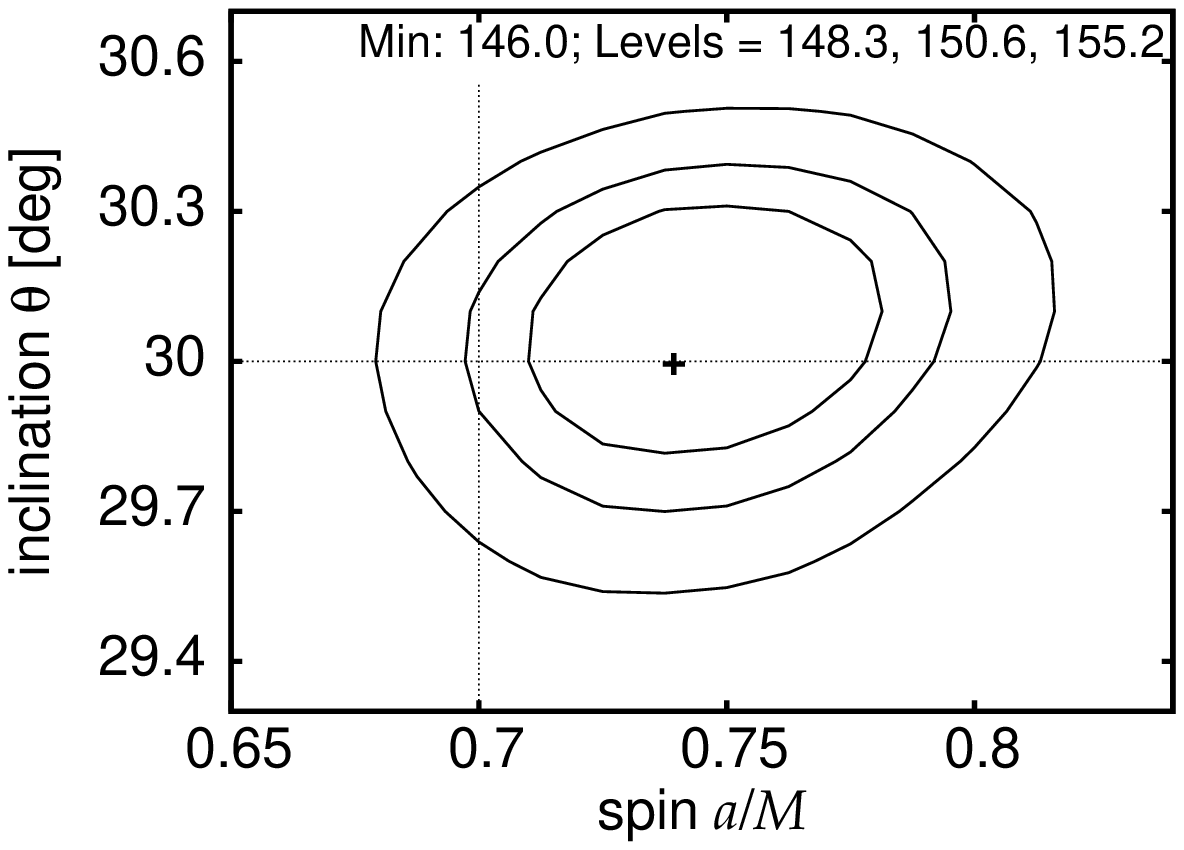}\\
\end{tabular}
\caption{Test results of the theoretical fits with 
$a_{\rm f}=0.7$, $\theta_{\rm f}=30\deg$ and three different profiles
of the emission directionality - \textposown{left}: limb brightening (Case~1 in eq.~\ref{case123}), 
\textposown{middle}: isotropic (Case~2), \textposown{right}: limb darkening (Case~3). 
The simulated data were generated using the 
\textscown{powerlaw} + \textscown{kyrline} model
with isotropic directionality.
\textposown{Top}: dependence of the best fit $\chi^2$ values on the spin value.
The horizontal (dashed) line represents the 90\% confidence level.
\textposown{Bottom}: contour graphs of $a$ versus $\theta_{\rm o}$. The contour lines correspond to $1$, $2$,
and $3$ sigma. The position of the minimal value of $\chi^{2}$ is marked
with a small cross. The values of $\chi^{2}$ corresponding to the minimum
and to the contour levels are shown at the top of each contour graph.
The large cross indicates the position of the fiducial values of 
the angular momentum and the emission angle.
Besides the values of spin, inclination and normalisation constants of the model, 
the other parameters were kept fixed at their default values: 
$\Gamma=1.9$, $E_0=6.4$~keV, $r_{\rm in}=r_{\rm ms}$, $r_{\rm out}=400$,  $q=3$, 
(default values of normalisation constants: $K_\Gamma=10^{-2}$, $K_{\rm line}=10^{-4}$).}
\label{fig7}
\end{center}
\end{figure}

\begin{figure}[tbh!]
\begin{center}
\begin{tabular}{ccc}
%  \multicolumn{3}{c}{\textbf{a=0.6}}\\
  \includegraphics[width=0.315\textwidth]{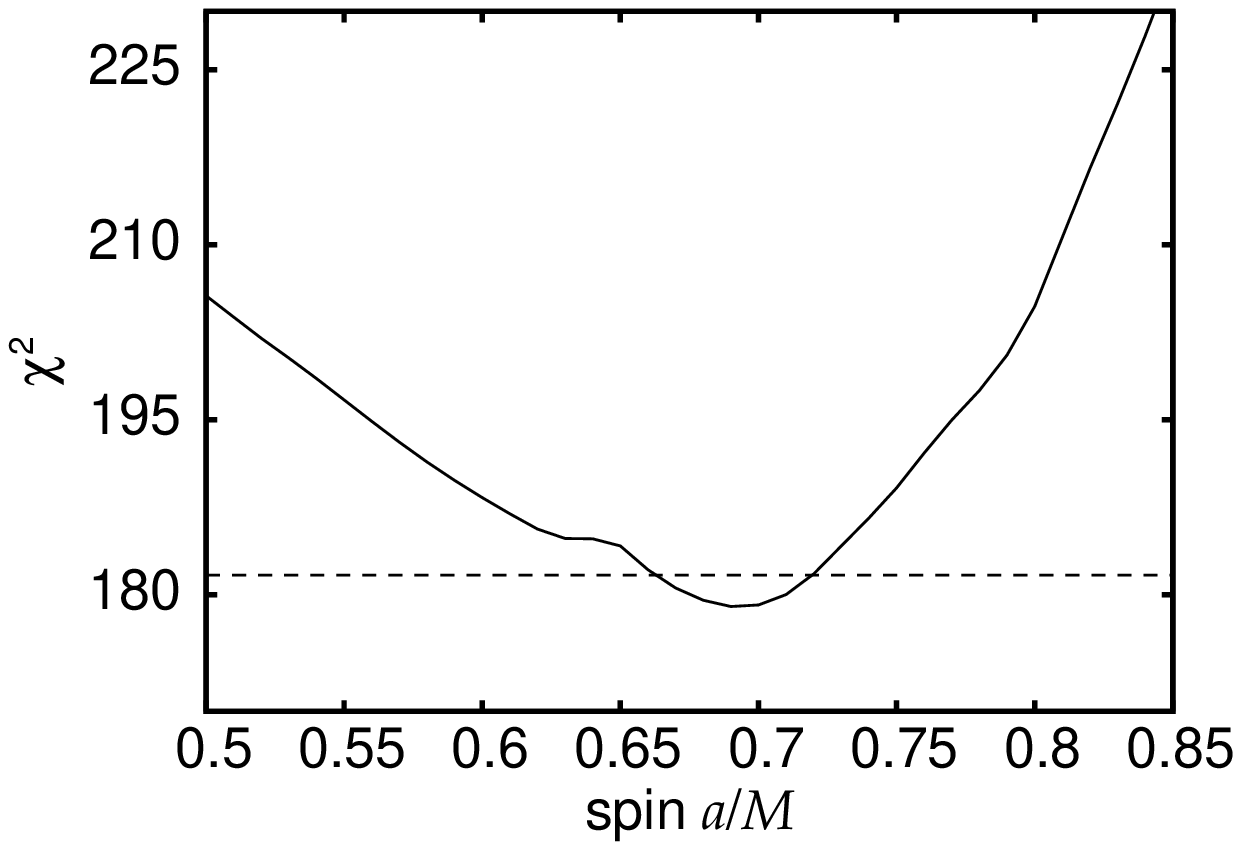} &
  \includegraphics[width=0.315\textwidth]{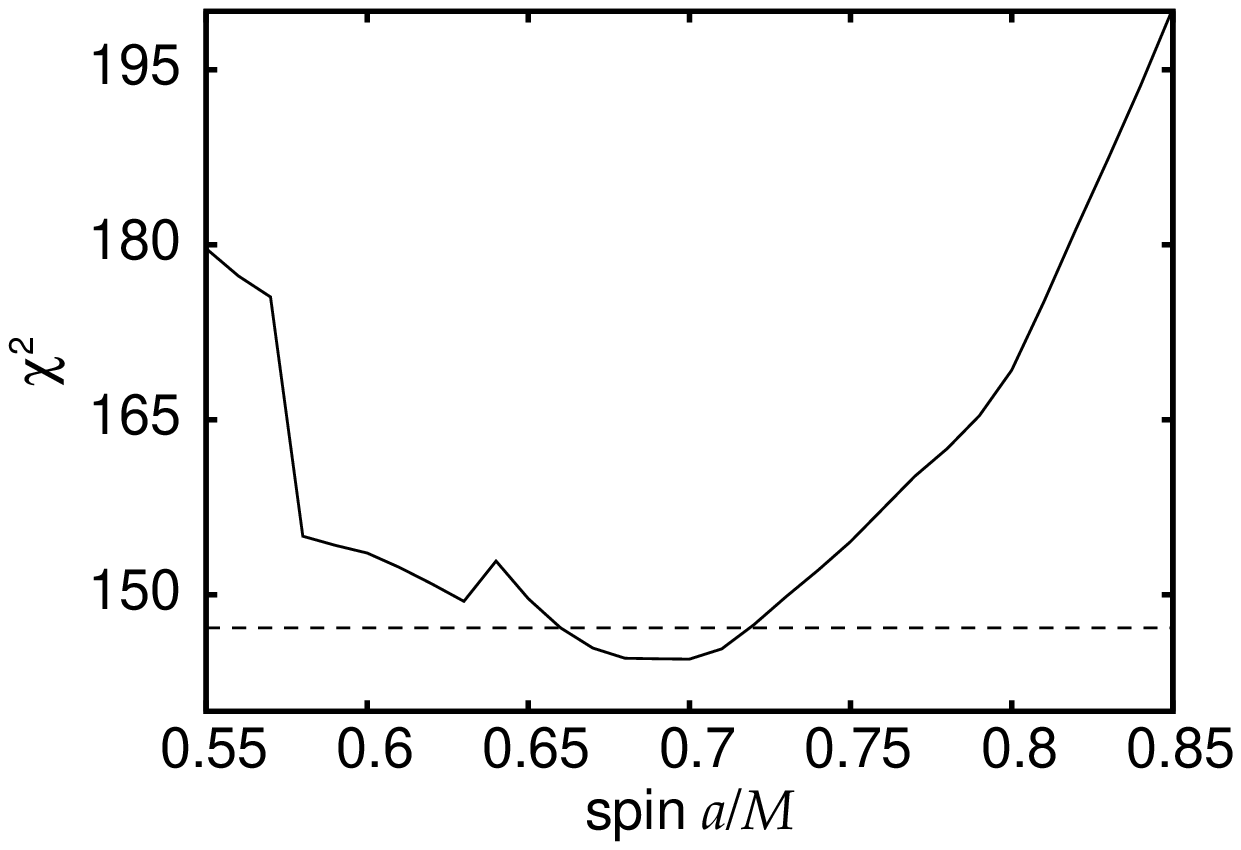} &
  \includegraphics[width=0.315\textwidth]{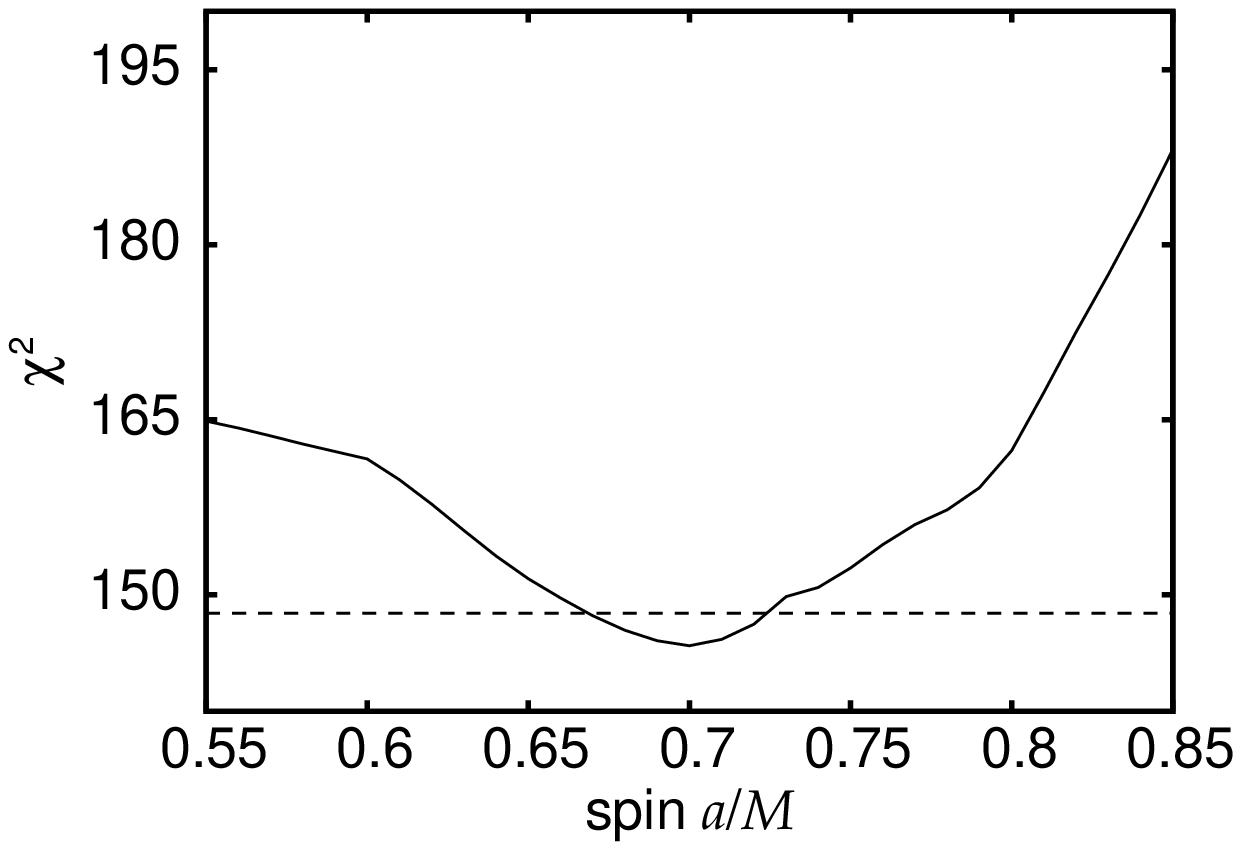}\\
\end{tabular}
\begin{tabular}{ccc}
  \includegraphics[width=0.315\textwidth]{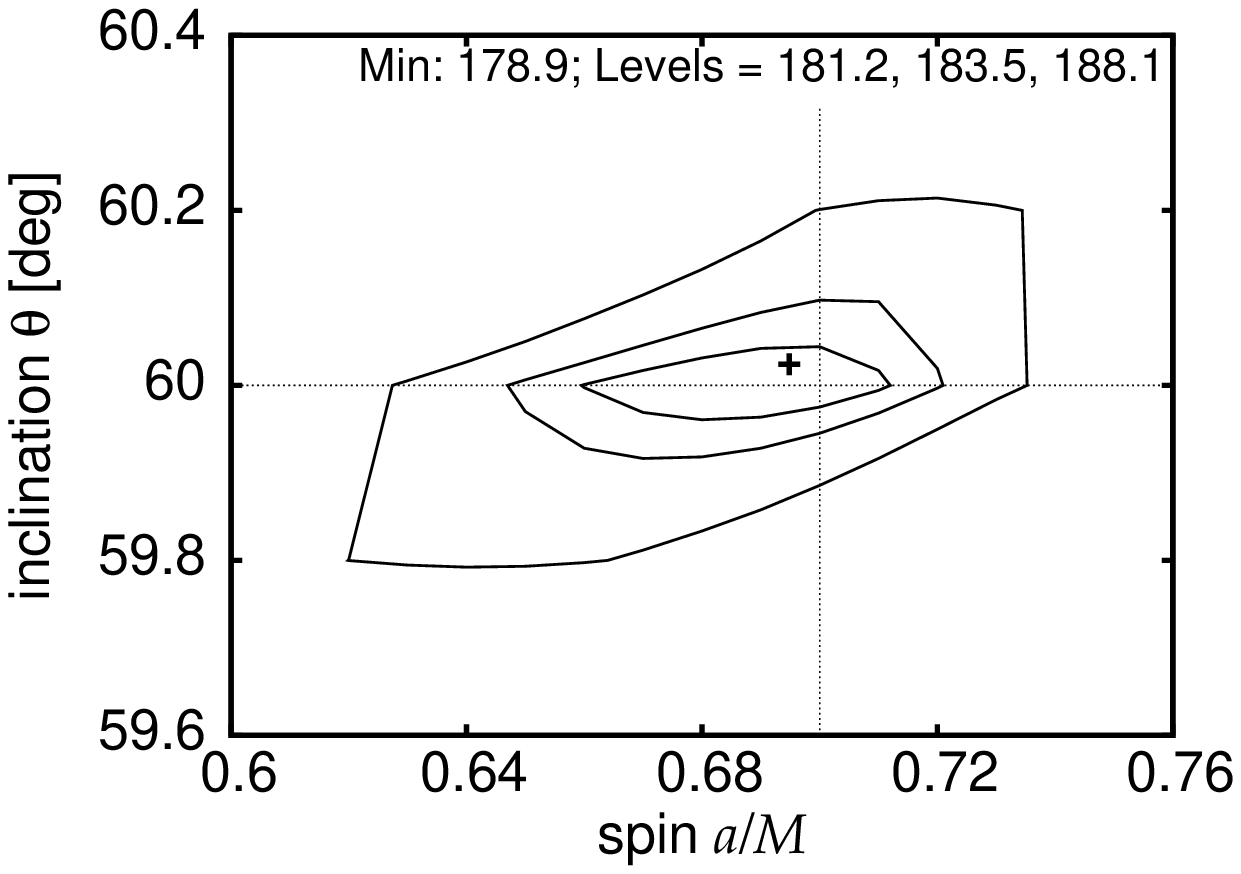} &
  \includegraphics[width=0.315\textwidth]{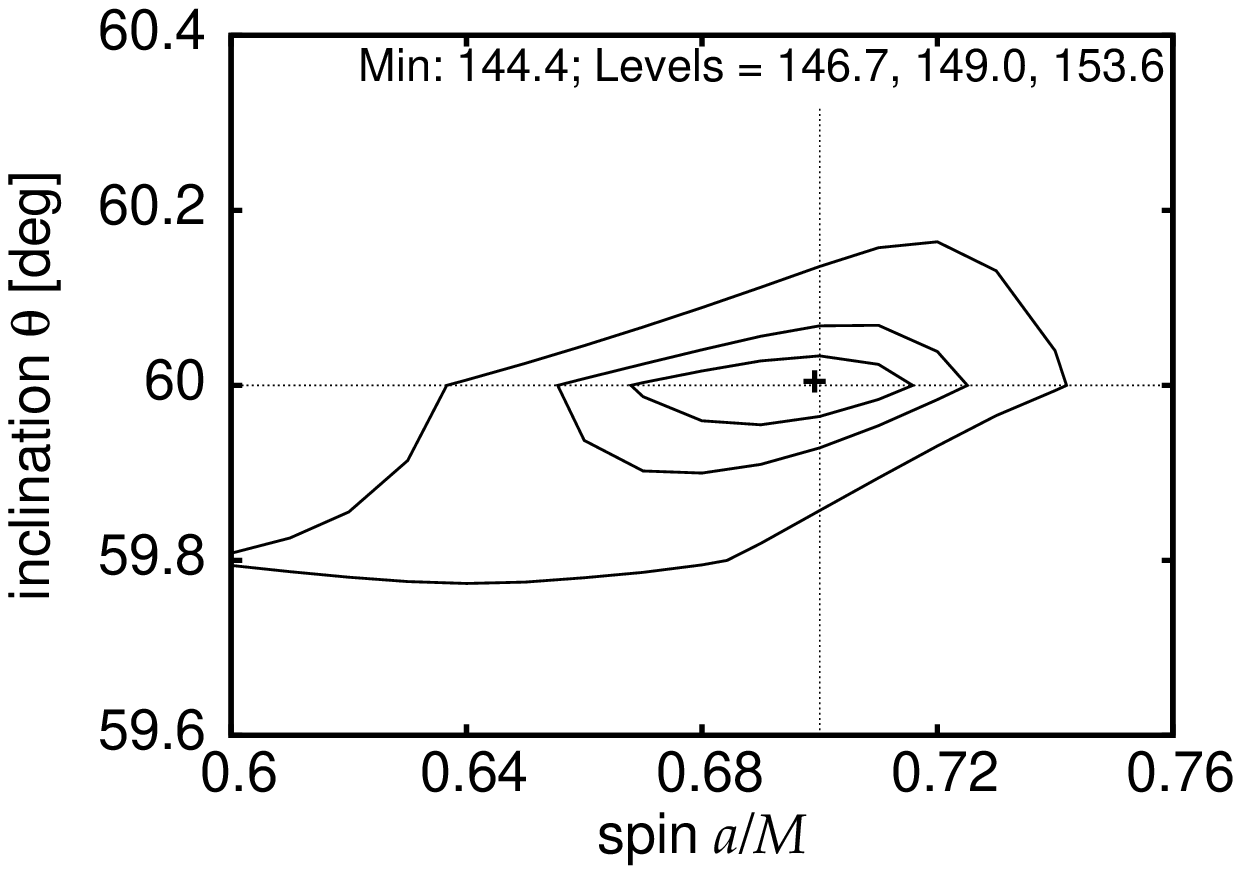} &
  \includegraphics[width=0.315\textwidth]{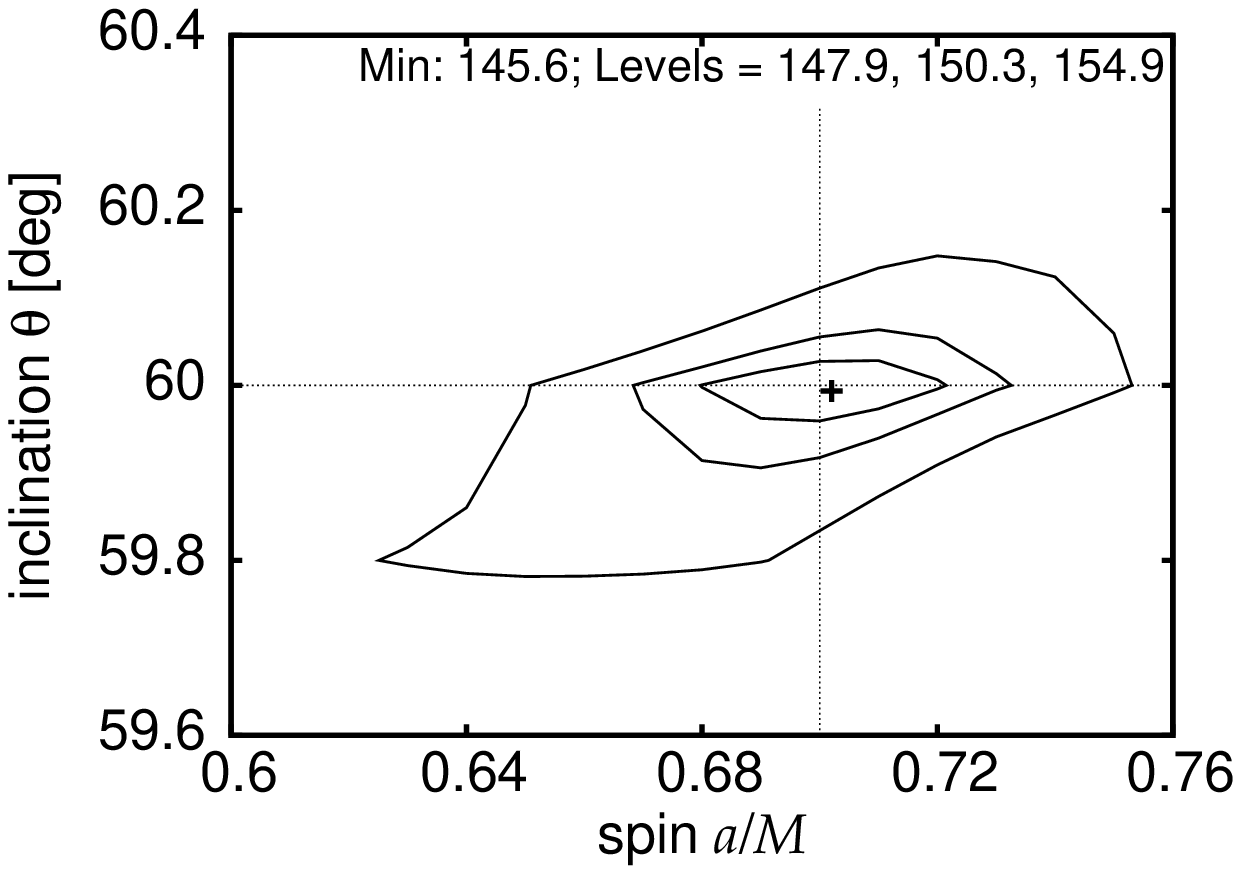}\\
\end{tabular}
\caption{The same as in Figure~\ref{fig7}, but for 
$a_{\rm f}=0.7$ and $\theta_{\rm f}=60\deg$.}
\label{fig8}
\end{center}
\end{figure}

\begin{figure}[tbh!]
\begin{center}
\begin{tabular}{ccc}
%  \multicolumn{3}{c}{\textbf{a=0.6}}\\
  \includegraphics[width=0.315\textwidth]{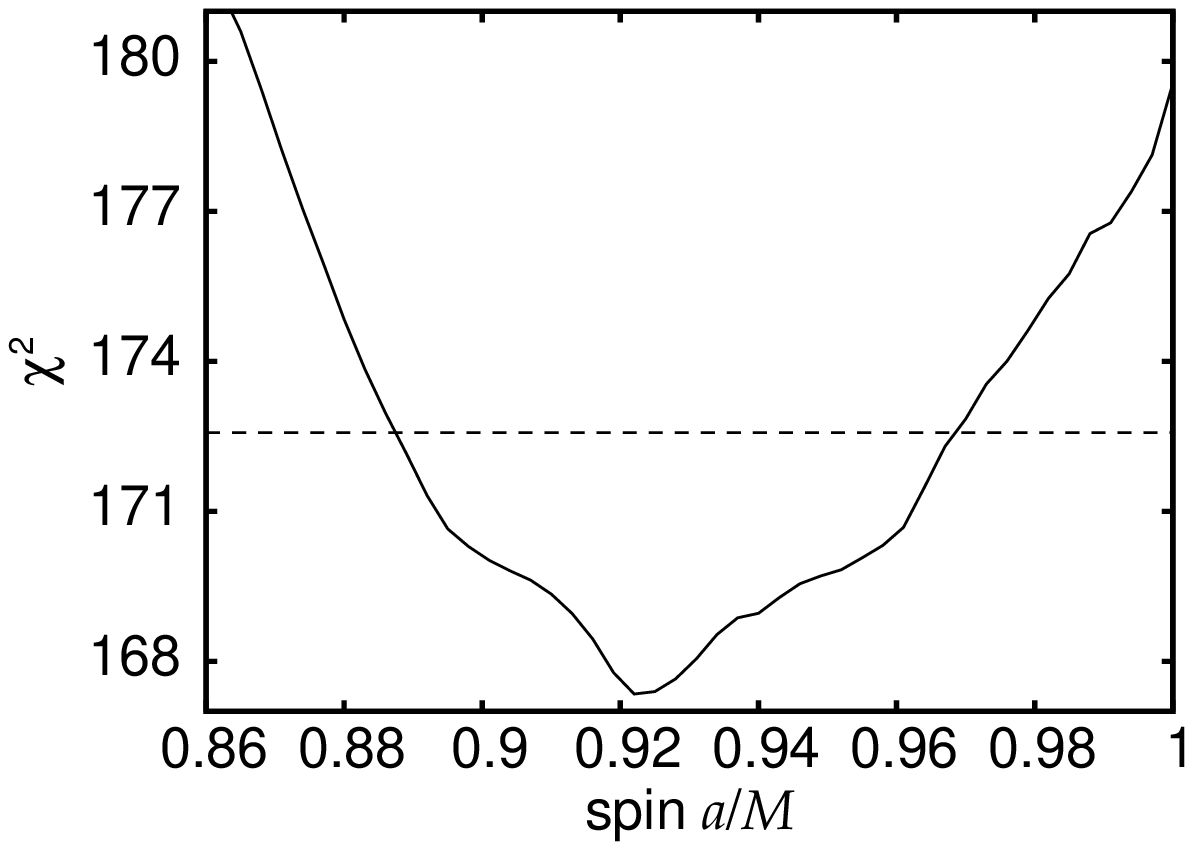} &
  \includegraphics[width=0.315\textwidth]{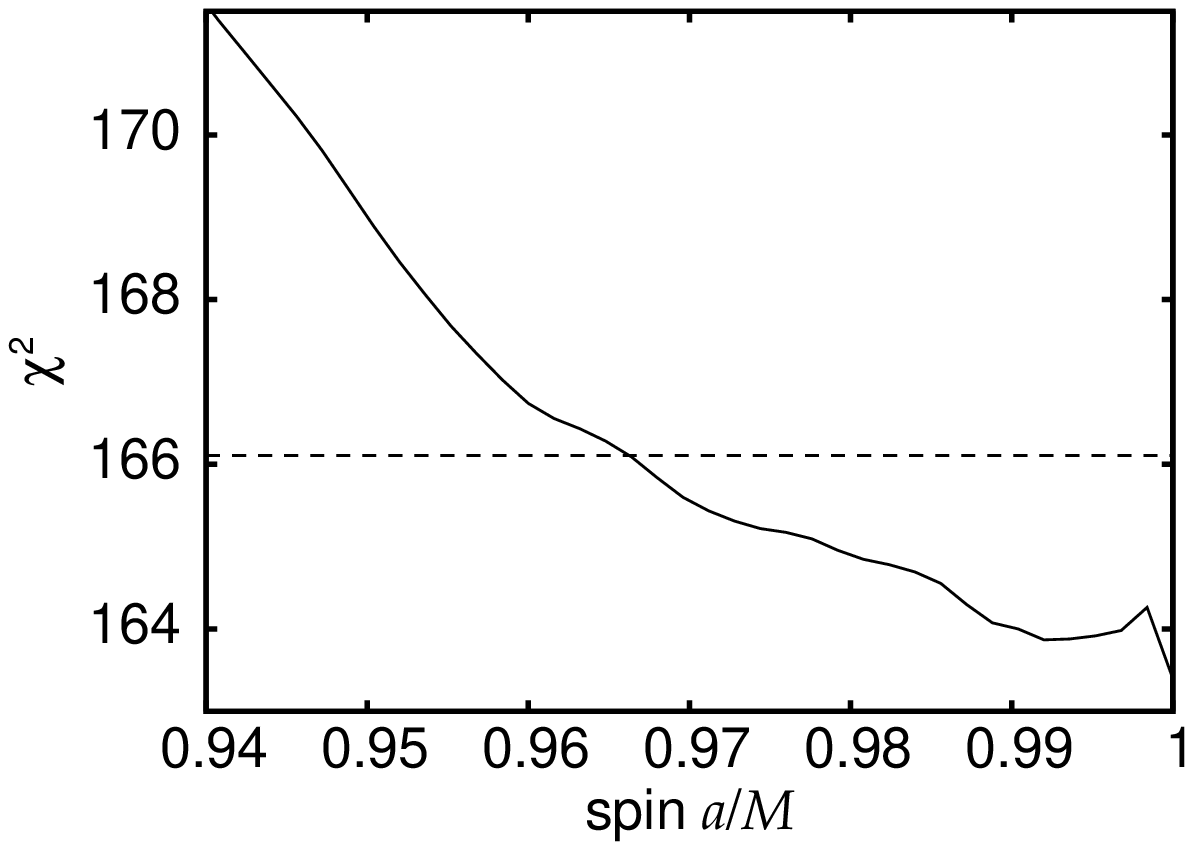} &
  \includegraphics[width=0.315\textwidth]{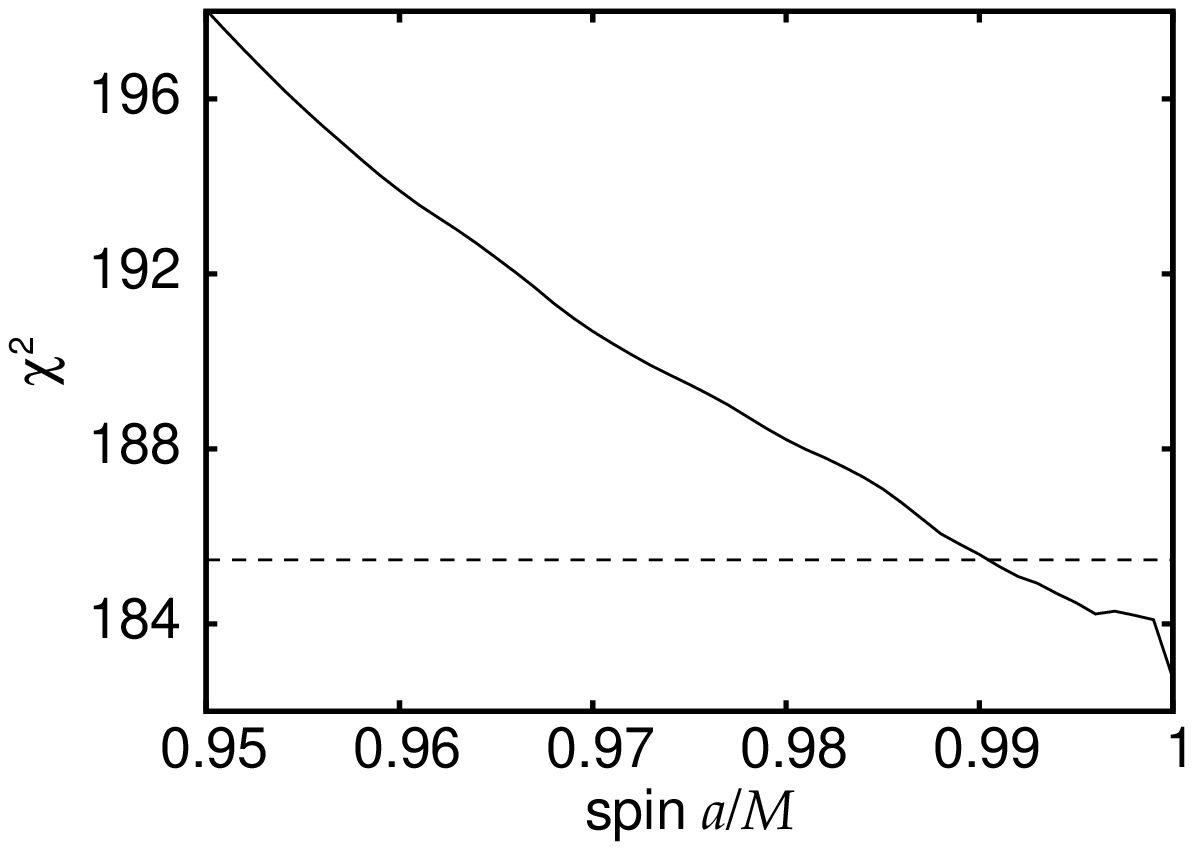}\\
\end{tabular}
\begin{tabular}{ccc}
  \includegraphics[width=0.315\textwidth]{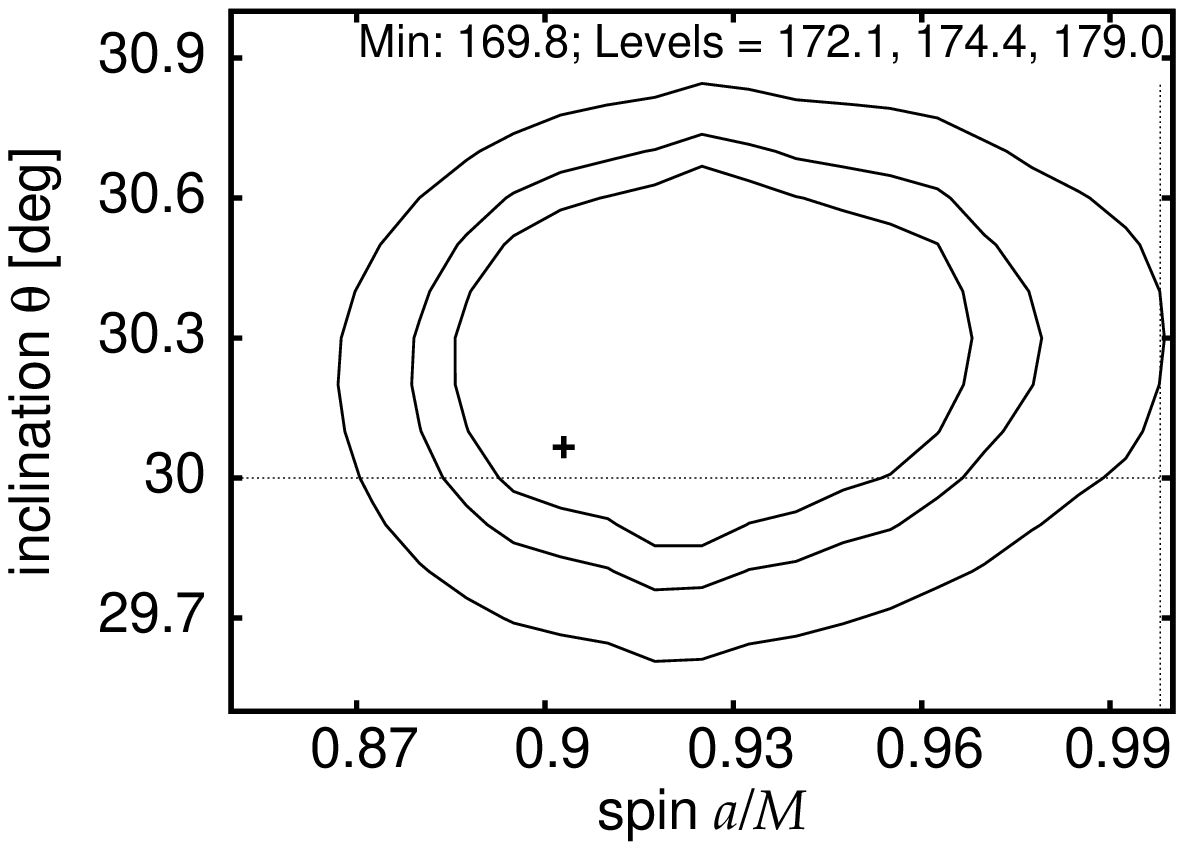} &
  \includegraphics[width=0.315\textwidth]{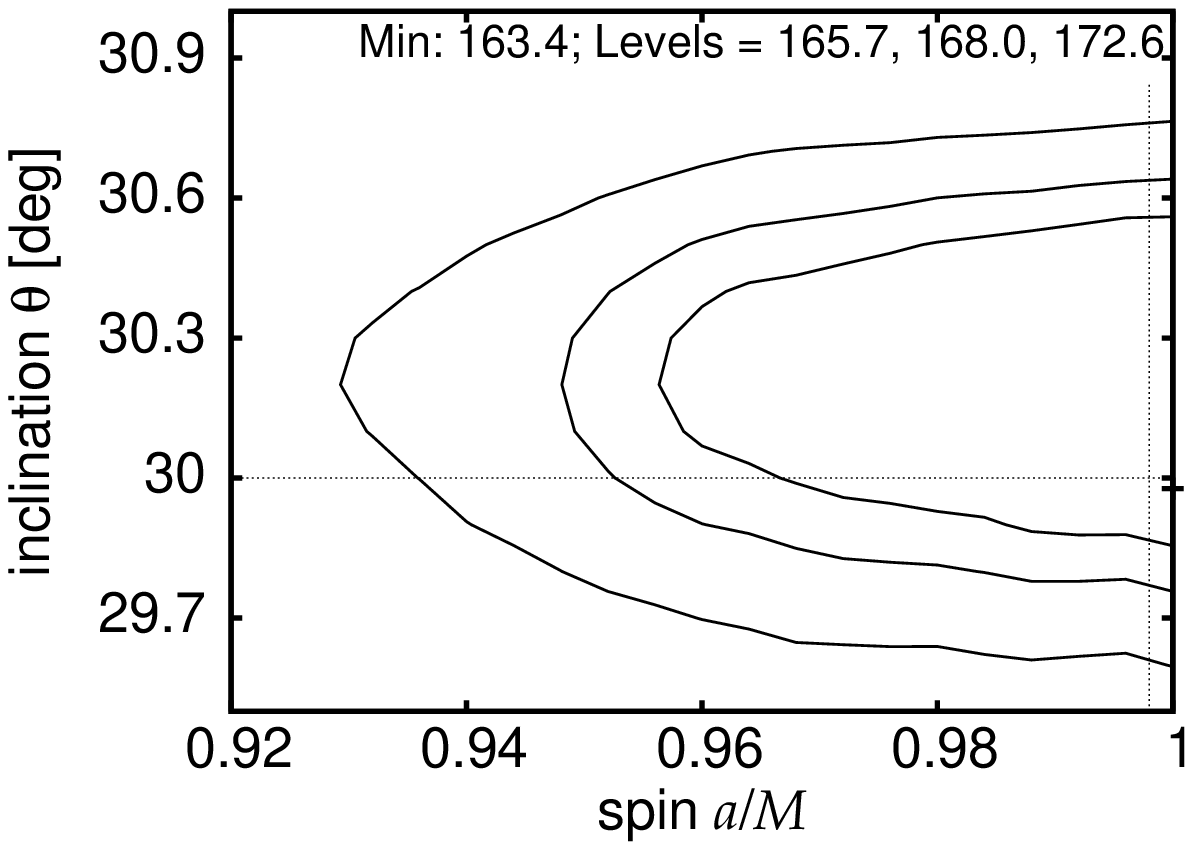} &
  \includegraphics[width=0.315\textwidth]{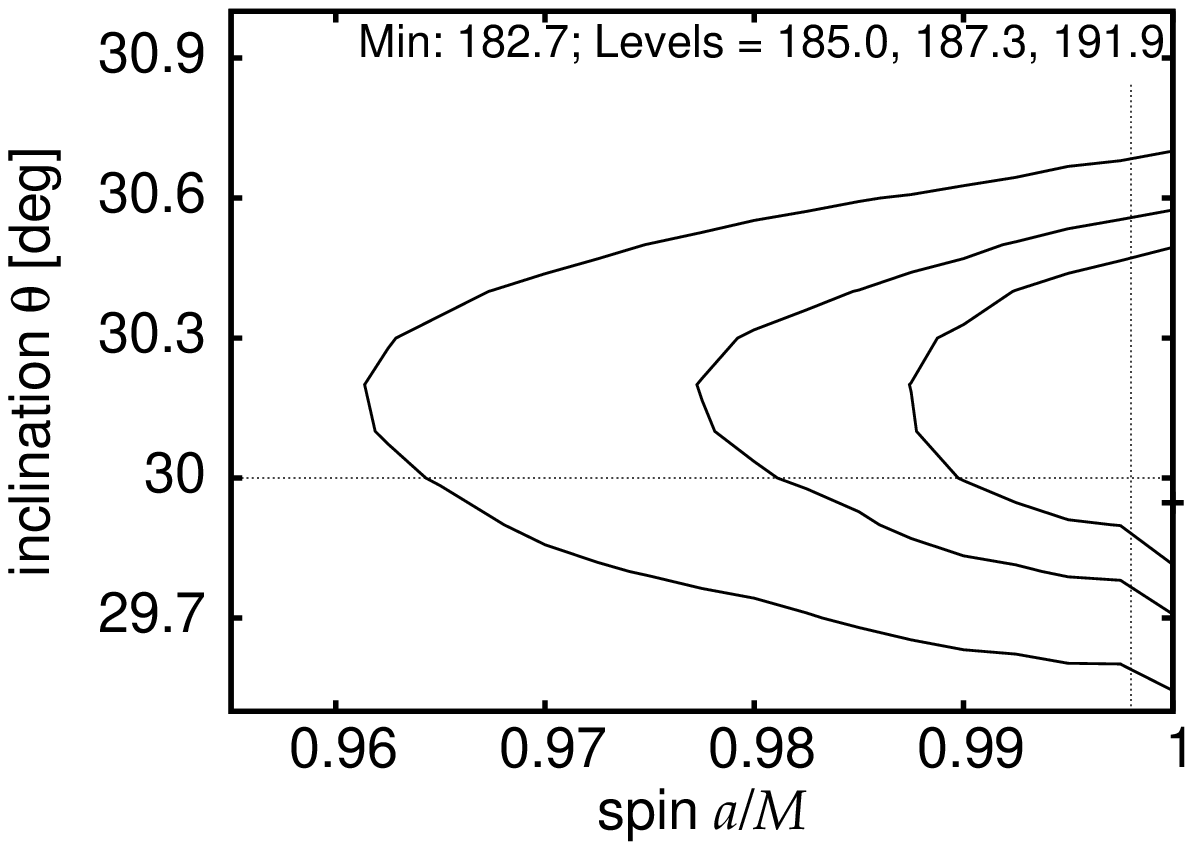}\\
\end{tabular}
\caption{The same as in Figure~\ref{fig7}, but for $a_{\rm f}=0.998$ and $\theta_{\rm f}=30\deg$.}
\label{fig9}
\end{center}
\end{figure}

\begin{figure}[tbh!]
\begin{center}
\begin{tabular}{ccc}
%  \multicolumn{3}{c}{\textbf{a=0.6}}\\
  \includegraphics[width=0.315\textwidth]{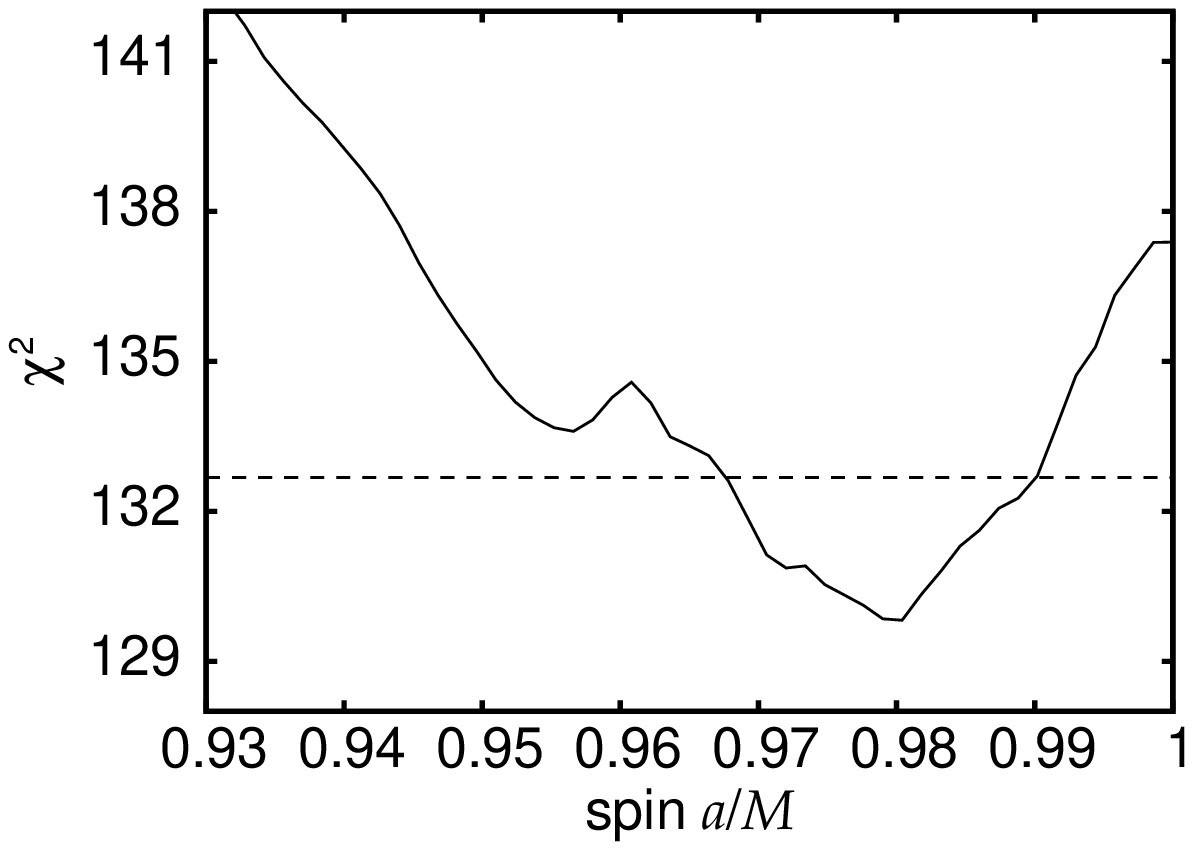} &
  \includegraphics[width=0.315\textwidth]{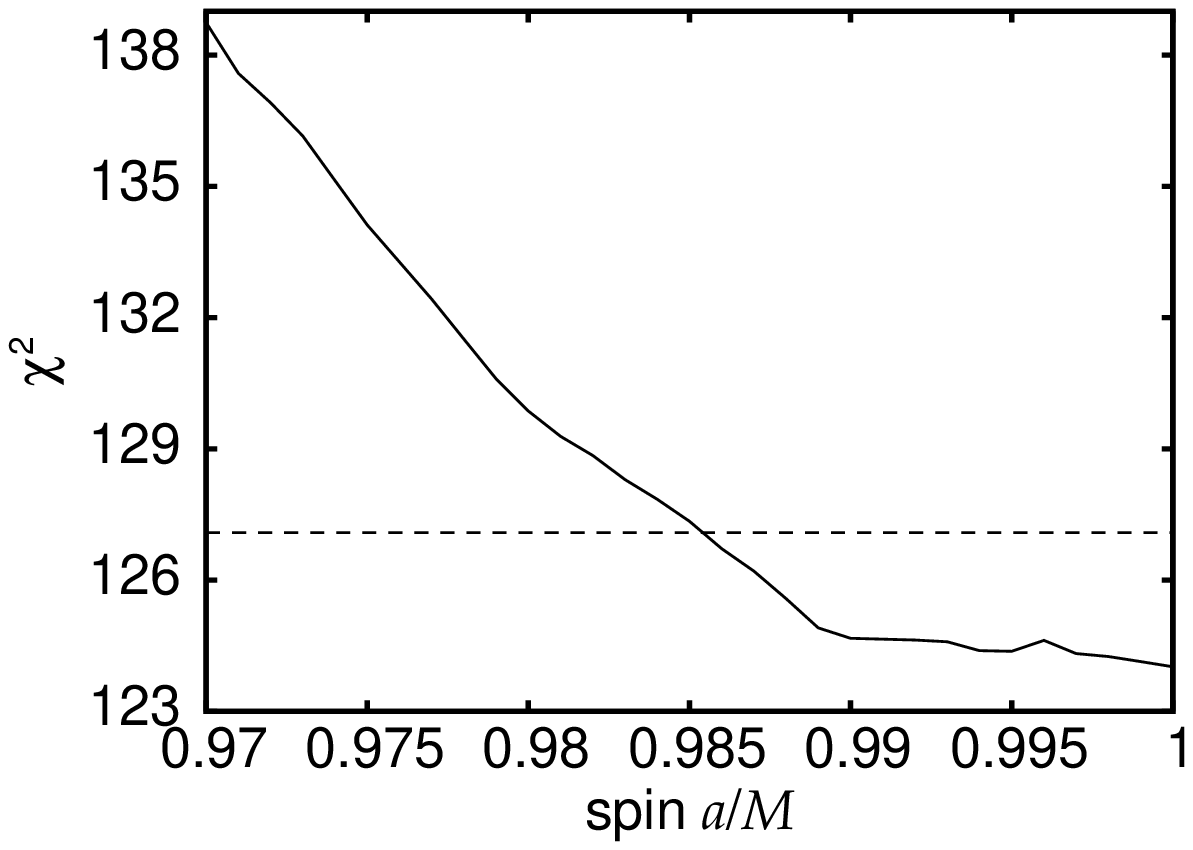} &
  \includegraphics[width=0.315\textwidth]{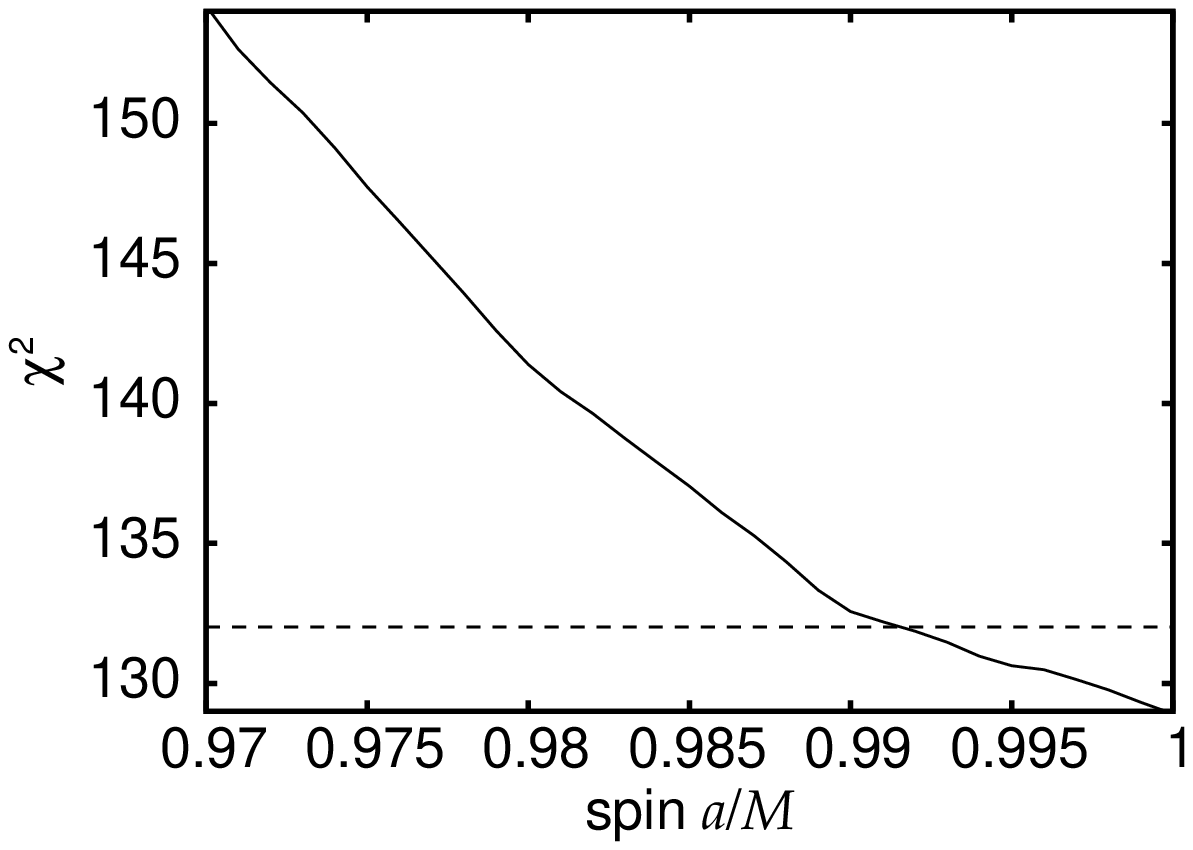}\\
\end{tabular}
\begin{tabular}{ccc}
  \includegraphics[width=0.315\textwidth]{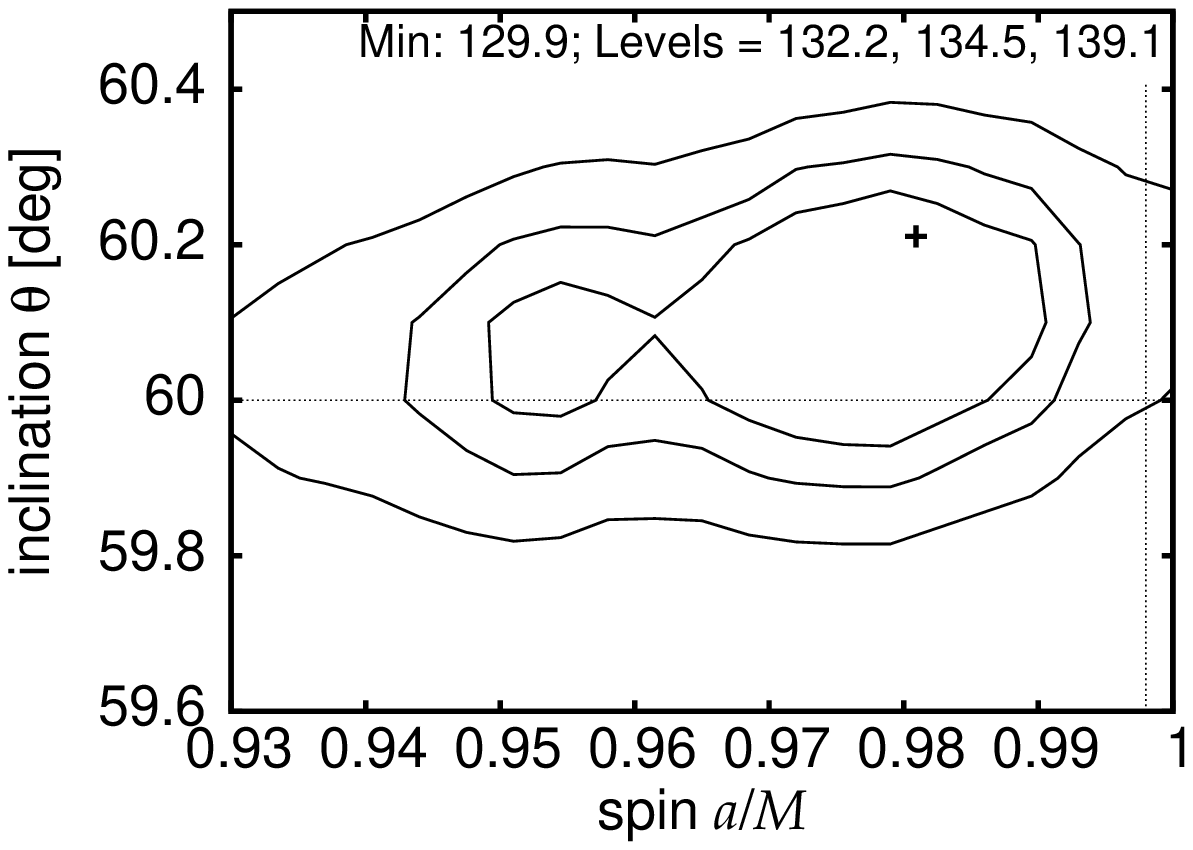} &
  \includegraphics[width=0.315\textwidth]{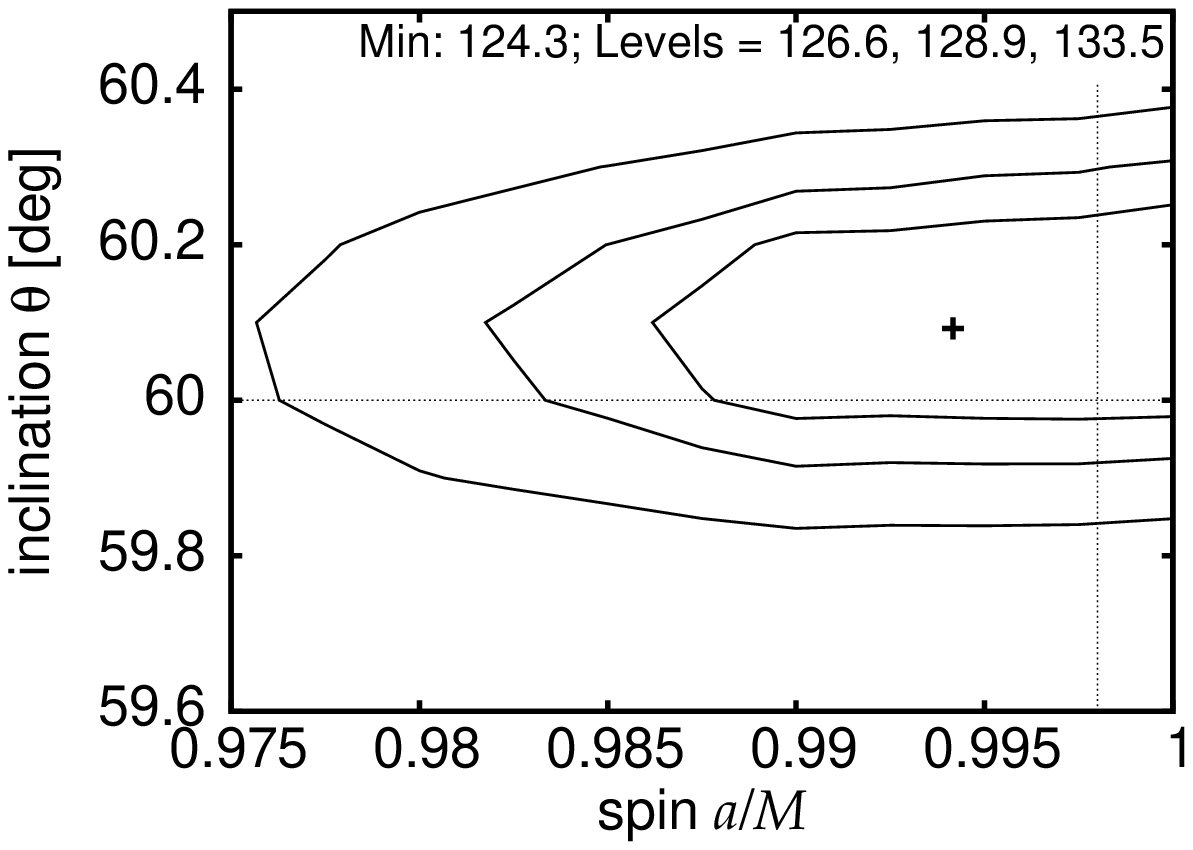} &
  \includegraphics[width=0.315\textwidth]{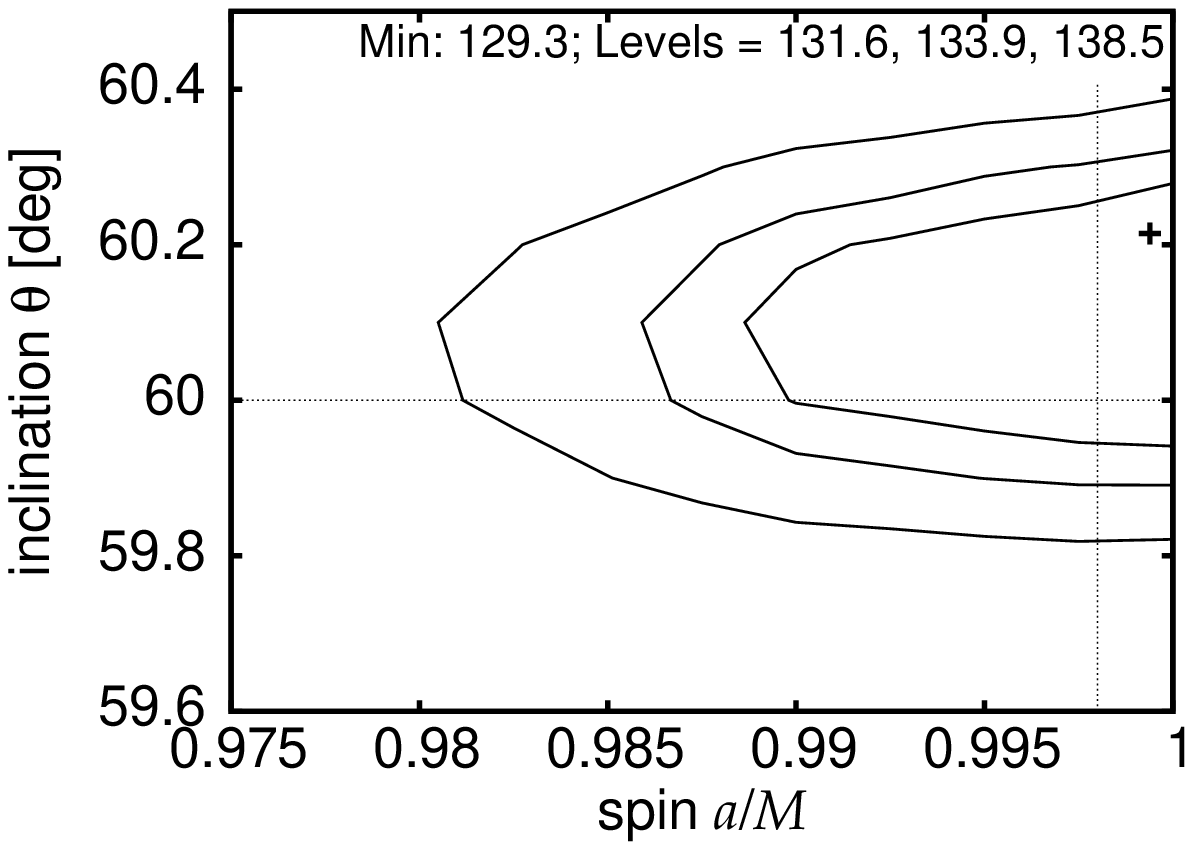}\\
\end{tabular}
\caption{The same as in Figure~\ref{fig7}, but for $a_{\rm f}=0.998$ and $\theta_{\rm f}=60\deg$.}
\label{fig10}
\end{center}
\end{figure}

We generated a set of ``fake'' spectra (i.e., artificial spectra in the XSPEC 
terminology). These spectra were produced in a grid of angular 
momentum values while assuming isotropic directionality, Case~2 in eq.~({\ref{case123}).
We call the assumed angular momentum of the black hole the fiducial spin
and we denote it as $a_{\rm f}$. We performed the fitting loop to these data points using 
each of the three angular emissivity profiles. Once the fit reached convergence, 
we recorded the {\em inferred spin $a$}. 
Figures \ref{fig7}--\ref{fig10} show the results in terms of best-fit 
$\chi^2$ profiles and the confidence contours
for two different {\em fiducial values of the spin}
(we assumed $a_{\rm f}=0.7$, and $0.998$).
We summarise the values for the inferred spin
for two inclination angles $i=30$\,deg and
$i=60$\,deg in Table~\ref{tab1}.

The fitting procedure was performed in two different ways -- having the rest
of the parameters free or keeping them frozen. Obviously the former approach
results in an extremely complicated $\chi^2$ space. Therefore, for simplicity of
the graphical representation, we plot only the results of the second approach which, 
however, gives broadly consistent results (though it misses some 
local minima of $\chi^2$). In other words, the plots have the parameters 
of the power law continuum, the energy of the line and the radial 
dependence parameter fixed at $\Gamma=1.9$, $E_0=6.4$~keV, and $q=3$. 

The conclusion from this analysis is that 
the determination of $a$ indeed seems to be sensitive within certain 
limits to the assumed directionality of the intrinsic emission. 
The suppression of the flux of
the reflection component at high values of $\theta_{\rm e}$ may
lead to overestimating the spin, and vice versa.
The middle panels of Figs.\ \ref{fig7}--\ref{fig10} show the fit results
for isotropic directionality, which was also the seed model
used to generate the test data, and so these contours illustrate
the magnitude of combined dispersion due to the simulated noise 
and the degeneracy between the spin and the inclination.. The fiducial values 
are well inside the $1\sigma$ confidence contour in all the graphs
in the middle panels.

However, systematically lower values of the angular momentum are
obtained for the limb brightening profile and, vice versa, higher values
are found for the limb darkening profile. The magnitude of the difference
is larger for higher values of angular momentum.

\subsection{Angular emission profile of the detailed reprocessing model}
\label{sec-titan-noar-modelling}

\begin{figure}[tbh!]
\begin{center}
\includegraphics[width=0.48\textwidth,angle=0]{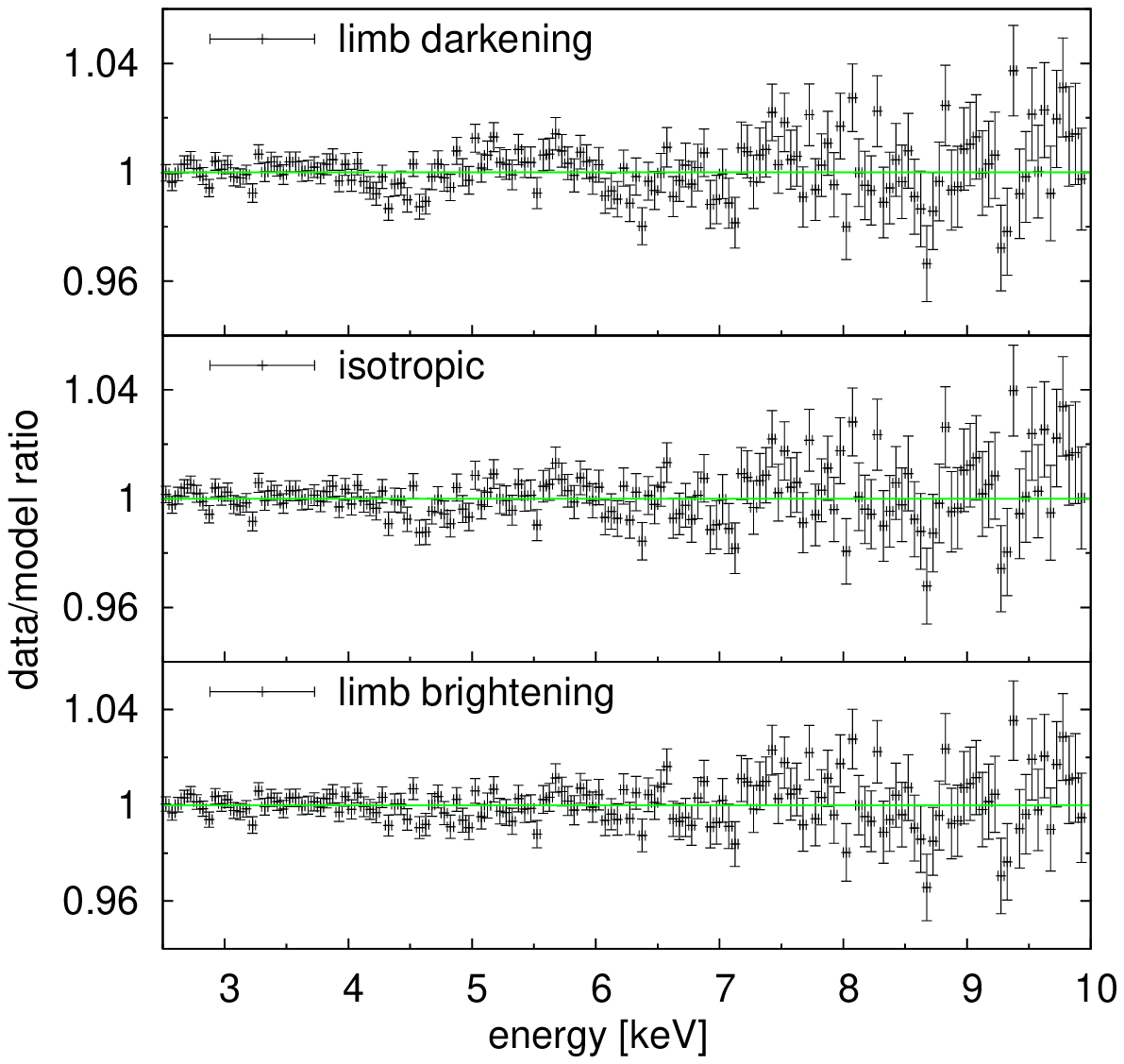}
\hfill
\includegraphics[width=0.48\textwidth,angle=0]{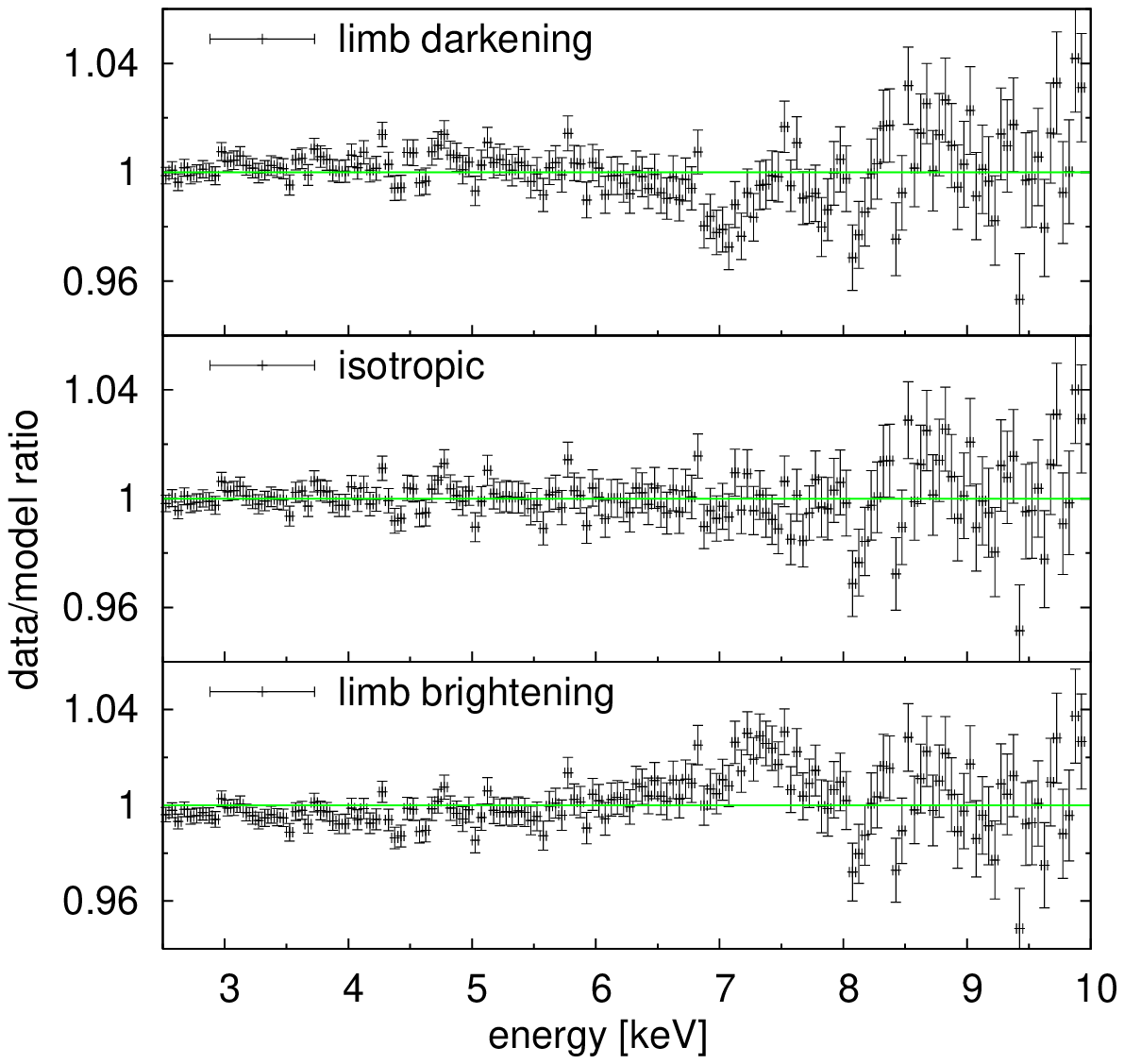}
\caption{Plots of data/model ratio, where the data are simulated as
\textscown{powerlaw} + \textscown{kyl2cr} and the model applied to the data
is \textscown{powerlaw} + \textscown{kyl3cr} with a particular analytical approach
of the directionality. The default angular momentum value is $a_{\rm f}=0.7$,
and the emission angle $\theta_{\rm f}=30^{\circ}$ (\textposown{left}), 
and $\theta_{\rm f}=60^{\circ}$ (\textposown{right}). These parameters 
and the normalisation of the reflection component were allowed 
to vary during the fitting procedure.
The plotted results correspond to the terminal values 
of the parameters obtained during the $\chi^{2}$ minimisation process.
Other parameters of the model were kept frozen at their default
values: $\Gamma=1.9$, $r_{\rm in}=r_{\rm ms}$, $r_{\rm out}=400$, 
$q=3$ and normalisation of the power law $K_{\Gamma} = 10^{-2}$.}
\label{fig_a07_ratio}
\end{center}
\end{figure}

\begin{figure}[tbh!]
\begin{center}
\includegraphics[width=0.48\textwidth,angle=0]{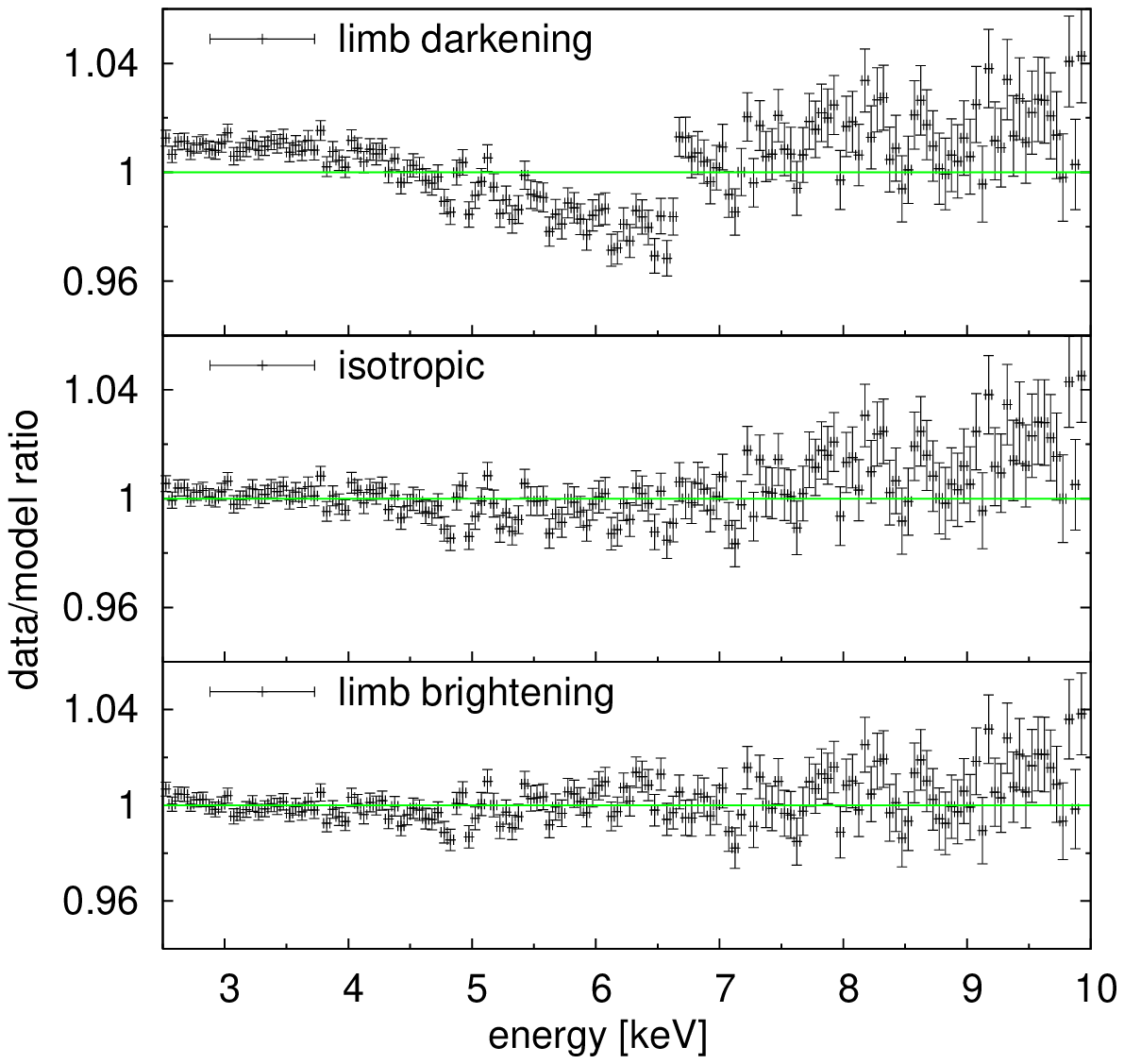}
\hfill
\includegraphics[width=0.48\textwidth,angle=0]{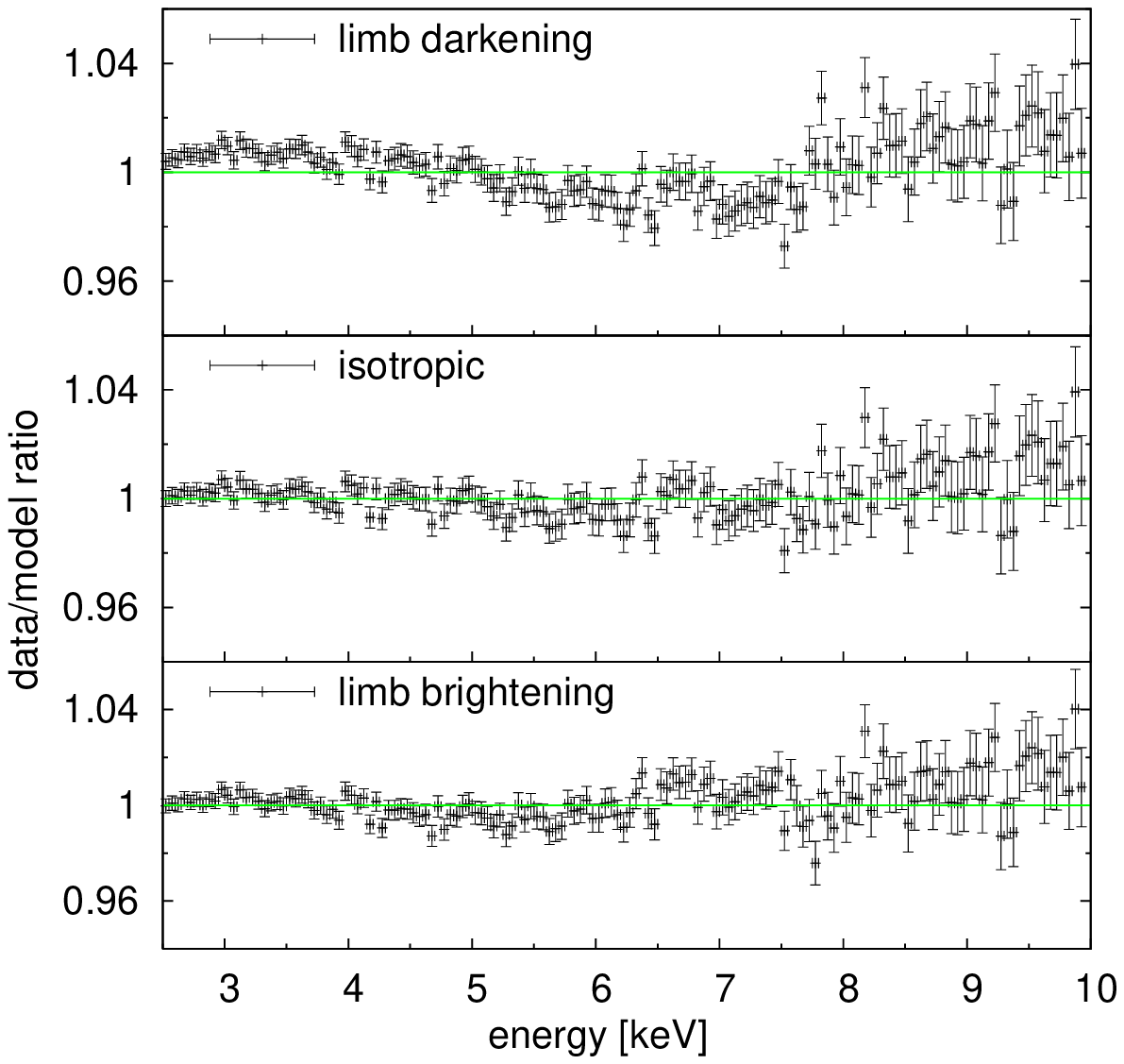}
\caption{The same plots of data/model ratio, but for $a=0.998$
(\textposown{left}: $\theta_{\rm f}=30^{\circ}$, \textposown{right}: $\theta_{\rm f}=60^{\circ}$). }
\label{fig_a0998_ratio}
\end{center}
\end{figure}

\begin{figure}[tbh!]
\begin{center}
\begin{tabular}{ccc}
\includegraphics[width=0.315\textwidth]{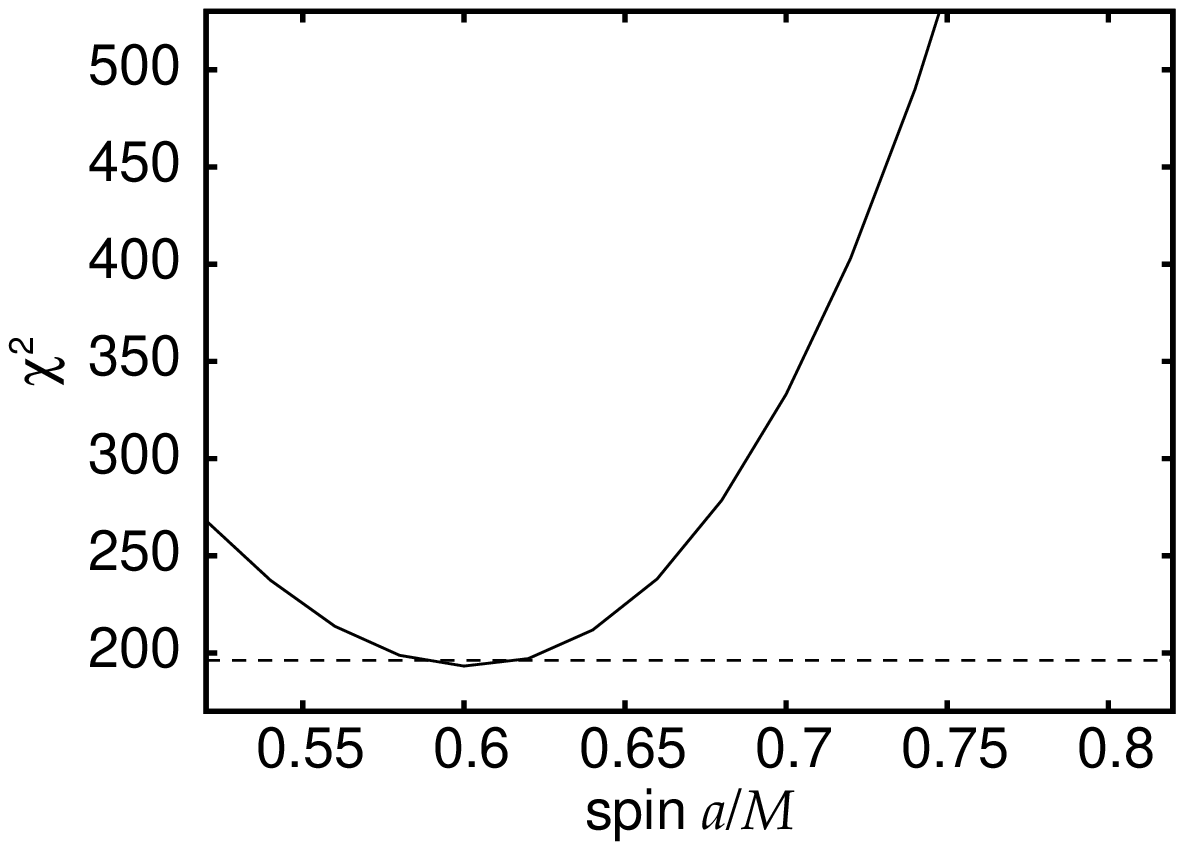} &
\includegraphics[width=0.315\textwidth]{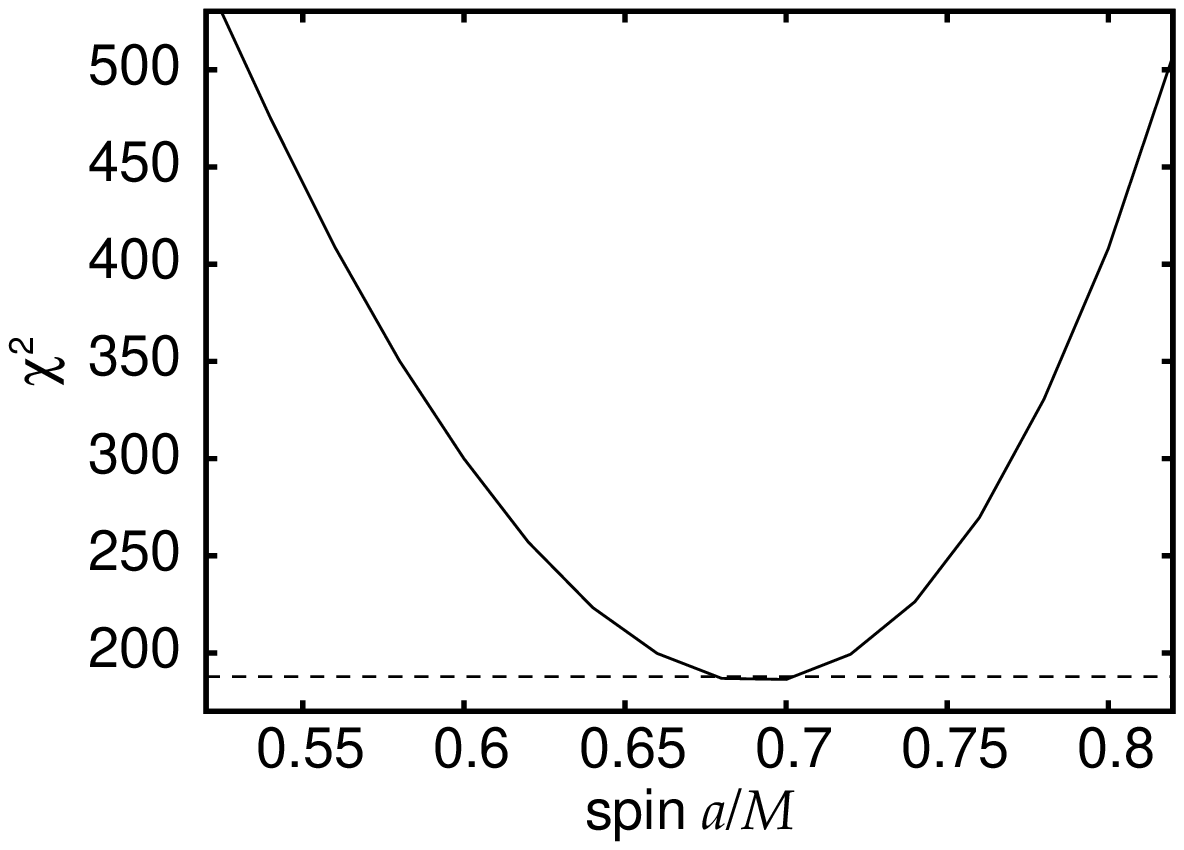} &
\includegraphics[width=0.315\textwidth]{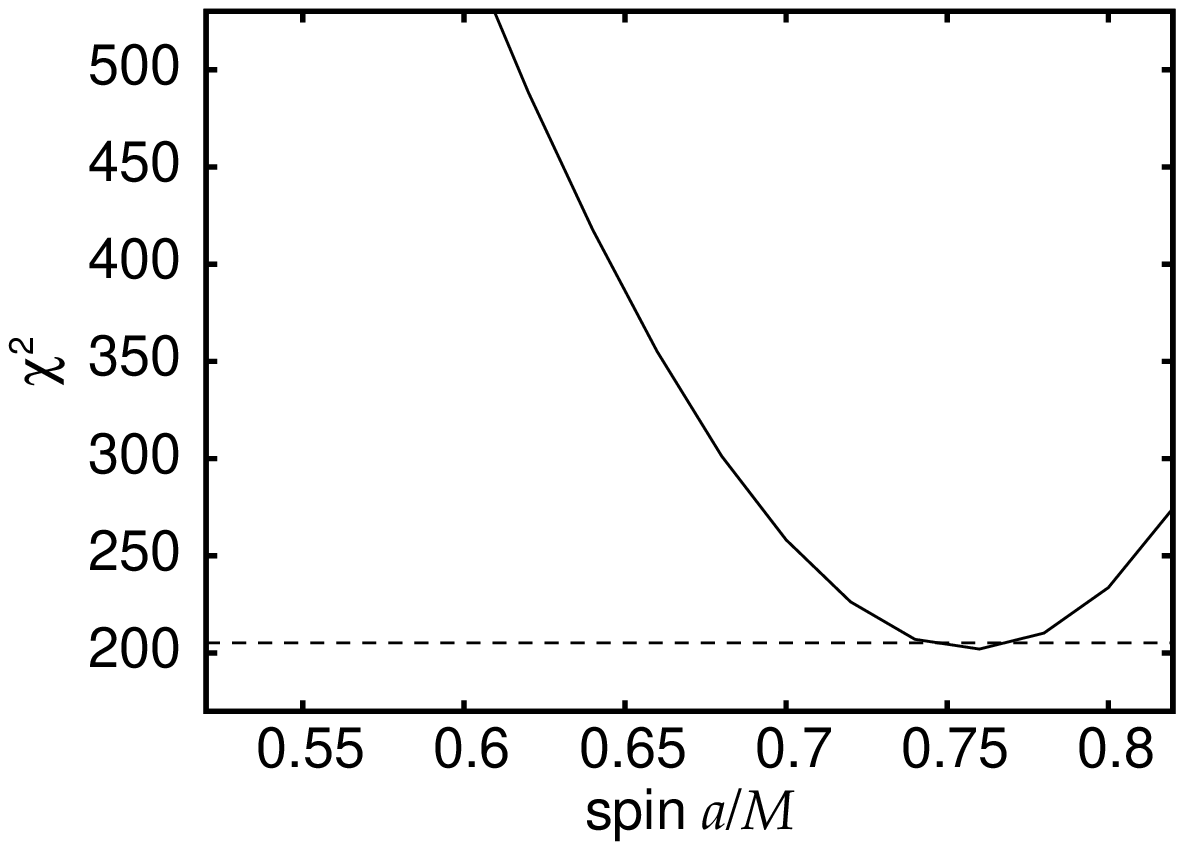} \\
\end{tabular}
\begin{tabular}{ccc}
\includegraphics[width=0.315\textwidth]{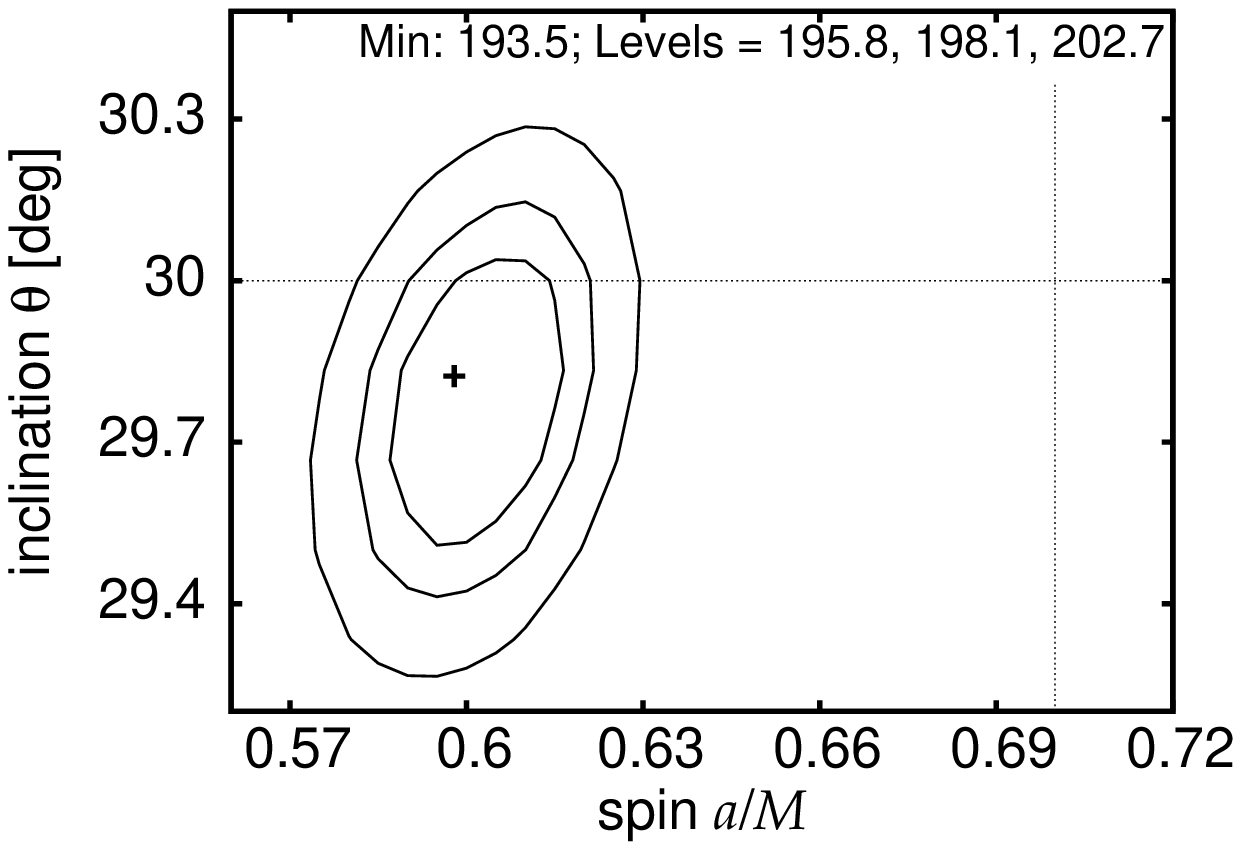} &
\includegraphics[width=0.315\textwidth]{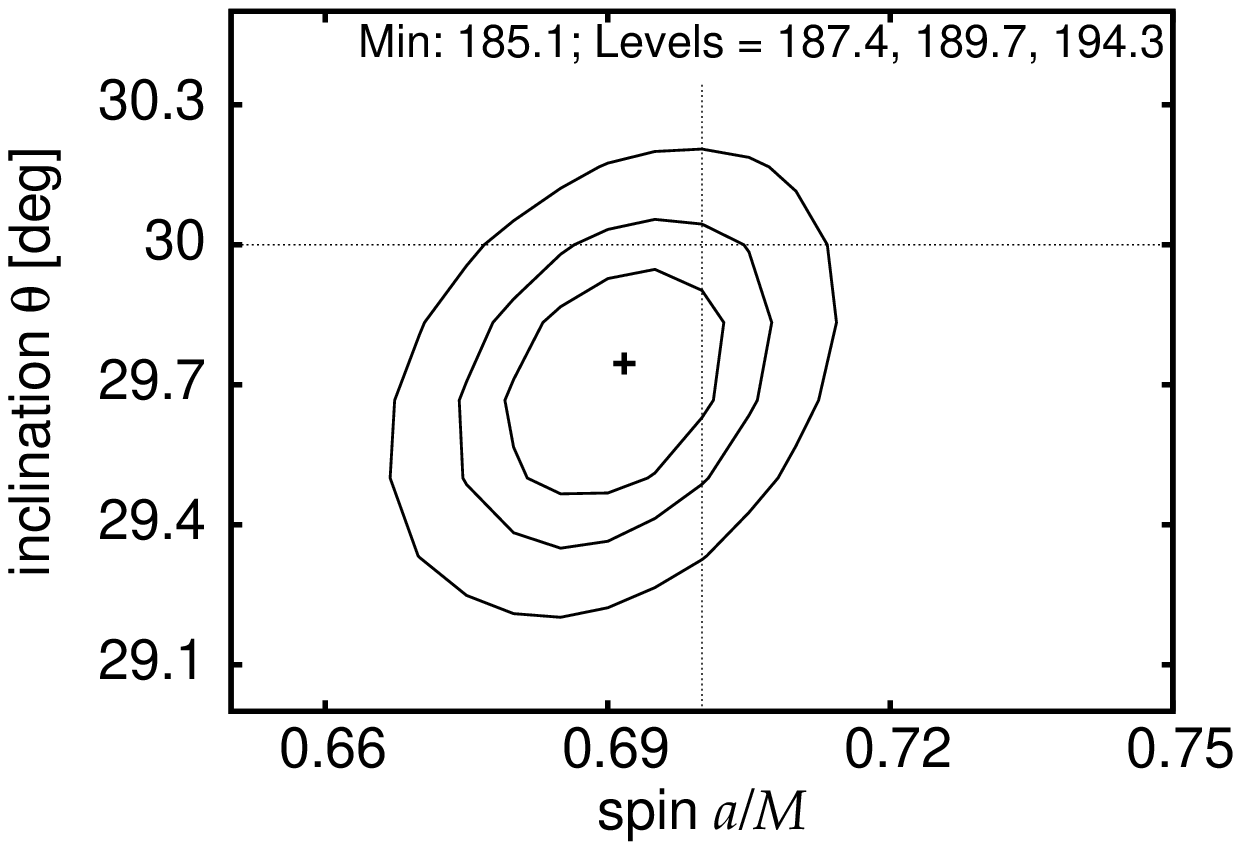} &
\includegraphics[width=0.315\textwidth]{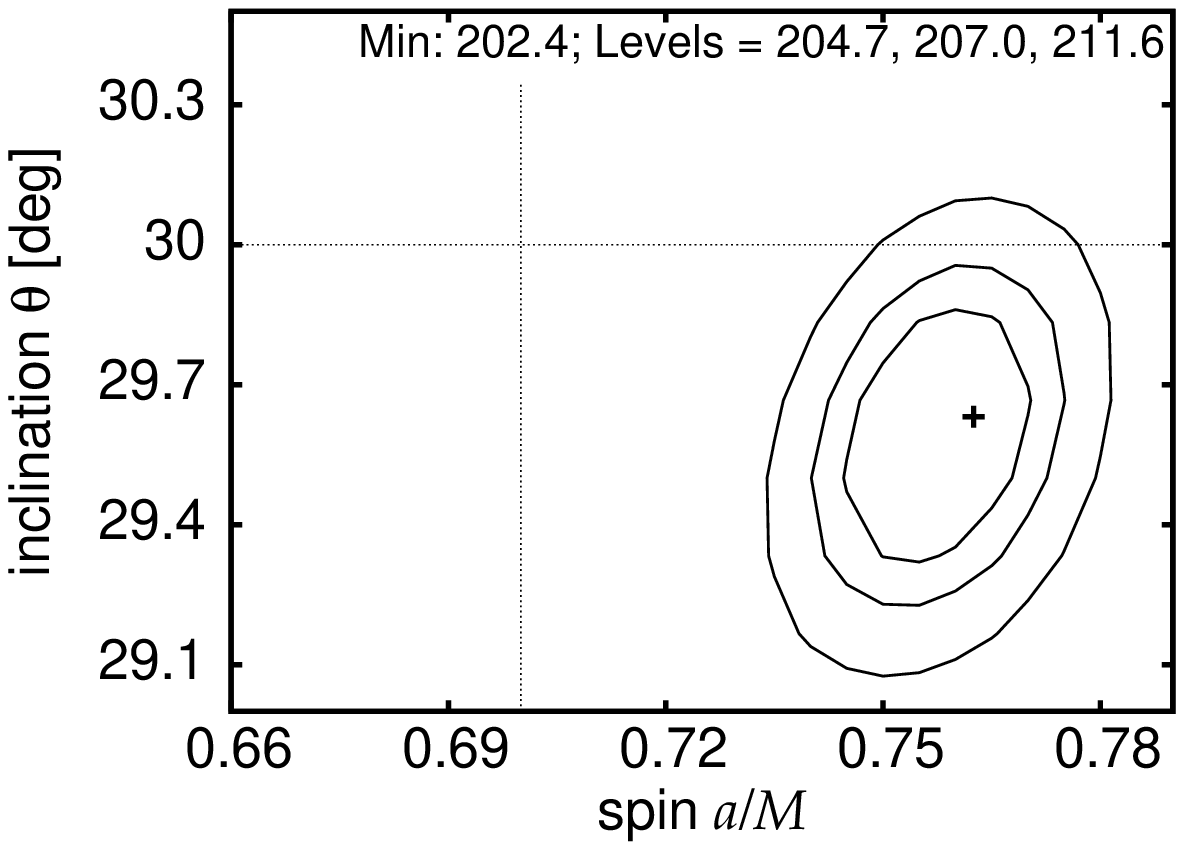}\\
\end{tabular}
\caption{Results from the test fits with $a_{\rm f}=0.7$ and $\theta_{\rm f}=30\deg$,
using the \textscown{powerlaw} + \textscown{kyl3cr} model applied to the data simulated
with \textscown{powerlaw} + \textscown{kyl2cr}. Three different profiles
of the emission directionality are shown in columns -- \textposown{left}:
limb brightening, \textposown{middle}: isotropic, \textposown{right}: limb darkening. 
\textposown{Top}: dependence of the best fit $\chi^2$ values on the fiducial spin value.
The horizontal (dashed) line represents the 90\% confidence level.
\textposown{Bottom}: contour graphs of $a$ versus $\theta_{\rm o}$. 
The contour lines refer to $1$, $2$, and $3$ sigma levels.
The position of the minimal value of $\chi^{2}$ is marked
with a small cross. The values of $\chi^{2}$ corresponding to the minimum
and to the contour levels are shown at the top of each contour graph.
The large cross indicates the position of the fiducial values of 
the angular momentum and the emission angle.
Other parameters of the model were kept fixed at default
values: $\Gamma=1.9$, $r_{\rm in}=r_{\rm ms}$, $r_{\rm
out}=400$,  $q=3$ and normalisation of the power law
$K_\Gamma=10^{-2}$.}
\label{fig_a07i30}
\end{center}
\end{figure}

\begin{figure}[tbh!]
\begin{center}
\begin{tabular}{ccc}
\includegraphics[width=0.315\textwidth]{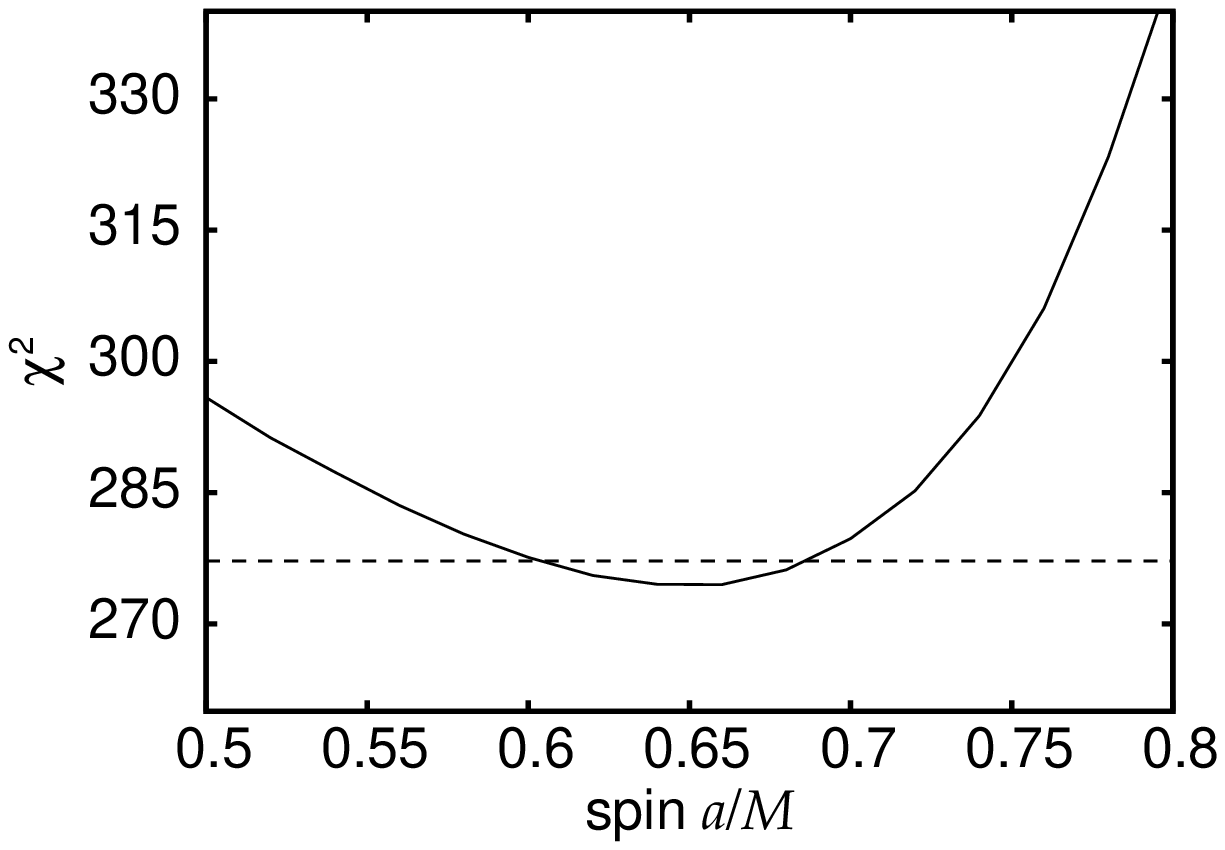} &
\includegraphics[width=0.315\textwidth]{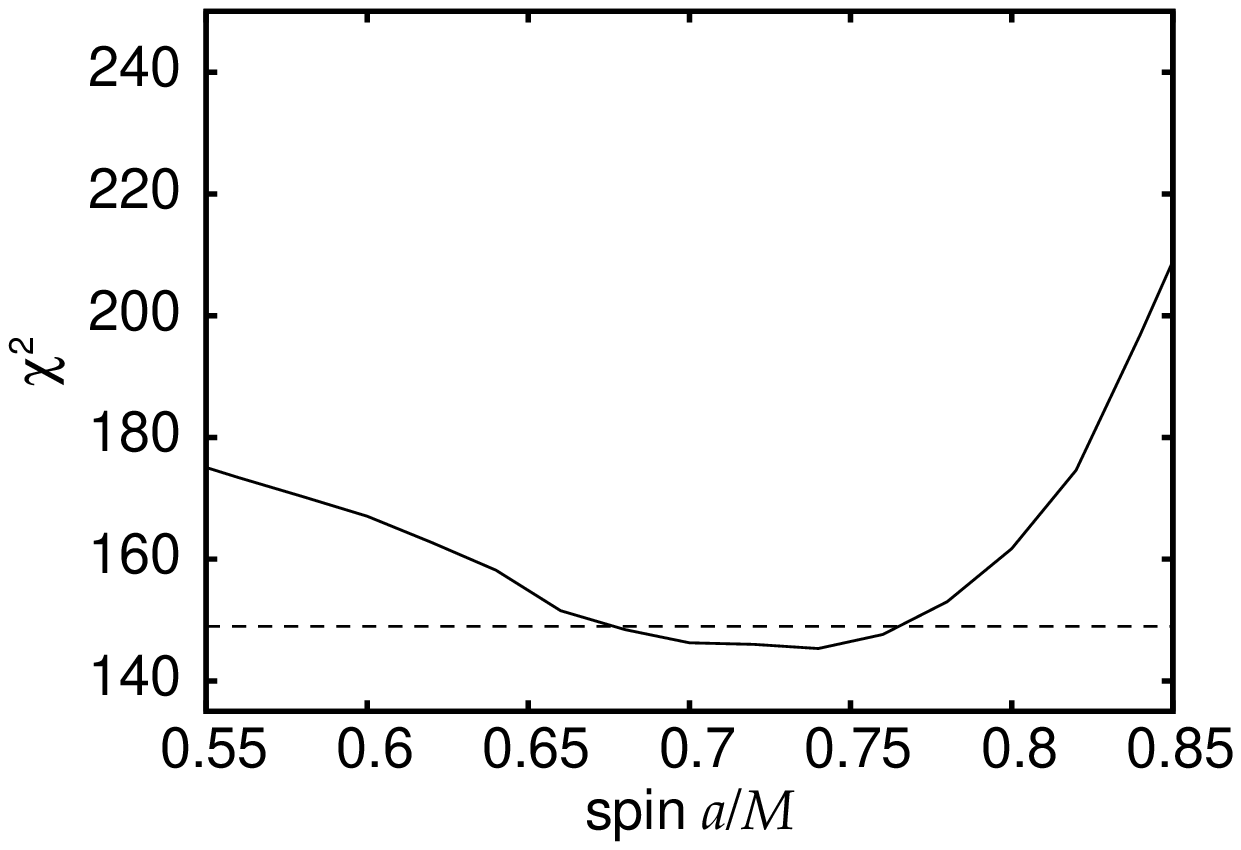} &
\includegraphics[width=0.315\textwidth]{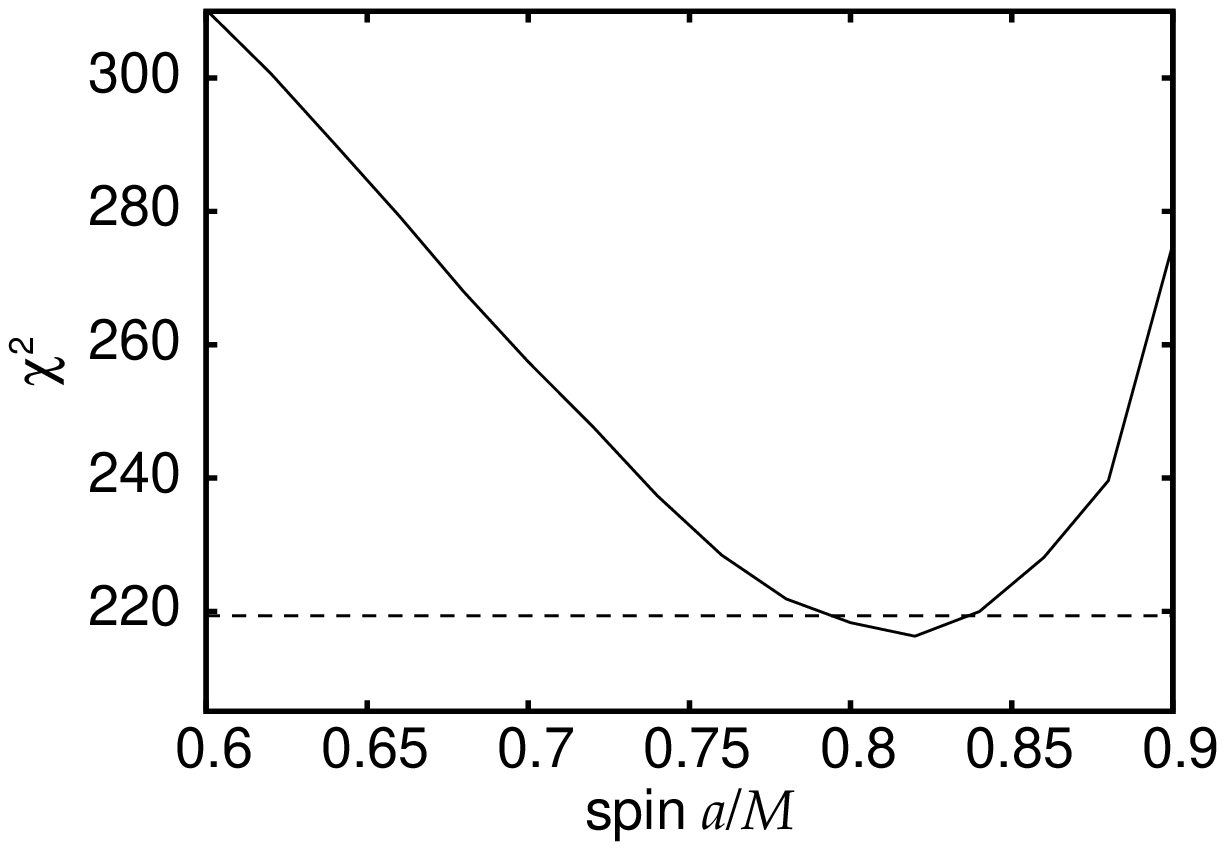} \\
\end{tabular}
\begin{tabular}{ccc}
  \includegraphics[width=0.315\textwidth]{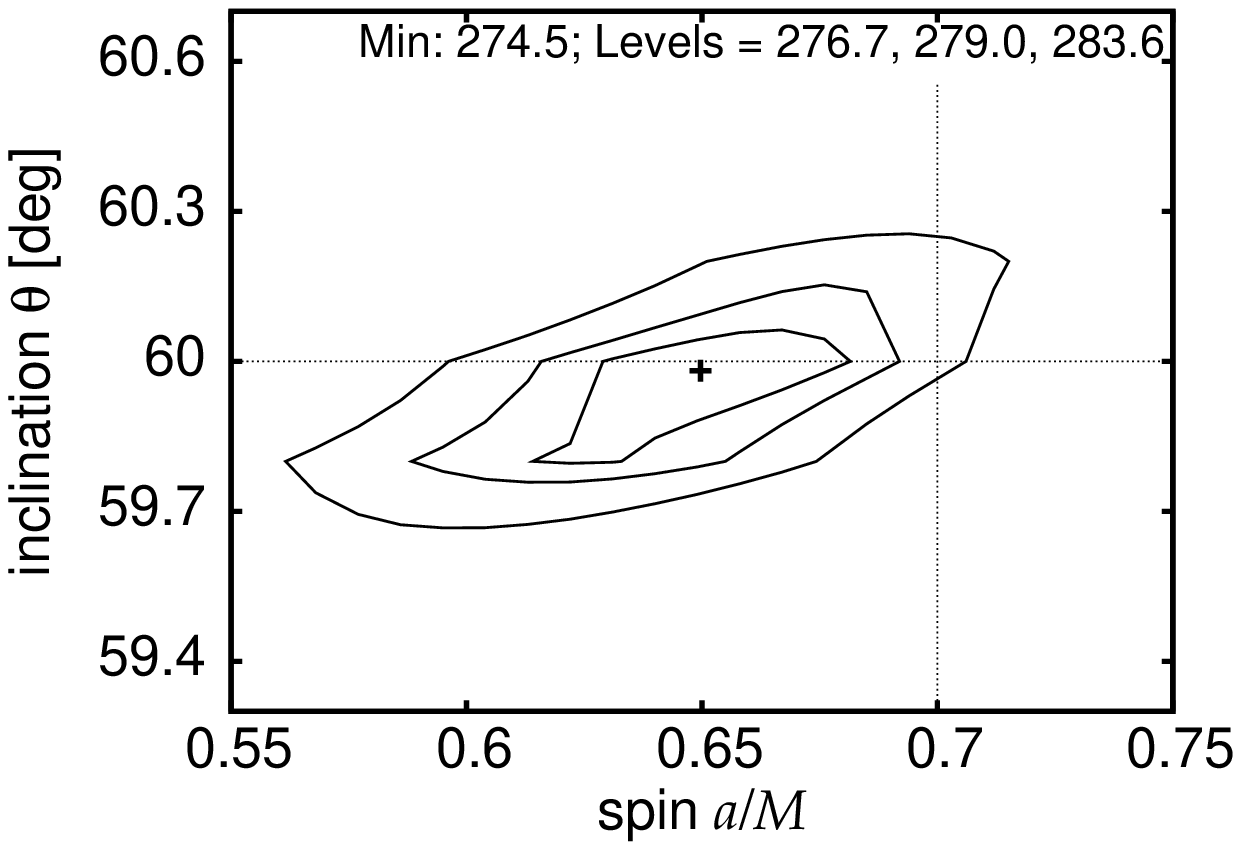} &
  \includegraphics[width=0.315\textwidth]{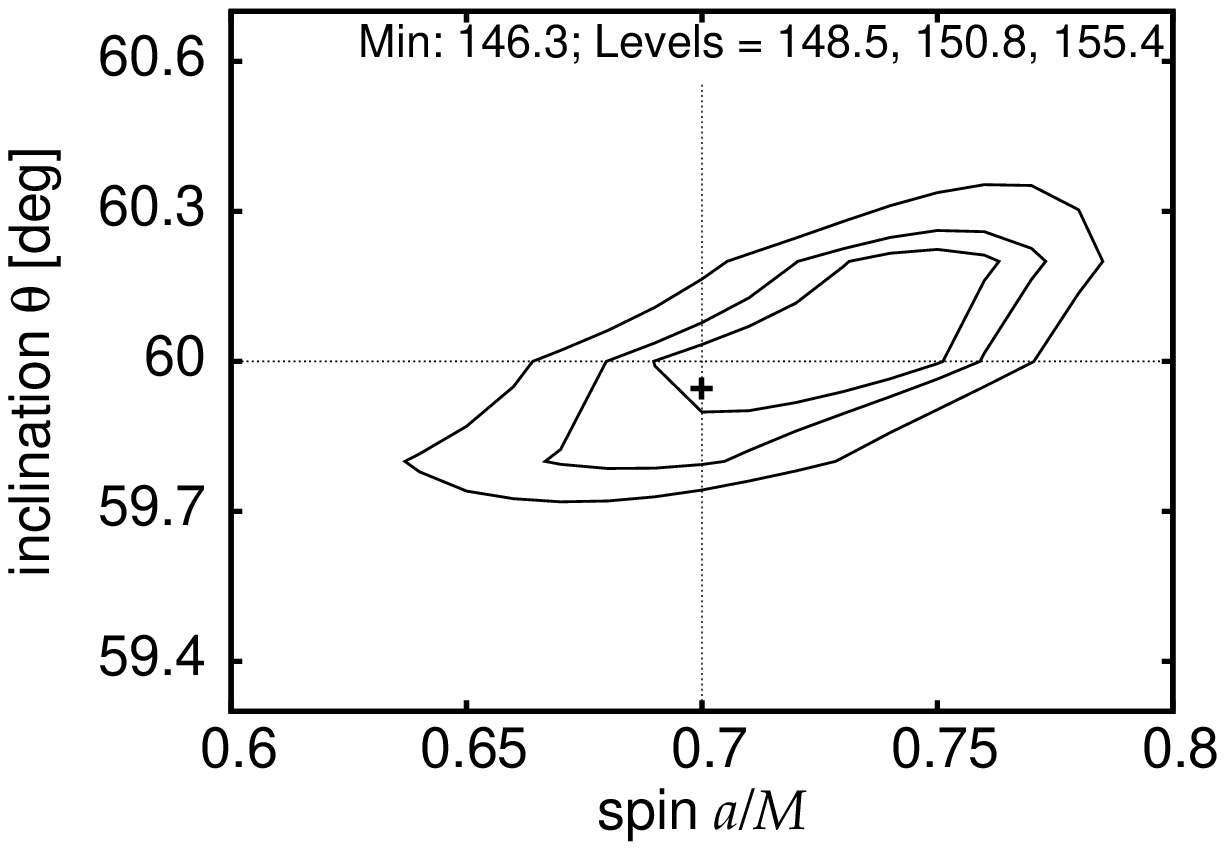} &
  \includegraphics[width=0.315\textwidth]{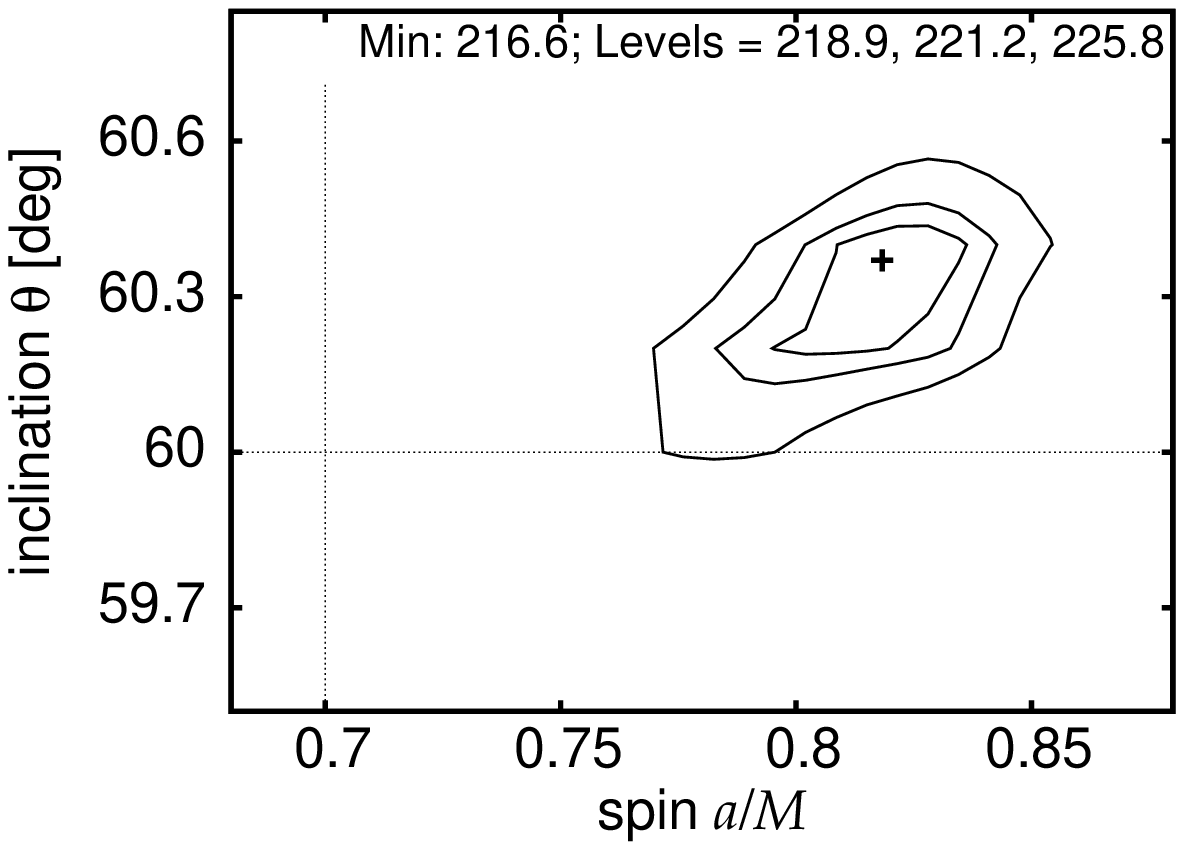}\\
\end{tabular}
\caption{The same as in Figure~\ref{fig_a07i30}, but for 
$a_{\rm f}=0.7$ and $\theta_{\rm f}=60\deg$.}
\label{fig_a07i60}
\end{center}
\end{figure}

\begin{figure}[tbh!]
\begin{center}
\begin{tabular}{ccc}
\includegraphics[width=0.315\textwidth]{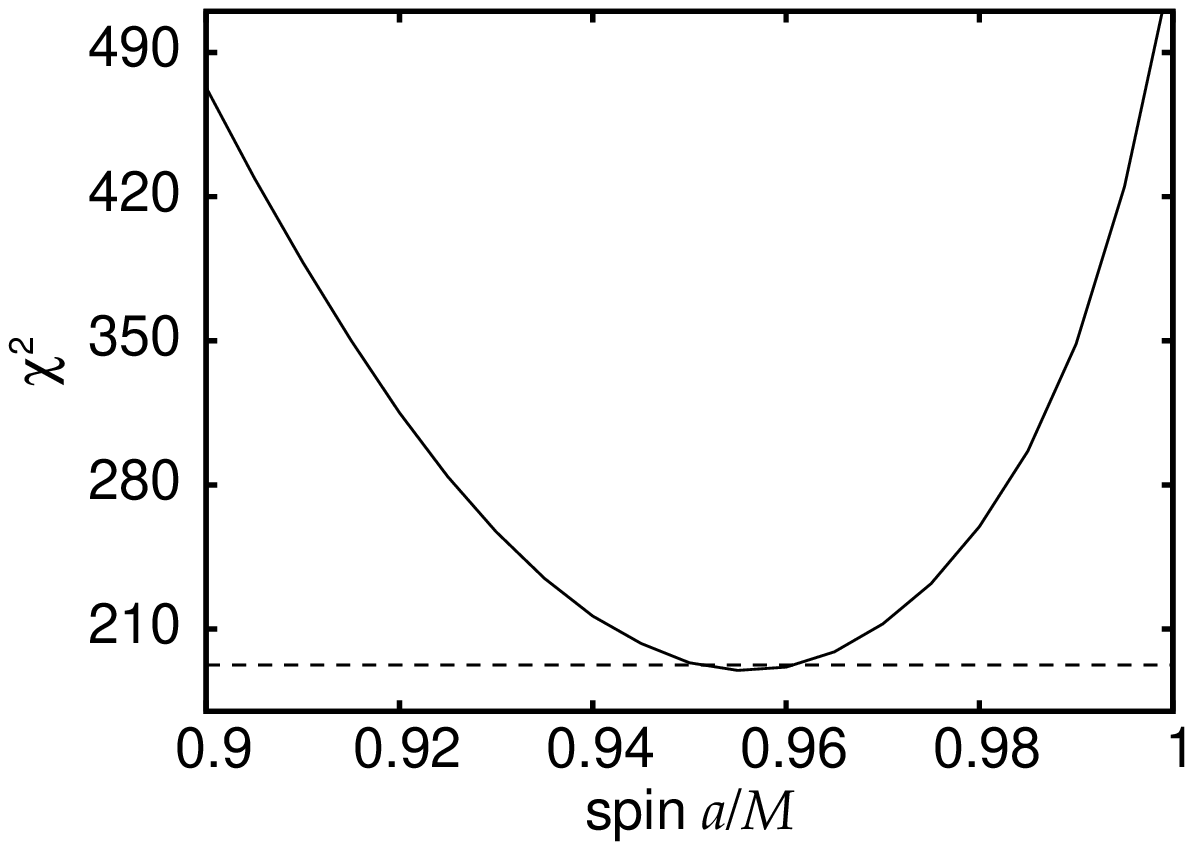} &
\includegraphics[width=0.315\textwidth]{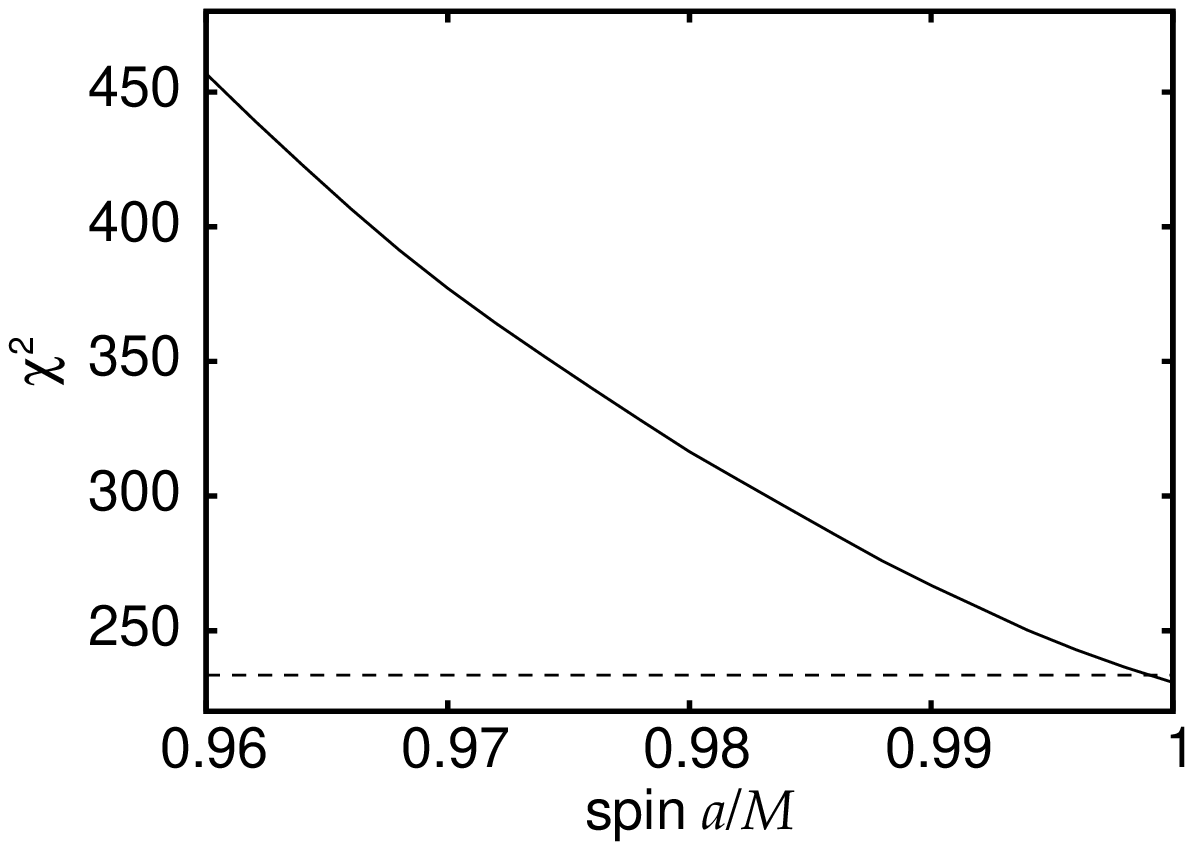} &
\includegraphics[width=0.315\textwidth]{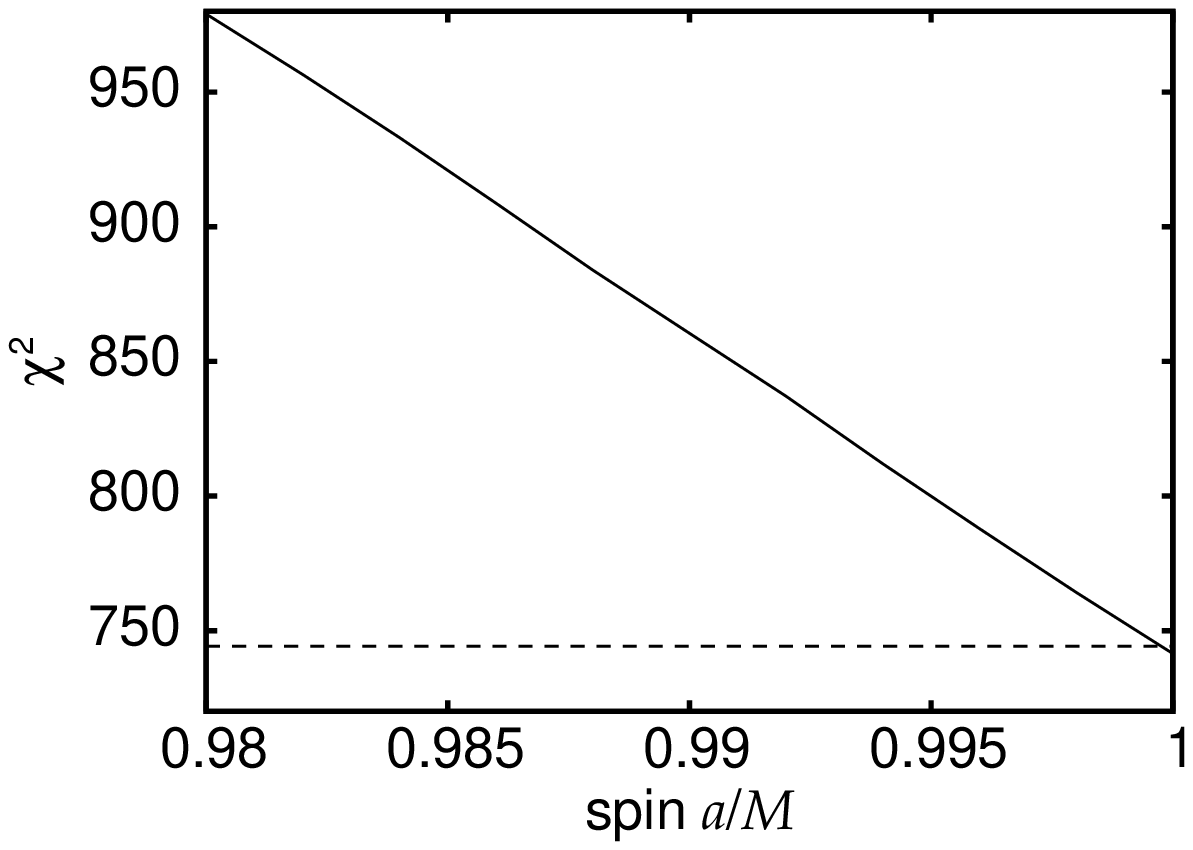} \\
\end{tabular}
\begin{tabular}{ccc}
  \includegraphics[width=0.315\textwidth]{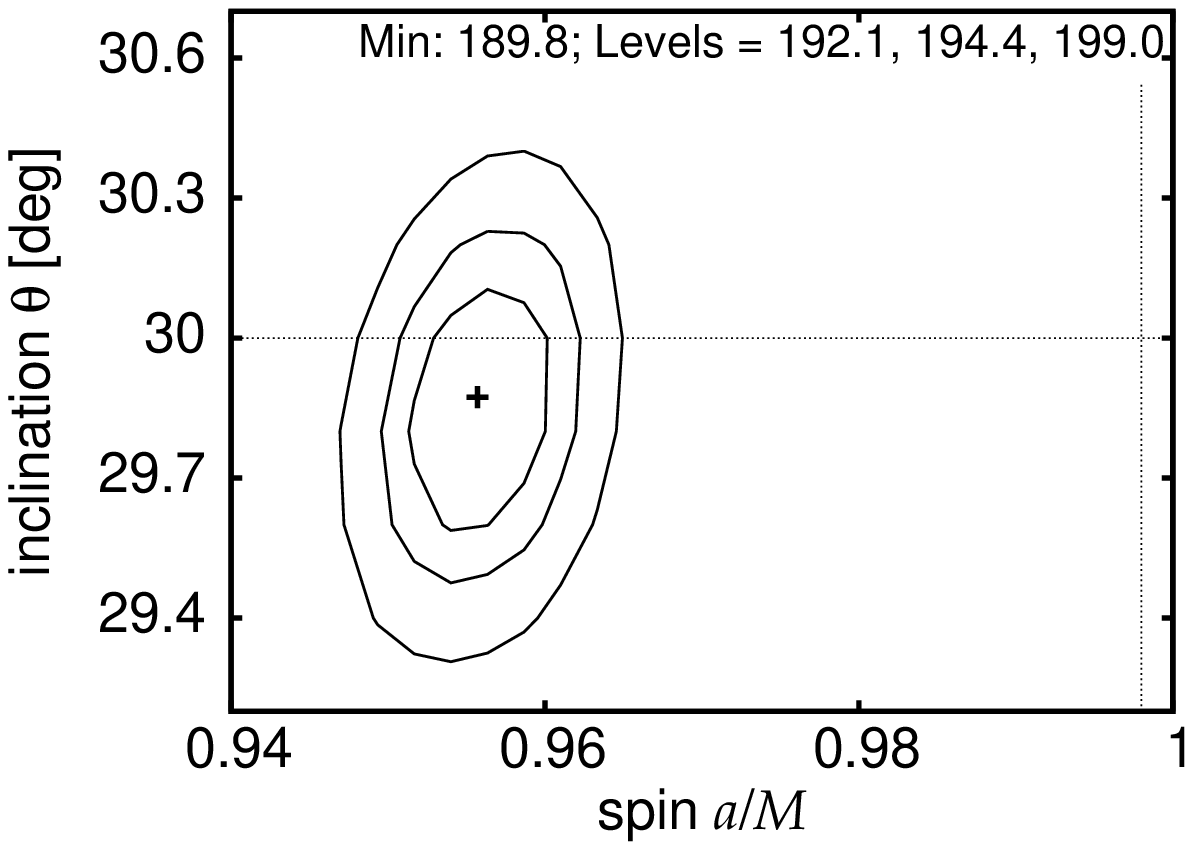} &
  \includegraphics[width=0.315\textwidth]{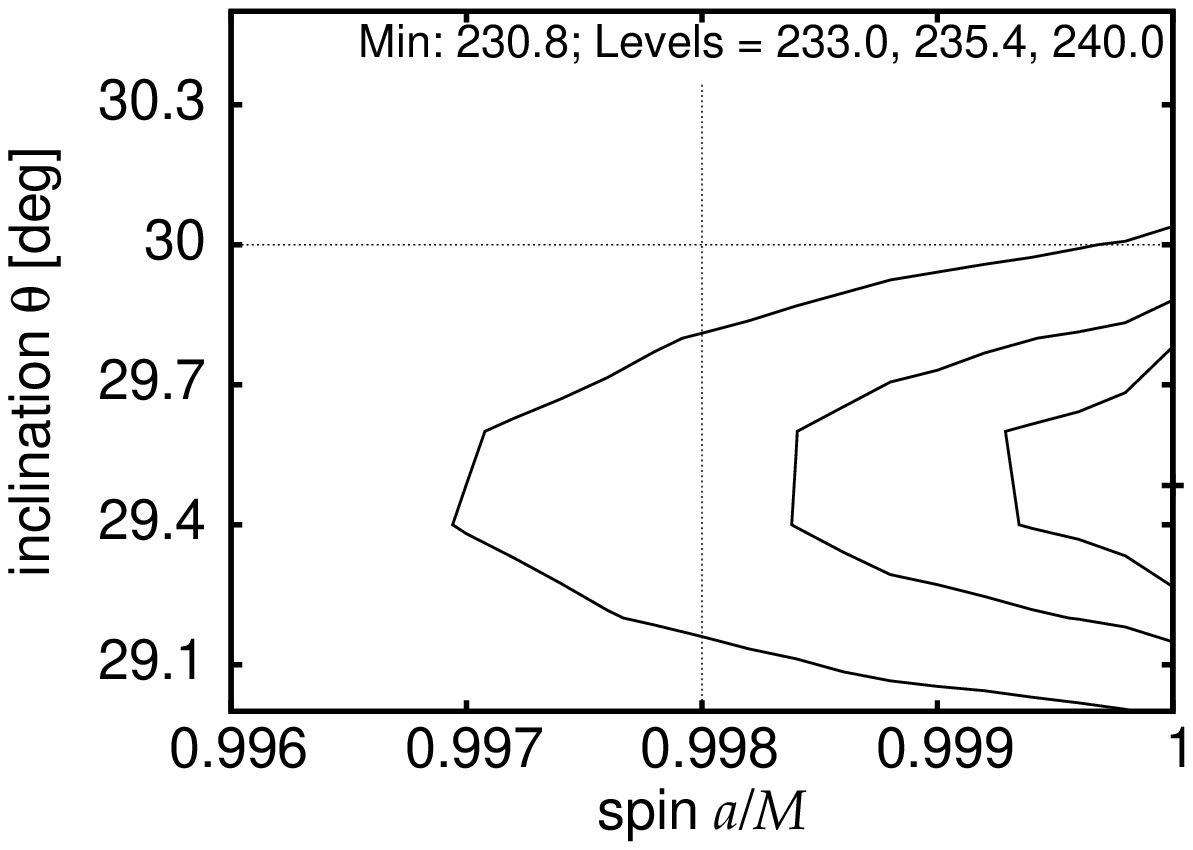} &
  \includegraphics[width=0.315\textwidth]{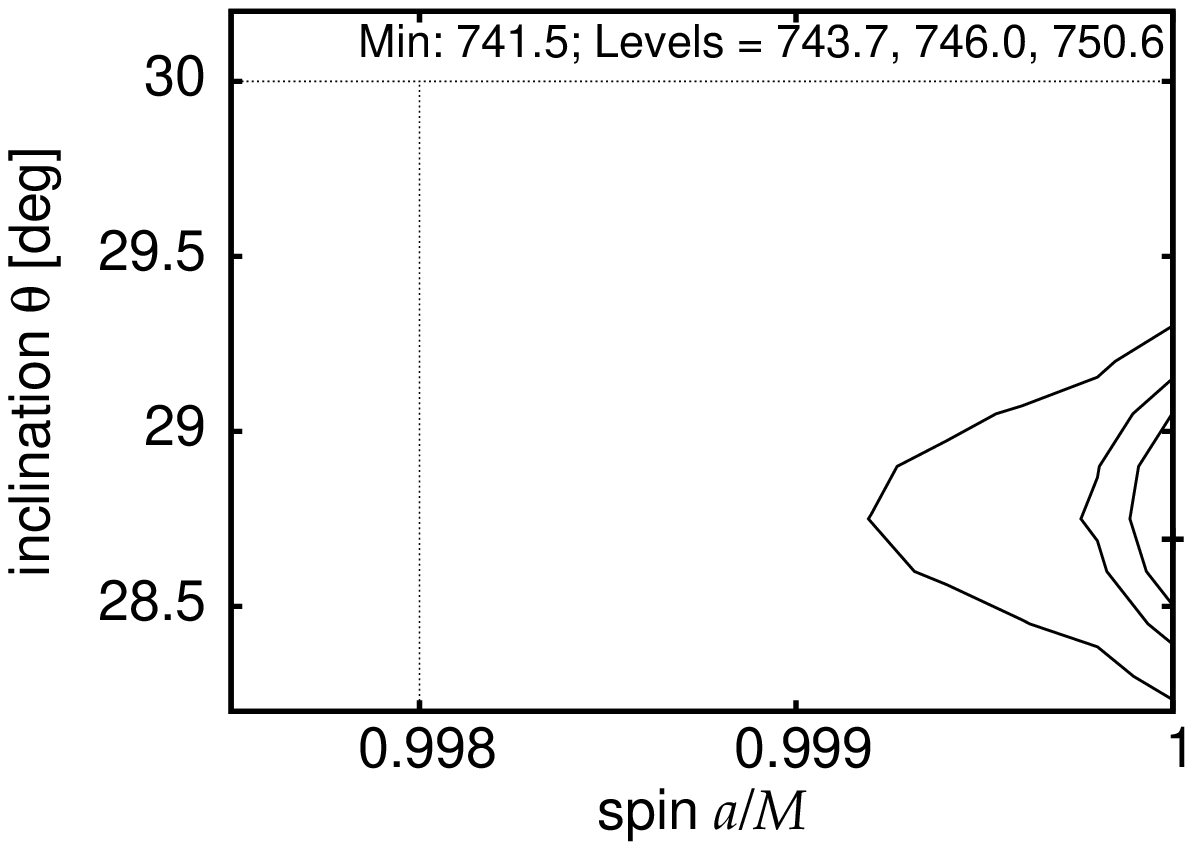}\\
\end{tabular}
\caption{The same as in Figure~\ref{fig_a07i30}, but for 
$a_{\rm f}=0.998$ and $\theta_{\rm f}=30\deg$.}
\label{fig_a0998i30}
\end{center}
\end{figure}

\begin{figure}[tbh!]
\begin{center}
\begin{tabular}{ccc}
\includegraphics[width=0.315\textwidth]{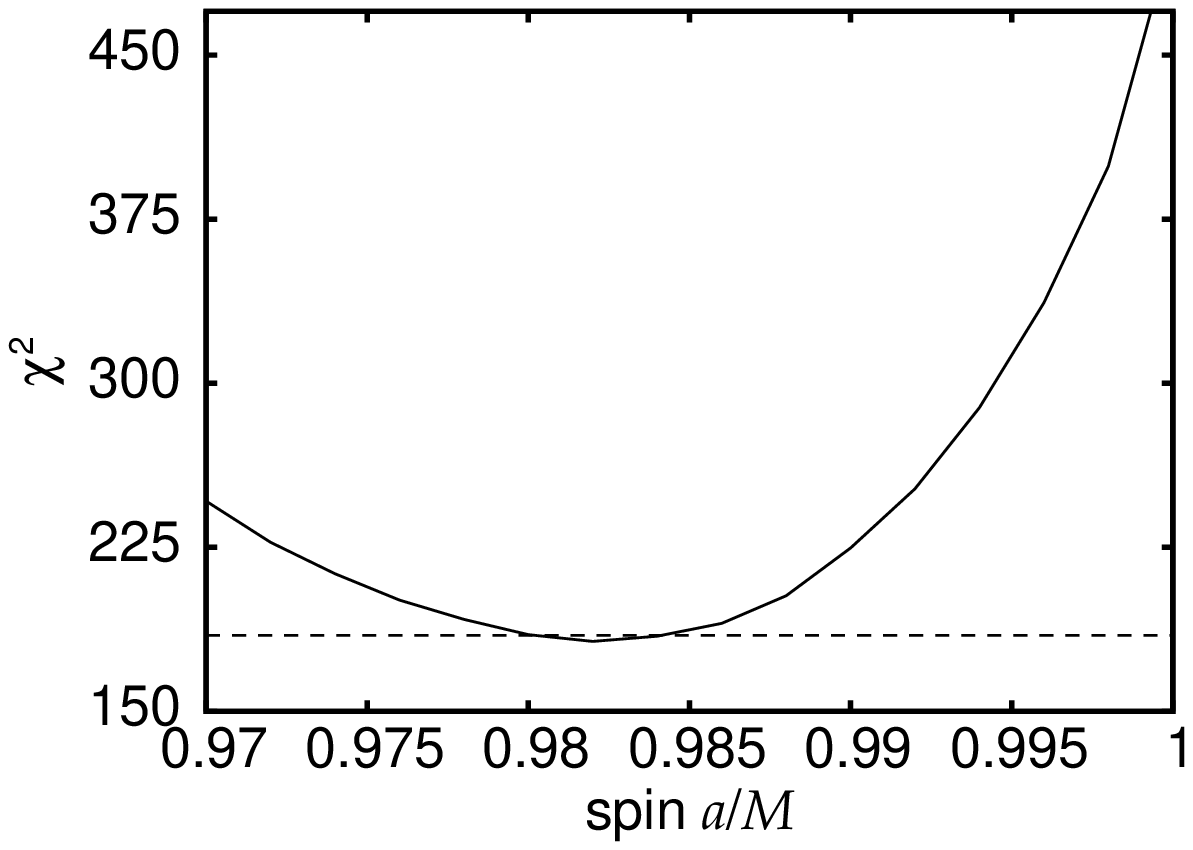} &
\includegraphics[width=0.315\textwidth]{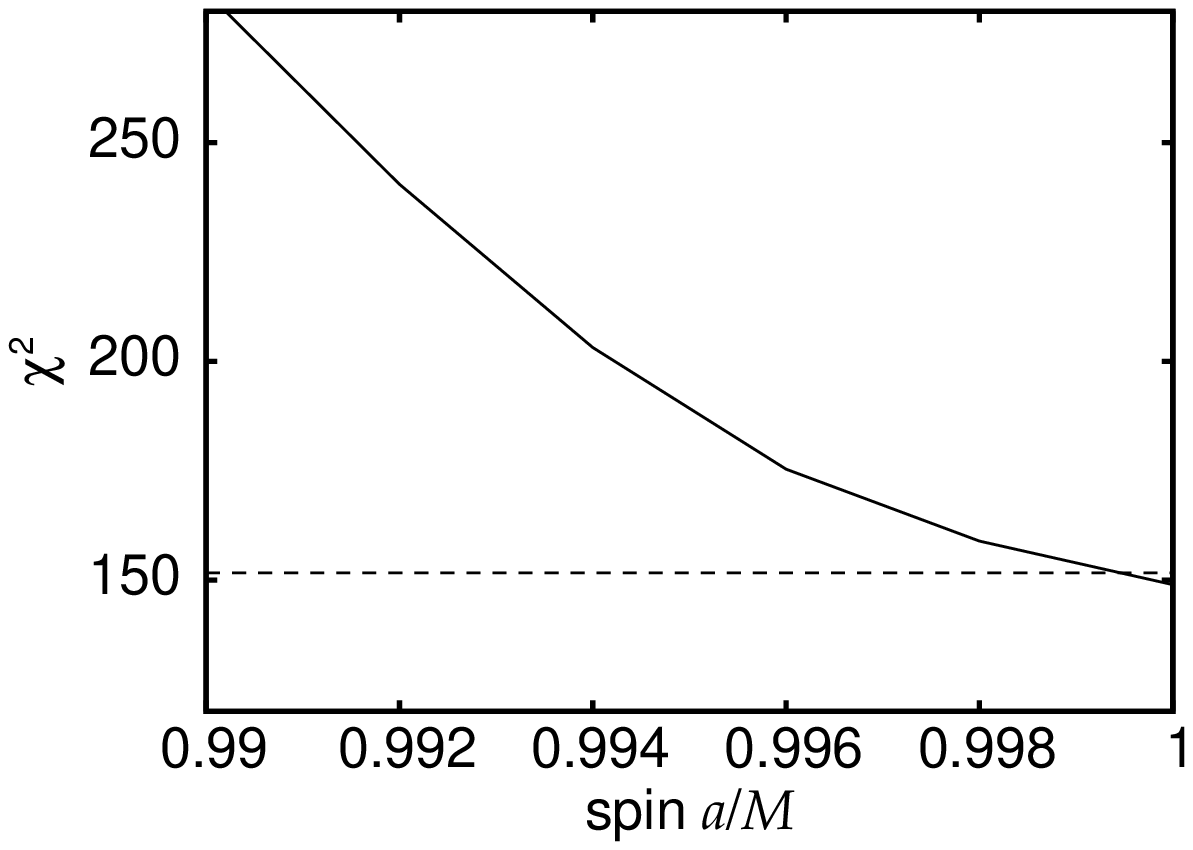} &
\includegraphics[width=0.315\textwidth]{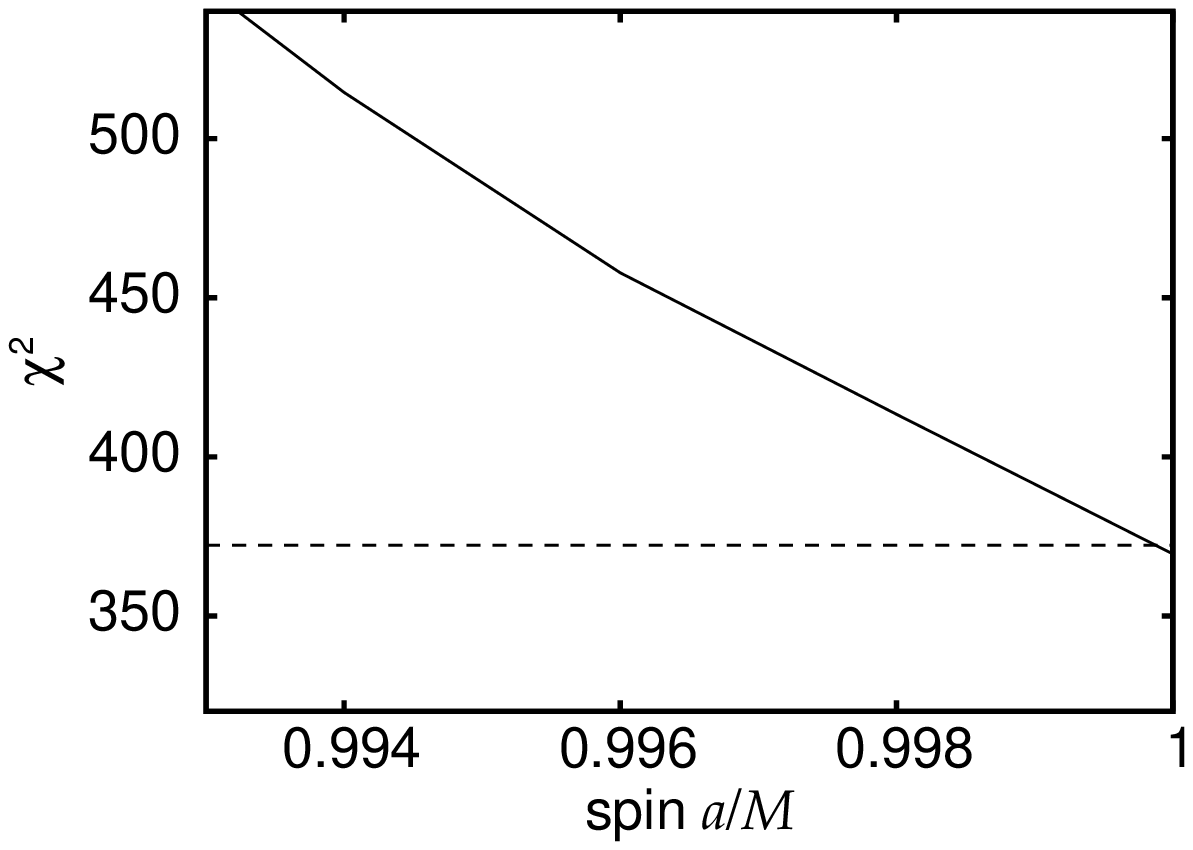} \\
\end{tabular}
\begin{tabular}{ccc}
  \includegraphics[width=0.315\textwidth]{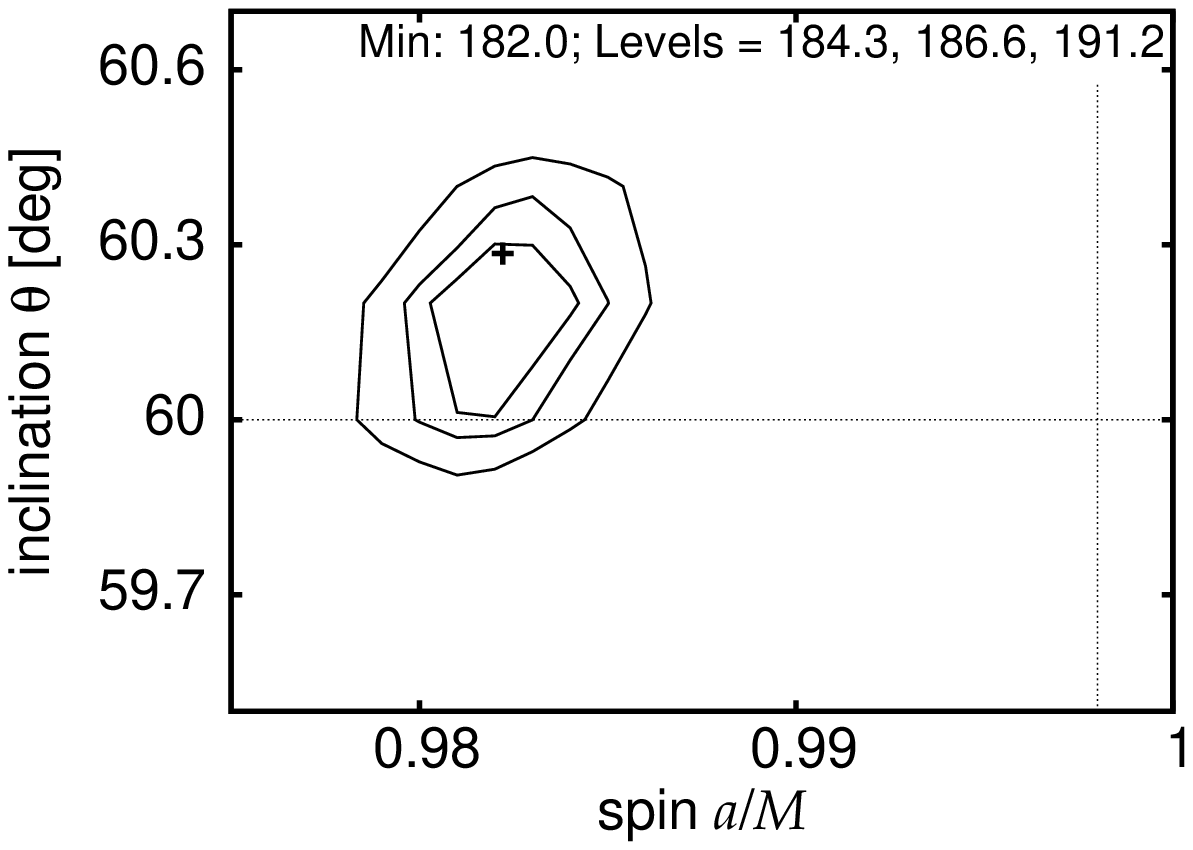} &
  \includegraphics[width=0.315\textwidth]{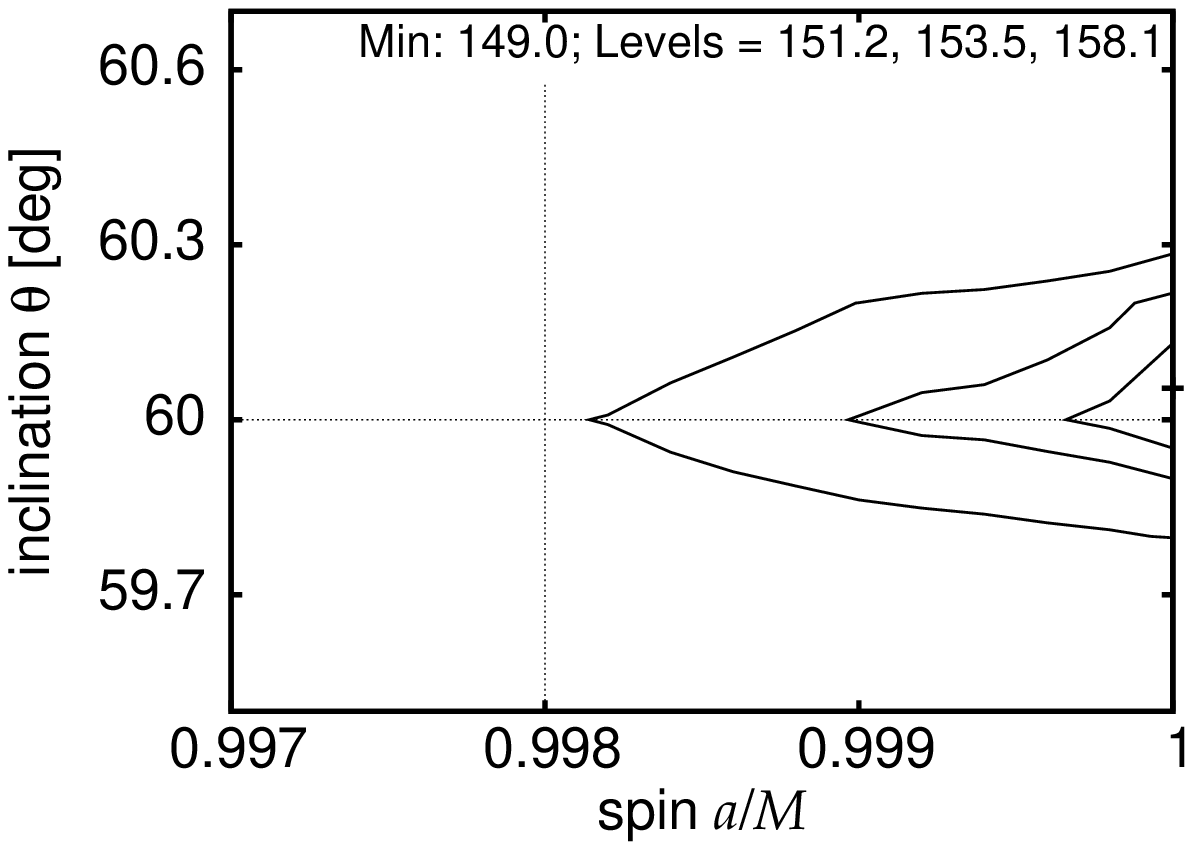} &
  \includegraphics[width=0.315\textwidth]{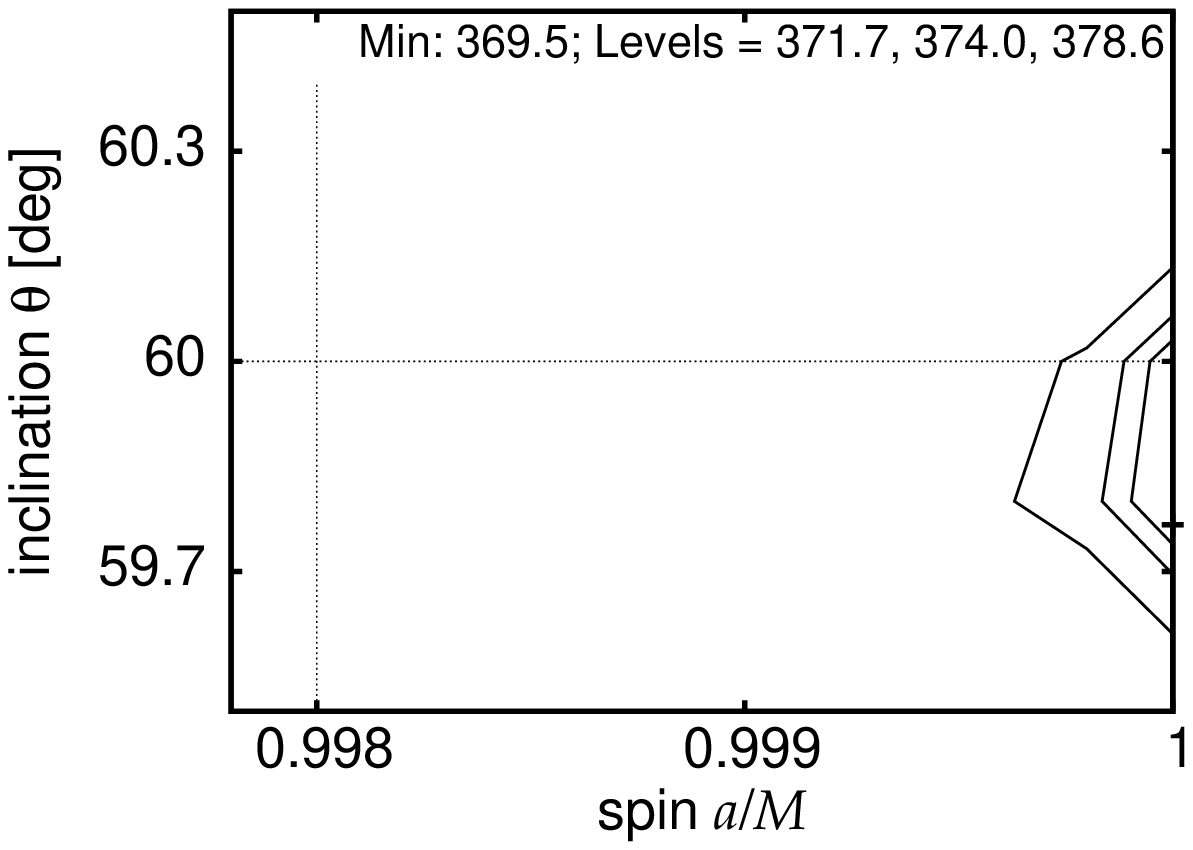}\\
\end{tabular}
\caption{The same as in Figure~\ref{fig_a07i30}, but for 
$a_{\rm f}=0.998$ and $\theta_{\rm f}=60\deg$.}
\label{fig_a0998i60}
\end{center}
\end{figure}

\begin{table}
\caption{The best-fit spin and inclination angle values inferred for the three cases of the
limb darkening/brightening law in the \textscown{kyl3cr} model.}
\begin{center}
\begin{tabular}{c|ccc} 
\hline \hline
\rule[-0.7em]{0pt}{2em}  & Case 1 & Case 2 & Case 3 \\ \hline
\rule[-1.3em]{0pt}{3em} & \multicolumn{3}{c}{$a_{\rm f}=0.7$, $\theta_{\rm f}=30^{\circ}$}\\
%\hline
%\rule[-0.7em]{0pt}{2em}  & Case 1 & Case 2 & Case 3 \\ \hline
\rule[-0.7em]{0pt}{2em} $a$ & $0.60^{+0.02}_{-0.01}$ & $0.69^{+0.01}_{-0.01}$ & $0.76^{+0.01}_{-0.02}$ \\ 
\rule[-0.7em]{0pt}{2em} $\theta_{\rm{}o}$ [deg] & $29.8^{+0.2}_{-0.3}$ & $29.7^{+0.3}_{-0.3}$ & $29.6^{+0.3}_{-0.3}$ \\
%\hline
%$\chi^{2}$ & 202		   & 185		    & 194 \\
\rule[-0.7em]{0pt}{2em} $\chi^{2}/\nu$ & 	1.33	   & 1.27	    & 1.39 \\
\hline
\hline
\rule[-1.3em]{0pt}{3em}  & \multicolumn{3}{c}{$a_{\rm f}=0.7$, $\theta_{\rm f}=60^{\circ}$}\\
%\hline
%\rule[-0.7em]{0pt}{2em}  & Case 1 & Case 2 & Case 3 \\ \hline
\rule[-0.7em]{0pt}{2em} $a$ & $0.65^{+0.03}_{-0.05}$ & $0.73^{+0.03}_{-0.04}$ & $0.82^{+0.02}_{-0.02}$ \\ 
\rule[-0.7em]{0pt}{2em} $\theta_{\rm{}o}$ [deg] & $60.0^{+0.1}_{-0.2}$ & $60.0^{+0.2}_{-0.1}$ & $60.3^{+0.1}_{-0.1}$ \\
%\hline
%$\chi^{2}$ & 202		   & 185		    & 194 \\
\rule[-0.7em]{0pt}{2em} $\chi^{2}/\nu$ &	1.87	   & 1.00	    & 1.48  \\
\hline
\hline
\rule[-1.3em]{0pt}{3em} & \multicolumn{3}{c}{$a_{\rm f}=0.998$, $\theta_{\rm f}=30^{\circ}$}\\
%\hline
%\rule[-0.7em]{0pt}{2em} & Case 1 & Case 2 & Case 3 \\ \hline
\rule[-0.7em]{0pt}{2em} $a$ & $0.956^{+0.005}_{-0.005}$ & $1^{+0}_{-1{\rm E}-3}$ & $1^{+0}_{-8{\rm E}-5}$ \\ 
\rule[-0.7em]{0pt}{2em} $\theta_{\rm{}o}$ [deg] & $29.9^{+0.3}_{-0.3}$ & $29.5^{+0.3}_{-0.3}$ & $28.7^{+0.4}_{-0.2}$  \\
%\hline
\rule[-0.7em]{0pt}{2em} $\chi^{2}/\nu$ & 	1.30	   & 1.58	    & 5.08 \\
\hline
\hline
\rule[-1.3em]{0pt}{3em} & \multicolumn{3}{c}{$a_{\rm f}=0.998$, $\theta_{\rm f}=60^{\circ}$}\\
%\hline
%\rule[-0.7em]{0pt}{2em}  & Case 1 & Case 2 & Case 3 \\ \hline
\rule[-0.7em]{0pt}{2em} $a$ & $0.982^{+0.002}_{-0.002}$ & $1^{+0}_{-3{\rm E}-4}$ & $1^{+0}_{-6{\rm E}-5}$ \\ 
\rule[-0.7em]{0pt}{2em} $\theta_{\rm{}o}$ [deg] & $59.9^{+0.1}_{-0.2}$ & $60.1^{+0.1}_{-0.2}$ & $60.2^{+0.1}_{-0.2}$ \\
%\hline
\rule[-0.7em]{0pt}{2em} $\chi^{2}/\nu$ & 	1.24	   & 1.02	    & 2.51 \\
\hline
\end{tabular}%$
\label{tab_angdirfp}
\end{center}
 
Data were generated using the \textscown{kyl2cr} model. See the main text for details. 
The quoted errors correspond to the 90$\%$ confidence level.
\end{table}

The results presented in the previous sections 
show that using different emission directionality approaches
leads to a different location of the $\chi^{2}$ minimum
in the parameter space. 
Doubt about the correct prescription for the emission directionality
thus brings some non-negligible inaccuracy into the evaluation of 
the model parameters. The magnitude of this error cannot be easily 
assessed as a unique number because other parameters are also involved.

A possible way to tackle the problem is to derive the intrinsic
spectrum from self-consistent numerical computations. This has the potential
of removing the uncertainty about the emission directionality (although,
to a certain degree this uncertainty is only moved to a different level 
of the underlying model assumptions).
In this section, we present such results from modelling the artificial
data generated by numerical simulations,
i.e.\ independently of an analytical approximation 
of the emission directionality presented in the previous sections.

Let us remind the reader that the orbital speed within the inner
$\lesssim10r_{\rm g}$ reaches a considerable fraction of the speed of
light (Fig.~\ref{v_orbit}). Beaming, aberration, and the light-bending 
all affect the emitted photons very significantly in
this region. Less energetic photons come
from the outer parts where the motion slows down and the relativistic
effects are of diminished importance. This reasoning suggests that the
analysis of the previous section may be inaccurate because the adopted
analytical approximations (\ref{case123}) neglect any dependence on
energy and distance. 

We applied the Monte-Carlo radiative transfer code NOAR \citep[see
Section~5 in][]{2000A&A...357..823D} for the case of ``cold''
reflection, i.e.\ for neutral or weakly-ionised matter. 
The NOAR code computes absorption cross sections in each layer. 
Free-free absorption and the recombination continua of hydrogen-
and helium- like ions are taken into account, as well as the 
direct and inverse Compton scattering. The NOAR
code enables us to obtain the angle-dependent intensity for the
reprocessed emission. The cold reflection case serves as a reference
point that can be later compared against the
models involving  stronger irradiation and higher ionisation of the disc
medium.

The directional distribution of the intrinsic emissivity of the
reprocessing model is shown in Fig.~\ref{fig4} (right panel).
The continuum photon index $\Gamma=1.9$ is considered
and the energies are integrated over the $2$--$100$ keV range.
Although the results of the radiation transfer computations do 
show the limb-brightening effect, it is a rather mild one,
and not as strong as the Case~1.
In the same plot we also show the angular profile of the emissivity 
distribution in different energy ranges: (i)~2--6~keV (i.e.\ below the iron 
K$\alpha$ line rest energy); (ii)~6--15~keV (including the iron K$\alpha$ 
line); (iii)~15--100~keV (including the Compton hump). We notice that
the energy integrated profile is dominated by the contribution from
the Compton hump, where much of the emerging flux originates.
However, all of the energy sub-ranges indicate the limb-brightening
effect, albeit with slightly different prominence.

We implemented the numerical results of NOAR modelling of a reflected
radiation from a cold disc as \textscown{kyl2cr} in the \textscown{ky}
collection of models. Furthermore, we produced an averaged model,
\textscown{kyl3cr}, by integrating \textscown{kyl2cr} tables over all angles.
Therefore, \textscown{kyl3cr} lacks information about the detailed 
angular distribution of the intrinsic local emission from the disc
surface. On the other hand, it has the advantage of increased
computational speed and the results are adequate if the 
emission is locally isotropic. Furthermore, \textscown{kyl3cr} can be {\em
a posteriori} equipped with an analytical prescription for the angular
dependence (Cases 1--3 in eq.~\ref{case123}), which brings the angular 
resolution back into consideration. This approach allows us to switch
between the three prescriptions for comparison and rapid evaluation. 

In order to constrain the feasibility of the aforementioned approaches, we
generated the artificial data using the \textscown{powerlaw + kyl2cr} model.
The parameters of the model are:
photon index of the power law $\Gamma=1.9$ and its normalisation
$K_\Gamma=0.01$, spin of the black hole $a$, inclination angle
$\theta_{\rm o}$, the inner and outer radii of the disc $r_{\rm in}=r_{\rm
ms}$ and $r_{\rm out}=400$, the index of the radial dependence of the
emissivity $q=3$, and the normalisation of the reflection component
$K_{\rm kyl2cr}=0.1$. We simulated the data for two different values of
the spin, $a=0.7$ and $a=0.998$, and for inclination angles
$\theta_{\rm o}=30^{\circ}$ and  $\theta_{\rm o}=60^{\circ}$.
The simulated flux of the primary power law component is the same
as in the previous section. However, now an important fraction
of the primary radiation is reflected from the disc. The total
flux depends on the extension of the disc and its inclination. 
Its value is $4.9-5.6\times10^{-11}$\,erg\,cm$^{-2}$\,s$^{-1}$ for our
choice of the parameters.

As a next step, we replaced \textscown{kyl2cr} by \textscown{kyl3cr} and
searched back for the best-fit results using the latter model. In this
way, using \textscown{kyl3cr} we obtained the values of the spin and the
inclination angle for different directionalities. The fitting results are
summarised in Table~\ref{tab_angdirfp}. Besides the spin and the
inclination angle, only normalisation of the reflection component was
allowed to vary during the fitting procedure. The remaining parameters
of the model were kept frozen at their default values. 

The resulting data/model ratios are shown in
Figures~\ref{fig_a07_ratio} and \ref{fig_a0998_ratio}. For  $a=0.7$
and $\theta_{\rm o}=30^{\circ}$ the graphs look very similar in all
three cases. However, the inferred spin value differs from the fiducial
value with which the test data were originally created.
The dependence of the best-fit $\chi^{2}$ statistic on the spin
and the corresponding graphs of the confidence contours 
for spin versus inclination angle are shown
in Figures \ref{fig_a07i30}--\ref{fig_a0998i60}, again for the three cases of
angular directionality. These figures confirm that for the limb-brightening
profile the inferred spin value comes out somewhat lower than the
correct value, whereas it is higher if the limb-darkening profile is
assumed. 

In each of the three cases the error of the resulting $a$ determination
depends on the inclination angle and the spin itself. However,
we find that the isotropic directionality reproduces our data to the
best precision. The limb darkening profile is not accurate at higher values
of the spin, such as $a=0.998$, when the resulting  $\chi^{2}/\nu$ value
even exceeds $2$. The limb darkening profile is characterised by an 
enhanced blue peak of the line while the height of the red peak is
reduced (see Figs.~\ref{fig4a}--\ref{fig4b}). Consequently, the model
profile is too steep to fit the data. This is clearly visible in the
data/model ratio plots for $a=0.998$ shown in Figures~\ref{fig_a0998_ratio}.
The flux is underestimated by the model below a mean energy value $E_{\rm mean}$ of the
line (for $a=0.998$ and $i=30$\,deg $E_{\rm mean} \approx 5$\,keV) 
and overestimated above $E_{\rm mean}$. This fact leads to
a noticeable jump in the data/model ratio plot.

A noteworthy result appears in comparing 
of the contours produced by the model with 
limb brightening and limb darkening for 
$a=0.7$ and $i=60$\,deg (Fig.~\ref{fig_a07i60}). 
Although the former model (limb brightening) gives 
a statistically worse fit with $\chi^{2}/\nu=1.87$ 
than the limb darkening case ($\chi^{2}/\nu=1.48$),
the inferred values of the spin and the inclination 
angle are consistent with the fiducial values within
the $3\sigma$ level. On the other hand, the spin value 
inferred from the limb darkening model
is far from the fiducial (i.e., the correct) value.

\begin{figure}[tbh!]
\begin{center}
\begin{tabular}{ccc}
\includegraphics[width=0.31\textwidth]{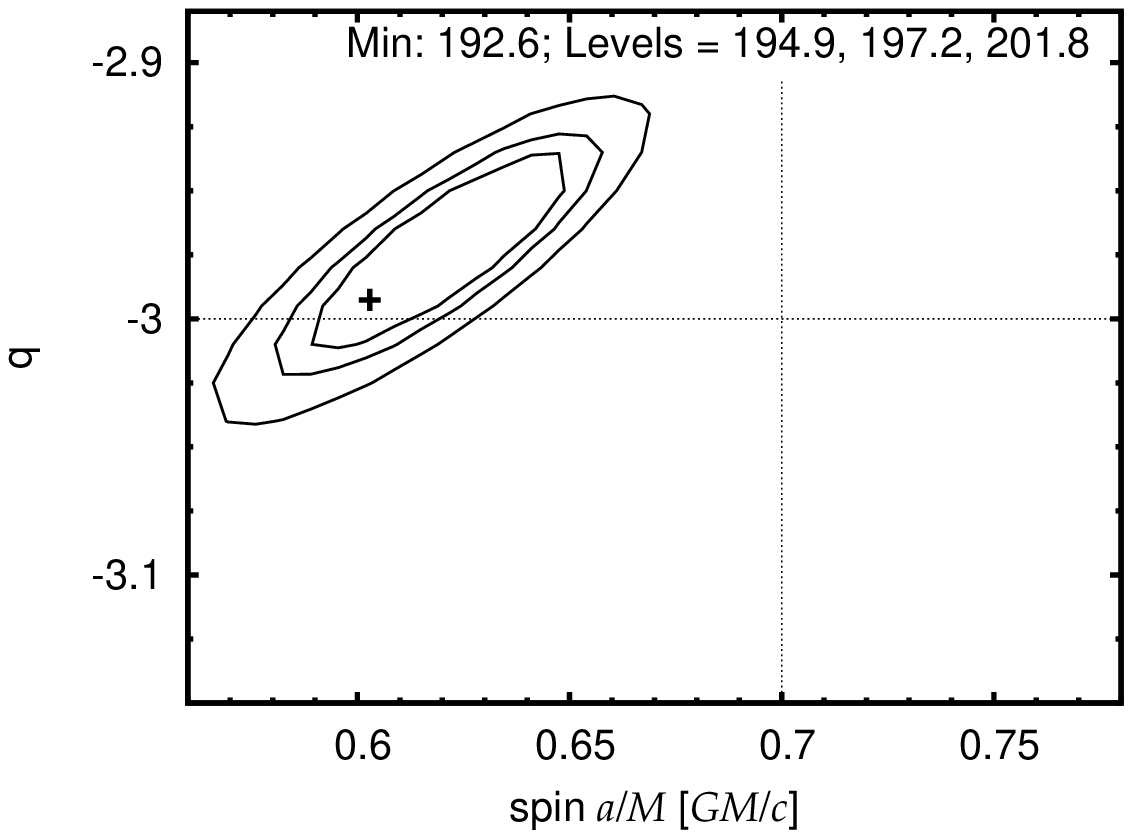} &
\includegraphics[width=0.31\textwidth]{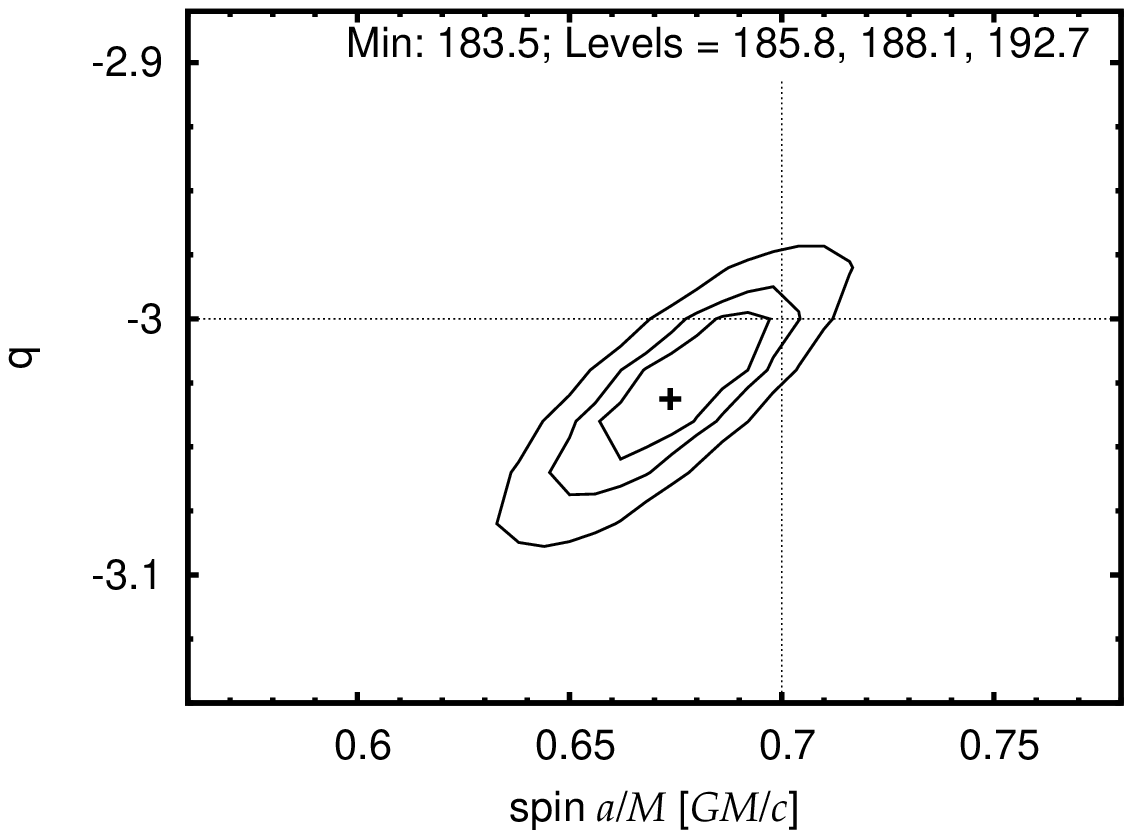} &
\includegraphics[width=0.31\textwidth]{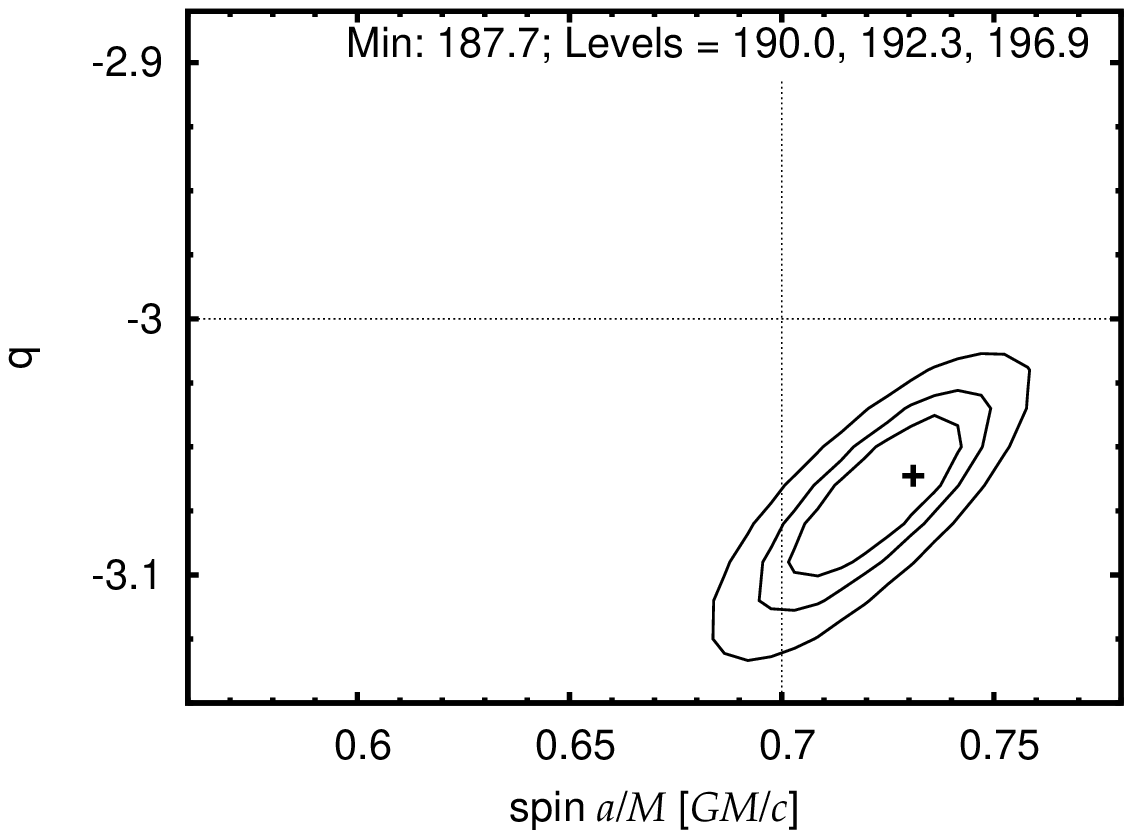} \\
\end{tabular}
\begin{tabular}{ccc}
\includegraphics[width=0.31\textwidth]{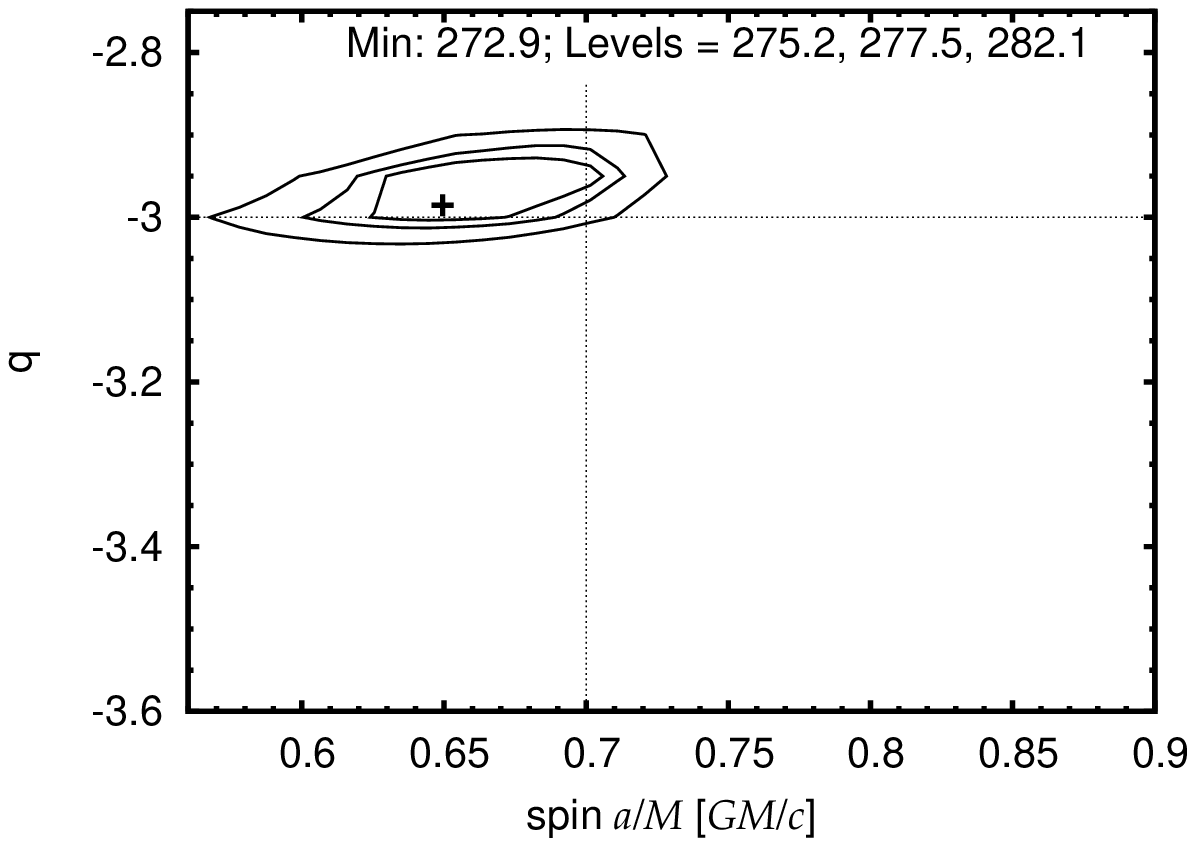} &
\includegraphics[width=0.31\textwidth]{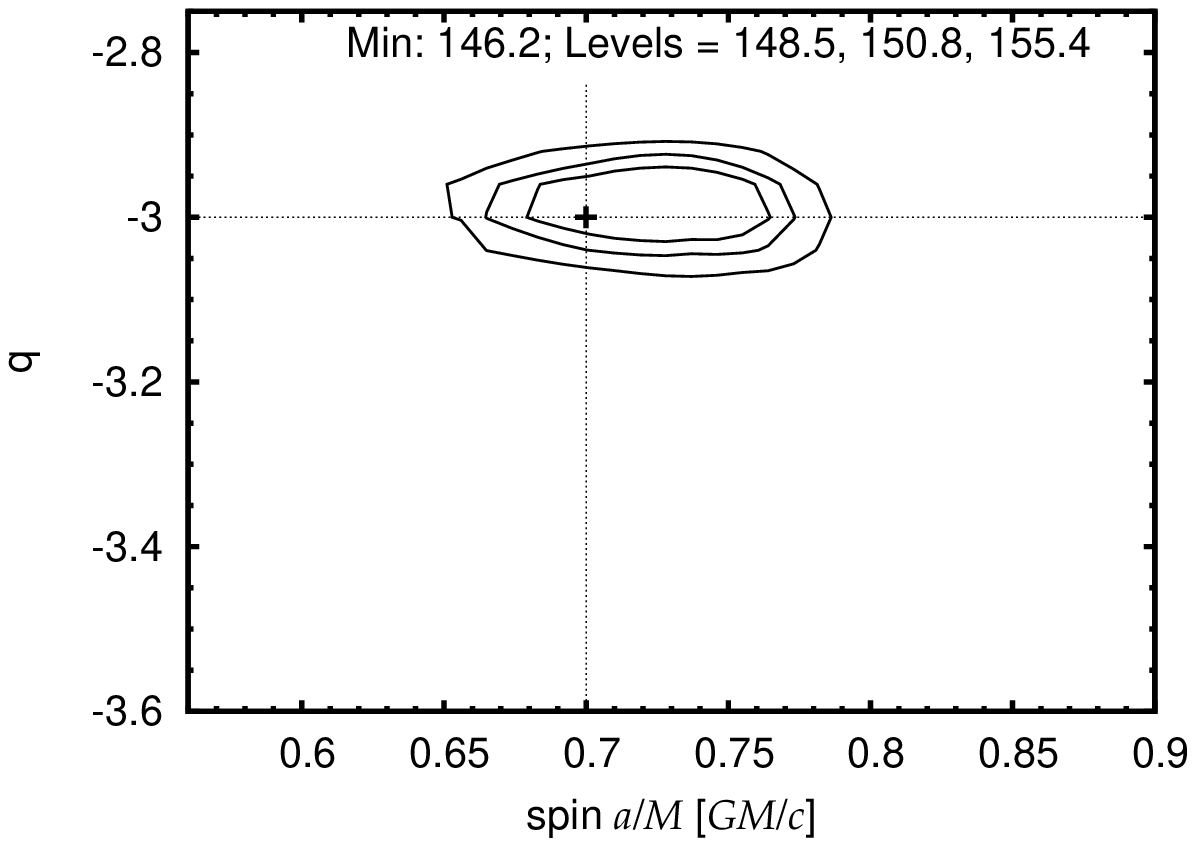} &
\includegraphics[width=0.31\textwidth]{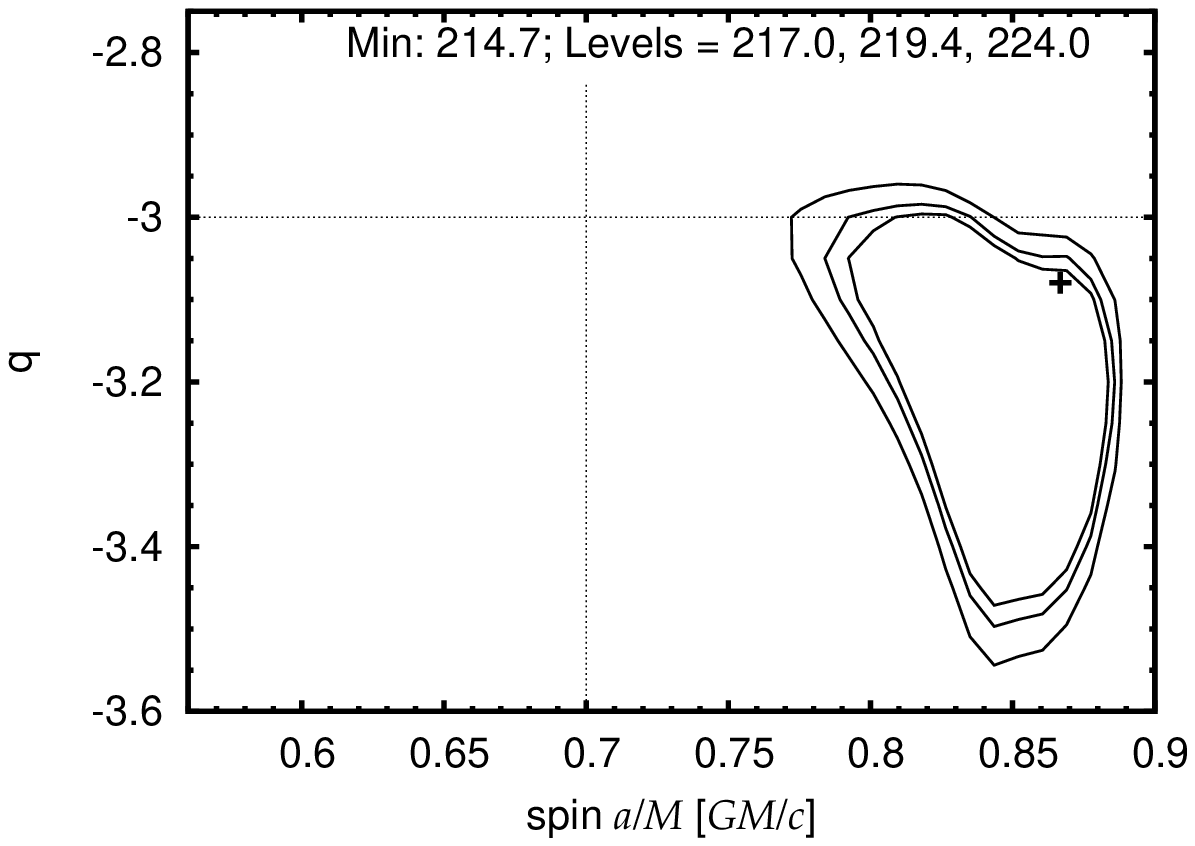} \\
\end{tabular}
\begin{tabular}{ccc}
\includegraphics[width=0.31\textwidth]{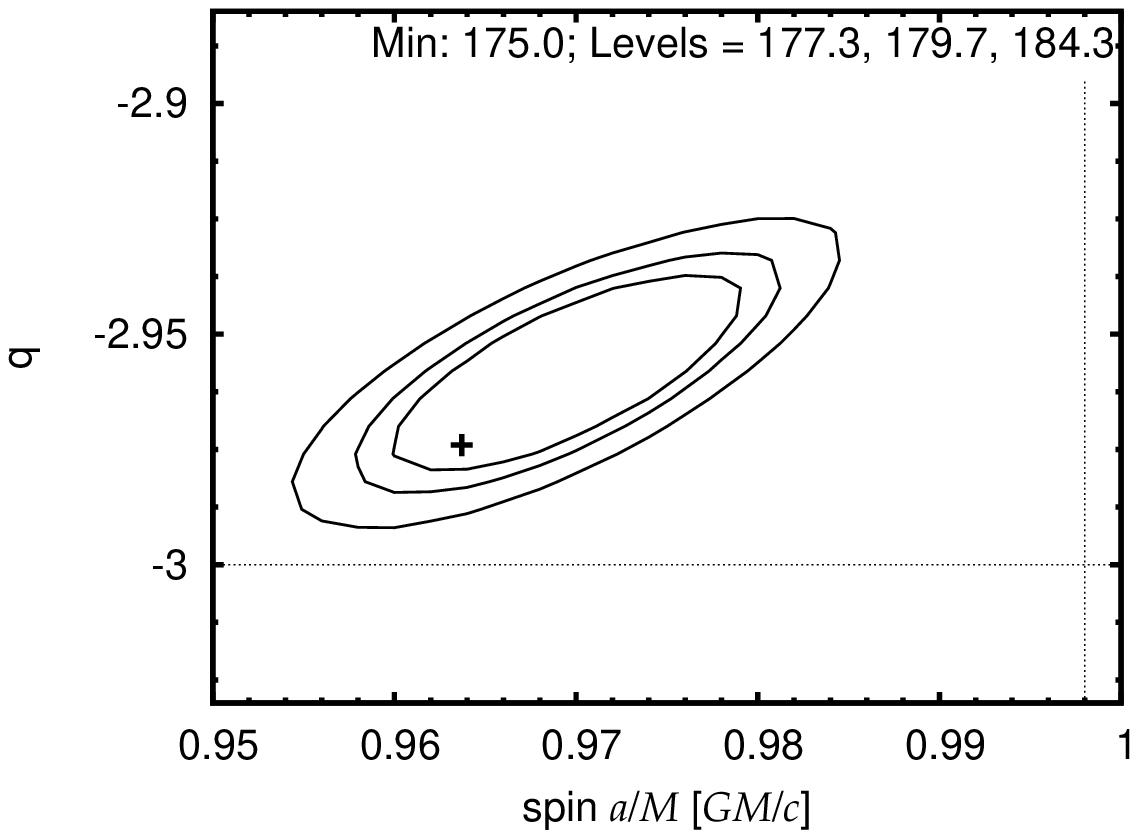} &
\includegraphics[width=0.31\textwidth]{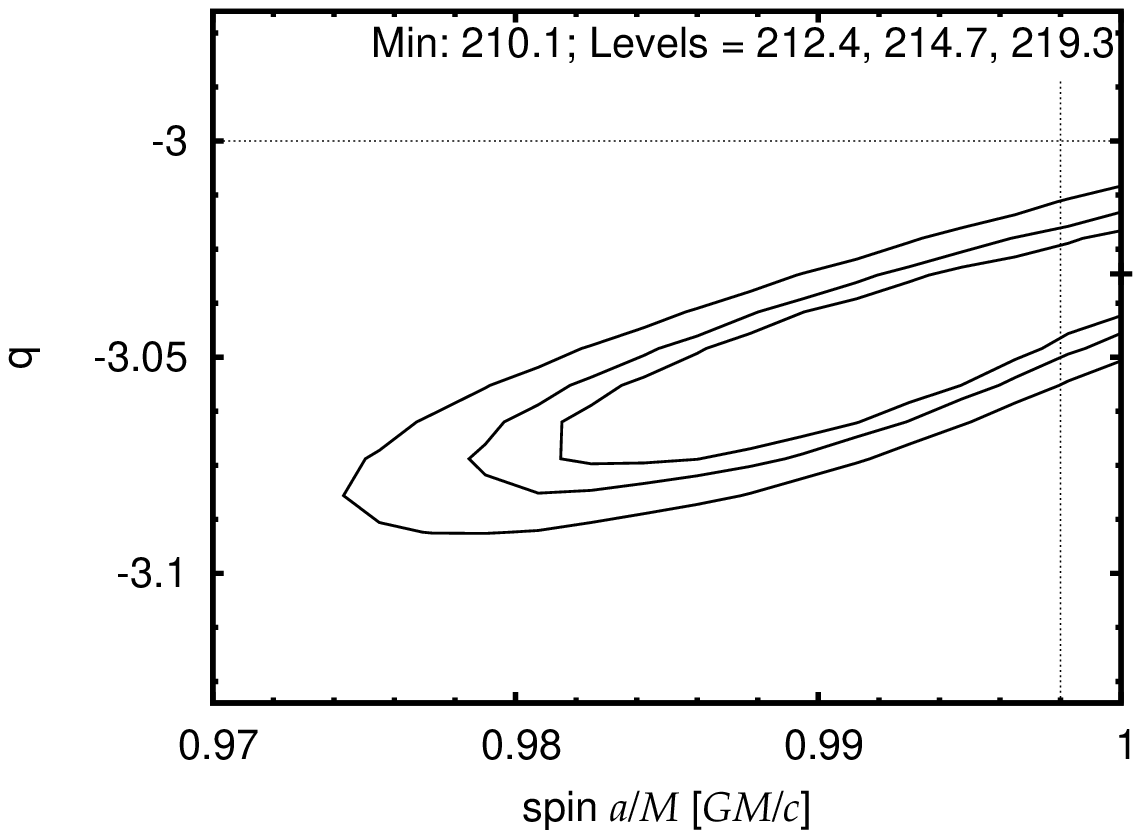} &
\includegraphics[width=0.31\textwidth]{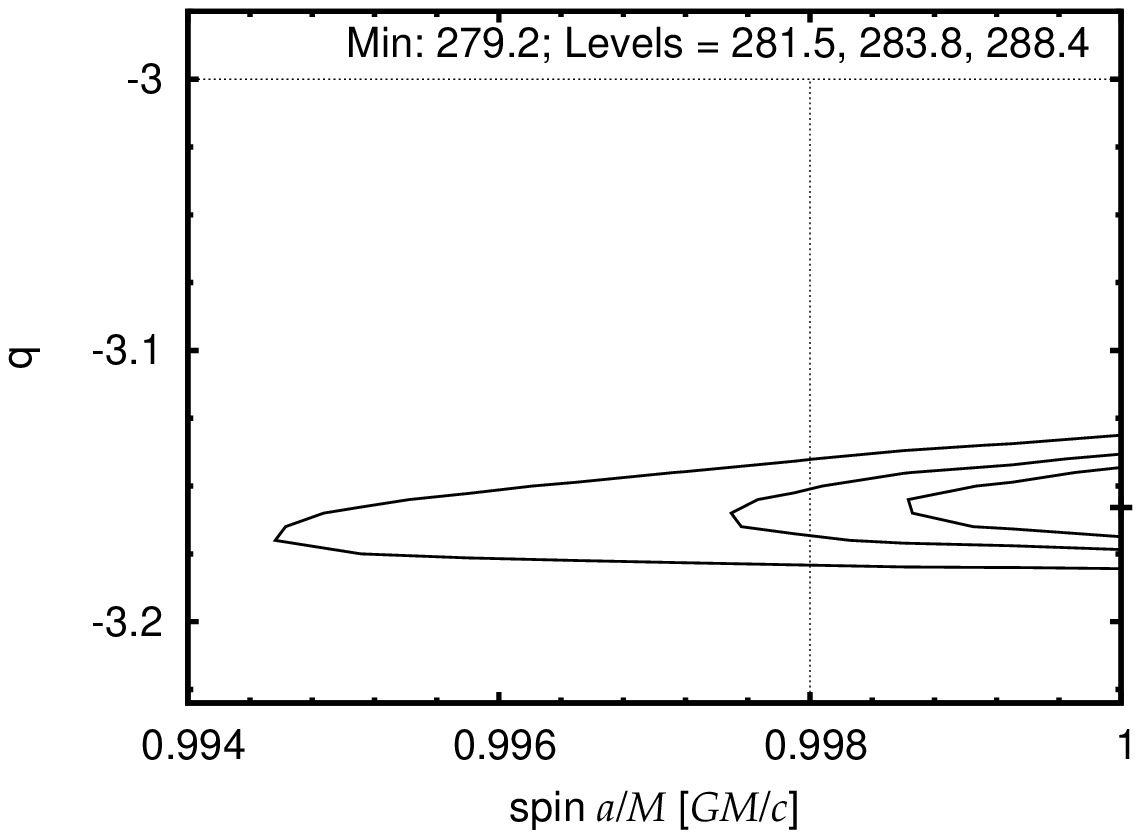} \\
\end{tabular}
\begin{tabular}{ccc}
\includegraphics[width=0.31\textwidth]{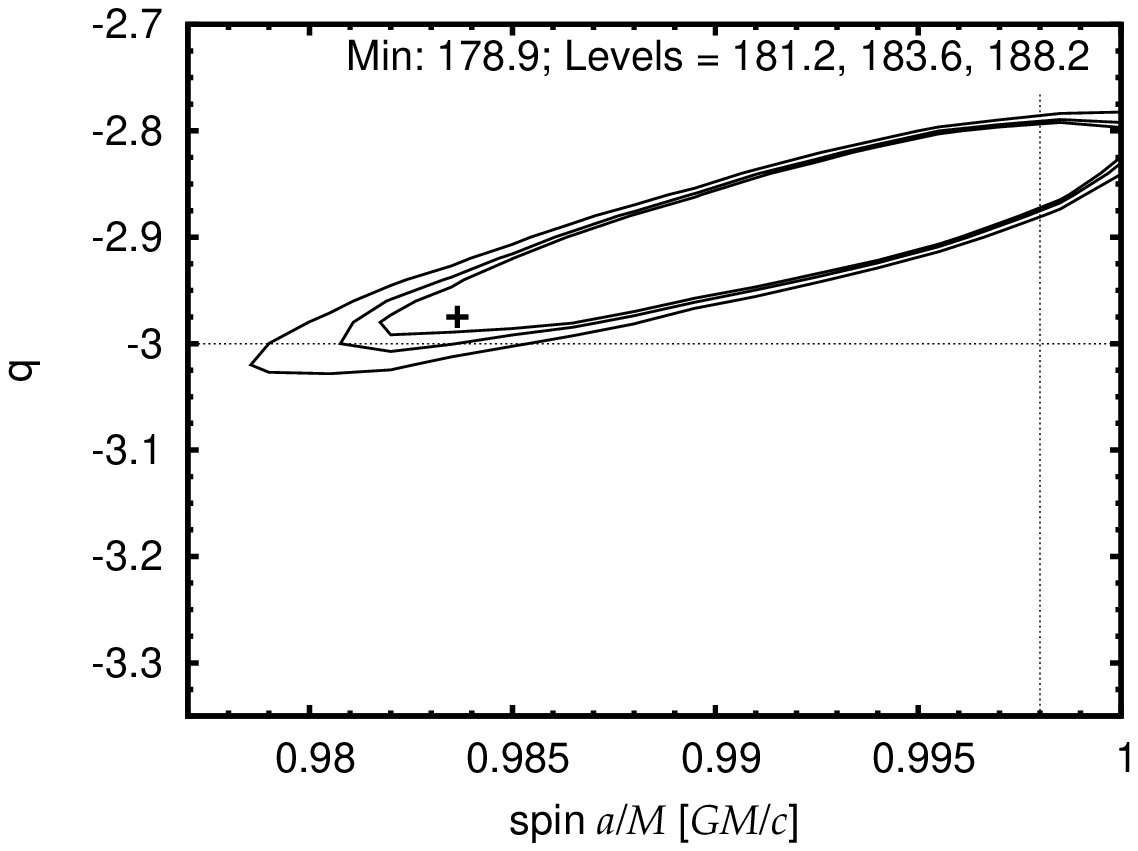} &
\includegraphics[width=0.31\textwidth]{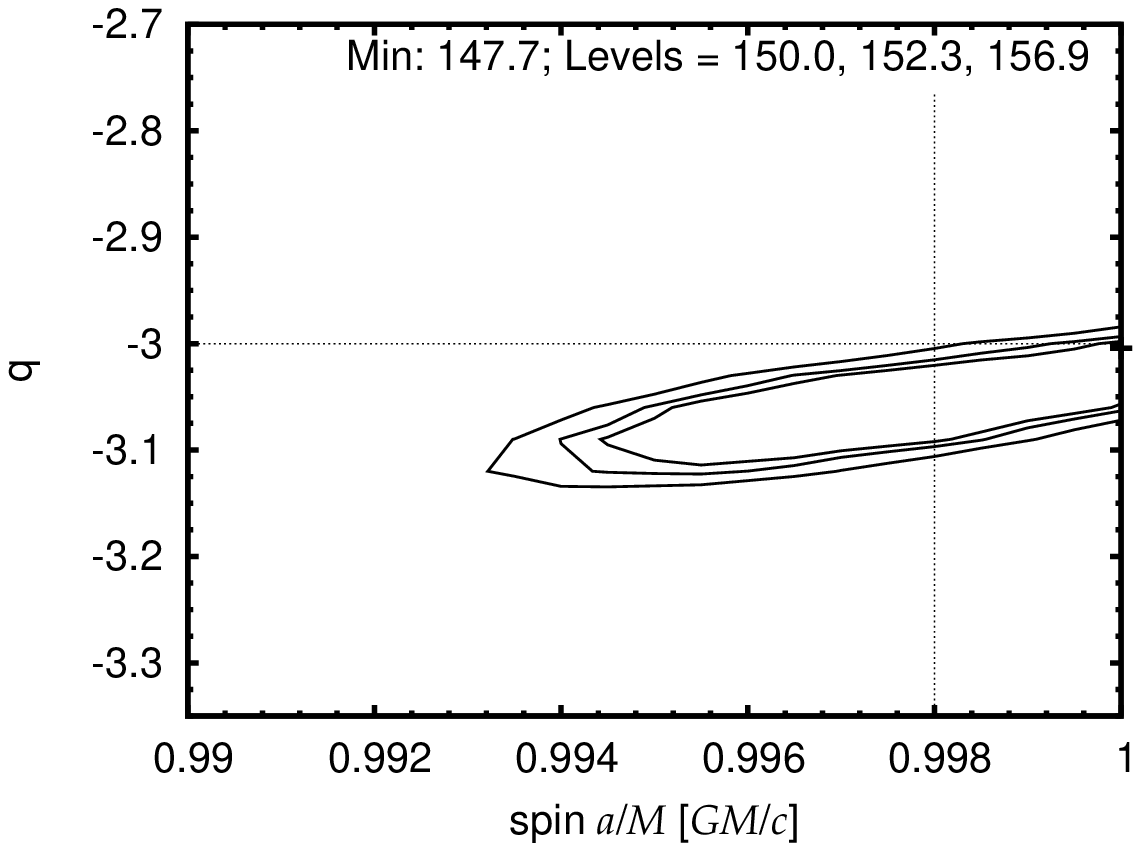} &
\includegraphics[width=0.31\textwidth]{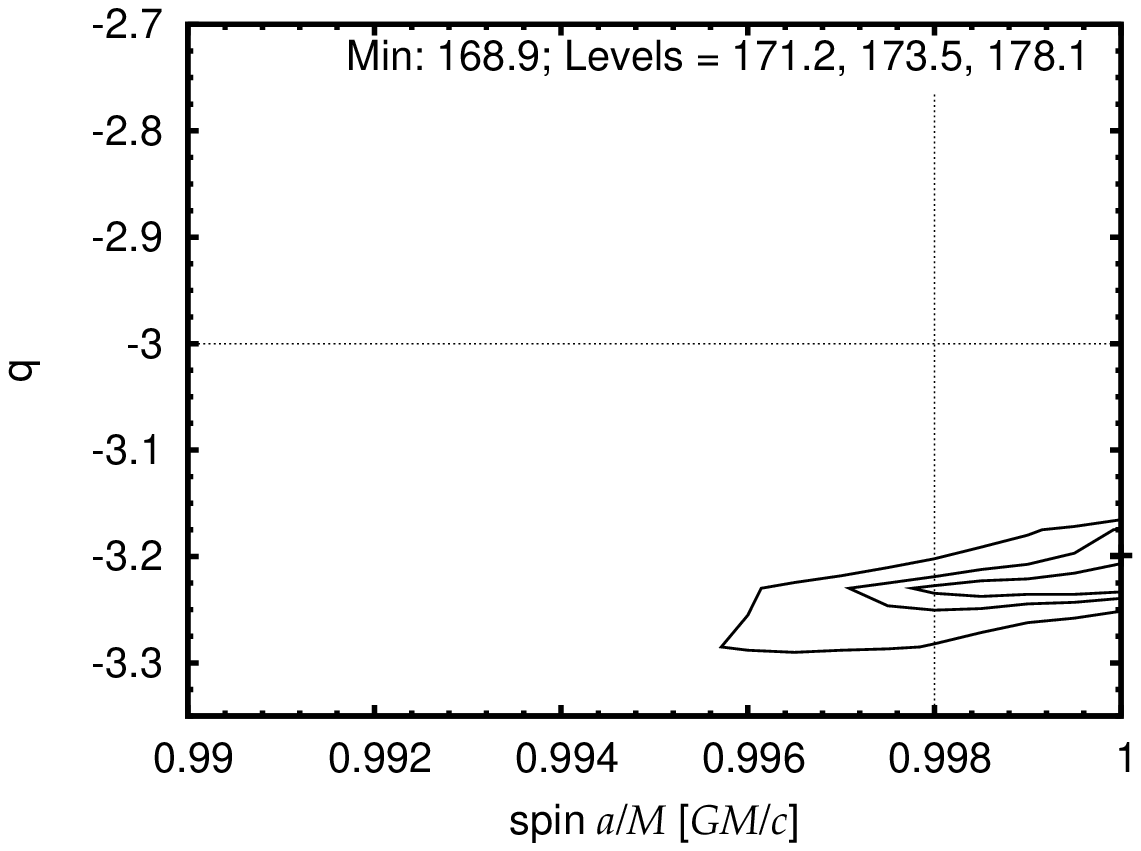} \\
\end{tabular}
\caption{The contour graphs of the spin $a$ and the parameter $q$ of the radial emissivity
for different fiducial values of the spin and inclination.
From \textposown{top} to \textposown{bottom}: $\{a=0.7$, $i=30$\,deg$\}$, $\{a=0.7$, $i=60$\,deg$\}$,
$\{a=0.998$, $i=30$\,deg$\}$, $\{a=0.998$, $i=60$\,deg$\}$. The fiducial value
of the radial-emissivity parameter was always $q=3$. 
Three different profiles of the emission directionality are shown in columns -- \textposown{left}:
limb brightening, \textposown{middle}: isotropic, \textposown{right}: limb darkening. 
The contour lines refer to $1\,\sigma$, $2\,\sigma$, and $3\,\sigma$ levels.
The position of the minimal value of $\chi^{2}$ is marked
with a small cross. The values of $\chi^{2}$ corresponding to the minimum
and to the contour levels are shown at the top of each contour graph.
The large cross indicates the position of the fiducial values of 
the angular momentum and the emission angle.
Other parameters of the model were kept fixed at default
values: $\Gamma=1.9$, $r_{\rm in}=r_{\rm ms}$, $r_{\rm
out}=400$ and normalisation of the power law
$K_\Gamma=10^{-2}$.}
\label{contaq}
\end{center}
\end{figure}

It was shown for the case of XMM-Newton data of MCG\,-6-30-15
that the radial emissivity can mimic different angular emissivity
(Section~\ref{direxammcg}). Therefore, as a next step, 
we allowed $q$ parameter to vary during the fitting procedure
of the artificial data. The contour graphs of the spin and the 
$q$ parameter are shown in Figure~\ref{contaq}. 
It is clear from the figures that the correct value of the spin
can be reached by adjusting of the $q$ parameter -- for limb
brightening by decreasing steepness of radial emissivity,
and vice versa for limb darkening. However, this is not
valid for each set of parameters. For $a=0.7$, and $i=60$\,deg,
the model with limb darkening overestimates the spin value
for any value of $q$ within $3\,\sigma$ confidence level.
Oppositely, for $a=0.998$, and $i=30$\,deg,
the model with limb brightening underestimates the spin value
for any value of $q$ within $3\,\sigma$ confidence level.

\section{Discussion of the results}
\label{sec:discussion}

We investigated whether the spin measurements of accreting black holes
are affected by the uncertainty of the angular emissivity law, 
${\cal M}(\mu_{\rm e},E_{\rm e})$,
in the relativistically broadened iron K$\alpha$ line models.
We employed three different approximations of the angular emissivity
profiles, representing limb brightening, isotropic and limb darkening emission profiles.
For the radius-integrated line profile of the disc emission,
and especially for higher values of the spin, the
broadened line has a triangle-like profile. The
differences among the considered profiles concern
mainly the width of the line's red wing.
However, the height of the individual peaks,
also affected by the emission directionality,
is important for the case of an orbiting spot 
(or a narrow ring), which produces a characteristic 
double-horn profile.

We re-analysed an XMM-Newton observation of MCG\,-6-30-15 
to study the emission directionality effect on
the broad iron line, as measured by current X-ray instruments.
We showed the graphs of $\chi^{2}$ values as a function
of spin for different cases of directionality. 
We can conclude that the limb darkening law favours higher values of
spin and/or steeper radial dependence of the line emissivity;
vice versa for the limb brightening profile.
Both effects are comprehensible after examining the 
left panel of Fig.~\ref{fig4a}. The limb darkening
profile exhibits a deficit of flux in the red wing compared
with the limb brightening profile. Both higher spin value
and steeper radial profile of the intensity can compensate for this deficit.

The higher spin value has the effect of shifting the 
inferred position of the marginally 
stable orbit (ISCO) closer to the black hole, in accordance 
with eq.~(\ref{a_rms}). Consequently, the accretion disc
is extended closer to the black hole.
The radiation comes from shorter radii 
and it is affected by the extreme gravitational redshift.
Hence, the contribution to the red wing of the total 
disc line profile is enhanced. Naturally, these considerations
are based on the assumption that the inner edge
of the line emitting region coincides with the ISCO.
Likewise, the steeper radial dependence of the emissivity means
that more radiation comes from the inner parts of
the accretion disc than from the outer parts, and
this produces a similar effect to decreasing the 
inner edge radius. With the limb brightening law 
the above-given considerations work the other way around.

We further simulated the data with the \textscown{powerlaw} + \textscown{kyrline} 
model. The simple model allows us to keep better control 
over the parameters
and to evaluate the differences in the spin determination.
We used one of the preliminary response matrices for
the IXO mission
and we chose the flux at a level similar
to bright Seyfert 1 galaxies observable
by current X-ray satellites \citep{2007MNRAS.382..194N}.
The simulations with the \textscown{kyrline} model confirm that the
measurements would overestimate the spin for the limb 
darkening profile and, vice versa, they tend to the lower
spin values for the limb brightening profile. 

Although the interdependence of the model parameters is essential
and it is not possible to give the result of our analysis in terms of a 
single number, one can very roughly estimate that the uncertainties
in the angular distribution of the disc emission will produce an uncertainty
of the inferred inner disc radius of about $20$\% for the high quality 
data by IXO.\footnote{Here, we refer to the inner disc radius instead of the spin
because the latter is not as uniform a quantity as the corresponding
ISCO radius to express the uncertainty simply as a percentage value.
However, we still suppose that the inner edge of the line emitting region of the 
disc coincides with the marginally stable orbit.} 
We consider this value as realistic.

In the next step, we applied the NOAR radiation transfer 
code to achieve a self-consistent simulation of the outgoing
spectrum without imposing an ad hoc formula
for the emission angular distribution. We assumed 
a cold isotropically illuminated disc with a constant 
density atmosphere.
We created new models for the \textscown{ky} suite.
The results of NOAR computations of the cold disc are implemented
in the \textscown{kyl2cr} model, while the \textscown{kyl3cr} model
uses the angle-integrated tables over the entire range of emission
directions. This enables us to include, a posteriori, the analytical 
formulae for directionality and check how precisely they reproduce the
original angle-resolved calculations.
We simulated the data using the preliminary IXO response matrix
with the seed model \textscown{powerlaw} + \textscown{kyl2cr}.
Then we replaced the \textscown{kyl2cr} model by \textscown{kyl3cr} to test
the analytical directionality approaches on these artificial data.

We found that, for $a=0.998$, none of the three assumed cases 
of the directionality profile covered
the fiducial values for the spin and the inclination angle
within the $3\sigma$ contour line 
(see Figs. \ref{fig_a0998i30}--\ref{contaq}).
The suitability of the particular directionality prescription
depends on the fiducial values of the spin and the inclination
angle. The limb brightening profile successfully minimises the 
$\chi^{2}$ values for $a=0.998$ and $i=30$\,deg (though with slightly 
different parameter values),
but for $a=0.7$ and $i=60$\,deg it gives the worst fit
of all studied cases of the directionality laws.

On the whole, we found that the isotropic angular 
dependence of the emission intensity fits best.
The model with the limb darkening profile was not able to reproduce the data,
especially for higher values of the black hole spin.
The best fit $\chi^{2}/\nu$ value exceeds $5$ 
for $a=0.998$ and $i=30$\,deg, which means 
a more than three times worse fit than using 
isotropic or limb brightening directionality. 
The inclination angle was underestimated by more than $1$\,deg,
the $q$-parameter was overestimated by almost $10\%$.

This is an important result because much
of the recent work on the iron lines, both in
AGN and black hole binaries, has revealed
a significant relativistic broadening near
rapidly rotating central black holes \citep[see e.g.][]{2007ARA&A..45..441M}. 
In some of these works, the limb darkening law was employed and different 
options were not tested. The modelled broad lines are typically
characterised by a steep power law in the
radial part of the intensity across the inner region
of the accretion disc, as in the mentioned MCG\,-6-30-15 
observation. This behaviour has been interpreted as a case of
a highly spinning compact source where the black hole rotational energy
is electromagnetically extracted \citep{2001MNRAS.328L..27W}.
We conclude that the significant steepness of the radial part of the intensity 
also persists in our analysis, however, the exact values depend partly 
on the assumed angular distribution of the emissivity 
of the reflected radiation.

It should be noted that, in reality, the angular distribution
of the disc emission is significantly influenced by the vertical
structure of the accretion disc. However, our comprehension 
of accretion disc physics is still evolving. In recent years, 
several detailed models have been developed for irradiated black hole 
accretion discs in hydrostatic equilibrium 
\citep[see e.g.][]{2000ApJ...540L..37N, 2001MNRAS.327...10B, 2002MNRAS.332..799R}. 
These models combine radiative transfer simulations with calculations of the
hydrostatic balance in the stratified disc medium. Aside from
reprocessing spectra, the models provide solutions for the vertical
disc profile of the density, temperature, and ionisation fractions. 
In \citet{2007A&A...475..155G} the effects of general relativity and
advection on the disc medium were added.

In \citet{2000ApJ...540L..37N}, \citet{2007A&A...475..155G}, and
\citet{2008MNRAS.386.1872R} the reprocessed spectra are evaluated at
different local emission angles. The shape and
normalisation of these spectra depend on various model
assumptions. Until we know in more detail how accretion discs
work, it is hard to choose which is the ``correct'' reprocessing
model. One could always argue that more physical processes and properties
should be included in the radiative transfer simulations, such as the impact
of magnetic fields, macroscopic turbulence, or different chemical
compositions of the medium. Using the above-mentioned models for an
accretion disc in hydrostatic equilibrium may then easily become
computationally intense.

For the practical purpose of data analysis, however, the computation
of large model grids is necessary, which requires sufficiently
fast methods. This is why simple constant density models are most often
used to analyse the observational data. In fact,
\citet{2001MNRAS.327...10B} have shown that their reprocessing
spectra for a stratified disc medium in hydrostatic equilibrium can be
satisfactorily represented by spectra that are computed for irradiated
constant density slabs. Therefore, we included in our present analysis
the angular emissivity obtained from the modelling of neutral
reprocessing in a constant density slab. 
We have not discussed in any further detail the dependence on the
ionisation parameter, which we expect to be rather important, and it will
be possibly addressed in a future work.

We emphasise that the main strategy of our presented research
is not intended to find the ``correct'' angular emissivity, as this is 
still beyond our computational abilities and understanding of all the physical 
processes shaping the accretion flow. Instead, we examined the three 
different prototypical dependencies which are mutually disparate 
(i.e., the limb-darkening, isotropic, and limb-brightening cases), 
applied in current data analysis, and which presumably reflect the range of possibilities. 
By including these different cases we mimic various uncertainties, 
such as those in the vertical stratification, and we estimate 
the expected error that these uncertainties can produce in the spin determinations. 
Further detailed computations of reprocessing models 
and the angular emissivity are needed in the future 
in order to understand its role in different spectral states of
accreting black holes.

% ##########################################################################

\chapter{Data reduction and spectral analysis}
%\chaptermark{Reduction and spectral analysis \ldots}
 \label{xmm_analysis}
 \thispagestyle{empty}
\section{Preliminaries}
 \subsection{XMM-Newton satellite}
\label{xmm_newton}

X-ray astronomy is essential to study neutron stars and accreting black holes.
These objects would not be revealed on the sky without X-ray detectors
which were first used in the early 1960s. 
It was quite a surprise at that time that the X-ray sky was so different from the optical
sky. The first detected X-ray source besides our Sun, Scorpius X-1 
(discovered by Aerobee~150 rocket launched on 12th June 1962),
was found to be 10,000 times brighter in X-rays than in its optical
emission. 
X-ray astronomy is not possible from the Earth surface
because our atmosphere blocks out all X-rays. Only telescopes
above the atmosphere can detect X-ray radiation. Following the
first ``rockoons'' experiments (hybrid of a rocket and a balloon) 
many X-ray satellites have been launched.

The XMM-Newton satellite (X-ray Multi Mirror) belongs to the largest
scientific satellites ever launched and it represents a cornerstone
mission of the European Space Agency's Horizon 2000 programme
\citep{2001A&A...365L...1J}. It was launched by the Ariane~504 rocket 
on 10th December 1999, and it is planned to operate still several more years.
Its large effective area ($4700$\,cm$^{2}$) is its major advantage 
compared to other recent satellites (like Chandra). The satellite is
in a highly eccentric orbit (between $7000$ and $114,000$\,km) allowing
long uninterrupted exposure times ($\approx 40$\,hours).
Observing time on XMM-Newton is being made available to the scientific community, 
applying for observational periods on a competitive basis.
The data are stored in the publicly available archive\footnote{http://xmm.esac.esa.int/xsa}. 
%new data are accessible for public typically after one year
%long period reserved for the principal investigator of the project}.

The XMM-Newton satellite carries three CCD cameras for X-ray
spectroscopy and imaging -- European Photon Imaging Camera, EPIC,
which includes two MOS detectors \citep{mos} 
and one PN detector \citep{pn}, 
two spectrometers for high resolution X-ray spectroscopy --
Reflection Grating Spectrometer, RGS \citep{rgs},
and one optical/UV imaging and grism spectroscopy instrument --
Optical Monitor, OM \citep{2001A&A...365L..36M}.
All these instrument are able to operate simultaneously. The main
properties of the instruments are summarised in Table~\ref{xmm_instruments}.

\begin{table}[tb]
\caption{Capabilities of detectors on-board XMM-Newton satellite.}
 \begin{center}
\small{
  \begin{tabular}{l|c|c|c|c}
    \hline
    \hline
\rule[-0.7em]{0pt}{2em} Instrument &	EPIC PN	&	EPIC MOS	&	RGS	&	OM \\
\hline
\rule[-0.7em]{0pt}{2em} Energy bandpass & $0.15-12$\,keV & $0.15-12$\,keV & $0.35-2.5$\,keV & $180-600$\,nm \\
\rule[-0.7em]{0pt}{2em} Orbital target visibility\,$^1$ & $5-135$\,ks & $5-135$\,ks & $5-135$\,ks & $5-145$\,ks \\
\rule[-0.7em]{0pt}{2em} Sensitivity\,$^2$ & $\approx10^{-14}$ & $\approx10^{-14}$ & $\approx8\times10^{-5}$ & $20.7$\,mag 	\\
\rule[-0.7em]{0pt}{2em} Field of view & $30'$ & $30'$ & $\approx5'$ & $17'$ \\
\rule[-0.7em]{0pt}{2em} Spectral resolution\,$^3$ & $\approx80$\,eV & $\approx70$\,eV & $0.4/0.025$\,\AA & $350$ \\
\hline
%\rule[-0.7em]{0pt}{2em}
  \end{tabular}
 \label{xmm_instruments}
}
 \end{center}
\small{
\textposown{Notes:} $^1$\,The orbital target visibility represents the total time
available for scientific measurement per orbit. The value depends on what time the satellite
is outside van Allen radiation belts which could damage the detectors.\\
$^2$\,The sensitivity is a) for EPIC detectors: in units of erg\,s$^{-1}$\,cm$^{-2}$ 
 measured in $0.15-15$\,keV range after 10\,ks exposure,
b) for RGS detectors: in units of counts\,s$^{-1}$\,cm$^{-2}$ measured
in O\,VII~0.57\,keV line flux with a background of $10^{-4}$ counts\,s$^{-1}$\,cm$^{-2}$,
c) for OM: 5\,$\sigma$ detection of an A0 star in 1\,ks exposure.\\
$^3$\,The spectral resolution of EPIC detectors is at 1\,keV (at the energy of Fe~K$\alpha$
line $E=6.4$\,keV the energy resolution of both detectors is $\approx 150$\,eV.
The resolution of OM is the resolving power ($\lambda/\Delta\lambda$).
}
\end{table}

\subsubsection{Observation modes of EPIC detectors}
\label{xmm_modes}

The EPIC detectors are suitable for our research of broad iron lines
with their energy range (see Table~\ref{xmm_instruments}) and high 
effective area. There are several observation modes which differ 
from each other mainly in the covered field of view
and the read-out time. The basic characteristics of the PN and MOS observation
modes are summarised in Tables~\ref{xmm_pnmodes}-\ref{xmm_mosmodes}.
The ``Full Frame'' or ``Extended Full Frame''
modes are useful for faint sources where long read-out time is necessary
to reach a reasonable signal to noise ratio.
Oppositely, ``Timing'' mode or ``Burst'' mode 
are dedicated for observation of very bright sources to avoid the problem
with pile-up (see Section~\ref{pileup}).
The duty cycle in the ``Burst'' mode is only 3$\%$ 
which means that the rest 97$\%$ of counts are lost.

\begin{table}[tbh]
\caption{EPIC PN observation modes.}
 \begin{center}
\small{
  \begin{tabular}{l|c|c|c}
    \hline
    \hline
\rule[-0.7em]{0pt}{2em} Modes & Time resolution [ms]	& Live time [$\%$]	&	Max. count rate* [cts\,s$^{-1}$] \\
\hline
\rule[-0.7em]{0pt}{2em} Extended Full Frame & $200$ & $100$ & $2$  \\
\rule[-0.7em]{0pt}{2em} Full Frame & $73.4$ & $99.9$ & $6$  \\
\rule[-0.7em]{0pt}{2em} Large Window & $48$ & $94.9$ & $10$  \\
\rule[-0.7em]{0pt}{2em} Small Window & $6$ & $71$ & $100$  \\
\rule[-0.7em]{0pt}{2em} Timing & $0.03$ & $99.5$ & $800$ \\
\rule[-0.7em]{0pt}{2em} Burst & $0.007$ & $3$ & $60000$ \\
\hline
%\rule[-0.7em]{0pt}{2em}
  \end{tabular}
 }
\label{xmm_pnmodes}
\end{center}
\end{table}

\begin{table}[tbh]
\caption{EPIC MOS observation modes.}
 \begin{center}
 \small{ 
\begin{tabular}{l|c|c|c}
    \hline
    \hline
\rule[-0.7em]{0pt}{2em} Modes & Time resolution [ms]	&	Live time [$\%$]	&	Max. count rate* [cts\,s$^{-1}$] \\
\hline
\rule[-0.7em]{0pt}{2em} Full Frame  & $2600$ & $100$ & $0.7$ \\
\rule[-0.7em]{0pt}{2em} Large Window  & $900$ & $99.5$ & $1.8$ \\
\rule[-0.7em]{0pt}{2em} Small Window & $300$ & $97.5$ & $5$ \\
\rule[-0.7em]{0pt}{2em} Timing  & $1.75$ & $100$ & $100$ \\
\hline
%\rule[-0.7em]{0pt}{2em}
  \end{tabular}
}
\label{xmm_mosmodes}
 \end{center}
*{\small The maximal count rate values are shown for a point source. 
Only observations of sources with lower flux avoid problems with 
an irremovable pile-up.}
\end{table}

\subsubsection{SAS - tool for reduction of the XMM-Newton data}
\label{sas}

The Science Analysis System (SAS) is a collection of tasks, scripts and libraries, 
specifically designed to reduce and analyse data collected by the XMM-Newton observatory
\citep{2004ASPC..314..759G}.
It is publicly available at \textit{http://xmm.esac.esa.int/sas},
where also the instructions for install and usage are described.
In this Section, we briefly summarise what is needed in preparation
of the X-ray data before the spectral analysis and which is all 
accessible with the SAS software:

\begin{enumerate}
 \item The raw XMM-Newton data, called Observation Data Files (ODF), 
are provided in FITS (Flexible Image Transport System) format. 
The SAS software allows
to create a calibrated and concatenated event list where events from all detector
chips are included in one single file. The information about the instrumental 
calibration is necessary to properly interpret the data.
The researchers are encouraged to use the most updated calibration data set.
\item The SAS software helps its users to create scientific products 
according to their own taste and data screening criteria.
High X-ray background flare events are standardly removed
by creating of good-time intervals of the observation
where the background level is under a certain criterion.
\item After choosing good-time intervals, an X-ray image might
be created. The source spectrum is extracted from the region on the chip where
the point spread function (PSF) is the brightest (it is usually a circle around
the centre of the PSF). The background spectrum is extracted from the outer
regions of the same chip where the PSF is weaker.
\item Each X-ray instrument has a typical response on the incoming radiation
and has a different sensitivity in different wavelengths. The instrumental
spectral transfer functions (response matrices) are needed for the quantitative
spectral analysis. The SAS software allows the users to generate it
for the individual observation.
\item The SAS software contains also a tool for estimation
of the pile-up fraction in the data (see Section~\ref{pileup}),
and many other useful tasks (for more information see \citet{2004ASPC..314..759G}
or look at the SAS web pages).
\end{enumerate}

\subsection{Pile-up}
\label{pileup}

Pile-up is one of the biggest problems of the detection technique
using a CCD camera when the observed source is too bright.
Any CCD camera is composed from individual pixels
which are collecting photons. 
The energy of an incoming photon invokes a charge in the
photo-active region 
(in typical CCDs, epitaxial layers of silicon are used).
The charge is then read-out during the duty cycle.
Problem with pile-up occurs when two or even more photons
deposit charge packets in a single pixel (``photon pile-up''),
or in neighbouring pixels (``pattern pile-up'') during one
read-out cycle.

In general, the pile-up makes spectra harder 
because two or more soft photons are detected as only one hard photon.
In addition, when the summed energy is above a rejection threshold,
it leads to a more or less pronounced depression of counts in the central
part of the point spread function implying undesirable flux loss.
The Tables~\ref{xmm_pnmodes}-\ref{xmm_mosmodes} contain the values of 
the maximal count rates of the observed source fluxes for different 
observation modes to avoid the problem with pile-up.
In general, the MOS detector is more susceptible to pile-up
than the PN camera.

If the pile-up is only moderate, it might be removed by:
\begin{enumerate}
 \item using only single events, 
i.e. events where no signal is detected in neighbouring pixels. 
The single events are less sensitive to the pile-up
than any multiple events.
 \item excising of the PSF core. However, this procedure may cause a drastic loss of counts.
\end{enumerate}

% detection of pile through pattern pile-up

A SAS task \textit{epatplot} is a suitable tool
for constraining the pile-up level. 
It measures the ``pattern pile-up''
which is proportional to the ``photon pile-up''.
The pattern pile-up occurs when two single events 
are interpreted as one double event with twice the energy. 
The probability of single and double events fractions
can be modelled for a given flux. If there is a pile-up
the double events are more frequent than predicted.
The rate of diversion of the data from the theoretical model
curves corresponds to the pile-up level.

We encountered the pile-up problem in the observation 
of a stellar-mass black hole binary, GX\,339-4, 
performed by the XMM-Newton satellite in 2004, 
see Section~\ref{gx_lhs} for details.

\subsection{Re-binning of the data}
\label{sec_grouping}

There is not a unique way of the treatment with the data after their reduction
described in Section~\ref{sas}. Some astronomers use the data for the
spectral analysis without any re-binning. Some re-bin the data but
with different conditions. The common condition for grouping
is to have at least a certain number of counts per bin, e.g. 20.
The reason is that the commonly used statistics for fitting the data is 
the Gaussian one with $\chi^2$ values describing the fit goodness,
and the derived formulae of the Gaussian statistics are based on
the assumption of a sufficient number of counts (see Section~\ref{spec_analysis}). 
However, when the number of counts per bin is not adequate
C-statistics \citep{1976A&A....52..307C} may be used, instead.
Further, we discuss re-binning of the data because
of the energy resolution rather than an insufficient number
of counts. 

The counts are intrinsically distributed in many channels 
of the instrument according to the detected energy of photons. 
However, the energy resolution of currently used X-ray instruments 
is not so brilliant and it is largely exceeded by the total number of channels.
For instance, the PN detector has the energy resolution of order 
of $\approx 100$\,eV (see Table~\ref{xmm_instruments})
and the energy range $\approx 10$\,keV which is about
$\approx 100\times$ larger than the energy resolution.
Thus, the corresponding number of energy bins of the width
equal to the order of the energy resolution is $\approx 100$.
However, the PN detector has 4095 intrinsic channels in total,
i.e. more than one order of magnitude more.

Hence, we find necessary to re-bin (group) the data 
with respect to the intrinsic resolution of the instrument.
We find sufficient re-binning which over-samples the energy resolution
by a factor of three. We use the \textit{pharbn} script 
provided us by M.~Guainazzi for this purpose.
The energy resolutions of the instruments are typically approximated
as power laws of the energy (with an exponent $e=-0.5$ for
the PN detector and $e=-0.46$ for the MOS detector),
and with a known energy resolution at some reference energy
($2.26\%$ at 6\,keV for PN, $6.31\%$ at 1\,keV for MOS).

We emphasise that the interpretation of spectral results
can be influenced by the adopted procedure of the data re-binning,
and so different authors may reach different conclusions.
We will illustrate this fact on an X-ray spectrum
of GX\,339-4 (see the details in Section~\ref{gx_vhs}).

\subsection{Goodness of the fit}
\label{spec_analysis}

In this section, we briefly describe how to constrain
the goodness of the fit. Readers are referred to e.g. 
\citet{2002nrc..book.....P} for more information about
different statistical techniques to analyse the data.

X-ray spectra are measured with spectrometers which do not obtain 
the actual spectrum but rather photon counts $C$ within
specific instrument channels $I$.
This observed spectrum is related to the actual
spectrum of the source $f(E)$ by:
\begin{equation}
 C(I) = \int_0^\infty f(E)\,R(I,E)\,dE,
\end{equation}
where $R(I,E)$ is the instrumental response
which is proportional to the probability that an
incoming photon of energy $E$ will be detected in channel $I$. 

The instrumental spectral transfer functions (response matrices)
are not generally invertible \citep[see e.g.][]{1979MNRAS.186...45B}, 
and it is therefore insecure  
to reconstruct the original spectrum (so called unfolded spectrum).
Instead of deriving $f(E)$ for a given set of $C(I)$, 
model spectra $f_M(E)$ are calculated 
how they would look like when
passing through the detector. The predicted counts $C_p(I)$ are:
\begin{equation}
 C_p(I) = \int_0^\infty f_M(E)\,R(I,E)\,dE,
\end{equation}
and then, they are compared with the data counts $C(I)$. 

The quality of the fit, i.e. how the model is able to interpret
the data, is usually characterised by the $\chi^2$ value given by:
\begin{equation}
 \chi^2 = \sum_I \frac{\left(C\left(I\right)-C_p\left(I\right)\right)^2}{\left(\sigma\left(I\right)\right)^2},
\end{equation}
where $C(I)$ is the number of counts in the energy bin $I$ (DATA),
$C_p(I)$ is the number of predicted counts in the energy bin $I$ (MODEL),
and $\sigma(I)$ is the (generally unknown) error for the channel $I$.
The error $\sigma(I)$ is usually estimated as $\approx \sqrt{C(I)}$.

The $\chi^2$ statistics provides a goodness-of-fit criterion for
a given number of degrees of freedom $\nu$, which is calculated as the number of
channels minus the number of model parameters, and for a given confidence level.
If $\chi^2$ value exceeds a critical value 
one can conclude that the fit $f_M(E)$ is not an adequate model for $C(I)$. 
As a general rule, one wants to have a reduced $\chi_{\rm red}^2 \equiv \chi^2/\nu$ 
to be approximately equal to one ($\chi^2 \approx \nu$). 
A reduced $\chi^2$ value that is much greater than one,
$\chi_{\rm red}^2 >> 1$, indicates a poor fit, 
while $\chi_{\rm red}^2 << 1$ indicates that the errors on 
the data have been over-estimated. 

If $\chi_{\rm red}^2 \approx 1$, so the best-fit model $f_M(E)$ 
pass the ``goodness-of-fit'' test, 
one still cannot say that $f_M(E)$ is the only acceptable model. 
For example, if the data used in the fit are
not particularly good, one may be able to find many different models for which
adequate fits can be found. In such a case, the choice of the correct model to fit is a
matter of scientific judgement.

There are several software packages to analyse X-ray spectra
-- XSPEC \citep{1996ASPC..101...17A}, SPEX \citep{1996uxsa.conf..411K}, 
ISIS \citep{2000ASPC..216..591H}.
For all the spectral analysis presented in this Thesis,
we used the XSPEC fitting package. The best-fit values
of the model parameters were found
using the modified Levenberg-Marquardt algorithm \citep{1969drea.book.....B},
which is the default one in the XSPEC.
This algorithm is local rather than a global one, so 
we were aware of the possibility to get stuck
in a local minimum instead of the global minimum
by starting of the fitting process from reasonable
initial values of model parameters. Therefore,
we standardly repeated fitting procedures for several different initial
parameter values.
\newpage
\section{GX 339-4}
 \label{gx339}

GX\,339-4 belongs to the Galactic black hole candidates \citep{2006ARA&A..44...49R}.
Dynamical properties of the system are still not completely known. The companion star
stays unresolved by the present measurements even during the quiescent state 
of the object. However, the optical spectroscopy reveals periodic 
behaviour of He~II and Bowen blend N~III lines in spectra  
\citep{2003ApJ...583L..95H, 2003MNRAS.342..105B}. 
\citet{2003ApJ...583L..95H} determined the orbital period of the system as 
$P \approx 1.7d$. From the given period the radial velocity curve $K_{2}$ 
of the secondary can be fitted. The mass function $f(M)$ is given by
eq.~(\ref{mf_binary}).
For GX\,339-4: $f(M) = 5.8 \pm 0.5$.
The mass of the black hole can be determined from the knowledge 
of the inclination of the system and the ratio 
of the mass of the companion to the mass of 
the black hole (see eq.~\ref{mf_binary}).

However, these quantities are not yet well constrained. 
\citet{2003ApJ...583L..95H} estimated the mass ratio $q \leq 0.08$ 
from a small modulation of the wings of the He~II line.
The inclination angle is unknown. However,  
a jet is resolved from radio and infrared observations. 
The derived inclination from these measurements is $i < 26^{\circ}$ 
\citep{2004MNRAS.347L..52G}, but this need not be necessarily 
related with the inclination of the orbital plane.

We chose this object for our analysis because it exhibited a relativistic 
broadened iron line profile in the X-ray spectra 
in both high/soft state \citep{2004ApJ...606L.131M}
and low/hard state \citep{2006ApJ...653..525M},
see Figure~\ref{fig_miller}.
This fact rather contrasts with the interpretation 
of the low/hard state via a truncated disc (Section~\ref{accretion_flows}).
The extremely skewed line profiles suggest 
that the radiation comes from the innermost parts
of the accretion disc within a few gravitational radii.
Assuming accreting material radiates only above the marginal stable orbit 
it implies a high value of the spin of the black hole (Section~\ref{rellinemod}). 
\citet{2008ApJ...679L.113M} determined the value of the spin 
$a = 0.93 \pm 0.01$ (statistical) $\pm 0.03$ (systematic) by joint fitting 
of two XMM-Newton and one Suzaku observations.
The XMM-Newton spectra were re-analysed by \citet{2008MNRAS.387.1489R}
who measured the spin value as $a=0.935 \pm 0.01$ (statistical) $\pm 0.02$ (systematic).

\subsection{Low/hard state observation}
\label{gx_lhs}

The low/hard state observations of GX\,339-4
were performed by the XMM-Newton satellite
in March 2004 (Obs. ID\,\#0204730201 and \#0204730301).
The MOS detectors were operating in the ``Full Frame'' mode.
Although the object was in the low/hard state, 
it was still bright with the count rate $\gtrsim 20$\,cts\,s$^{-1}$, 
i.e. more than one order of magnitude greater
than the maximal recommended value to avoid the pile-up problem 
($0.7$cts\,s$^{-1}$, see Table~\ref{xmm_mosmodes}).
Indeed, we found a substantial level of the pile-up in the data
using the \textit{epatplot} tool (see Section~\ref{pileup}).

\begin{figure}[tbh]
\begin{center}
 \includegraphics[width=0.6\textwidth]{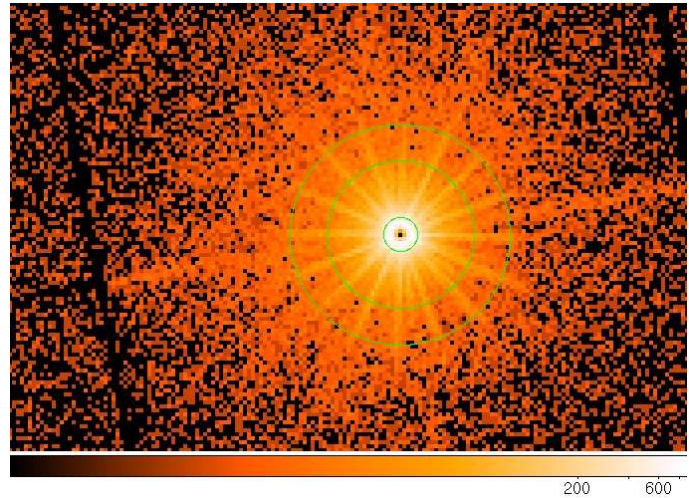}
 \caption{Central chip of the MOS\,1 detector (zoomed on PSF) of the observation ID\,\#0204730201.
Radii of the green circles are 18, 80 and 120\,arcsec.}
\label{mos_chip}
\end{center}
\end{figure}

\begin{figure}[tbh]
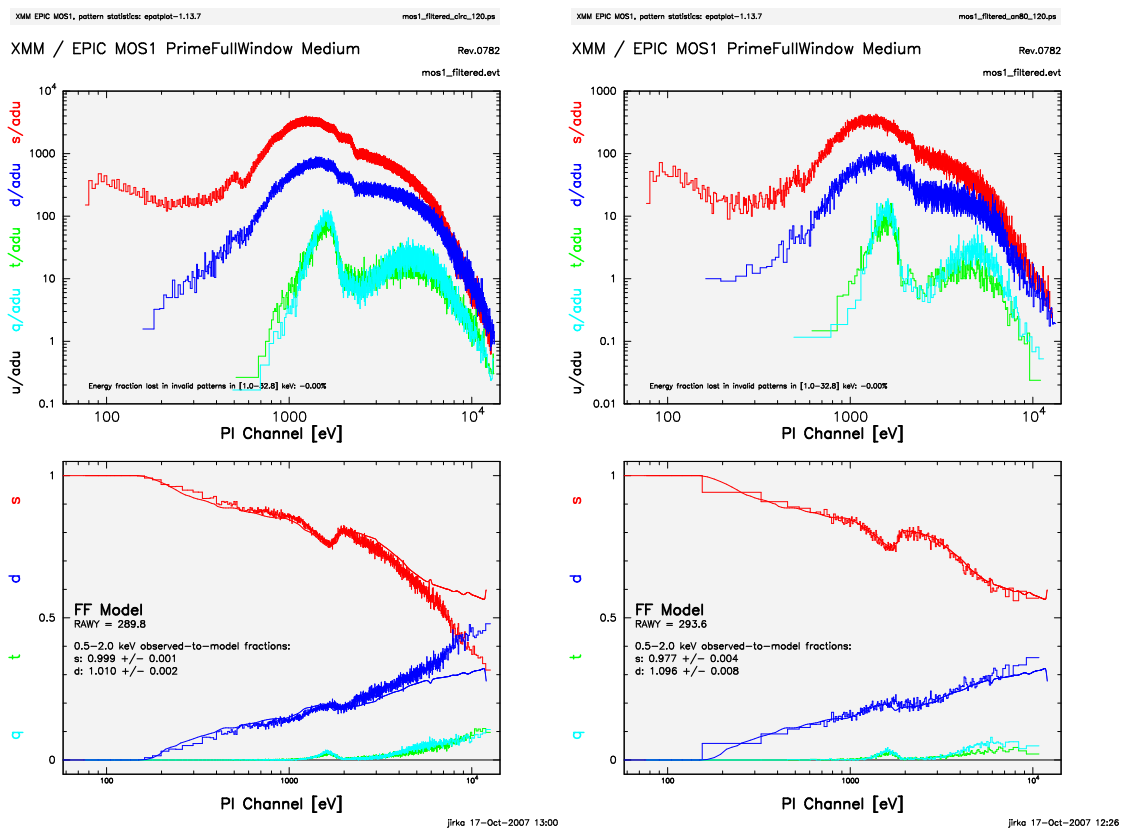

\begin{center}
  \includegraphics[angle=270,width=0.49\textwidth]{mos1_filtered_circ_120}
  \includegraphics[angle=270,width=0.49\textwidth]{mos1_filtered_an80_120}
  \caption{Quantitative pile-up diagnostics in the MOS\,1 camera of the observation
ID\,\#0204730201 using the SAS task \textit{epatplot}. 
The data points correspond to single events $s$ (red), double events $d$ (dark blue),
triple events $t$ (green), and quadruple events $q$ (light blue).
\textposown{Up:} Spectra related to the individual patterns.
\textposown{Bottom:} Pattern fractions. The solid lines represent expected fractions
if there is no pile-up effect.
\textposown{Left:} Data extracted from a circle with radius of 120\,arcsec.
\textposown{Right:} Data extracted from an outer annulus between 80 and 120\,arcsec.}
\label{pileup_fractions_201}
\end{center}
\end{figure}

\begin{figure}[tbh]
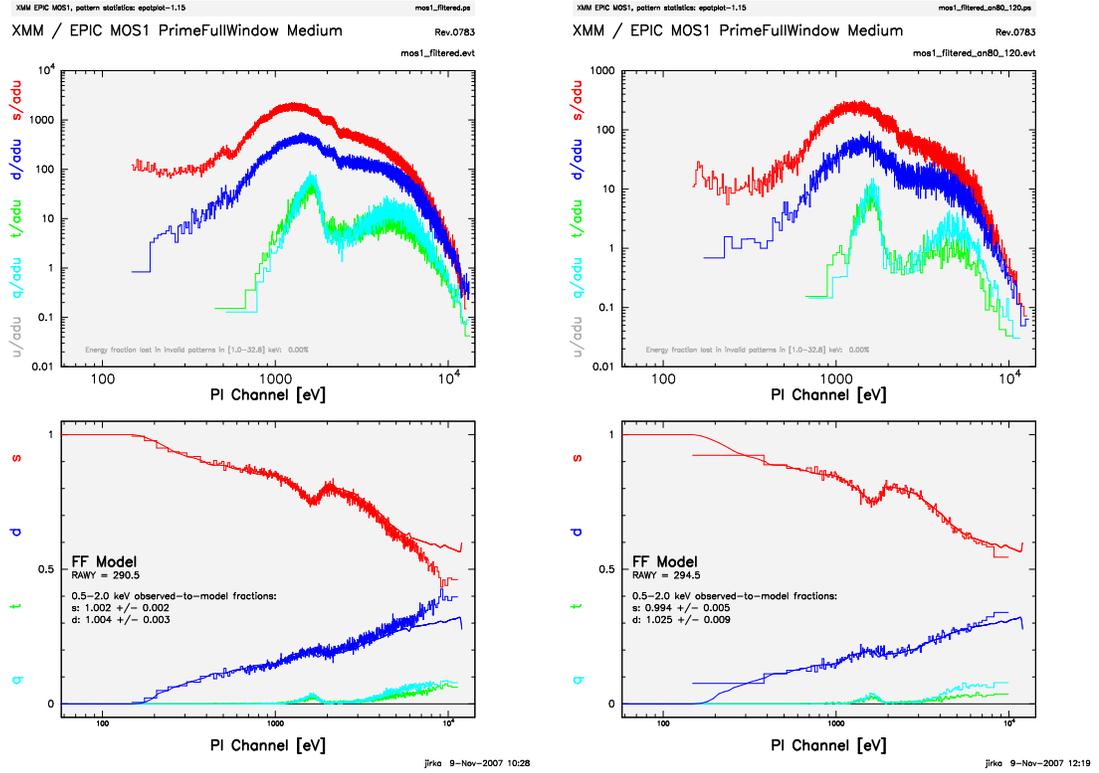

\begin{center}
  \includegraphics[angle=270,width=0.49\textwidth]{mos1_301_filtered}
  \includegraphics[angle=270,width=0.49\textwidth]{mos1_301_filtered_an80_120}
  \caption{The same quantitative pile-up diagnostics as in Figure~\ref{pileup_fractions_201}
but for the second observation (ID\,\#0204730301). 
\textposown{Left:} Data extracted from a circle with radius of 120\,arcsec.
\textposown{Right:} Data extracted from an outer annulus between 80 and 120\,arcsec.}
\label{pileup_fractions_301}
\end{center}
\end{figure}

\begin{figure}[tbh]
\begin{center}
  \includegraphics[angle=270,width=0.49\textwidth]{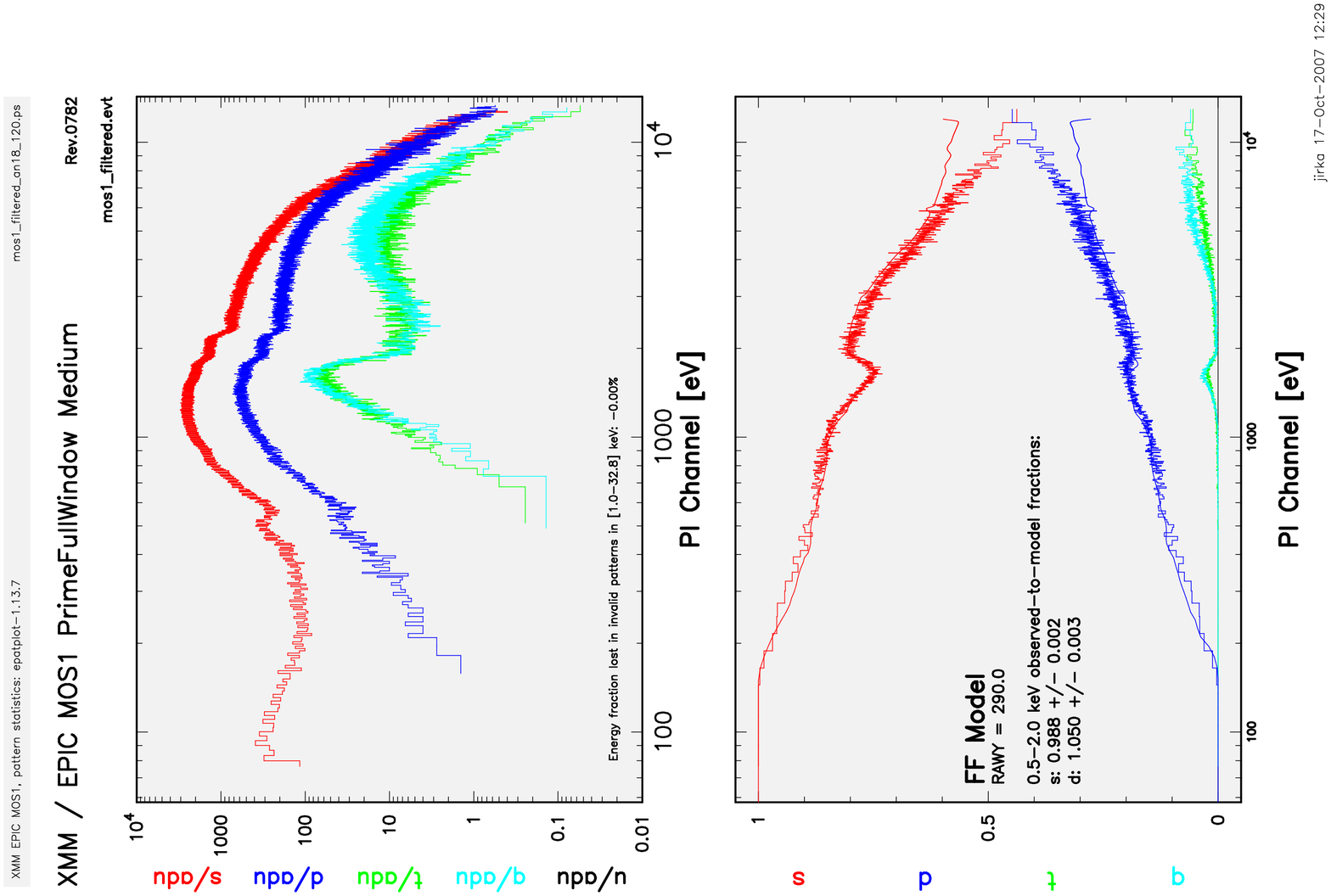}
  \includegraphics[angle=270,width=0.49\textwidth]{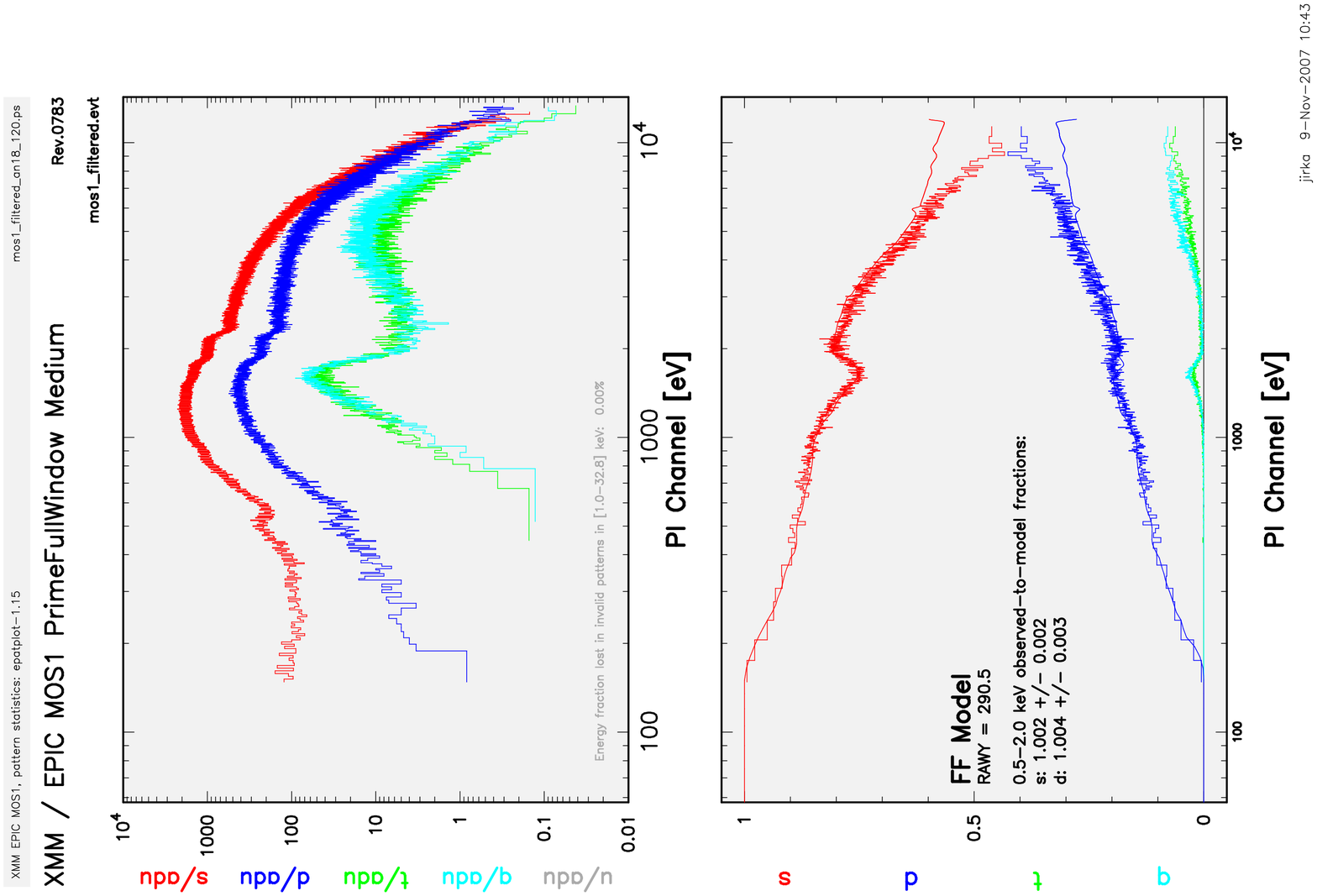}
  \caption{The same quantitative pile-up diagnostics 
as in Figure~\ref{pileup_fractions_201} 
and Figure~\ref{pileup_fractions_301} 
but for the same extracted region of the source as in \citet{2006ApJ...653..525M},
i.e. an $18-120$\,arcsec annulus.
\textposown{Left:} Observation \#0204730201.
\textposown{Right:} Observation \#0204730301.}
\label{pileup_fractions_annulus}
\end{center}
\end{figure}

Figure~\ref{mos_chip} shows the central chip of the MOS~1 detector 
of the first observation (it is similar in the second observation). 
The black core itself indicates the over-exposition of the detector.
Figures~\ref{pileup_fractions_201} and \ref{pileup_fractions_301} 
show pattern distribution with the energy
for two distinctly extracted regions from the central 
MOS chip of the two observations. 
The first extracted region is a circle around the core of 
point spread function (PSF) which is the standard
way of the region extraction to obtain the source spectra if there is no pile-up
problem. The radius was chosen as $120$\,arcsec.
The second region is an outer annulus between 80 and 120\,arcsec. 
In this case, however, the most of the flux is lost,
and the signal to noise ratio becomes very low. %(see Figure~\ref{mos_chip}).

The more deviated from the expected curves the pattern fractions are 
(bottom panels of the figures)
the more significant the effect of pile-up is.
It is clear that the pile-up is highly presented 
when the extracted region is the whole circle including the core of PSF.
However, the pile-up is not completely removed nor for a faraway outer
annulus though the level of the pile-up is  much lower.
Hence, the spectral analysis might be possible
in combination with the use of only single events. 
However, the drastic excision
of such a large region causes that the signal to noise is so low
that the quantitative spectral analysis gives very poor constraints
about the spectrum and is not suitable for a detailed investigation
of any model parameters (not speaking of a very sensitive spin parameter).

Besides the two extreme extraction regions, we checked several annuli
with the inner radius between 0 and 80\,arcsec.
We found that the pile-up
is significant up to $\approx 80$\,arcsec.
\citet{2006ApJ...653..525M} were satisfied with an annulus $18-120$\,arcsec.
However, Figure~\ref{pileup_fractions_annulus} shows very strong
indications of the pile-up in both observations for this extraction region.
Because of the pile-up problem, we left these data sets as unsuitable 
for our purposes, regarding any spectral results obtained with these data
as untrustworthy.

Meanwhile, the data were re-analysed by \citet{2009arXiv0911.3243D}
who also found the pile-up level significantly high. 
Moreover, they performed a more comprehensive
analysis including PN data, as well.
They showed that the PN data are less affected by the pile-up than MOS data, %\footenote{The previous
%authors did not consider the PN data because they did not know how
and so they can be used for the spectral analysis if only single
events are considered and the brightest core of PSF is excised
(in the PN ``Timing mode'', it means to excise several central columns of the source image).
They concluded that the PN data do not support the presence
of a broad iron line as suggested from the piled-up MOS data
(presented by \citealp{2006ApJ...653..525M} and \citealp{2008MNRAS.387.1489R}).
%\citep[presented by][]{2006ApJ...653..525M,2008MNRAS.387.1489R}.
They argued 
that missing relativistic smearing supports the truncated
disc interpretation of the low/hard state. 
Similar issues with the pile-up
were also reported by \citet{2009ApJ...707L.109Y} 
with Suzaku data of GX\,339-4.

Evidently, the pile-up issues and the possible ways of removing its
degradation of the spectra are subject of discussions. Different researchers
seem to agree on the general conclusion that the pile-up can be a serious
problem. However, in individual observations the impact of the effect is
often estimated differently. It appears that the problem can be definitively
avoided only by future high throughput detectors.

\subsection{Very high state observation}
\label{gx_vhs}

We also re-analysed the observation of GX\,339-4 in the very high state (VHS).
The PN detector was operating in the ``Burst mode'' 
which avoids the pile-up problem (see Table~\ref{xmm_pnmodes}).
Using the SAS task \textit{epatplot} we found
the pile-up to be insignificant with these data.
Therefore, we used these data for the further spectral analysis
and also for the comparison of the relativistic line models in Section~\ref{gx339_vhs_spectrum}.
%
%We reduced the data in the same way as described in \citet{2004ApJ...606L.131M}.
However, we obtained an unacceptable fit with $\chi^{2}_{\rm reduced} \approx 6.9$
when we applied the same model as in \citet{2004ApJ...606L.131M}
to the data in the energy range $0.7-9$\,keV. 
The difference of the results is likely due to 
the different re-binning of the instrumental energy channels (see Section~\ref{grouping}).
%We found that the spectral analysis by \citet{2004ApJ...606L.131M} 
%(and also re-analysis by \citealp{2008MNRAS.387.1489R})
%was performed on practically un-binned data.

\begin{figure}[tbh]
\begin{center}
  \includegraphics[width=0.65\textwidth]{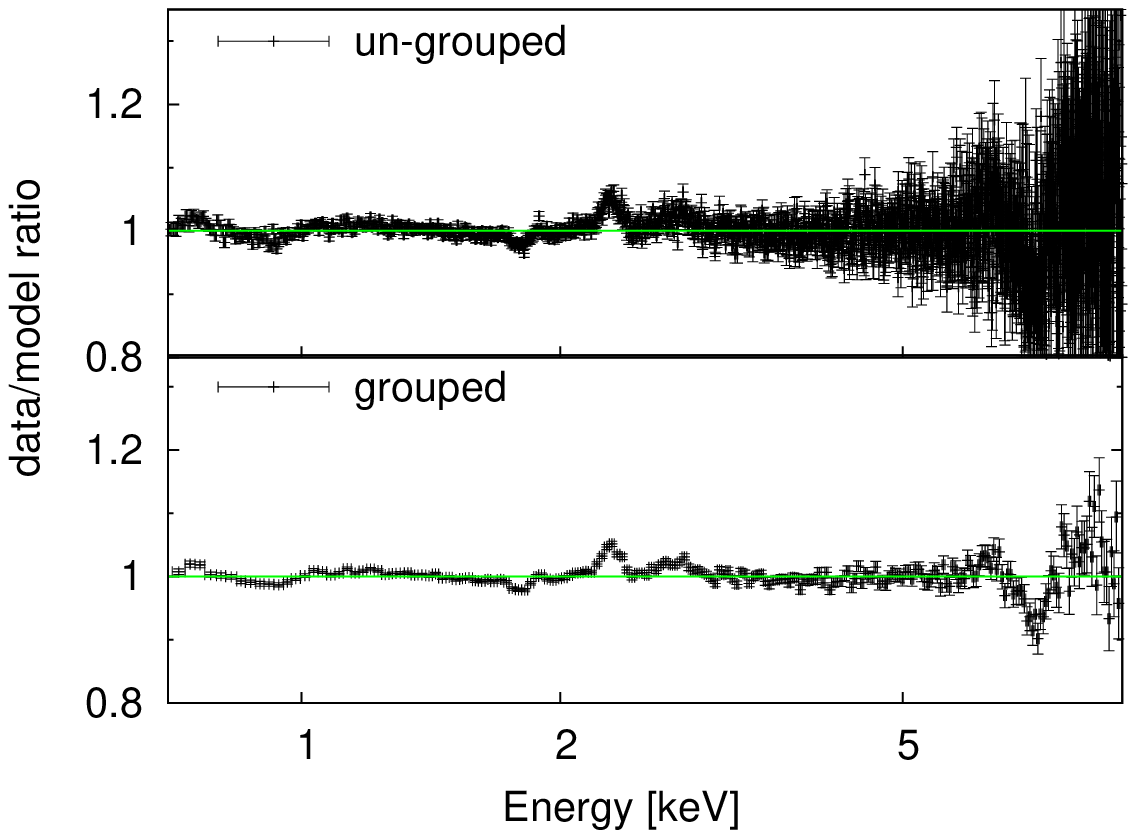}
  \caption{Ratios of the PN data of GX\,339-4 to the model used
in \citet{2004ApJ...606L.131M} and \citet{2008MNRAS.387.1489R}
with ungrouped data (\textposown{top}) and grouped data (\textposown{bottom})
as described in the main text.}
\label{grouping}
\end{center}
\end{figure}

Figure~\ref{grouping} shows ratios of differently binned
data to the same model as used in \citet{2004ApJ...606L.131M}
and \citet{2008MNRAS.387.1489R}. 
The upper panel shows the data
re-binned with the only condition to have at least 20 counts per bin,
which was used in the mentioned works. 
This condition is, however, very weak with respect to the total number of 
counts $N_{\rm counts} \approx 1.0\times 10^{7}$ and the total 
number of energy channels $N_{\rm chan} = 1.5\times 10^{3}$ 
in the $2-10$\,keV energy range. As a result, the data stayed practically un-binned
and the energy resolution was excessively over-sampled. 

In the lower panel of Figure~\ref{grouping}, 
the data are re-binned with respect to the 
instrumental energy resolution (the energy resolution
is over-sampled by a factor of three). 
An absorption feature at high energies which was concealed
in large error bars of the ungrouped data is now clearly revealed.
The discrepancy is even more visible in the quantitative 
spectral analysis, i.e. after constraining the fit-goodness 
in the $0.7-9$\,keV energy range
by the reduced $\chi^2$ values (see Section~\ref{spec_analysis}).
While $\chi_{\rm red}^2 \equiv \chi^2/\nu \doteq 2963/1659 \doteq 1.8$
for the ungrouped data 
$\chi_{\rm red}^2 \doteq 1427/208 \doteq 6.86$
for the grouped data.
After adding a Gaussian line to model a spectral feature at $2.3$\,keV,
possibly being some calibration issue only \citep{2004ApJ...606L.131M}, 
$\chi_{\rm red}^2 \doteq 2453/1657 \doteq 1.48$
for the ungrouped data, and $\chi_{\rm reduced}^2 = 921/206 \doteq 4.47$
for the grouped data in the energy range $0.7-9$\,keV.
So, the model would still have not been acceptable if applied
to the appropriately grouped data.

\begin{figure}[tbh!]
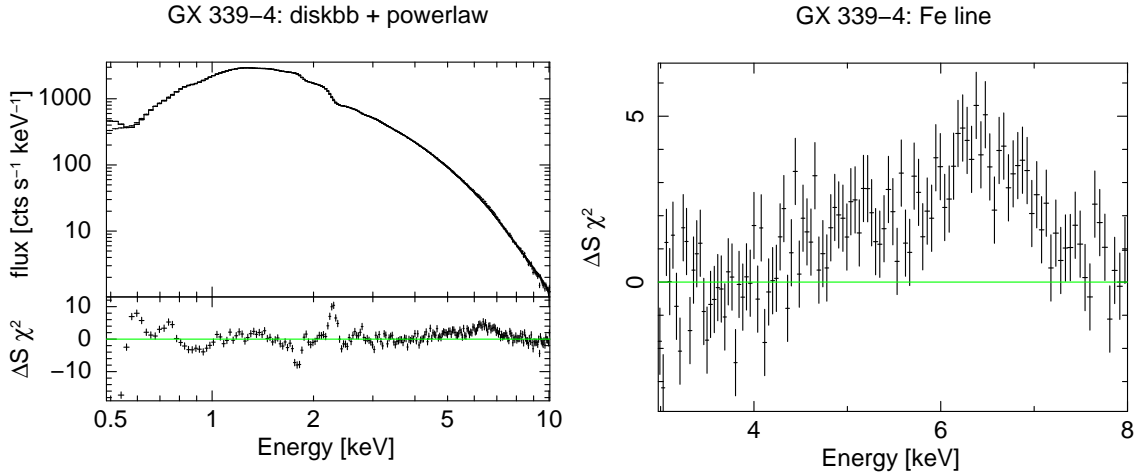

\begin{center}
\begin{tabular}{ccc}
%\epsfbox{fm_line_ratio.eps}
\includegraphics[angle=270,width=0.5\textwidth]{gx_cont.eps}
\includegraphics[angle=270,width=0.5\textwidth]{gx_ironlineregion.eps}
\end{tabular}
%\end{center}
\caption{\textposown{Left}: The X-ray spectrum of GX 339-4. 
The residuals from the model are plotted in the bottom panel 
of the graph (see the main text for further details). 
\textposown{Right}: More detailed view of the iron line band.}
\label{gx}
\end{center}
\end{figure}

\begin{table}[tbh!]
\begin{center}
\caption{Results for PN spectrum of GX\,339-4 (VHS) in 0.7--9\,keV.}
	\vspace{0.35cm}
\begin{tabular}{c|c||c|c|c}
	\hline \hline
\rule{0pt}{1.5em}	model  & 	parameter   &	\citet{2004ApJ...606L.131M} &	\citet{2008MNRAS.387.1489R}	& our fit	 \\[2pt]
	\hline 
\rule{0pt}{1.5em} {\phabs} & $n_{\rm H} \, [10^{22}$\,cm$^{-2}]$ & $0.51 \pm 0.01$	& $0.53$ & $0.63 \pm 0.01$	\\[2pt]
\hline
\rule{0pt}{1.5em} {\powerlaw} & $\Gamma$ & $2.60_{-0.05}$	& $2.553 \pm 0.003$ & $3.08 \pm 0.05$	\\
\rule{0pt}{1.5em}		&	normalisation	& $2.2^{+0.3}_{-0.1}$	&	$2.52^{+0.01}_{-0.02}$	& $5.64^{+0.12}_{-0.18}$ \\[2pt]
\hline
\rule{0pt}{1.5em} {\diskbb} & $kT$\,[keV] & $0.76 \pm 0.01$ & $0.726^{+0.001}_{-0.001}$ & $0.87 \pm 0.01$ \\
\rule{0pt}{1.5em}		&	normalisation	& $2300^{+100}_{-200}$	&	$2545 \pm 20$	& $1450^{+31}_{-25}$ \\[2pt]
\hline
\rule{0pt}{1.5em} {\laor}	& $E$\,[keV]	& $6.97_{-0.2}$	&	$6.88^{+0.09}_{-0.008}$	& $6.97_{-0.2}$ \\
\rule{0pt}{1.5em} & $r_{\rm in}$	& $2.1^{+0.2}_{-0.1}$	&	$1.91^{+0.02}_{-0.005}$	& $3.0^{+0.9}_{-0.6}$ \\
\rule{0pt}{1.5em} & $\theta$\,[deg]	& $11^{+5}_{-1}$	&	$18.8 \pm 0.5$	& $17 \pm 4$ \\
\rule{0pt}{1.5em} & $q$	& $5.5^{+0.5}_{-0.1}$	&	$6.88^{+0.09}_{-0.008}$	& $3.3 \pm 0.1$ \\
\rule{0pt}{1.5em} & normalisation	&	$0.077^{+0.005}_{-0.015}$	&	$0.113 \pm 0.002$	& $0.006^{+0.004}_{-0.002}$ \\[2pt]
%\rule{0pt}{1.5em} & EW\,[eV]	&	$200^{+10}_{-40}$	&	$0.113 \pm 0.002$	& $0.006^{+0.004}_{-0.002}$ \\[2pt]
\hline
\rule{0pt}{1.5em} {\smedge} & $E$\,[keV] & $7.9^{+0.1}_{-0.4}$	&	$7.1 \pm 0.1$	& - \\
\rule{0pt}{1.5em} & $\tau$ & $0.6^{+0.4}_{-0.1}$	& $2.5^{+0.2}_{-0.1}$	& - \\
\rule{0pt}{1.5em} & $W$\,[keV] & $1.0 \pm 0.3$	& $4.5^{+0.5}_{-2.0}$	& - \\[2pt]
\hline
\rule{0pt}{1.5em} & $\chi^{2}/v$	&	$3456.5/1894$	&	$2344.6/1652$ 	&	$277/198$ 	\\[2pt]
\hline
\end{tabular}
\label{table_gx_fit}
\end{center}

\small{\textposown{Note}: \citet{2004ApJ...606L.131M} required the photon index of the power law
to lie within $\Delta\Gamma \leq 0.1$ of the value from the simultaneous RXTE
measurements. The value $\Gamma = 2.6$ is their upper limit.}
\end{table}

Further, we analysed the spectrum of the re-binned data.
First, we removed an artificial {\smedge} model which was
used by \citet{2004ApJ...606L.131M} for fitting a smeared edge.
As a next step, we allowed the model parameters of the continuum and
also the iron line to float. As a result, we found a better fit 
with $\chi^{2}/\nu = 277/202 \doteq 1.7$ in 0.7--9\,keV.
The model parameters are shown in Table~\ref{table_gx_fit}
and they are compared there with the values found by
\citet{2004ApJ...606L.131M} and \citet{2008MNRAS.387.1489R}.

Our fit is characterised by a steeper power law with $\Gamma = 3.08 \pm 0.05$,
a slightly hotter disc with $kT_{\rm in} = \left(0.87 \pm 0.01 \right)$\,keV,
and a less prominent broad iron line. 
The column density $n_{\rm H} = \left(0.63 \pm 0.01\right)\,\times 10^{22}$\,cm$^{-2}$
is slightly higher than the value $n_{\rm H} = 0.374\times 10^{22}$\,cm$^{-2}$ 
from the Leiden/Argentine/Bonn H\,I measurements \citep{2005A&A...440..775K},
which may be explained by an additional local absorption. 
We tried also to add a reflection component into the model, as {\pexrav} ({\pexriv}) 
or {\refsch} model, but without any improvements of the fit. 
\citet{2008MNRAS.387.1489R} used the {\textscown{refhidden}} model
for fitting of the reflection component from the hot disc. However, we did not have
the opportunity to use it because it is not a publicly available model.

The spectrum of GX 339-4 is shown in Figure~\ref{gx} with a detailed view
of the iron line band in the right panel. 
A broadened iron line feature is still present. However, due to 
the different slope of the underlying power law
the line is much weaker than the one presented in \citet{2004ApJ...606L.131M}.
The equivalent width found by \citet{2004ApJ...606L.131M} is $200^{+10}_{-40}$\,eV
while our value is $154^{+21}_{-17}$\,eV. The main difference is in the
determination of the inner disc radius inferred from the iron line modelling.
While \citet{2004ApJ...606L.131M} and \citet{2008MNRAS.387.1489R}
concluded that the black hole in GX\,339-4 is very rapidly rotating
with $a \approx 0.9$ we found an intermediate value for the spin from our fitting,
$a \approx 0.7$ (see also Section~\ref{gx339_vhs_spectrum}).

%\newpage
\section{MCG\,-6-30-15}
  \label{mcg}

MCG\,-6-30-15 is a nearby Seyfert galaxy ($z = 0.00775$) of the type I 
(see Section~\ref{section_agn}).
This object is one of the most studied objects in X-ray domain
since it exhibited a very broad iron line in the spectrum of the 
ASCA observation \citep{1995Natur.375..659T}. The broadness of the line
has been attributed to its origin from the relativistic accretion
disc around a rapidly spinning black hole. 
%where the innermost edge
%of the disc coinciding with the marginally stable is allowed to be
%very close to the black hole horizon. 
Further investigations of this object with following X-ray missions
confirmed the presence of the relativistically broadened iron K$\alpha$ line:
BeppoSAX \citep{1999A&A...341L..27G}, XMM-Newton \citep{2001MNRAS.328L..27W,
2002MNRAS.335L...1F,2002A&A...383L..23M,2004MNRAS.348.1415V,2006ApJ...652.1028B},
Chandra + RXTE \citep{2002ApJ...570L..47L}, 
and Suzaku \citep{2007PASJ...59S.315M}.
All cited authors concluded that the extension of the red wing can be reasonably
interpreted only with models employing a very high value of the black hole spin.
However, \citet{2008A&A...483..437M} showed in their analysis that
the substantial X-ray data set for MCG\,-6-30-15, comprising 
all the mentioned missions, could be equally well fitted 
by an absorption-dominated model with no relativistically smeared emission.
The main idea of this work is that the source is obscured by several
clumpy absorption zones (``partial covering'' absorption). 

A possible way how to distinguish between the two different models 
is to study the time variability. The lack of variability in the energy range
of the broad iron line was noticed by \citet{2003MNRAS.340L..28F} in the
XMM-Newton spectrum and explained by the light bending model \citep{2004MNRAS.349.1435M}.
They constrained the inner edge of the disc to be within two
gravitational radii (the spin value $a > 0.94$). 
However, \citet{2008MNRAS.386..759N} pointed out that the line profile
has always a pronounced blue peak in the light bending model,
whereas the blue peak is missing in the deep minimum state
observation \citep{1996MNRAS.282.1038I}. Hence, \citet{2008MNRAS.386..759N} concluded that
the innermost region of the disc is not closer than 2--3 gravitational radii.
It has been also shown that a possible warm absorber may have an 
important effect on the spectral variability \citep{2008A&A...483..437M, 2009PASJ...61.1355M}.  
Within these models, the variability is explained 
by changing of the ionisation state of the warm absorber.

The uncertainty which model is more realistic pertains to all similar sources.
A strong argument against partial-covering warm absorber model was pronounced by
\citet{2009Natur.459..540F}. In the X-ray spectrum of an extragalactic source, 1H0707-495,
the authors detected an iron L line as well (thanks to extraordinarily high abundances of iron), 
and measured a time reverberation lag
between the direct X-ray continuum and its reflection from matter falling into black hole.
The time lag of 30\,s is comparable with the light travel time within one gravitational radius,
implying that the radiation must come from the closest neighbourhood around
a maximally rotating black hole.

\subsubsection{Long XMM-Newton observation of MCG\,-6-30-15}

A long XMM-Newton observation took place in summer 2001 (31st July - 5th August)
with the acquired exposure time was about 350\,ks (Obs. \#0029740101, \#0029740701, \#0029740801).
The EPN and both MOS cameras were operating in the ``Small Window'' mode
(see Section~\ref{xmm_modes}). 
The spectra were analysed by several authors \citep[etc.]{2002MNRAS.335L...1F, 
2003MNRAS.342..239B, 2004MNRAS.348.1415V, 2006ApJ...652.1028B,2008A&A...483..437M}.
We re-analysed the data and then, we used them for analyses presented
in the previous sections of the Thesis (Section~\ref{laky} and \ref{direxammcg}). 
In this section, we describe the details of our re-analysis.

We reduced the EPN data from three sequential revolutions (301, 302, 303)
using the SAS software version 7.1.2. %\footnote{http://xmm.esac.esa.int/sas}
Intervals of high particle background were removed by applying count rate thresholds
on the field-of-view (EPIC, single events) light curves of 0.35\,cts/s for the PN, 
and 0.5\,cts/s for the MOS.
The patterns 0--12 were used for both MOS cameras, and patterns 0--4 (i.e.\ single and double events) for the PN camera. 
The source spectra were extracted from a circle of 40 arcsec in radius defined around the centroid position 
with the background taken from an offset position close to the source. 
The two MOS spectra and the related response files were joined 
into a single spectrum and response matrix.

Using the FTOOL MATHPHA (part of the HEASOFT) we merged together 
the three spectra to improve the statistical significance. 
Further, we used the XSPEC version 12.2 for the spectral analysis.
We re-binned all the data channels to over-sample 
the instrumental energy resolution maximally by a factor of 3 
and to have at least 20 counts per bin. The first condition is much stronger 
with respect to the total flux of the source -- $4\times10^{-11}$\,erg\,cm$^{-2}$\,s$^{-1}$ 
($1.1\times10^{6}$\,cts) in the 2--10\,keV energy interval.

\begin{figure}[tb]
\begin{center}
%\epsfbox{fm_line_ratio.eps}
\includegraphics[angle=270,width=0.7\textwidth]{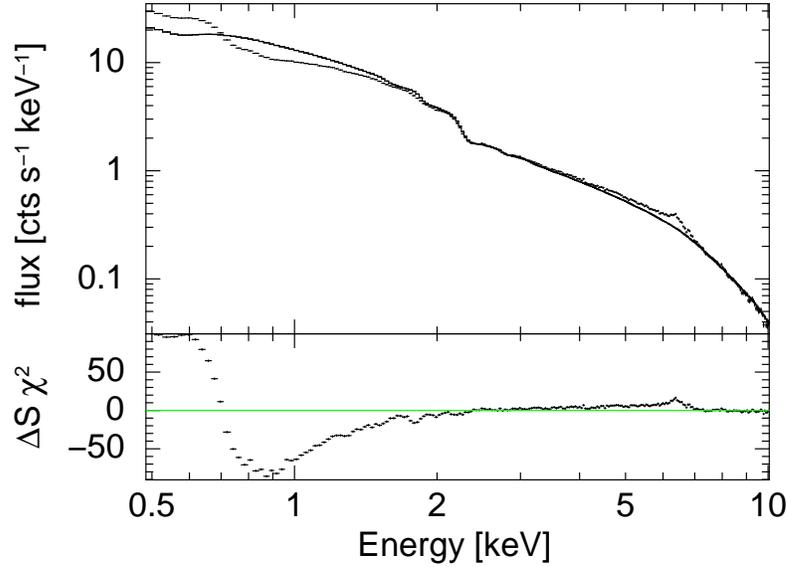}
\caption{The X-ray spectrum of MCG-6-30-15 
observed by the XMM-Newton satellite. 
The continuum is characterised by a power law with the photon index $\Gamma=1.9$ 
absorbed by Galactic gas
matter along the line of sight with column density 
$n_{\rm H} = 0.41\times 10^{21}$\,cm$^{-2}$. The residuals from 
the model are plotted in the bottom panel clearly revealing features of 
a local absorption and soft excess in the soft X-ray band,
and a feature at around $6$\,keV which can be explained by the presence
of a broad iron line.}
\label{mcg_spec}
\end{center}
\end{figure}

\subsubsection{X-ray continuum}

The resulting X-ray spectrum of MCG\,-6-30-15 %from 0.5\,keV to 10\,keV
is shown in  Figure~\ref{mcg_spec}. Above $E \approx\,2.5\,$keV, 
the X-ray continuum can be described by a power law component 
with the photon index $\Gamma =1.9$
absorbed by Galactic gas
matter along the line of sight with column density
$n_{\rm H} = 0.41\times 10^{21}$\,cm$^{-2}$
revealing a broad iron line at $E\approx 3-7$\,keV \citep{2002MNRAS.335L...1F}.
It is clear from the figure that other components 
need to be added into the model to describe the spectrum
at lower energies as well. 

The shape of the data residuals in lower energies 
suggests a combination of two effects:
an additional absorption and a soft excess. The local additional absorption
is usually attributed to a warm absorber, 
i.e.\ absorption by totally or partially ionised matter; see, e.g.,
\citet{2003ApJ...599..933N,blustin05,2007ApJ...659.1022K} for more 
information about warm absorbers in Seyfert galaxies.
The soft excess can be caused by reflection on the 
ionised surface of the accretion disc \citep{2006MNRAS.365.1067C}.
Alternative explanations for the soft excess are black-body radiation
or complex partially ionised absorption (see Section~\ref{similarity}
for brief discussion of the origin of the soft excess). 

\citet{2006ApJ...652.1028B} included warm absorber and ionised reflection in their model
and they showed that the extreme relativistic smearing is
still required. However, in a spectral analysis by 
\citet{2008MNRAS.483..437}, the spectrum is characterised by a complex
partial covering absorption in four different zones which affects 
the higher energy band more significantly implying that even 
no relativistic smearing was needed in the model. 
Our aim is not to discuss the appropriateness of the particular models.
Instead, we relied on the model of a broad line which we used in Section~\ref{laky}
for a comparison of two iron line models, {\kyrline} and {\laor},
and in Section~\ref{direxammcg}, for a 
comparison of different angular emissivity laws.

\begin{figure}[tb]
\begin{center}
\includegraphics[angle=270,width=0.65\textwidth]{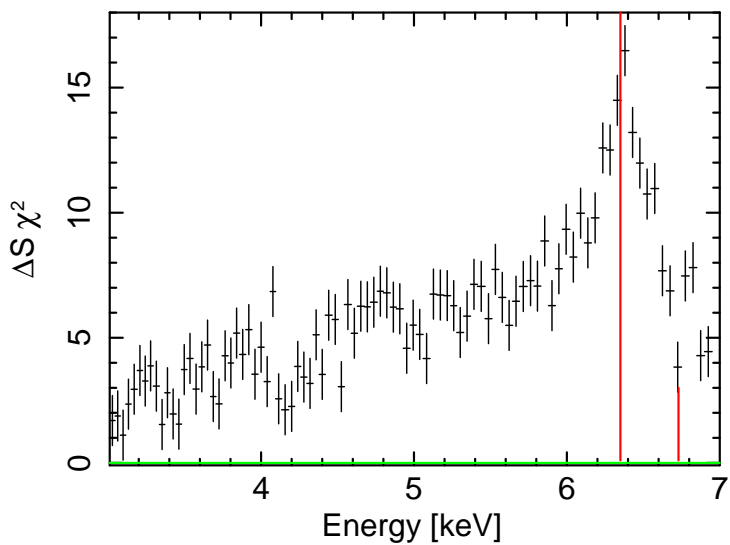}
\caption{A detailed view of the iron line band:
the data residuals from the power law model are shown
in terms of sigmas with error bars of size $1\,\sigma$.
Energies of narrow components considered in the model
are indicated by red lines:
$E=6.4$\,keV (observed at $E=6.35$\,keV due to the cosmological
redshift) of the neutral K\,$\alpha$ emission line,
and $E=6.77$\,keV (observed at $E=6.72$\,keV) of 
a mildly ionised absorption line. See the main text
for further details.}
\label{mcg_line}
\end{center}
\end{figure}

\subsubsection{Iron line complex}

The iron line complex (see Figure~\ref{mcg_line}) consists of three components: 
\begin{enumerate}
 \item a narrow line at $E=6.4$\,keV possibly originating by reflection 
on a distant torus and/or outer parts of an extended accretion disc  
 \item a broad relativistic line
 \item an additional narrow line which can be either
an emission line $E \approx 6.97$\,keV (of hydrogen-like ionised iron atoms),
or an absorption line at $E \approx 6.7$\,keV
\end{enumerate}

It is impossible to distinguish from the spectral analysis
the statistical preference between the models 
with the additional emission line of fully ionised iron atoms and with the absorption
line of mildly ionised iron atoms.
The values of the broad iron line parameters obviously depend on the 
parameters of the underlying model. However,
it was discussed by \citet{2002MNRAS.335L...1F} that the energy of the broad 
relativistic line is consistent with neutral iron K$\alpha$ line
originating in the innermost regions of cold disc independently 
of the inclusion of the narrow emission or absorption line. 
This was confirmed
in the analysis by \citet{2006ApJ...652.1028B} who preferred the emission
line at $E \approx 6.97$\,keV because they found it to be more consistent
with a Chandra observation \citep{2005ApJ...631..733Y}. However, they
considered also the model with the absorption line. They found the equivalent width
of the absorption line $EW=-21.3$\,eV much less than the one 
derived by \citet{2002MNRAS.335L...1F}, where $EW=-138$\,eV,
and so, they concluded that it is not robustly required by the fit.

We considered both cases and realised that the parameters of the broad
iron line are more sensitive to the underlying model than it was
suggested from the previous works, including the broad line 
energy, i.e. the ionisation state of the disc.

\begin{figure}[tb]
\begin{center}
 \includegraphics[width=0.49\textwidth]{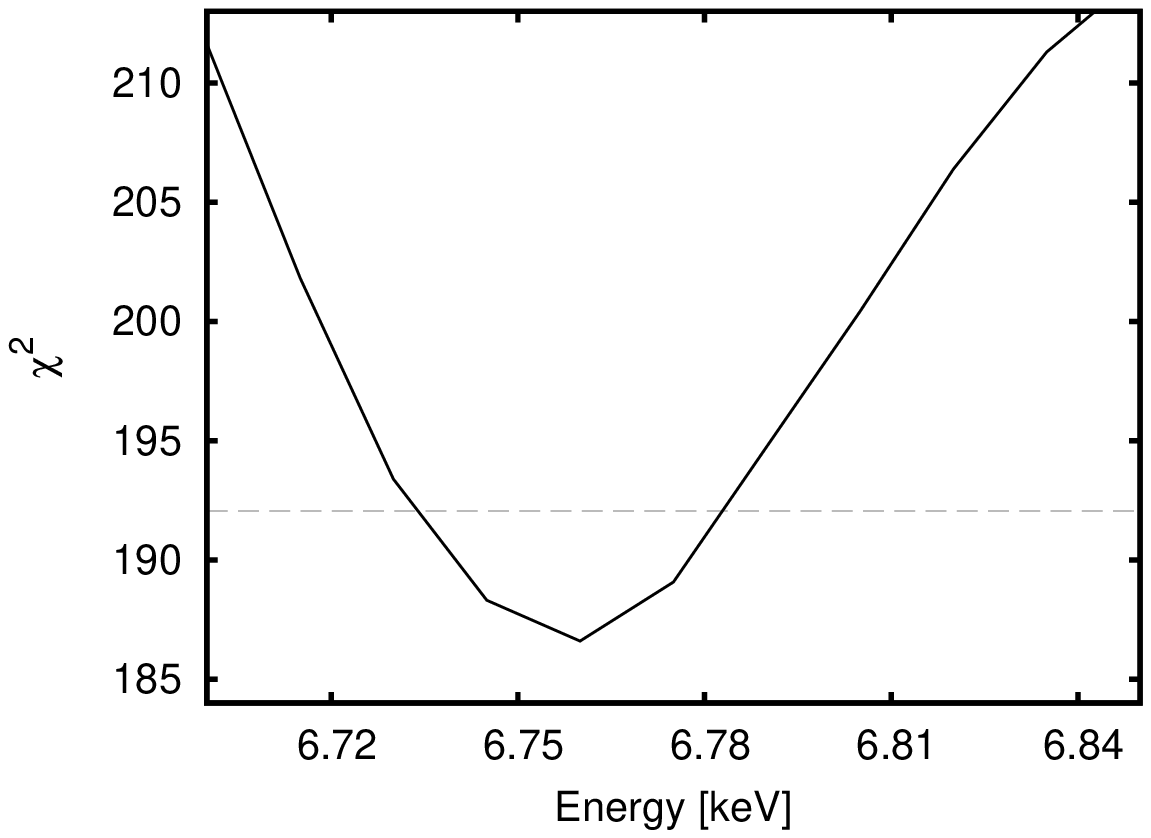}
 \includegraphics[width=0.49\textwidth]{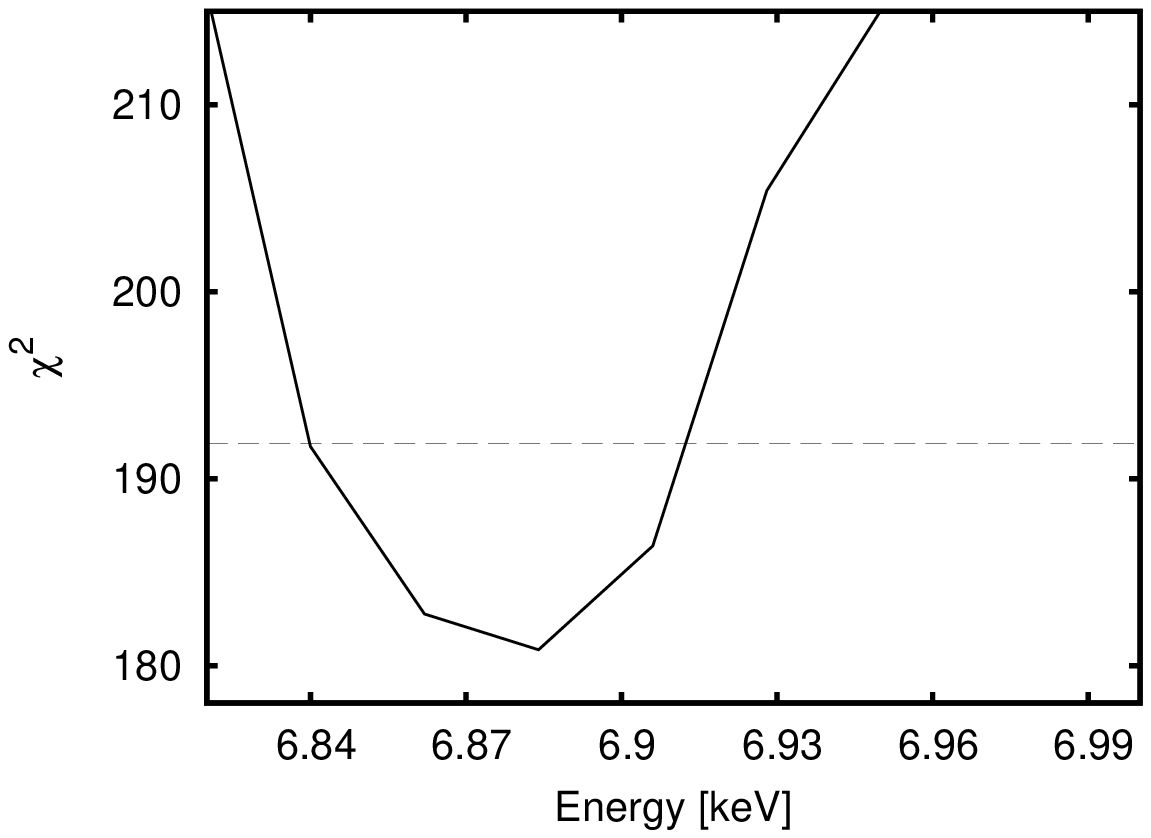}
\caption{Confidence plot of the energy value of 
a considered narrow absorption (\textposown{left})
or emission (\textposown{right}) spectral line. 
%The model further consists from an absorbed power law, 
%a broad iron line and a narrow emission line at 6.4\,keV.
The dashed line represents 90\% confidence level.}
\label{narrowlines}
\end{center}
\end{figure}

\begin{figure}[h!]
\begin{center}
 \includegraphics[width=0.49\textwidth]{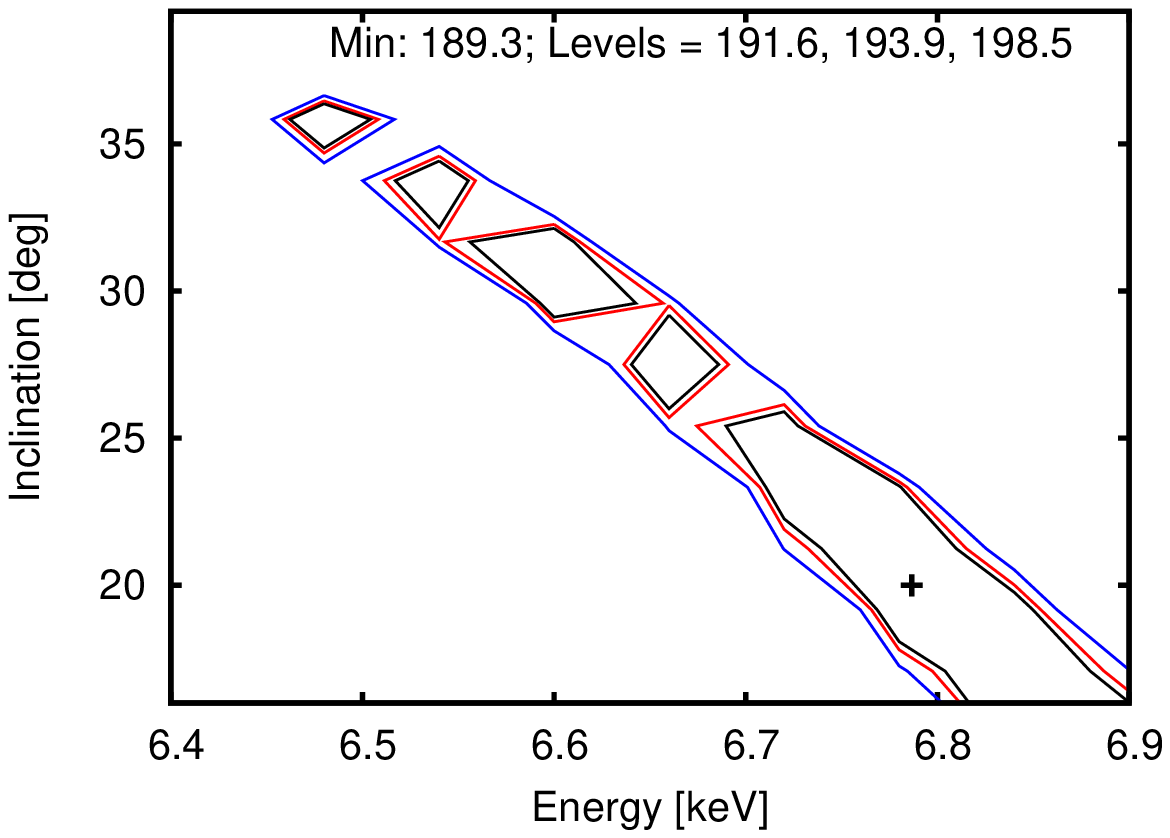}
 \includegraphics[width=0.49\textwidth]{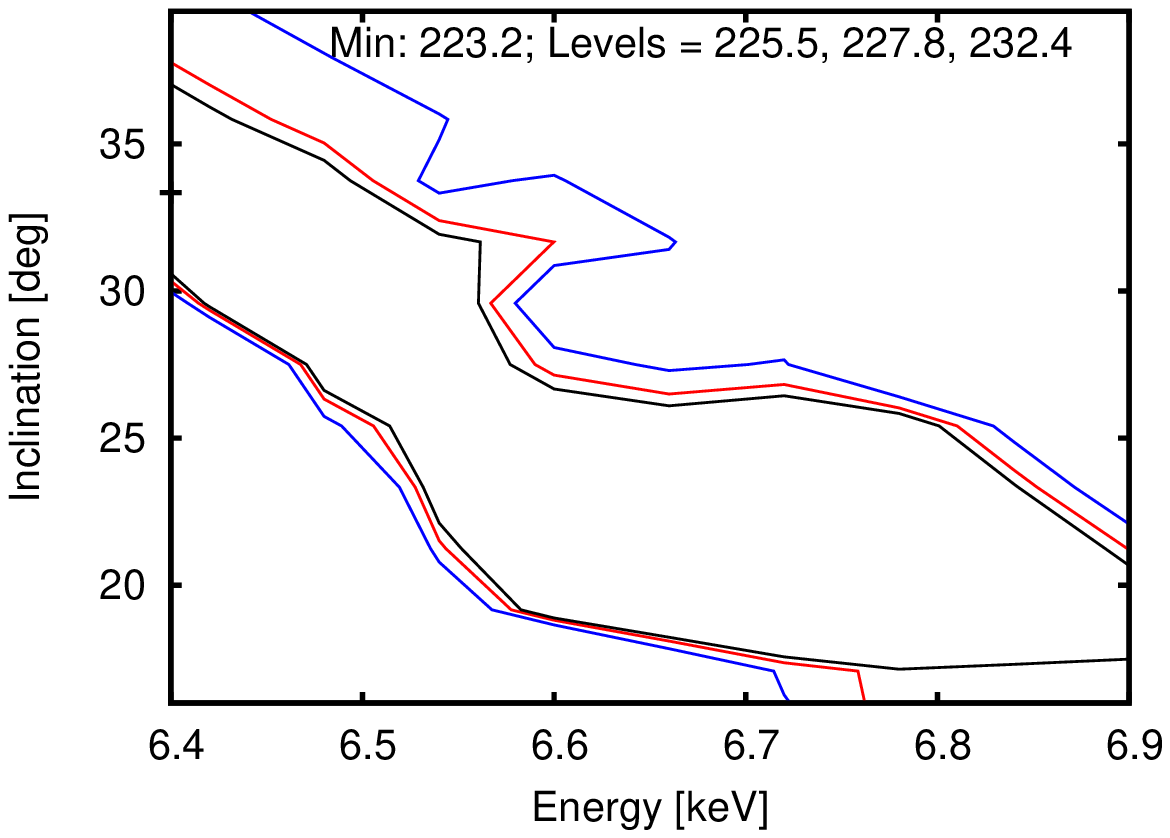}
\caption{Contour plots of the energy of the broad iron line
and inclination angle for two model alternatives. \textposown{Left}: narrow 
absorption line at $E=6.77$\,keV is included.
\textposown{Right}: narrow emission line at $E=6.97$\,keV is included, instead.
The individual curves correspond to $1\,\sigma$, $2\,\sigma$,
and $3\,\sigma$, respectively. The best fit values are indicated
with a small cross (in the right panel, the energy has the lowest
possible value, $E=6.4$\,keV, so the cross is situated in the 
graph boundary).}
\label{conie}
\end{center}
\end{figure}

The energy of the narrow absorption line was constrained to be $E=6.77^{+0.01}_{-0.03}$\,keV,
see the left panel of Figure~\ref{narrowlines}, where the dependence of the 
goodness of the fit is plotted against the energy of the absorption line.
The photon index of the power law, spin, inclination, energy and
radial emissivity of the broad iron line, and all normalisations
were allowed to vary during the fitting procedure.
%The energy level of the considered 
%absorption line is marked by a small red line in the Figure~\ref{mcg_line} 
%where the data residuals related to the iron line complex are shown.
We measured the equivalent width of the absorption line
$EW=-(21 \pm 5)$\,eV, consistently with the result obtained by
\citet{2006ApJ...652.1028B}.

The alternative model employs the emission line, instead. 
The energy of the narrow
emission line was constrained to be $E=6.88 \pm 0.02$\,keV,
a rather lower value than $6.97$\,keV considered by \citet{2002MNRAS.335L...1F},
but consistently with \citet{2006ApJ...652.1028B}.
The total goodness of the fit against the energy of the emission line
is shown in the right panel of Figure~\ref{narrowlines}.
The equivalent width is $EW=(28 \pm 4)$\,eV, i.e. about twice
larger than the value by \citet{2006ApJ...652.1028B}.

Further, we investigated how the energy of the broad iron line
depends on the choice of one of the model alternatives. 
We constructed contour plots for the energy of the broad line and inclination angle for the two model
alternatives - see Figure~\ref{conie}. The parameters of the narrow
lines modelled as Gaussian profiles were fixed. We used
$E=6.77$\,keV for the energy of the absorption line,
and $E=6.97$\,keV for the energy of the emission line.
The photon index of the power law, spin, inclination, energy
and radial emissivity of the broad iron line, and all normalisations
were allowed to vary during the fitting procedure.

The values for the
broad line energy are relatively consistent, similarly
depending on the inclination angle value. Nevertheless, in the alternative
with the absorption line (left panel of Figure~\ref{conie}), 
the ionised case is preferred rather than the case of a cold disc.

%proc jsme pouzili absorption line model, proc jsme pouzili 6.97..?
%\newpage
\section{IRAS~05078+1626}
  \label{iras05078}

%\subsection{Introduction}

IRAS 05078+1626 is a nearby Seyfert 1.5 galaxy. %observed by the Infrared Astronomical Satellite (IRAS); 
Before its identification as an infrared source
it was also known as CSV 6150 (Catalogue of Suspected Variables). 
Its position on the sky is $l=186.1$ and $b=-13.5$ in the Galactic coordinates. 
The cosmological redshift of this galaxy is $z\approx 0.018$ \citep{takata94}. %(Takata et al., 1994). 
It had never been spectroscopically examined in X-ray prior to the
observation discussed in this Thesis.
However, it was detected in a number of X-ray surveys, such as the
all-sky monitoring of the INTEGRAL IBIS/ISGRI instrument \citep{sazonov07}, 
the SWIFT BAT instrument \citep{ajello08,2008ApJ...681..113T}, and the RXTE Slew Survey (XSS) \citep{2004A&A...418..927R}. 

The X-ray spectroscopic properties of intermediate Seyferts are rather elusive:
both obscured Type 1 and unobscured Type 2 active galactic nuclei (AGN) have been reported
\citep[e.g.,][]{2006A&A...446..459C,2008MNRAS.390.1241B,2009ApJ...695..781B}.
It has been suggested that intermediate Seyfert galaxies are seen at intermediate 
inclination angles between pure ``face-on'' Seyfert~1s and pure ``edge-on'' Seyfert~2s,
which follows directly from the orientation-based AGN unification scenarios 
\citep{1985ApJ...297..621A,1993ARA&A..31..473A,1995PASP..107..803U}.
For this reason, X-ray spectroscopy of type 1.5 Seyferts may provide clues
to the nature and geometrical distributions of optically thick gas surrounding
the active nucleus, the latter being the fundamental ingredient behind 
the unification scenarios.

IRAS~05078+1626 is included in the FERO project 
(``Finding Extreme Relativistic Objects''; \citeauthor{2008MmSAI..79..259L}
\citeyear{2008MmSAI..79..259L}) 
with the aim of establishing the fraction of a relativistically broadened 
K$\alpha$ iron lines in the spectrum of a complete flux-limited sample (2--10\,keV flux $>$ 1 mCrab).

\begin{figure}[tb]
%\begin{tabular}{c}
% \includegraphics[height=6cm,angle=270]{powerlaw.eps} 
\centering
\includegraphics[angle=0,width=0.7\textwidth]{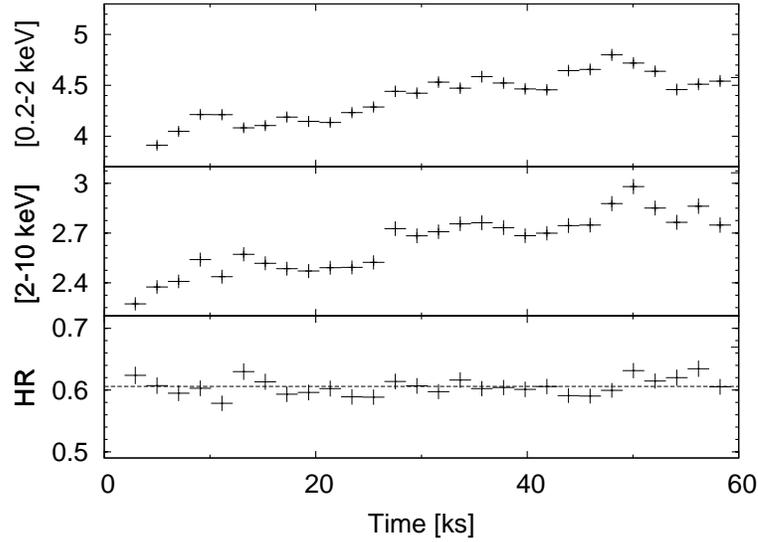} 
% \includegraphics[height=5.3cm,angle=270]{pm_onlypowerlaw.eps}
%\includegraphics[angle=270,width=0.48\textwidth]{pm_onlypowerlaw.eps}
%\end{tabular}
\caption{EPIC-PN light curves in the 0.2--2\,keV band (upper panel) and 2--10\,keV band (middle). 
The hardness ratio HR is defined as the ratio of the counts at 2--10\,keV to the counts at 0.2--2\,keV
and presented as a function of time in the lower panel. The bin time is as 2048\,s.}
\label{lc}
\end{figure}

\begin{figure}[tb]
\centering
\includegraphics[angle=0,width=0.7\textwidth]{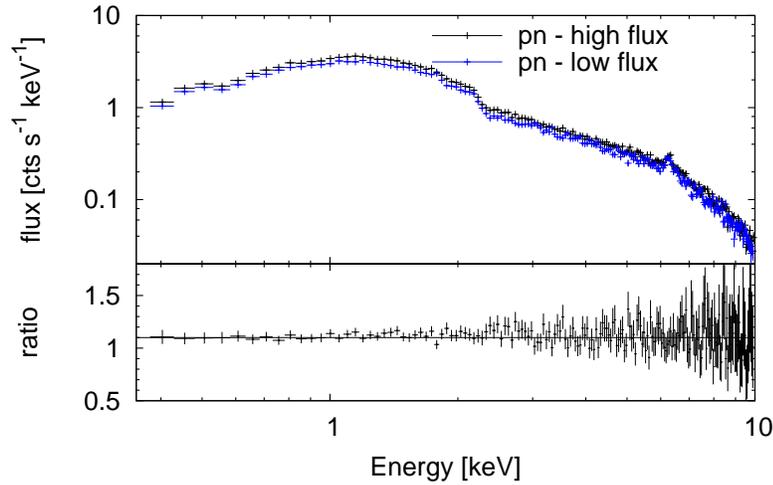} 
\caption{PN spectrum extracted from the first half of the observation with
the lower source flux (blue) and from the second half of the observation with
the higher source flux (black). The ratio of the two spectra is presented
in the lower panel.}
\label{pndata}
\end{figure}

\begin{figure}[tb]
\centering
\includegraphics[width=0.7\textwidth]{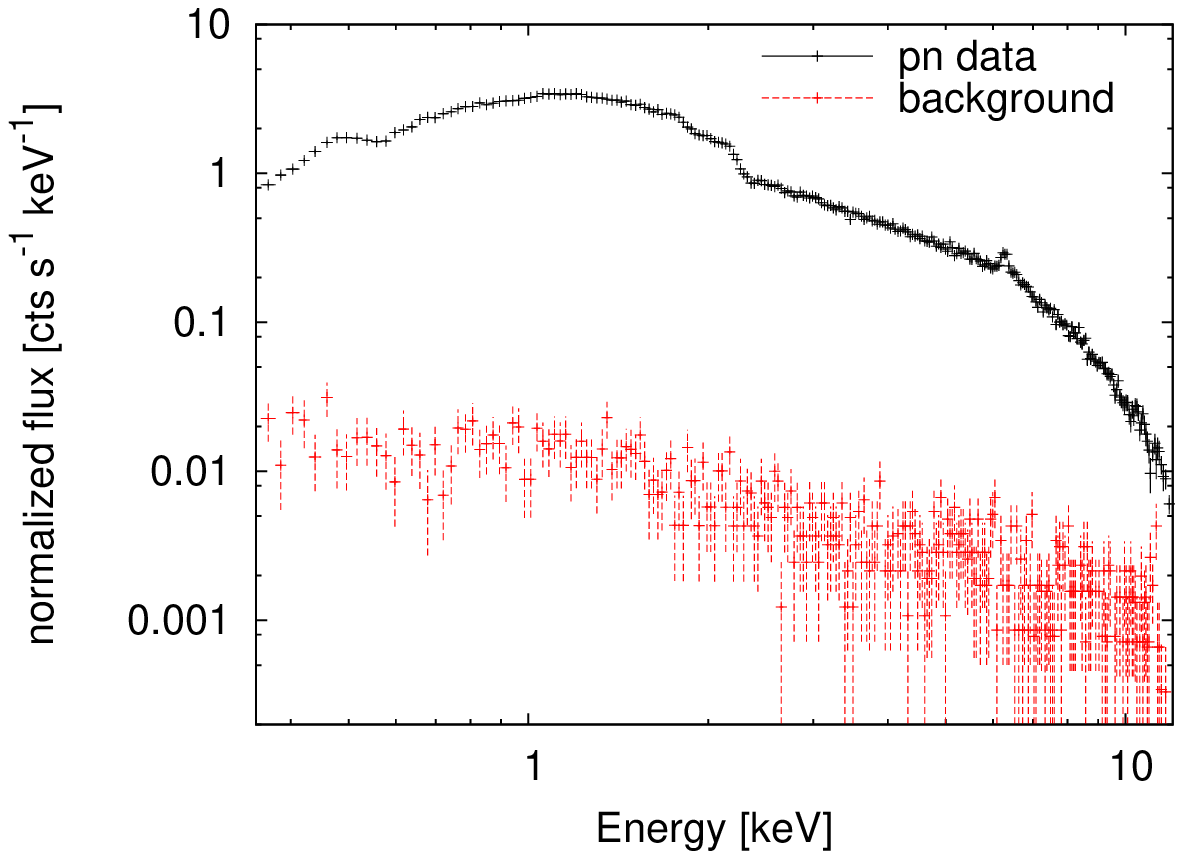}
\caption{Total PN spectrum with the background level showing 
that the signal-to-noise ratio is very good up to high energies.}
\label{pndata_back}
\end{figure}

\subsection{Results from XMM-Newton observation in 2007}
\subsubsection{Observations and data reduction}
\label{sec:observations}

The XMM-Newton observation of IRAS 05078+1626 was performed between 2007 August 21 
UT 22:24:49 and 22 UT 15:35:43 (Obs. \#0502090501).
The EPN and both MOS cameras were operating in the ``Small Window'' mode
(see Section~\ref{xmm_modes}). The RGS cameras were operating in the ``Spectroscopic'' mode \citep{rgs}.
The spectra were reduced with the SAS software version 9.0.0 \citep{2004ASPC..314..759G}. 
Intervals of high particle background were removed by applying count rate thresholds
on the field-of-view (EPIC, single events) and CCD\,\#9 (RGS) light curves of 0.35\,cts/s for the PN, 
0.5\,cts/s for the MOS and 0.15\,cts/s for the RGS. The exposure time after data
screening is $\approx 56$\,ks for MOS, $\approx 40$\,ks for PN and $\approx 58$\,ks for RGS, respectively.
The patterns 0--12 were used for both MOS cameras, and patterns 0--4 (i.e.\ single and double events) for the PN camera. 
The source spectra were extracted from a circle of 40 arcsec in radius defined around the centroid position 
with the background taken from an offset position close to the source. 
The two MOS spectra and the related response files were joined 
into a single spectrum and response matrix.
Finally, the PN and MOS spectra were re-binned in order to have 
at least 25 counts per bin and to over-sample the energy resolution 
of the instrument maximally by a factor of three, while the RGS spectra
were left un-binned. Consequently, different statistics were used in fitting
the spectra -- the traditional $\chi^2$ statistics to fit the PN and MOS spectra
and the C-statistics \citep{1976A&A....52..307C} for all fits including RGS data.
For the spectral analysis, we used XSPEC 
version 12.5, which is part of the HEASOFT software package version~6.6.%\footnote{http://heasarc.gsfc.nasa.gov}

\subsubsection{Timing properties}
\label{time}

The PN light curve of the source is shown in Figure~\ref{lc}. We have divided the energy range
into two bands and checked the light curve behaviour in each of them, 
as well as a hardness ratio, which
we defined as the ratio of the counts at 2--10\,keV to the counts at 0.2--2\,keV.
The energy ranges were chosen for sampling different spectral components,
as indicated by the energy where the continuum starts deviating from a power law model
that describes the hard X-ray spectrum (see Sect.~\ref{msp}).
The hardness ratio stays almost constant during the observation, 
suggesting that no significant spectral variations occur, 
although the source flux increased by around 20\%. 

To confirm this conclusion also for narrow spectral features,
such as the iron emission line, we compared the PN spectra extracted during the
first and the second halves of the observation (see Figure~\ref{pndata}).
The spectra correspond to the lower/higher source flux %(respectively to the higher source flux),
because the flux is increasing nearly monotonically during the observation.
We calculated the ratio values of the two data sets and fit them with a simple
function $f(E)=a \,E+b$ using the least square method. The fitting results are
$a=-0.004 \pm 0.003$ and $b=1.12 \pm 0.01$ with the sum of the residuals $\chi^{2}=111$ 
for 190 degrees of freedom. When we set $a=0$, the fitting results
are comparably good with $b=1.11 \pm 0.01$ and $\chi^{2}=113$. 
The ratio of the spectra is plotted in the lower panel of Figure~\ref{pndata}. 
Because no significant spectral differences are evident\footnote{Meanwhile, 
\citet{2009A&A...507..159D} reported
study of variability of FERO-sample galaxies,
including IRAS\,05078+1626.
They also found that the spectral variability
of this source is diminutive.},
we analyse the time-averaged spectra hereafter.

\begin{figure}[tb]
\centering
\includegraphics[angle=0,width=0.75\textwidth]{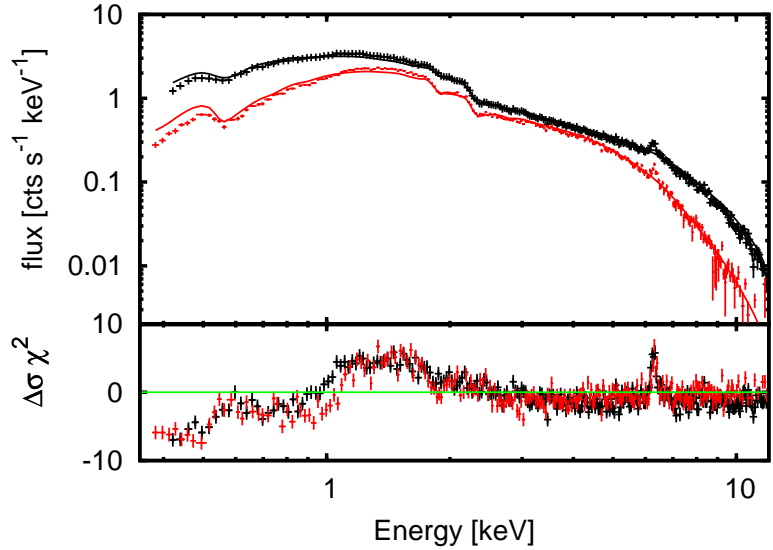} 
\caption{XMM-Newton PN (black) and joint MOS (red) spectrum of IRAS05078+1626 described 
by a simple power law model absorbed by Galactic neutral hydrogen in the line of sight 
with $n_{\rm H}=0.188 \times 10^{22}$\,cm$^{-2}$. The photon index of the power law is $\Gamma = 1.49$. 
%The model reveals an apparent excess at $E=6.4$\,keV associated with the iron line K$\alpha$ 
%and some wiggle-like residuals at lower energies. A more detailed view of the data residuals 
%in these parts of the spectrum is shown in Figure~\ref{pm_resid}.}
}
\label{powerlaw}
\end{figure}

\begin{figure}[tb]
%\begin{tabular}{cc}
% \includegraphics[height=6cm,angle=270]{powerlaw.eps} 
%\includegraphics[angle=0,width=0.99\textwidth]{phapo.eps} 
% \includegraphics[height=5.3cm,angle=270]{pm_onlypowerlaw.eps}
\includegraphics[width=0.48\textwidth]{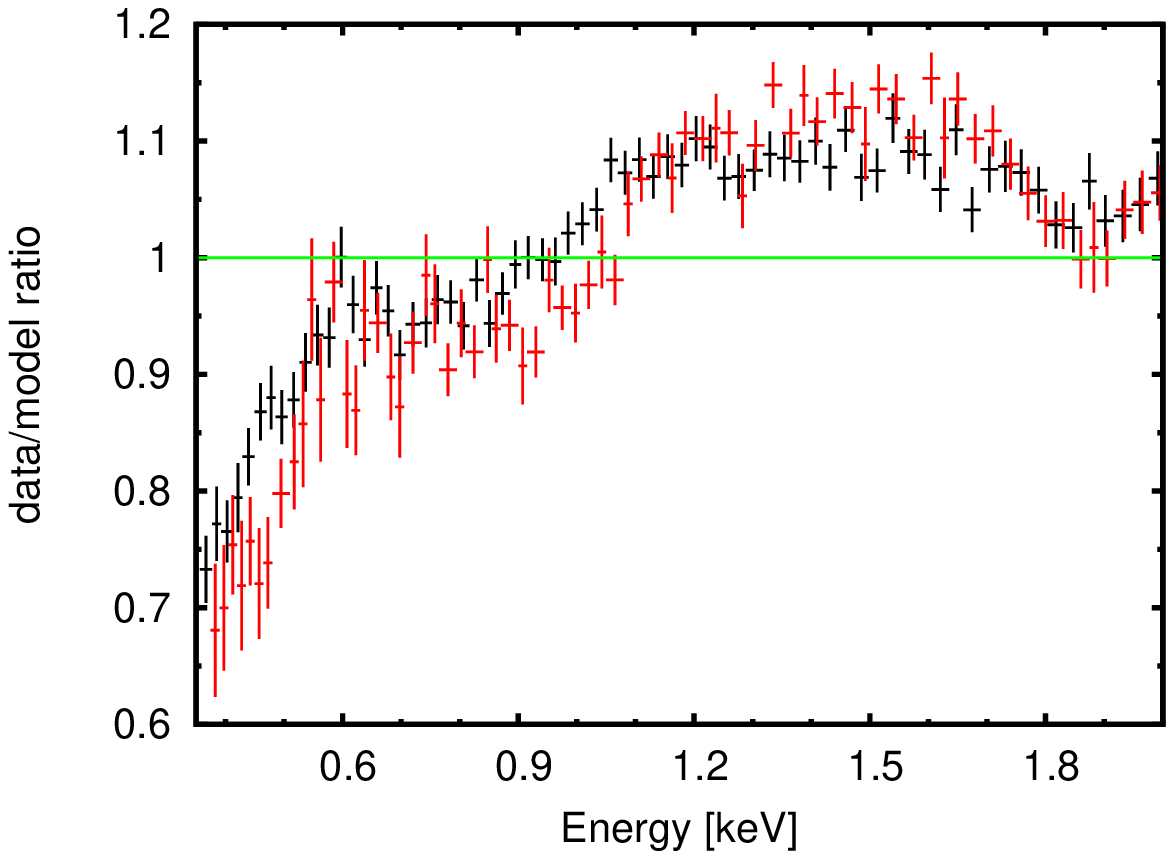}
\includegraphics[width=0.48\textwidth]{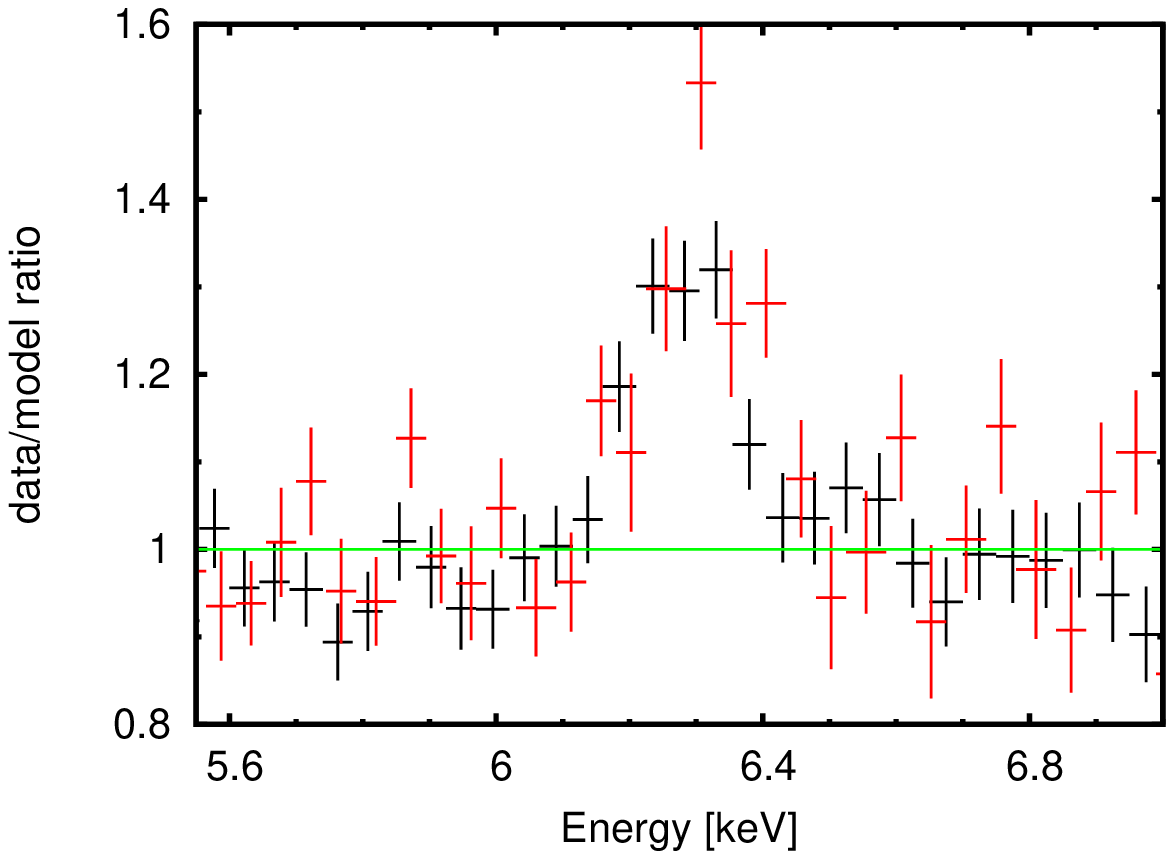}
%\end{tabular}
\caption{Ratios of the simple power law model (the same as in Figure~\ref{powerlaw}) to the data
in different energy bands: \textposown{left:} at lower energies, 
\textposown{right:} in the iron line band, where the narrow K$\alpha$ line
 at the rest energy $E=6.4$\,keV is prominent (observed at $E=6.29$\,keV due to
the cosmological redshift). Black crosses correspond to the PN data points,
while the red crosses correspond to the MOS data points.}
%An extra emission feature at slightly higher energies $E\approx 6.6$\,keV is suggested from the spectrum.} 
%and an eventual absorption line at $E\approx 6.7$\,keV are suggested from the spectrum.}
\label{pm_resid}
\end{figure}

\subsubsection{Mean spectral properties}
\label{msp}

%The X-ray spectrum of IRAS 05078+1626 is quite hard for this type of object, 
%with the hardness ratio around $HR \approx 0.6$.
%The luminosity of the source is $L = 2.6 \times 10^{43} $erg\,s$^{-1}$ in the 0.35-12\,keV energy range,
%$L = 2.3 \times 10^{43} $erg\,s$^{-1}$ in the 0.54-10\,keV energy range.
The signal-to-noise ratio is very good up to high energies (Figure~\ref{pndata_back}), 
so we fit the EPIC spectra spectra in the full energy range where they are
well calibrated (0.35--12\,keV).
The X-ray continuum is described by a power law model at energies above 2~keV, 
although the iron line at $E=6.4$\,keV is present (Figure~\ref{powerlaw}). 
The photon index of the power law is $\Gamma \simeq 1.49(1)$.
In this and all subsequent models we included absorption by Galactic gas
matter along the line of sight with column density $n_{\rm H}=0.188 \times 10^{22}$\,cm$^{-2}$. 
This value is from the Leiden/Argentine/Bonn H\,I measurements \citep{2005A&A...440..775K}. 
We used the {\tbabs} model \citep{wilms00} to fit the absorption produced by the Galactic interstellar matter. 

We applied the simple {\tbabs}*{\powerlaw} model to both PN and MOS spectra.
The $\chi^{2}$ value is 3557 with 528 degrees of freedom ($\chi^{2}/\nu=6.7$) 
in the $0.35-12.0$\,keV energy range. 
The spectra differ from the power law model not only around $E=6.4$\,keV
but also at lower energy band $0.35-2.0$\,keV (adding a Gaussian line model
to fit the iron line improves the fit only to $\chi^{2}/\nu = 3384/524 \doteq 6.5$). 
Residuals against this model are shown in Figure~\ref{pm_resid}. They may be 
due to a warm absorber and/or soft excess.

The spectral residuals reveal certain discrepancies between the PN and MOS spectra
(see Figure~\ref{pm_resid} with the data/model ratios of both spectra, PN and MOS, 
with the identical model parameters of the spectra). The level
of discrepancy is, however, comparable to the level of systematic uncertainties
in the cross-calibration between the EPIC cameras \citep{2006ESASP.604..937S}.
%(http://xmm2.esac.esa.int/docs/documents/CAL-TN-0052.ps.gz).
Nonetheless, we conservatively analyse the EPIC spectra separately. 
We use the same models for both spectra but allow the values 
of the model parameters to be different. %The precedence was yielded to the PN spectrum. 
The values of the photon index using the simple power law model differ from each other 
when fitting the spectra independently, resulting in a harder PN spectrum with $\Gamma = 1.60(1)$ 
compared to the MOS spectrum with $\Gamma = 1.54(1)$, ignoring the energies 
below 2\,keV and also between 5.5-7.5\,keV.
Although the absolute value of these spectral index
measurements does not have a direct physical meaning,
given the simplicity of the model applied on a small 
energy band, the comparison between them is illustrative 
of the quality of the cross-calibration between the EPIC cameras.
Differences of the order of $\Delta \Gamma \simeq$ 0.06 
in the hard X-ray band are consistent with current systematic 
uncertainties \citep{2006ESASP.604..937S}.

\subsubsection{RGS spectrum}
\label{rgs}

We jointly fit the un-binned first-order spectra of the two RGS cameras 
with the same model's parameter values except the overall normalisations.
The continuum is well-fitted by the simple power law model with the photon index $\Gamma = 1.57$.
We searched further for narrow emission and absorption lines in the spectrum using
several {\zgauss} models with the intrinsic width $\sigma$ set to zero. We calculated the 90\,\% confidence 
interval for a blind search, as $P=P_{0}/N_{\rm trial}$, where $N_{\rm trial}=N_{\rm bins}/3 = 3400/3$
and $P_{0}=0.1$. For the RGS data $P \doteq 8.8\times 10^{-5}$, to which $\Delta C = 22.4$
corresponds assuming the Student probability distribution.
The only line fulfilling this criterion by improving the fit about $\Delta C=31.7$ 
is an emission line at the energy $E=0.561\pm0.001$\,keV ($22.10 \pm 0.04\,\AA$) 
and the equivalent width $7^{+5}_{-3}$\,eV.
We identified it with the forbidden line of the O~VII triplet ($E_{\rm LAB} = 0.561$\,keV).

%\subsection{Effects of warm absorber and disc reflection}
\subsubsection{EPIC spectrum}
\label{res}

The forbidden line of the OVII triplet is clear signature of a photoionised plasma. 
No significant features were detected that may be expected alongside the O~VII~(f) line, 
if it were produced in a collisionally ionised plasma, such as the resonance 
line in the OVII triplet or the OVIII~Ly$\alpha$. 
This led us to try and explain the residuals against a power law model
in the soft X-rays as effect of intervening ionised absorption gas.
We used the {\xstar} model version 2.1ln7c 
\citep{2001ApJS..133..221K}\footnote{http://heasarc.gsfc.nasa.gov/docs/software/xstar/xstar.html} 
to calculate a grid of tabular models with the input parameters constrained 
from the preliminary data analysis with simple models whenever possible
(photon index $\Gamma \approx 1.7$, density $\rho \leq 10^{14}$\,cm$^{-3}$,
luminosity $L \leq 10^{44}$\,erg\,s$^{-1}$, column density 
$10^{19}$\,cm$^{-2} \leq n_{\rm H} \leq 10^{25}$\,cm$^{-2}$, and ionisation parameter
$-5 \leq \log \xi \leq 5$).

\begin{sidewaystable}
\begin{center}
{\small
\caption{Parameters of the `baseline' and `final' models.}
\begin{tabular}{c|c|c|c|c|c|c} 
	\hline \hline \rule{0cm}{0.5cm}
 Model 	&	Model		&\multicolumn{2}{c|}{`baseline' model}	&	\multicolumn{3}{c}{`final' (`double reflection') model}	 \\
  component	&	 parameter	&	PN	&	MOS &	PN	&	MOS   &  PN+MOS+RGS  \\
	\hline
\rule[-0.5em]{0pt}{1.6em} \zphabs	& $n_{\rm H} [10^{22}$\,cm$^{-2}]$	& $0.104^{+0.005}_{-0.007}$	&	$0.129^{+0.007}_{-0.007}$	&	$0.102^{+0.009}_{-0.005}$	& $0.120^{+0.008}_{-0.005}$	& $0.106^{+0.004}_{-0.004}$ \\
\hline
\rule[-0.5em]{0pt}{1.6em} \xstar	& $n_{\rm H} [10^{22}$\,cm$^{-2}]$	&	$130^{+20}_{-10}$	& 	$170^{+20}_{-20}$	&	$120^{+30}_{-30}$	&	$150^{+70}_{-20}$ 	&	$130^{+20}_{-20}$	\\
\rule[-0.5em]{0pt}{1.6em}	&	$\log\,\xi$			&$2.3^{+0.1}_{-0.1}$	&	$2.4^{+0.1}_{-0.1}$	&	$2.2^{+1.4}_{-0.6}$	&	$2.5^{+1.0}_{-0.5}$  &	$2.5^{+1.0}_{-0.4}$	\\
%\rule[-0.7em]{0pt}{2em}	&	z				&	$0.018$	(f)	&	$0.018$	(f) &	$0.018$	(f) \\
\rule[-0.5em]{0pt}{1.6em}	&	He/He$_{\rm Solar}$ - Ca/Ca$_{\rm Solar}$ 	&	$1$ (f)	
&	$1$ (f)	&	$1$ (f)	&	$1$ (f) &	$1$ (f)	\\
\rule[-0.5em]{0pt}{1.6em}	&	Fe$/$Fe$_{\rm Solar}$ - Ni$/$Ni$_{\rm Solar}$		&	$0.2^{+0.1}_{-0.2}$	&	$0.1^{+0.1}_{-0.1}$		&	$1.2^{+0.3}_{-0.3}$	&	$0.9^{+0.2}_{-0.2}$	&	$1.1^{+0.2}_{-0.2}$	\\
\hline
\rule[-0.5em]{0pt}{1.6em} \powerlaw &	$\Gamma$		&	$1.81^{+0.03}_{-0.05}$		&	$1.80^{+0.05}_{-0.05}$	&	$1.75^{+0.10}_{-0.03}$		&	$1.74^{+0.07}_{-0.03}$	&	$1.76^{+0.04}_{-0.02}$	\\
\rule[-0.5em]{0pt}{1.6em}	&	normalisation		&	$\left(7 \pm 1\right) \times 10^{-4}$		&	$(7\pm1) \times 10^{-4}$	&	$(6\pm1) \times 10^{-4}$		&	$(7\pm2) \times 10^{-4}$	& .....	\\
\hline
\rule[-0.5em]{0pt}{1.6em} \reflion &	$\Gamma$	&	$1.81$ (b)	&	$1.80$ (b)	&	$1.75$	(b)	&	$1.74$ (b)	&	$1.76$ (b)	\\
\rule[-0.5em]{0pt}{1.6em}	&	$\log\,\xi$		&	$3.0^{+0.2}_{-0.2}$		&	$3.2^{+0.2}_{-0.2}$	&	$3.0^{+0.2}_{-0.2}$		&	$3^{+2}_{-3}$	&	$3.0^{+0.1}_{-0.2}$	\\
\rule[-0.5em]{0pt}{1.6em}	&	Fe$/$Fe$_{\rm Solar}$ - Ni$/$Ni$_{\rm Solar}$	& 	$0.2$ (b)	&	$0.1$ (b)	&	$1.2$ (b)	&	$0.9$ (b)	&	$1.1$ (b)	\\
\rule[-0.5em]{0pt}{1.6em}	&	normalisation 		&	$(3 \pm 2) \times 10^{-9}$	&	$(3 \pm 2) \times 10^{-9}$	&	$(2\pm1) \times 10^{-9}$	&	$(1\pm1) \times 10^{-9}$		& .....		\\
\hline
\rule[-0.5em]{0pt}{1.6em} \reflion~2&	normalisation 		&	-	&	-	&	$3\pm1 \times 10^{-7}$	&	$4\pm1 \times 10^{-7}$		& .....		\\
\hline
\rule[-0.5em]{0pt}{1.6em} \zgauss &	E [keV]		&	$6.40^{+0.01}_{-0.01}$	&	$6.44^{+0.02}_{-0.02}$	& -  & -  & -	\\
\rule[-0.5em]{0pt}{1.6em}	&	$\sigma$ [keV]		&	$0.06^{+0.03}_{-0.04} $	&	$0.02^{+0.04}_{-0.02} $	& -  & - & - \\
\rule[-0.5em]{0pt}{1.6em}	&	z	&	$0.018$	(f)	&	$0.018$	(f)  & - & - & -\\
\rule[-0.5em]{0pt}{1.6em}	&	normalisation 		&	$(3\pm1) \times 10^{-5}$ &	$(3\pm1) \times 10^{-5}$	& -  & - & -	\\
\hline
\hline
\rule[-0.5em]{0pt}{1.6em} $\chi^2/\nu$	& &	246/264		&	405/243		&	256/266		&	404/244		
&	$C/\nu$ = 1551/1347  \\
\end{tabular}
\label{model}
}
\end{center}
{\small
Note: The sign (f) after a value means that it was fixed during the fitting procedure. 
The sign (b) means that the parameter was bound to the value of the corresponding 
parameter of the previous model component. The sign ``-'' means that the model component
was not included while dots only mean that there
are more values related to the individual spectra which are not necessary to be all 
shown in the table.
}
\end{sidewaystable} 

\begin{figure}[tb]
\begin{center}
\includegraphics[width=0.75\textwidth]{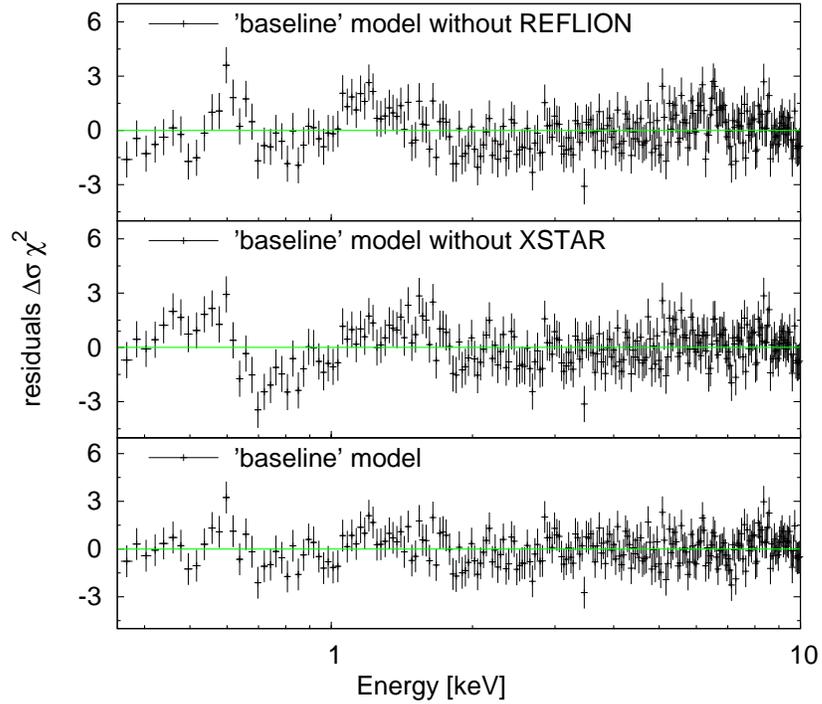}
\caption{Residuals of the PN data from the `baseline' model
including both ionised reflection and absorption (\textposown{lower}),
only reflection (\textposown{middle}), and only absorption (\textposown{upper}).}
% See the main text for further details.}
\label{del_chi}
\end{center}
\end{figure}

\begin{figure}[tb]
\begin{center}
\includegraphics[width=0.75\textwidth]{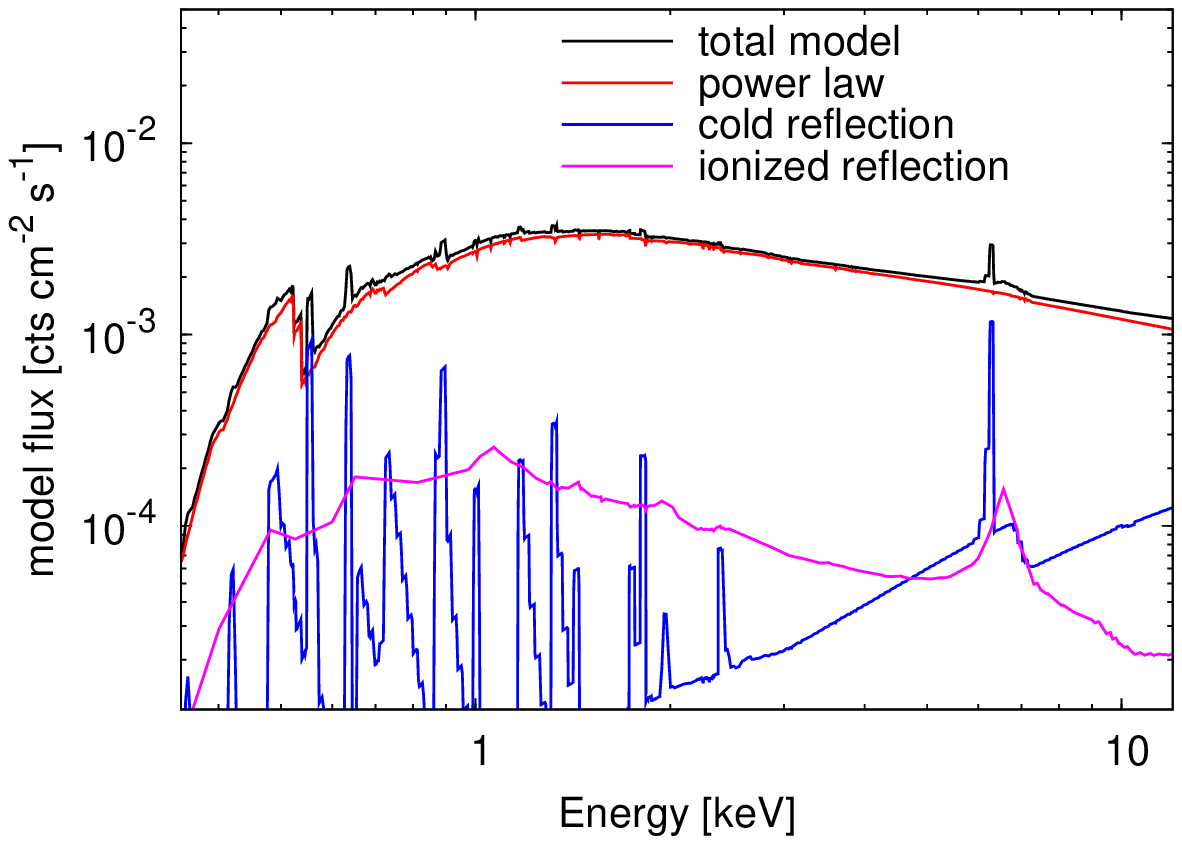}
\caption{The `final' model. The total model is shown in black (solid line), the primary radiation
is red (dashed), the {\reflion} components are blue (dotted) for cold reflection 
and magenta (dot-dashed) for ionised reflection. 
}
\label{emo}
\end{center}
\end{figure}

A single-zone warm absorber component modifying the power law continuum dramatically
improved the fit from $\chi^2 / \nu = 1850/270 \doteq  6.9$ 
to $\chi^2 / \nu = 402/270 \doteq  1.5$ for the PN spectrum.
The ionisation parameter converged to a very low value, and we found 
that this almost neutral absorption can be successfully reproduced with {\zphabs}, 
which is a simpler model than {\xstar}, so we preferred this possibility.
The addition of another warm absorber zone improves the fit 
to $\chi^2 / \nu = 320/266 \doteq 1.2$ for the PN spectrum, and it requires
the ionisation parameter $\log \xi \cong 3.9$.
We checked that adding another warm absorber zones does not improve
the fit significantly.
%The resulting `warm absorber' model is in XSPEC notation: 
%{\tbabs} $\times$ {\zphabs} $\times$ {\xstar} ({\powerlaw} + {\zgauss}),
%and the values of model parameters are presented in the Table~\ref{model}.
The residuals from the model (see Figure~\ref{del_chi}, upper panel)
reveal an extra emission that remains at low energies, as well as 
around the iron K$\alpha$ line band.

These features can come from reflection of the primary radiation on the surface 
of the accretion disc, so we added the {\reflion} model \citep{2005MNRAS.358..211R}, 
which calculates the ionised reflection for an optically thick atmosphere with constant density.
We examined the significance of the addition of the reflection component 
into the complex model of the PN spectrum by the statistical F-test. 
The low value of the F-test probability ($5 \times 10^{-15}$) strongly favours this additional model component.
The best fit was now $\chi^2/\nu = 246/264 \doteq 0.95$ for the PN spectrum.
We hereafter call this model the `baseline' model; in the XSPEC notation:
{\tbabs} $\times$ {\zphabs}$_{\,\rm N}$ $\times$ {\xstar} $\times$ ({\powerlaw} + {\reflion} + {\zgauss}).

%The MOS spectrum is still affected by the complexities that cannot be described in terms of the warm absorber zones 
%and/or disc reflection; in this case the best-fit $\chi^2/\nu = 1.6$. 
The parameter values of the `baseline' model are presented in the Table~\ref{model}.
The quoted errors of the parameters represent a 90~$\%$ confidence 
level for a single interesting parameter. The measurement is obviously 
affected by a much larger systematic error, which, however, 
could be properly quantified only if we knew the ``right'' model.
The value of the power law photon index increased to $\Gamma \approx 1.8$
compared to the simple model applied to the data in Sect.~\ref{msp},
because we included the additional local absorption in the model.
The data residuals from the model are shown in the lower panel of Figure~\ref{del_chi}.
In the same figure, we also show residuals from the best fit
performed with the `baseline' model, excluding the ionised absorption
(middle panel) and the ionised reflection component (upper panel). %baseline bez xstaru ma nejlepsi fit X2/v = 319/266=1.2

The narrow iron K$\alpha$ line with the rest energy $E=6.40 \pm 0.01$\,keV,
the width $\sigma=0.06 \pm 0.03$\,keV, and the equivalent width $EW = 82 \pm 15$\,eV
evidently represents cold reflection. This suggests an origin of this spectral component 
in the outer part of the disc, or from the torus. 
The cold reflection is also supposed to contribute to the soft part of the spectrum
with the individual emission lines. For this reason, we replaced the Gaussian profile
in the `baseline' model with another {\reflion} component (called as {\reflion}~2 
in the Table ~\ref{model}) with the same values for the photon index 
and abundances as the {\reflion}~1 model component.
The ionisation parameter was kept free during the fitting procedure, but 
it very quickly converged to its lowest value $\xi = 30$ ($\log \xi = 1.477$). 
The advantage of the {\reflion} model compared to the other available 
reflection models is that it also includes the soft X-ray lines, 
with the disadvantage in this case that the ionisation parameter cannot be set to zero.

\begin{figure}[tb]
\begin{center}
\includegraphics[width=0.7\textwidth]{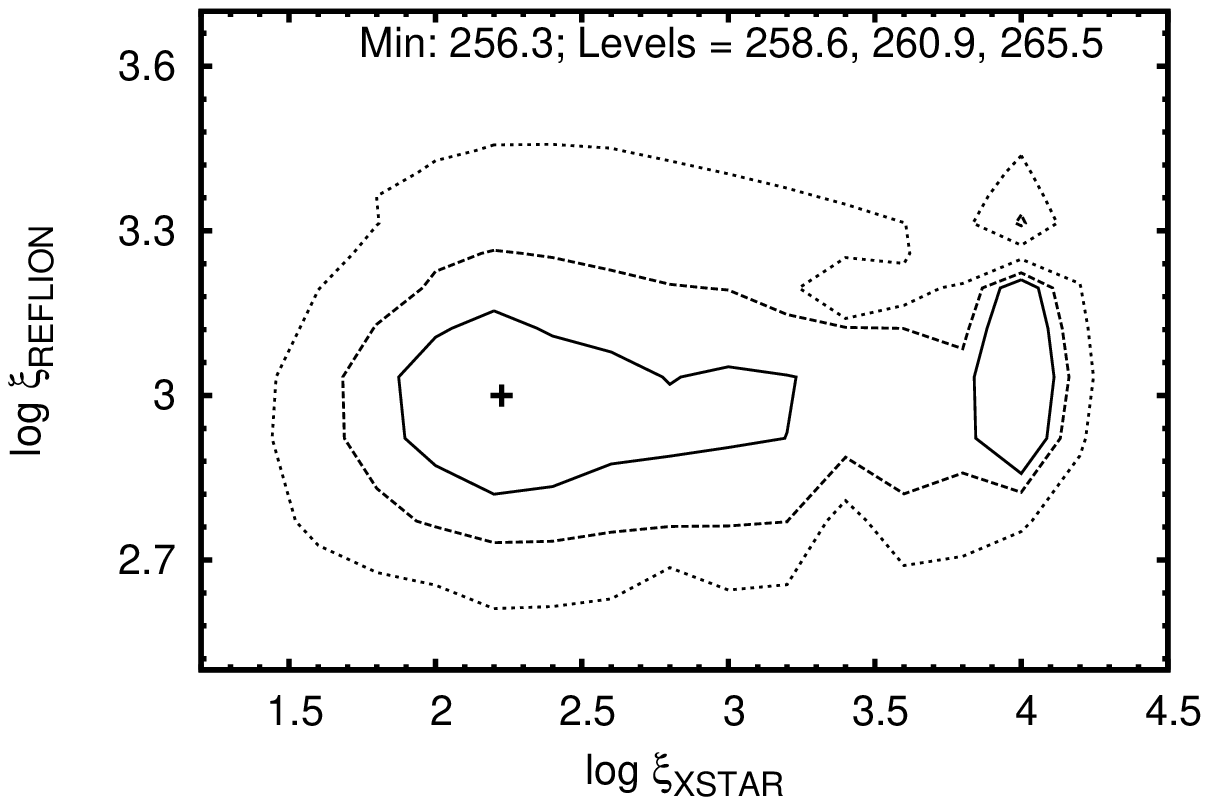}
\caption{The contour plot of the ionisation parameters of the {\reflion} model,
representing the ionised accretion disc, and of the {\xstar} model, representing
the warm absorber in the `final' model. 
The individual curves correspond to the 1\,$\sigma$, 2\,$\sigma$, and 3\,$\sigma$
levels. The position of the minimal value of $\chi^{2}$ found by the fitting 
procedure is marked by a cross. The corresponding $\chi^2$ values are given
in the plot.}
\label{cont_ion}
\end{center}
\end{figure}

\begin{table}[tbh]
\caption{Flux values of the `final' model and its individual components.} 
\begin{center}
\begin{tabular}{c|cc|cc} 
	\hline \hline
\rule[-0.6em]{0pt}{1.7em} Model	&	\multicolumn{2}{c|}{Flux at $0.5-2\,$keV} &	\multicolumn{2}{c}{Flux at $2-10\,$keV}	\\
\rule[-0.6em]{0pt}{1.7em} component	&	\multicolumn{2}{c|}{$[10^{-12}$erg\,cm$^{-2}$\,s$^{-1}$]}	&	\multicolumn{2}{c}{$[10^{-12}$erg\,cm$^{-2}$\,s$^{-1}$]}	 \\
\rule[-0.6em]{0pt}{1.7em} 	& 	PN	&	MOS &	PN	&	MOS\\
\hline
\rule[-0.6em]{0pt}{1.7em} total	model	& $7.05^{+0.03}_{-0.03}$	&	$7.00^{+0.03}_{-0.03}$	& $25.0^{+0.1}_{-0.2}$	&	$25.4^{+0.1}_{-0.2}$	\\
\rule[-0.6em]{0pt}{1.7em} unabsorbed model	& $16.6^{+0.2}_{-0.2}$	&	$16.7^{+0.2}_{-0.2}$	& $25.7^{+0.2}_{-0.2}$	&	$26.5^{+0.2}_{-0.2}$	\\
\rule[-0.6em]{0pt}{1.7em} {\powerlaw}	& $13.6^{+0.1}_{-0.2}$	&	$14.3^{+0.1}_{-0.1}$	& $22.6^{+0.3}_{-0.2}$	&	$23.8^{+0.1}_{-0.1}$	\\
\rule[-0.6em]{0pt}{1.7em} {\reflion}$_{\rm ion}$	& $1.9^{+0.2}_{-0.2}$	&	$1.1^{+0.1}_{-0.1}$	& $1.6^{+0.1}_{-0.2}$	&	$0.9^{+0.1}_{-0.1}$	\\
\rule[-0.6em]{0pt}{1.7em} {\reflion}$_{\rm cold}$	& $1.1^{+0.1}_{-0.1}$	&	$1.3^{+0.1}_{-0.1}$	& $1.5^{+0.1}_{-0.1}$	&	$1.8^{+0.2}_{-0.1}$	\\
\hline
\rule[-0.6em]{0pt}{1.7em} $R_{\rm ion}$ $^*$		& $0.12^{+0.01}_{-0.01}$	&	$0.07^{+0.01}_{-0.01}$ & $0.06^{+0.01}_{-0.01}$	&	$0.03^{+0.01}_{-0.01}$\\
\rule[-0.6em]{0pt}{1.7em} $R_{\rm cold}$ $^*$	& $0.07^{+0.01}_{-0.01}$	&	$0.08^{+0.01}_{-0.01}$ & $0.06^{+0.01}_{-0.01}$	&	$0.07^{+0.01}_{-0.01}$\\
%\hline
\end{tabular}
\end{center}

{$^*$ the ratios of the reflection component flux values to the flux 
value of the total unabsorbed model (sum of the primary and reflected radiation).}
\label{rratio}

\end{table} 

This `double reflection' model, in the XSPEC notation
{\tbabs} $\times$ {\zphabs}$_{\,\rm N}$ $\times$ {\xstar} $\times$ ({\powerlaw} + {\reflion} + {\reflion}),
does not significantly improve the fit goodness over the `baseline' model 
(with $\chi^2/\nu = 256/265 \doteq 0.96$ for the PN spectrum), 
but it represents a more self-consistent astrophysical picture.
Therefore, we call the `double reflection' model as `final' model.
In contrast to the `baseline' model, it does not require sub-solar iron abundances, 
see the Table~\ref{model}, where the parameter values for this model are presented.
The `final' model with each component separately drawn is shown in Figure~\ref{emo}.
All the plotted components are absorbed by a warm absorber surrounding 
the central accretion disc and two kinds of cold absorber -- one from 
Galactic interstellar matter and one from local absorber in the host galaxy.

Some model parameters were not allowed to vary during the fitting procedure. 
The redshift of the ionised absorber was fixed to the source cosmological value, 
because leaving it free yields a negligible improvement in the quality of the fit.
Second, we used the same iron abundances across all the components in the model.

In the `final' model, the warm absorber ionisation parameter is consistent
with the ionised reflection component.
This result is also presented in Figure~\ref{cont_ion}, 
where the contour lines related to the $1\sigma$, $2\sigma$, and $3\sigma$ levels 
of $\chi^{2}$ between the ionisation parameters of the two model components are presented.

Table~\ref{rratio} summarises flux values of the individual components of the `final' model
for both PN and MOS spectra for two energy bands, $0.5-2$\,keV and $2-10$\,keV,
and also shows fractions of the reflection radiation to the total emission (sum of the
primary and reprocessed radiation). The flux ratio is almost equally 
shared between the cold and ionised reflection components, 
and its value is in total $R<0.2$.
The absorption--corrected luminosity values of the source in the same energy bands are
$L\,_{0.5-2\rm\,keV}=(1.21 \pm 0.02)\times10^{43}$\,erg\,s$^{-1}$
and $L\,_{2-10\rm\,keV}=(1.87 \pm 0.02)\times10^{43}$\,erg\,s$^{-1}$, respectively.

\begin{figure}[tb]
\begin{center}
\includegraphics[width=0.75\textwidth]{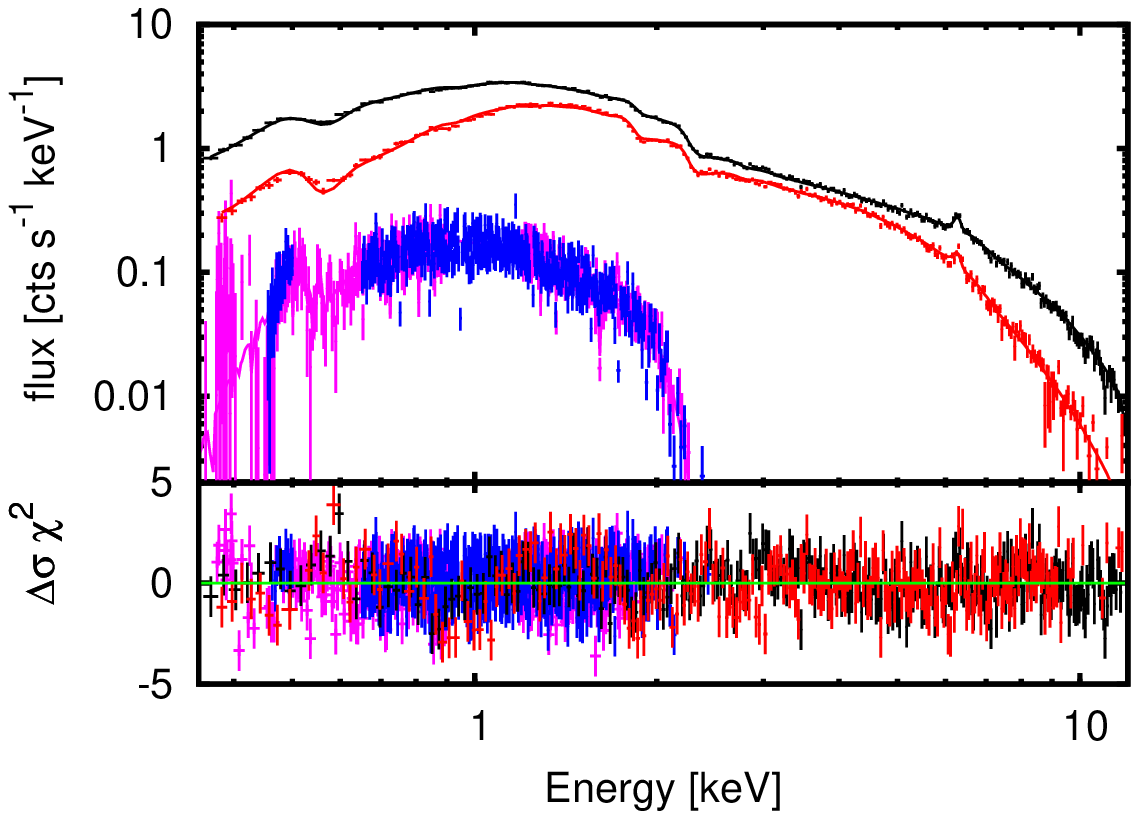}
\caption{The joint fit of all spectra of the XMM-Newton instruments -- 
PN (black), MOS (red), RGS\,1 (magenta), and RGS\,2 (blue), together with
the model residuals.}
\label{joint}
\end{center}
\end{figure}

We also used the `final' model for a joint fitting of all 
the XMM-Newton instruments (PN, MOS and both RGS spectra) together.
The parameter values were bound among all the spectra, only normalisation factors were allowed to vary.
The goodness of the joint fit is given in C-statistics because the RGS data are un-binned 
(and each individual bin contains only a few counts). The result is
$C = 1551$ for a number of degrees of freedom $\nu = 1347$.
All the spectra, together with the residuals, are shown in Figure~\ref{joint},
and the corresponding parameters in the last column of Table~\ref{model}.

\subsection{Discussion of the results}
\label{discussion}

\subsubsection{Constraints on the location of the absorbers}
\label{geometry}

In this section, we discuss a possible location of the absorber's system
in the `final' model. Photoelectric absorption is almost invariably
observed in Type~2 Seyferts \citep{1991ApJ...366...88A,1997ApJS..113...23T,2002ApJ...571..234R}
and generally attributed to an optically thick matter responsible for orientation-dependent 
classification in AGN unification scenarios \citep{1985ApJ...297..621A,1993ARA&A..31..473A}.

Because the IRAS 05078+1626 galaxy is probably viewed under an intermediate 
inclination between unobscured Seyfert 1s and obscured Seyfert 2s, 
the torus rim may also intercept the line of sight to the AGN
and absorb the radiation coming from the centre.
The cold absorption can, however, also be associated with the interstellar matter 
of the galaxy \citep{2006A&A...449..551L}.

Both reflection components are inside the ionised absorber in the `final' model.
The geometrical interpretation is that the cold reflection occurs on the outer
parts of the disc or the inner wall of the torus. Reflection on the nearer 
part of the torus is heavily absorbed by the torus itself, so only radiation
reflected on the farther peripheral part of the torus can reach the observer
after passing through the warm absorber.
However, an alternative scenario, in which the cold reflection is unaffected by the warm absorber, i.e.,
{\tbabs} $\times$ {\zphabs} $\times$ [{\reflion}$_{\rm cold}$ + {\xstar} $\times$ ({\powerlaw} + {\reflion}$_{\,\rm disc}$ )],
is also acceptable with  $\chi^{2}/\nu = 265/265$.

The lack of constraints on the variability in the warm absorbed features 
\citep{2007ApJ...659.1022K}, caused by the moderate dynamical range of the primary continuum, 
as well as statistical limitations in our spectra, prevents us from precisely 
constraining the location of the warm absorber.

\begin{figure}[tb]
\begin{center}
\includegraphics[width=0.75\textwidth]{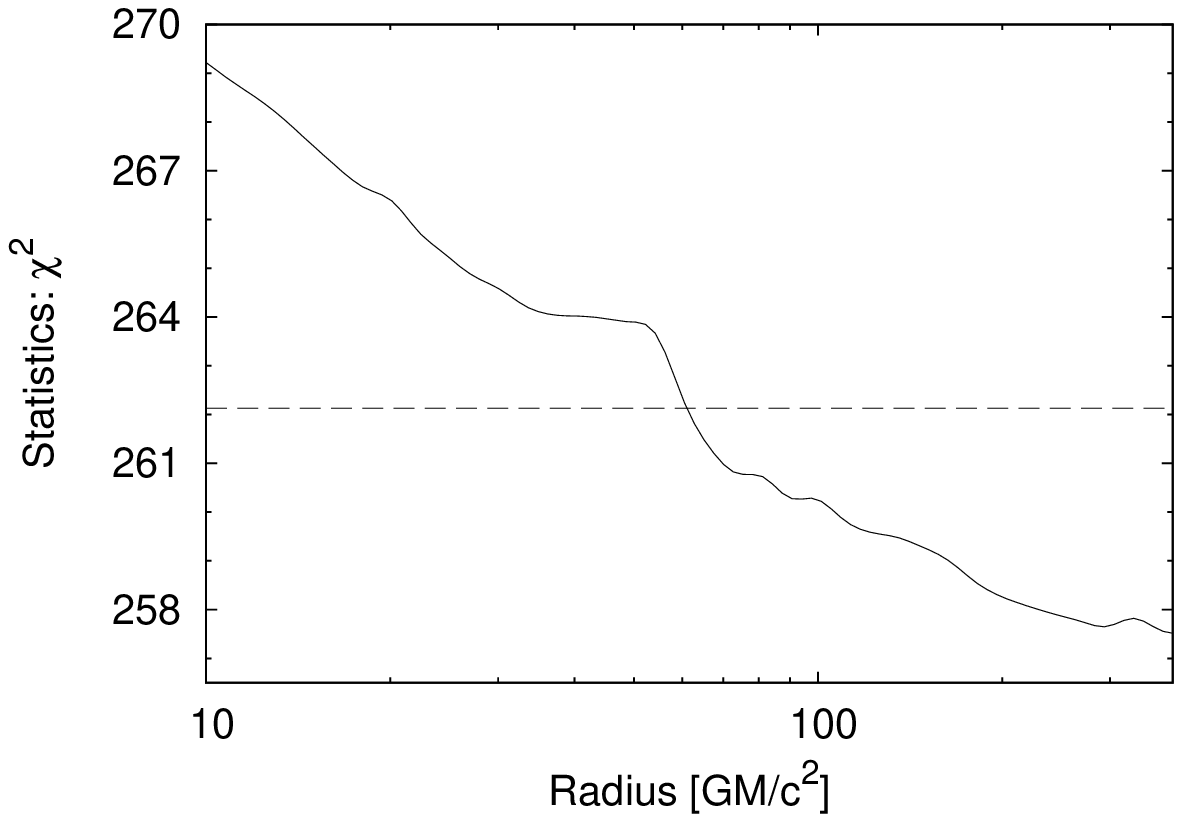}
\caption{The best-fit values of $\chi^2$ statistics for the inner disc radius parameter,
which we obtained by gradually stepping it from the horizon radius 
to the outer radius of the disc ($400 R_{\rm g}$). 
The dashed line is the 90\% confidence level for one interesting parameter. 
%Considering this confidence level,
%the accretion disc is not allowed to extend closer to the black hole than to $60 R_{\rm g}$.
}
\label{kyconvlevel}
\end{center}
\end{figure}

\subsubsection{Constraints on the location of the ionised reflector}
\label{disc}

The ionised reflection might occur either at the inner wall of a warm absorber cone
or on the accretion disc. Even in the latter case, the reflection cannot occur
arbitrarily close to the black hole. In this section we investigate 
the constraints of the accretion disc location and structure, which can be drawn 
from the lack of the significant relativistic blurring of the disc reflection component.

We convolved the ionised reflection component with the 
fully relativistic {\kyconv} model \citep{2004ApJS..153..205D}.
Two assumptions about the disc emissivity were considered.
First, the radial part of the intensity decreases with the power 
of the disc radius $q$ ($I \propto r^{-q}$),
where the value of $q$ was allowed to vary between $2$ and $3.5$. 
Second, the angular dependence was assumed to be isotropic
which seems to be appropriate approximation for our situation
of an X-ray irradiated accretion disc \citep{2009A&A...507....1S}.

We examined the expected confidence levels of the best-fit values
of the disc's inner radius
by stepping this parameter in the whole range of its possible
values -- from the horizon to the outer disc radius, which we set to 400
gravitational radii ($R_{\rm g} \equiv GM/c^{2}$). %$R_{\rm g}=\frac{GM}{c^{2}}$.
The results are shown in Figure~\ref{kyconvlevel}. At the 90~\% confidence level,
the accretion disc is not allowed to extend closer to the black hole than 
$60 R_{\rm g}$. 

The ``relativistic blurring method'' would be less appropriate
in looking for the imprints of the innermost parts of the accretion disc
if the disc were too highly ionised (log$~\xi \approx 4$) and the narrow
reflection features were not present \citep{2005MNRAS.358..211R}. However,
the ionisation parameter value of the reflection component is not so high
in the `final' model and the dominant feature is the intermediately 
ionised iron line ($E \approx 6.7$\,keV). If we assumed a stratified 
disc with the ionisation state decreasing with the radius from the centre, 
the hydrogen-like iron line would be also expected to appear in the spectrum
(as an intermediate stage between the over-ionised and mildly ionised contribution).
%owever, not observed in this source (see Figure~\ref{pm_resid}),
%and the accretion disc truncation provides therefore a more reasonable explanation.
Because it is not detected in the data,
the accretion disc truncation provides a more reasonable explanation
of missing signatures of the relativistic blurring.

\subsubsection{Mass accretion rate}

Disc truncation is expected in low--luminosity AGN where the inner
accretion flow is advection-dominated \citep[and references therein]{1994ApJ...428L..13N, 
1997ApJ...489..865E,2008NewAR..51..733N}.
The transition from the outer standard accretion disc may occur, e.g., via
the disc evaporation mechanism \citep{2000A&A...361..175M, 2009ApJ...707..233L}.
The observational evidence of a truncated accretion disc
in low--luminosity AGN was reported e.g. by \citet{1996ApJ...462..142L, 1999ApJ...525L..89Q}. 
However, its presence is also suggested in some observations of Seyfert galaxies 
\citep{2000ApJ...537L.103L, 2000ApJ...536..213D, 2003ApJ...586...97C, 2009ApJ...705..496M}
and even a quasar \citep{2005A&A...435..857M} where the luminosity value is estimated as a half
of the Eddington value. Generally, it is expected that the lower the luminosity, $L/L_{\rm Edd}$,
the larger transition radius \citep[see][and references therein]{2004ApJ...612..724Y}.
Furthermore, we investigate whether the disc truncation hypothesis is consistent 
with the IRAS~05078+1626 luminosity. To have these quantities in Eddington units, 
we first estimated the mass of the black hole. 

IRAS~05078+1626 belongs to the sample of the
infrared-selected Seyfert 1.5 galaxies observed by a 2.16~m optical telescope 
\citep{2006ApJ...638..106W} where, among others, the velocity dispersion 
in the O~III emission line was measured. 
The correlation between the O~III line width and the mass of active galactic
nucleus was discussed in \citet{2000ApJ...544L..91N} and \citet{2003ApJ...585..647B}. 
The value from the optical measurements, $\sigma_{\rm O\,III} \approx 130$\,km\,s$^{-1}$, %FWHM is 308, sigma=fwhm/2.35 
corresponds to the mass $M \approx 4\times10^{7} M_{\odot}$ using 
a correlation plot in \citet{2003ApJ...585..647B}. The scatter of the correlation
is somewhat large with the reported limit of a factor of 5 for an uncertainty
in the black hole mass determination, so the value only provides 
an order of magnitude estimation.

The value of the Eddington luminosity is
$L_{\rm Edd} \doteq 1.3 \times 10^{38} M/M_{\odot} $\,erg\,s$^{-1} \approx 5 \times 10^{45}$\,erg\,s$^{-1}$
for the given value of the mass.
We used luminosity-dependent corrections by \citet{2004MNRAS.351..169M} to estimate
the bolometric luminosity of IRAS~05078+1626 from the X-ray luminosity. Its value is 
$L \approx 5 \times 10^{44}$\,erg\,s$^{-1} \approx 10^{-1} L_{\rm Edd}$.
Correspondingly, the mass accretion rate, $\dot{M} = L/c^2$,
is sub-Eddington with $\dot{M} \approx 0.1 \dot{M}_{\rm Edd}$.
This value is typical of less luminous Seyfert galaxies \citep[see for example][]{2009A&A...495..421B},
and is consistent with the disc truncation hypothesis.

% ##########################################################################

%\chapter{Discussion}
%\chaptermark{Discussion}
% \thispagestyle{empty}

% ##########################################################################

\chapter{Conclusions}
\chaptermark{Conclusions}
 \thispagestyle{empty}
 \label{concl}

We employed the relativistic models of iron line
to probe the innermost regions of black hole accretion discs
with the current X-ray data by XMM-Newton, 
and with the simulated data for next generation X-ray
missions. We found the modelling of broad iron lines 
to be a suitable method to measure the angular momentum 
of black holes at all scales -- from stellar-mass
microquasars to giant black holes of billions solar masses 
of distant quasars.
In this section, we summarise the main conclusions
of our investigation presented in the previous chapters.

%%%%% LAOR - KYRLINE comparison %%%%%%%%%%%%%%%%%%%%%%%%%%%%%%%%%%%%%%%%%%%%%%%%%%%%%%%

\section{Relativistic line models}

In Section~\ref{rellinemod}, we described 
how the observed profile of an intrinsically narrow emission line 
is distorted due to the effects of rapid orbital motion and strong gravity. 
We investigated the iron line band of the X-ray spectrum 
for two representative sources -- 
MCG\,-6-30-15 (active galaxy) and GX 339-4 (X-ray binary).
The iron line is statistically better constrained in the active 
galaxy MCG\,-6-30-15 due to a significantly longer exposure
time of the available observations.
The spectra of both sources are well described by a continuum model 
plus a broad iron line model.
We determined the spin values using the {\kyrline} model as:
	  \begin{itemize}
	      \item MCG\,-6-30-15: $a = 0.9 - 1.0$, 
	      \item GX 339-4: $a = 0.56 - 0.85$.
	  \end{itemize} 

The value for MCG\,-6-30-15 is consistent with previously obtained
results \citep{2002MNRAS.335L...1F,2006ApJ...652.1028B}.
However, in the case of GX\,339-4, our best-fit value is lower than the one
by \citet{2004ApJ...606L.131M} and \citet{2008MNRAS.387.1489R}.
The model with their derived values does not provide 
an acceptable fit when applied to the data
re-binned with respect to the instrumental energy resolution.
We found that the iron line is not as extremely broad as
previous analyses suggested, and the spin value
has rather an intermediate value $a \approx 0.7$.

We compared two relativistic models of the broad 
iron line, {\laor} and {\kyrline}.
%which employ different ways to measure the spin.
In contrast to {\laor},
the {\kyrline} model has the spin value as a variable parameter.
However, the {\laor} model can still be
used for evaluation of the spin if one identifies the inner 
edge of the disc with the marginally stable orbit.
We realised that the discrepancies in the results between the 
{{\kyrline}} and {{\laor}} models
are within general uncertainties of the spin determination 
using the skewed line profile when applied to the current data
(for MCG\,-6-30-15: $a_{laor} = 0.94 - 0.9982$, and 
for GX 339-4: $a_{laor} = 0.63 - 0.87$).
This means that the spin is currently
determined entirely from the position of the marginally stable 
orbit, as it is done with the {\laor} model.

However, the results are apparently distinguishable for 
higher quality data, as those simulated for the next
generation X-ray missions, which will be sufficiently 
sensitive to resolve the slight variations in the overall
line shape due to the spin.
We found that the {\laor} model tends to over-estimate the spin value 
and furthermore, it has insufficient energy resolution 
which affects the correct determination of
the high-energy edge of the broad line.
The discrepancies in the overall shape of the line by the {\laor} model 
are more visible especially for lower values of the spin.

Another advantage of the {\kyrline} model over the
{\laor} model is that
the {{\kyrline}} model gives better pronounced minima of $\chi^{2}$ for the best-fit values.
The confidence contour plots for $a/M$ versus other model parameters are more regularly shaped.
This indicates that the {{\kyrline}} model has a smoother adjustment between the different
points in the parameter space allowing for more reliable constraints on spin.

%%%%% Emission directionality %%%%%%%%%%%%%%%%%%%%%%%%%%%%%%%%%%%%%%%%%%%%%%%%%%%%%%%

\section{Emission directionality}

In Section~\ref{directionality}, we investigated how 
the black hole spin measurements using X-ray spectroscopy
of relativistically broadened lines depend on the 
employed definition of the angular distribution of the disc emission.
We considered three different approximations of the angular
profiles, representing limb-brightening, isotropic and 
limb-darkening emission profiles.
We studied the role of the emission directionality
in the spin determination with the current data - we used
the XMM-Newton observation of MCG\,-6-30-15,
and with the artificial data simulated with energy
resolution and sensitivity of future X-ray missions, such as IXO
(International X-ray Observatory).

We realised that the use of an improper directionality profile could 
affect the other parameters inferred from the relativistic broad line model.
Especially for the frequently used case of limb darkening,
the radial steepness can interfere with the line parameters
of the best-fit model by enhancing the red wing of the line.
The limb darkening law favours higher values of
spin and/or steeper radial dependence of the line emissivity;
vice-versa for the limb brightening profile.

Steeper radial emissivity in the innermost region of an accretion
disc has been detected in several other sources,
both in AGN and black hole binaries \citep{2007ARA&A..45..441M}. The improperly
used limb darkening in the reflection model represents one of the possible 
explanations.  For this reason, constraining the ``correct'' 
directionality in more sophisticated models of reflection spectra 
of accretion discs is desirable.

We applied the NOAR radiation transfer code to achieve 
a self-consistent simulation of the outgoing 
spectrum reflected from a cold disc
without imposing any ad hoc formula
for the emission angular distribution.
We found that the isotropic directionality 
reproduces the simulated data to the best precision.
On the other hand, the model with the 
limb-darkening profile was not able to 
reproduce the simulated data especially 
for the higher values of black hole spin.

We consider this to be an important result since much
of the recent work on the iron lines has revealed
a significant relativistic broadening near
rapidly rotating central black holes. 
In some of these works, the limb darkening law was employed 
without testing different options. 
The modelled broad lines are typically
characterised by a steep power law in the
radial part of the intensity across the inner region
of the accretion disc, as in the mentioned MCG\,-6-30-15 
observation. 
We conclude that the significant steepness of the radial part of the intensity 
also persists in our analysis, however, the exact values depend partly 
on the assumed angular distribution of the emissivity 
of the reflected radiation.

%%%%% Spectral analysis of the data - GROUPING issue %%%%%%%%%%%%%%%%%%%%%%%%%%%%%%%%

%%%%%%%%%%%%%% IRAS 05078+1626 %%%%%%%%%%%%%%%%%%%%%%%%%%%%%%%%%%%%%%%%%%%%%%%%%%%%%%

\section{Data reduction}

We re-analysed XMM-Newton observations of MCG\,-6-30-15 and GX\,339-4.
According to the previous X-ray spectroscopy analyses,
these sources exhibited extremely broad iron lines.
The results of our spectral analysis are discussed in the context
of the aspects of iron line modelling in Sections~\ref{rellinemod}-\ref{directionality}.
We stressed that the correct re-binning of the data, 
which reflects the energy resolution of the used instrument,
is necessary to obtain statistically relevant results.
The un-grouped EPIC/PN or MOS data are characterised by large
error bars which implies 
that any applied model produces lower $\chi^2$ values 
when fitting such data (see Section~\ref{gx_vhs}).
As a consequence, a larger set of different models
may be accepted, including the incorrect models which would be
excluded if the data were properly grouped.

Photon pile-up is another problem which may affect the spectral analysis significantly.
It occurs during exposures of too bright sources, such as some Galactic black hole binaries.
We encountered a serious problem with the pile-up in the 
data re-analysis of the low/hard state observations of GX\,339-4
(see Section~\ref{gx_lhs}).
We found that the elimination of the pile-up by excising
the core of the point spread function of the source
is possible only at the cost of a drastic loss of counts
which makes any interpretation of the spectra rather disputable.

\section{X-ray spectrum of IRAS~05078+1626}

In Section~\ref{iras05078}, we presented a spectral analysis 
of the XMM-Newton observation of a Seyfert galaxy IRAS~05078+1626 
being the first X-ray spectroscopy study of this source.

The X-ray continuum spectrum of IRAS 05078+1626 
is dominated by a power law with a standard value of the photon index 
($\Gamma \cong 1.75$ in the `final model'). 
The residuals from the power-law continuum can be interpreted in terms 
of a warm absorber surrounding an accretion disc,
and a reflection of the primary radiation from an ionised matter and on a
cold torus. The outgoing radiation is absorbed by cold matter 
($n_{\rm H} \approx 1 \times 10^{21}$\,cm$^{-2}$), which can be either 
located in the inner side of the torus or caused by gas in the host galaxy. 
The type of the galaxy determined from the previous infrared and optical 
research is Seyfert 1.5, suggesting that the active nucleus 
could be seen at large inclination, consistent with either interpretation 
or even allowing a combination of both.

The ionised warm absorber occurs in the central part of the AGN.
Its column density was found to be $n_{\rm H} \geq 1 \times 10^{24}$\,cm$^{-2}$,
which is a rather high value compared to the warm absorbers detected
in other Seyfert galaxies \citep{blustin05}.
This may be because we are looking through a longer
optical path of a conical non-relativistic outflow due to the high
inclination of the system. Such a conical outflow is suggested
in the model by \citet{2000ApJ...545...63E} (see Figure~\ref{elvis}).
The ionisation parameter of the warm absorber is $\log \xi_{\rm WA} = 2.5 \pm 1.0$,
which is comparable to the value related to the ionised reflection
$\log \xi_{\rm reflection} = 3.0 \pm 0.2$, suggesting a link between them.

If the ionised reflection is associated to the warm absorber
(e.g. the inner walls of a conical outflow), the lack of spectral
features associated with the accretion disc is a natural consequence thereof.
If, instead, the ionised reflection occurs at the accretion disc, 
it cannot extend up to the marginally stable orbit. 
The lack of the significant relativistic blurring of this model component
requires the disc to be truncated (inner disc radius $R_{\rm in} \geq 60\,R_{g}$).
This idea is also supported by the low ratio of the reflection radiation 
to the primary one, $R < 0.2$, and also by the relatively low mass-accretion 
rate $\dot{M} \approx 0.1\dot{M}_{\rm Edd}$ determined
from the source luminosity.

% ##########################################################################

\chapter{Future perspectives}
\chaptermark{Future perspectives}
\label{futureperspectives}
\thispagestyle{empty}
 
X-ray spectroscopy is a feasible method to probe the innermost regions
of black hole accretion discs
in active galactic nuclei and stellar-mass black hole binaries.
Especially, it provides a unique way to measure black hole spin 
which plays an important role in black hole energetic balance and evolution.
The existence of powerful relativistic jets is frequently attributed 
to rapidly rotating black holes where the energy extracted from the black hole rotation
generates and maintains the far-reaching collimated ejections of plasma. 
But is it really so? Is there any correlation
between the spin and radio-loudness of galaxies and black hole binaries?
What is the statistical distribution of the black hole spin? 
Is the black hole spin natal or are black holes spun up via accretion? 
Such questions could be answered
if we reliably knew the values of the spin for a sufficiently large number of black holes.
Hence, precise measurements of the black hole spin 
is among the important future tasks of X-ray astronomy.

Black hole angular momentum has been measured 
since late 1980s \citep{1989MNRAS.238..729F}. 
At the beginning with a large uncertainty, but
following the fast development of X-ray detectors,
these measurements have become
increasingly precise. 
The original method dealt with a relativistically
broadened iron line.
This method is still one of the most suitable 
methods of the spin measurements. Its advantage is its
applicability to black holes at all mass-scales.
However, it depends on the accretion regime whether a broad 
relativistic line may occur in the spectrum or not. The necessary conditions
are that the accretion flow is in the ``thin-disc'' regime (the disc is geometrically
thin and optically thick) and not over-ionised ($\xi \leq 5000$).

The advantage of stellar-mass black hole binaries 
is that their spectral states evolve on relatively short time-scales 
so that it is possible to detect them in different states.
A relativistically broadened iron line should be common for stellar-mass black
hole binaries when observed in the appropriate accretion state \citep{2009ApJ...697..900M}.
On the other hand, X-ray spectra of the brightest Galactic black hole binaries,
such as Cyg\,X-1, GRS\,1915+105, GX\,339-4, etc.,
suffer significantly from the pile-up
when observed with current X-ray satellites. 
Instruments with a more frequent
and faster read-out cycle or with a different detection technology
would be more suitable for observations of such bright sources.

For stellar-mass black hole binaries, the spin measurements
with a relativistic iron line may be compared to the results
obtained by modelling of the disc thermal radiation \citep{2006ARA&A..44...49R}.
An agreement of the results obtained by both methods 
will enhance the credibility of the spin measurements.
Higher resolution X-ray detectors will allow other methods, such as 
X-ray reverberation or X-ray polarimetry, to independently
measure the spin. These will also be applied for active galaxies.

Many similarities between super-massive black holes in active galaxies 
and Galactic black holes were reported (see Section~\ref{similarity}).
Do they also have a similar distribution in the spin?
A statistical study of the presence of relativistically 
broadened iron lines in spectra of active galactic nuclei (AGN) 
was performed by \citet{2006AN....327.1032G}
on the sample of a hundred AGNs observed by the XMM-Newton satellite.
They estimated a fraction of $42 \pm 12\,\%$ of well exposed ($>\,10^4$\,cts) AGNs
that exhibit a relativistically broadened iron line. 
A continuation of this effort is
represented by \citet{2008MmSAI..79..259L} and de la Calle {\textit{et al.}}
(2009, submitted to {\textit{A}\&\textit{A}}).
Some other works have been done 
on different samples of galaxies 
\citep{2005A&A...432..395S, 2007MNRAS.382..194N,
2009ApJ...702.1367B}.

All of these researches have been done on well exposed observations of active 
galaxies in the local Universe. However, with more sensitive
instruments, such as those planned for IXO, the relativistic iron lines
will be detectable also in observations of more distant galaxies.
Constraints on cosmic density of relativistic lines
and its distribution on cosmological length-scales
will be available in near future. A pioneering work 
with the current data appeared recently \citep{2010ApJ...708L...1B}.

It is worth-mentioning that, 
like any other spectroscopic results, the spin measurements 
with a relativistic iron line are model-dependent. 
Models employing partial covering absorption
represent an alternative explanation of the characteristic spectral curvature
around the iron line energy \citep{2008MNRAS.483..437}. Observations with
future X-ray satellites may be helpful to distinguish 
between the two models thanks to an increased sensitivity
in combination with a broader energy range 
including the Compton hump as well.
Self-consistent reflection models are then desirable 
to model all the reflection components together. 
%\citep[see e.g.][]{2005MNRAS.358..211R}. 
%Observations with the Suzaku 
%satellite may improve precision of our results on iron lines in the individual sources.
If the source is considerably variable then  
the timing studies may be determinative between
the two scenarios. The first reverberation
measurement of 1H0707-495 strongly favours
the scenario with the relativistic line models \citep{2009Natur.459..540F}. 

Beyond the current sensitivity of X-ray detectors,
further details of accretion physics can be studied.
Not only spin will be measured by relativistic line models
but also the intrinsic assumptions inside the models will be tested.
One of the uncertainties in the relativistic line models
is due to the unknown emission directionality.
We discussed its effect on the spin value determination
for the case of an isotropically illuminated cold accretion disc
(see Section~\ref{directionality}).
This analysis can be enhanced in a future work by adopting different assumptions,
e.g. ionised disc, or an illuminating source localised above the black hole.
More comprehensive simulations of the irradiated accretion discs
taking the general relativistic effects on the radiation transfer into account
would be desirable. Detailed knowledge of the model assumptions
will be necessary to properly measure the spin.

Currently, the uncertainties in the precise position of the inner edge
represent an unknown error of the spin determination. However,
it appears that the expected magnitude of these errors does not
prevent us from setting interesting and realistic constraints on the
spin parameter. Our present treatment of the problem is incomplete
by neglecting the magnetohydrodynamical effects and their influence
on the ISCO location. Future improvements in our theoretical 
understanding of the inner edge location are desirable and 
will help to improve the confidence in the determination of the 
model parameters.

Higher resolution X-ray detectors will possibly reveal more complex
profiles of iron lines modified by a contribution of different 
non-axial patterns, for instance orbiting hot spots \citep{2004MNRAS.350..745D}
or spiralling waves \citep{2001PASJ...53..189K}. Some of the current
observations of AGNs already revealed 
periodic X-ray variability 
\citep[see e.g.][]{1998MNRAS.295L..20I, 2002ApJ...574L.123T}
or narrow emission lines in the iron line
band which may be due to orbiting hot spots 
\citep[see e.g.][]{2004MNRAS.350..745D, 2006A&A...453..839P}.
The study of non-axial patterns in the iron line profiles
is possible using the KY package \citep{2004ApJS..153..205D},
and will be promising with sensitive future X-ray missions.

%We believe that high-throughput X-ray telescopes, such as IXO,
%will produce amazing insights into the innermost accretion discs
%around black holes which will significantly enhance our knowledge
%of black holes in the Universe and accretion physics in strong gravity.

We would like to conclude by expressing our optimism regarding 
the future of X-ray research of black holes and accretion discs. 
New technologies that are currently being developed will 
enhance energy and time resolution of the observations 
and they will even open new channels of information (such as X-ray polarimetry). 
In addition, theoretical models and numerical codes are getting 
more advanced, thus allowing us to
make interpretation of the observed data 
closer to their real meaning.

% ##########################################################################

% \cleardoublepage
%\markright{}
\clearpage
\phantomsection\addcontentsline{toc}{chapter}{Bibliography}
\setlength{\bibsep}{0.2em}
\bibliographystyle{mynatbib}
\bibliography{svoboda_phd}
\end{document}